\documentclass[letterpaper,11pt]{report}

\pdfoutput=1

\title{Modular Audio Recognition Framework v.0.3.0.6 (0.3.0 final)\\and its\\Applications}

{\author
	{\bf
		{The MARF Research and Development Group}\\\hline\\
		Montr\'eal, Qu\'ebec, Canada\\\\\\
	}
}

\date{Sat 25 Jul 2009 00:12:48 EDT
}

\usepackage{graphicx}
\usepackage{latexsym}
\usepackage{makeidx}
\usepackage{url}

\makeindex

\newcommand{\marf}{\textsf{MARF}}

\newcommand{\xf}[1]{Figure~\ref{#1}}

\newcommand{\xs}[1]{Section~\ref{#1}}
\newcommand{\xa}[1]{Appendix~\ref{#1}}
\newcommand{\xc}[1]{Chapter~\ref{#1}}
\newcommand{\xt}[1]{Table~\ref{#1}}

\newcommand{\bestscore}[1]{\textsf{{\large #1\%}}}

\newcommand{\gnu}{{GNU\index{GNU}}}

\newcommand{\gipsy}{{GIPSY\index{GIPSY}}}

\newcommand{\C}{{C\index{C}}}
\newcommand{\cpp}{{C++\index{C++}}}
\newcommand{\perl}{{Perl\index{Perl}}}
\newcommand{\java}{{Java\index{Java}}}

\newcommand{\todo}[0]
{
	{\Large \[TODO\]}
}

\newcommand{\file}[1]{\texttt{#1}\index{Files!#1}}
\newcommand{\tool}[1]{\texttt{#1}\index{Tools!#1}}
\newcommand{\api}[1]{\texttt{#1}\index{API!#1}}

\begin{document}

	\begin{titlepage}
		\maketitle
	\end{titlepage}

	\pagenumbering{roman}
\tableofcontents
\clearpage
\pagenumbering{arabic}

\listoffigures
\listoftables

	\chapter{Introduction}
\index{MARF!Introduction}

$Revision: 1.22 $

\section{What is MARF?}

{\marf} stands for {\bf M}odular {\bf A}udio {\bf R}ecognition {\bf F}ramework.
It contains a collection of algorithms for Sound, Speech, and Natural Language
Processing arranged into an uniform framework to facilitate addition of new
algorithms for preprocessing, feature extraction, classification, parsing, etc.
implemented in Java.
{\marf} is also a research platform for various performance metrics of the
implemented algorithms.

\subsection{Purpose}
\index{MARF!Purpose}

Our main goal is to build a general open-source framework to allow developers in the
audio-recognition industry (be it speech, voice, sound, etc.) to choose and apply various methods,
contrast and compare them, and use them in their applications. As a proof of concept, a
user frontend application for Text-Independent (TI) Speaker Identification has
been created on top of the framework (the \api{SpeakerIdentApp} program). A variety
of testing applications and applications that show how to use various aspects of {\marf}
are also present. A new recent addition is some (highly) experimental NLP support, which is also
included in {\marf} as of 0.3.0-devel-20050606 (0.3.0.2). For more information on applications
that employ {\marf} see \xc{chapt:apps}.

\subsection{Why {\java}?}

We have chosen to implement our project using the Java programming language. This
choice is justified by the binary portability of the Java applications as well as
facilitating memory management tasks and other issues, so we can concentrate more on
the algorithms instead. Java also provides
us with built-in types and data-structures to manage
collections (build, sort, store/retrieve) efficiently \cite{javanuttshell}.

\subsection{Terminology and Notation}

$Revision: 1.4 $

The term ``{\marf}'' will be
to refer to the software that accompanies this
documentation.
An {\bf application programmer}
could be anyone who is using, or wants to use, any part of the
{\marf} system.
A {\bf {\marf} developer}
is a core member of {\marf} who is hacking away the
{\marf} system.

% EOF

\section{Authors, Contact, and Copyright Information}
\index{MARF!Contact}
\index{MARF!Copyright}

\subsection{COPYRIGHT}

$Revision: 1.14 $

{\marf} is Copyright $\copyright$ 2002 - 2009
by the MARF Research and Development Group and is distributed under
the terms of the BSD-style license below.

\vspace{15pt}
\noindent
Permission to use, copy, modify, and distribute this software and
its documentation for any purpose, without fee, and without a
written agreement is hereby granted, provided that the above
copyright notice and this paragraph and the following two paragraphs
appear in all copies.

\vspace{15pt}
\noindent
IN NO EVENT SHALL CONCORDIA UNIVERSITY OR THE AUTHORS BE LIABLE TO ANY
PARTY FOR DIRECT, INDIRECT, SPECIAL, INCIDENTAL, OR CONSEQUENTIAL
DAMAGES, INCLUDING LOST PROFITS, ARISING OUT OF THE USE OF THIS
SOFTWARE AND ITS DOCUMENTATION, EVEN IF CONCORDIA UNIVERSITY OR THE AUTHORS
HAVE BEEN ADVISED OF THE POSSIBILITY OF SUCH DAMAGE.

\vspace{15pt}
\noindent
CONCORDIA UNIVERSITY AND THE AUTHORS SPECIFICALLY DISCLAIM ANY WARRANTIES,
INCLUDING, BUT NOT LIMITED TO, THE IMPLIED WARRANTIES OF
MERCHANTABILITY AND FITNESS FOR A PARTICULAR PURPOSE.  THE SOFTWARE
PROVIDED HEREUNDER IS ON AN ``AS-IS'' BASIS, AND CONCORDIA UNIVERSITY AND THE
AUTHORS HAVE NO OBLIGATIONS TO PROVIDE MAINTENANCE, SUPPORT,
UPDATES, ENHANCEMENTS, OR MODIFICATIONS.

\subsection{Authors}
\index{MARF!Authors}

\noindent
Authors Emeritus, in alphabetical order:
\index{MARF!Authors Emeritus}

\begin{itemize}
	\item Ian Cl\'ement, \url{iclement@users.sourceforge.net}
	\item Serguei Mokhov, \url{mokhov@cs.concordia.ca}, a.k.a Serge
	\item Dimitrios Nicolacopoulos, \url{pwrslave@users.sourceforge.net}, a.k.a Jimmy
	\item Stephen Sinclair, \url{radarsat1@users.sourceforge.net}, a.k.a. Steve, radarsat1
\end{itemize}

\noindent
Contributors:
\index{MARF!Contributors}

\begin{itemize}
	\item Shuxin Fan, \url{fshuxin@gmail.com}, a.k.a Susan
\end{itemize}

\noindent
Current maintainers:
\index{MARF!Current maintainers}

\begin{itemize}
	\item Serguei Mokhov, \url{mokhov@cs.concordia.ca}
\end{itemize}

\noindent
If you have
some suggestions, contributions to make, or for bug reports, don't
hesitate to contact us :-)
For {\marf}-related issues please contact us at \url{marf-devel@lists.sf.net}.
Please report bugs to \url{marf-bugs@lists.sf.net}.

\section{Brief History of MARF}
\index{MARF!History}

$Revision: 1.13 $

The {\marf} project was initiated in September 26, 2002 by four students of
Concordia University, Montr\'eal, Canada as their course project for Pattern Recognition under
guidance of Dr. C.Y. Suen. This included Ian Cl\'ement, Stephen Sinclair,
Jimmy Nicolacopoulos, and Serguei Mokhov.

\subsection{Developers Emeritus}

\begin{itemize}
\item
Ian's primary contributions were the
LPC and Neural Network algorithms support with the \api{Spectrogram} dump.

\item
Steve has done an extensive
research and implementation of the FFT algorithm for feature extraction
and filtering and Euclidean Distance with the \api{WaveGrapher} class.

\item
Jimmy was focused on implementation of the WAVE file format
loader and other storage issues.

\item
Serguei designed the entire MARF framework and architecture,
originally implemented general distance classifier and its Chebyshev, Minkowski,
and Mahalanobis incarnations along with normalization of the sample data. Serguei
designed the Exceptions Framework of MARF and was involved into the integration
of all the modules and testing applications to use {\marf}.
\end{itemize}

\subsection{Contributors}

\begin{itemize}
\item
	Shuxin `Susan' Fan contributed to development and maintenance of some test applications
	(e.g. \api{TestFilters}) and initial adoption of the JUnit framework \cite{junit} within {\marf}.
	She has also finalized some utility modules (e.g. \api{marf.util.Arrays})
	till completion and performed {\marf} code auditing and ``inventory''.
	Shuxin has also added NetBeans project files to the build system of {\marf}.

\end{itemize}

\subsection{Current Status and Past Releases}

Now it's a project on its own, being maintained and developed as we have some spare time for it.
When the course was over, Serguei Mokhov is the primary maintainer of the
project. He rewrote \api{Storage} support and polished all of {\marf} during
two years and added various utility modules and NLP support and implementation of new
algorithms and applications. Serguei maintains this manual, the web site and most of
the sample database collection. He also made all the releases of the project
as follows:

\begin{itemize}
\item 0.3.0-devel-20060226 (0.3.0.5), Sunday, February 26, 2006
\item 0.3.0-devel-20050817 (0.3.0.4), Wednesday, August 17, 2005
\item 0.3.0-devel-20050730 (0.3.0.3), Saturday, July 30, 2005
\item 0.3.0-devel-20050606 (0.3.0.2), Monday, June 6, 2005
\item 0.3.0-devel-20040614 (0.3.0.1), Monday, June 14, 2004
\item 0.2.1, Monday, February 17, 2003
\item 0.2.0, Monday, February 10, 2003
\item 0.1.2, December 17, 2002 - Final Project Deliverable
\item 0.1.1, December 8, 2002 - Demo
\end{itemize}

\noindent
The project is currently geared towards completion planned TODO items on {\marf} and
its applications.

% EOF

%
%
%

\section{MARF Source Code}
\index{MARF!Source Code}

$Revision: 1.8 $

\subsection{Project Source and Location}
\index{MARF!Project Location}

Our project since the its inception has always been an open-source project.
All releases including the most current one should most of the time be
accessible via \verb+<http://marf.sourceforge.net>+ provided by \texttt{SourceForge.net}.
We have a complete API documentation as well as this manual and all the sources
available to download through this web page.

\subsection{Formatting}
\index{MARF!source code formatting}
\index{Source code formatting}

Source code formatting uses a 4 column tab spacing, currently with
tabs preserved (i.e. tabs are not expanded to spaces).

For Emacs, add the following (or something similar)
to your \file{\~/.emacs}
initialization file:

\begin{verbatim}
;; check for files with a path containing "marf"
(setq auto-mode-alist
  (cons '("\\(marf\\).*\\.java\\'" . marf-java-mode)
        auto-mode-alist))
(setq auto-mode-alist
  (cons '("\\(marf\\).*\\.java\\'" . marf-java-mode)
        auto-mode-alist))

(defun marf-java-mode ()
  ;; sets up formatting for MARF Java code
  (interactive)
  (java-mode)
  (setq-default tab-width 4)
  (java-set-style "bsd")      ; set java-basic-offset to 4, etc.
  (setq indent-tabs-mode t))  ; keep tabs for indentation
\end{verbatim}

For \tool{vim}, your
\file{\~/.vimrc} or equivalent file should contain
the following:

\begin{verbatim}
set tabstop=4
\end{verbatim}

    or equivalently from within vim, try

\begin{verbatim}
:set ts=4
\end{verbatim}

    The text browsing tools \tool{more} and
    \tool{less} can be invoked as

\begin{verbatim}
more -x4
less -x4
\end{verbatim}

\subsection{Coding and Naming Conventions}
\index{MARF!Coding Conventions}
\index{Coding and Naming Conventions}

For now, please see \texttt{http://marf.sf.net/coding.html}.

{\todo}

% EOF

%
%
%

\section{Versioning}
\index{Versioning}
\index{MARF!Versioning}

This section attempts to clarify versioning scheme employed
by the {\marf} project for stable and development releases.

In the 0.3.0 series a four digit version number was introduced
like 0.3.0.1 or 0.3.0.2 and so on. The first digit indicates
a {\em major version}. This typically indicates a significant
coverage of implementations of major milestones and improvements,
testing and quality validation and verification to justify a major
release. The {\em minor version} has to do with some smaller
milestones achieved throughout the development cycles.
It is a bit subjective of what the minor and major version bumps are, but
the TODO list in \xa{appx:todo} sets some tentative milestones
to be completed at each minor or major versions. The {\em revision}
is the third digit is typically applied to stable releases if there
are a number of critical bug fixes in the minor release, it's revision
is bumped. The last digit represents the {\em minor revision} of the
code release. This is typically used throughout development releases
to make the code available sooner for testing. This notion was first
introduced in 0.3.0 to count sort of increments in {\marf} development,
which included bug fixes from the past increment and some chunk of new
material. Any bump of major, minor versions, or revision, resets the
minor revision back to zero. In the 0.3.0-devel release series these
minor revisions were publicly displayed as dates (e.g. 0.3.0-devel-20050817)
on which that particular release was made.

All version as of 0.3.0 can be programmatically queried for
and validated against. In 0.3.0.5, a new \api{Version} class
was introduced to encapsulate all versioning information and
some validation of it to be used by the applications.

\begin{itemize}
\item
	the major version can be obtained from \api{marf.MARF.MAJOR\_VERSION}
	and \api{marf.Version.MAJOR\_VERSION} where as of 0.3.0.5 the former
	is an alias of the latter

\item
	the minor version can be obtained from \api{marf.MARF.MINOR\_VERSION}
	and \api{marf.Version.MINOR\_VERSION} where as of 0.3.0.5 the former
	is an alias of the latter

\item
	the revision can be obtained from \api{marf.MARF.REVISION}
	and \api{marf.Version.REVISION} where as of 0.3.0.5 the former
	is an alias of the latter

\item
	the minor revision can be obtained from \api{marf.MARF.MINOR\_REVISION}
	and \api{marf.Version.MINOR\_REVISION} where as of 0.3.0.5 the former
	is an alias of the latter
\end{itemize}

The \api{marf.Version} class provides some API to validate the version
by the application and report mismatches for convenience. See the API
reference for details. One can also query a full \api{marf.jar} or 
\api{marf-debug.jar} release for version where all four components
are displayed by typing:

\begin{verbatim}
    java -jar marf.jar --version
\end{verbatim}

% EOF

	\chapter{Build System}

$Revision: 1.1 $

\begin{itemize}
\item Makefiles
\item Eclipse
\item NetBeans
\item JBuilder
\end{itemize}

{\todo}

% EOF

	\chapter{MARF Installation Instructions}
\index{MARF!Installation}
\index{Installation}

$Revision: 1.11 $

\noindent
TODO: automatic sync with the \file{INSTALL} file.

\section{Requirements}
\index{Installation!Requirements}

In general, any modern platform should be able to run {\marf} provided
a Java Virtual machine (JRE 1.4 and above) is installed on it.
The following software packages are required for building {\marf} from sources:

\begin{itemize}

\item
	You need a Java compiler. Recent
	versions of \tool{javac} are recommended.
	You will need at least JDK 1.4 installed as JDKs lower than that
	are no longer supported (a patch can be made however if there is really
	a need for it). Additionally,
	you would need \tool{javadoc} and \tool{jar} (also a part
	of a JDK) if you want to build the appropriate API documentation
	and the \file{.jar} files.

\item
	On Linux and UNIX the {\gnu} \tool{make} \cite{gmake} is required; other
	\tool{make} programs will {\em not} work.
	{\gnu} \tool{make} is often installed under
	the name \file{gmake}; on some systems the GNU \tool{make} is the
	default tool with the name \tool{make}. This document will always
	refer to it by the name of ``make''. To see which make version you
	are running, type \texttt{make -v}.

\item
	If you plan to run unit tests of {\marf} or use the \file{marf-debug-*.jar}
	release, then you will also need the JUnit \cite{junit} framework's jar somewhere in
	your CLASSPATH.

% Obsolete:
%\item
%\api{NeuralNetwork} module requires the JAXP XML parser for JDK 1.3. You can get it at\\
%\url{http://java.sun.com/xml/downloads/javaxmlpack.html}.
%Click the ``Download now'' under the heading ``Java XML Pack - Summer 02 Update
%Release''. This should be the right one.

\end{itemize}

%MARF INSTALLATION
%=================

\section{{\marf} Installation}

There are several ways to ``install'' {\marf}.

\begin{itemize}
\item
Download the latest \file{marf-$<$ver$>$.jar}
\item
Build it from sources
	\begin{itemize}
	\item
	UNIXen
	\item
	Windows
	\end{itemize}
\end{itemize}

\section{Downloading the Latest Jar Files}
\label{sect:download-install-marf}
\index{Installation!Binary}
%-------------------------------------

Just go to \url{http://marf.sf.net}, and download an appropriate \file{marf-$<$ver$>$.jar}
from there.
To install it, put the downloaded \file{.jar} file(s) somewhere from within
the reach of your CLASSPATH or Java extensions directory, EXTDIRS.
Et voila, since now on you can try to write some mini mind-blowing
apps based on {\marf}. You can also get some demo applications,
such as \api{SpeakerIdentApp} from the same exact web site
to try out.
This ``install'' is ``platform-independent''.

As of MARF 0.3.0.5, there are several \file{.jar} files being released.
Their number may increase based on the demand. The different jars contain
various subsets of {\marf}'s code in addition to the full and complete
code set that was always released in the past. The below is the description
of the currently released jars:

\begin{itemize}

\item
\file{marf-$<$ver$>$.jar} contains a complete set of {\marf}'s code
excluding JUnit tests and debug information, optimized.

\item
\file{marf-debug-$<$ver$>$.jar} contains a complete set of {\marf}'s code
including JUnit tests and debug information.

\item
\file{marf-util-$<$ver$>$.jar} contains a subset of {\marf}'s code
corresponding primarily to the contents of the \api{marf.util} package, optimized.
This binary release is useful for apps which rely only on the quite comprehensive set
of general-purpose utility modules and nothing else. This jar is quite small
in size.

\item
\file{marf-storage-$<$ver$>$.jar} contains a subset of {\marf}'s code
corresponding primarily to the contents of the \api{marf.Storage} and some of the \api{marf.util} packages, optimized.
This binary release is useful for apps which rely only on the set
of general-purpose utility and storage modules.

\item
\file{marf-math-$<$ver$>$.jar} contains a subset of {\marf}'s code
corresponding primarily to the contents of the \api{marf.math} and some of the \api{marf.util} packages, optimized.
This binary release is useful for apps which rely only on the set
of general-purpose utility and math modules.

\item
\file{marf-utilimathstor-$<$ver$>$.jar} contains a subset of {\marf}'s code
corresponding primarily to the contents of the \api{marf.util}, \api{marf.math}, \api{marf.Stats}, and \api{marf.Storage} packages, optimized.
This binary release is useful for apps which rely only on the set
of general-purpose modules from these packages.

\end{itemize}

\section{Building From Sources}
\index{Installation!From Sources}
%---------------------

You can grab the latest tarball of the current CVS,
or pre-packaged \texttt{-src} release
and compile it yourself producing the \file{.jar} file, which
you will need to install as described in the \xs{sect:download-install-marf}.
The {\marf} sources can be obtained
from \url{http://marf.sf.net}. Extract:

\begin{verbatim}
    tar xvfz marf-src-<ver>.tar.gz
\end{verbatim}

\noindent
or

\begin{verbatim}
    tar xvfj marf-src-<ver>.tar.bz2
\end{verbatim}

\noindent
This will create a directory
\file{marf-$<$ver$>$} under the current directory
with the {\marf} sources.
Change into that directory for the rest
of the installation procedure.

\subsection{UNIXen}
\index{Installation!Linux}
\index{Installation!UNIXen}

We went with the makefile build approach. You will need
GNU \tool{make} (sometimes called `\tool{gmake}') to use it. Assuming
you have unpacked the sources, `cd' to
\file{src} and type:

\begin{verbatim}
    make
\end{verbatim}

\noindent
This will compile and build \file{marf-$<$ver$>$.jar} in the current directory.
(Remember to use {\gnu} \tool{make}.)
The last line displayed should be:

\begin{verbatim}
(-: MARF build has been successful :-)
\end{verbatim}

\noindent
To install {\marf} enter:

\begin{verbatim}
    make install
\end{verbatim}

\noindent
This will install the \file{.jar} file(s) in the pre-determined
place in the \file{/usr/lib/marf} directory.

\begin{verbatim}
    make apidoc
\end{verbatim}

\noindent
This will compile general API javadoc pages in \file{../../api}.

\begin{verbatim}
    make apidoc-dev
\end{verbatim}

\noindent
This will compile developer's API javadoc pages in \file{../../api-dev}.
Both APIs can be compiled at once by using \texttt{make api}.
Of course, you can compile w/o the makefile and use \tool{javac},
\tool{jar}, and \tool{javadoc} directly if you really want to.

\subsection{Windows}
\index{Installation!Windows}

We also used JBuilder \cite{jbuilder} from version 5 through to 2005, so there is a project file
\file{marf.jpx} in this directory. If you have JBuilder you can use
this project file to build \file{marf.jar}. There are also Eclipse \cite{eclipse} and
NetBeans \cite{netbeans} project files rooted at \file{.project} and \file{.classpath}
for Eclipse that you can import and \file{build.xml} and
\file{nbproject/*.*} for NetBeans.
Otherwise, you are stuck
with \tool{javac}/\tool{java}/\tool{jar} command-line tools for the moment.

\subsection{Cygwin / under Windows}
\index{Installation!Cygwin}

Follow pretty much the same steps as for UNIXen build above. You
might need to hack Makefiles or a corresponding environment variable
to support ``;'' directory separator
and quoting instead of ``:''.

\section{Configuration}

Typically, {\marf} does not really need much more configuration
tweaking for itself other than having a JDK or a JRE installed
to run it. The applications that use {\marf} should however be
able to find \file{marf.jar} by either setting the CLASSPATH
environment variable to point where the jar is or mention it
explicitly on the command line (or otherwise JVM argument)
with the \texttt{-cp} or \texttt{-classpath} options.

\subsection{Environment Variables}

\subsubsection{CLASSPATH}
\index{CLASSPATH}

Similarly to a commonly-used PATH variable, CLASSPATH tells where to
find \file{.class} or \file{.jar} files if they are not present
in the standard, known to the JVM default directories. This is a
colon-separated (``:'') for Linux/Windows and semicolon-separated (``;'')
for Windows/Cygwin list of directories with \file{.class} files
or explicitly mentioned \file{.jar}, \file{.war}, or \file{.zip}
archives containing \file{.class} files.

{\marf} itself does not depend on any non-default classes or
Java archives, so it requires no CLASSPATH by itself (except when used
with unit testing with JUnit \cite{junit}, the \file{junit.jar} must
be somewhere in CLASSPATH). However,
applications wishing to use {\marf} should point their CLASSPATH
to where find it unless it is in one of the default known to the JVM
places.

\subsubsection{EXTDIRS}
\index{EXTDIRS}

This variable typically lists a set of directories for Java
extensions where to find them. If you didn't place a pointer
in your CLASSPATH, you can place it in your EXTDIRS instead.

\subsection{Web Applications}

\subsubsection{Tomcat}
\index{Tomcat}

If for some reason you decide to use one of the {\marf}'s jars
in a web app, the web app servlet/JSP container such as Tomcat
\cite{tomcat} is capable of setting the CLASSPATH itself as long
as the jar(s) are in one of the \file{shared/lib} or 
\file{webapp/$<$your application here$>$/WEB-INF/lib} directories.

\subsection{JUnit}

If you intend to run unit tests or use \file{marf-debug*.jar} 
JUnit \cite{junit} has to be ``visible'' to {\marf} and the apps
through CLASSPATH.

%MARF UPGRADE
%============

\section{{\marf} Upgrade}
\index{upgrading}
\index{Installation!Upgrade}

Normally, an upgrade would simply mean just yanking old \file{.jar} out,
and plugging the new one in, UNLESS you depend on certain parts
of {\marf}'s experimental/legacy API.

It is important to note that {\marf}'s API is still stabilizing,
especially for the newer modules (e.g. NLP). Even older modules still
get affected as the {\marf} is still flexible in that. Thus,
the API changes do in fact happen every release insofar, for the
better (not always incompatible). This, however, also means that the serialized version
of the \api{Serializable} classes also change, so the corresponding
data needs to be retrained until an upgrade utility is available.
Therefore, please check the versions, \file{ChangeLog}, class revisions,
and talk to us if you are unsure of what can affect you. A great
deal of effort went into versioning each public class and
interface of {\marf}, which you can check by running:

\begin{verbatim}
    java -jar marf.jar --verbose
\end{verbatim}

\noindent
To ask about more detailed changes if they are unclear from
the \file{ChangeLog}, please write to us to \url{marf-devel@lists.sf.net}
or post a question in one of our forums at:

\begin{verbatim}
    https://sourceforge.net/forum/?group_id=63118
\end{verbatim}

\noindent
Each release of {\marf} also supplies the Javadoc API comments,
so you can and should consult those too. The bleeding edge
API is located at:

\begin{verbatim}
    http://marf.sourceforge.net/api/
\end{verbatim}

\section{Regression Tests}
\index{regression test application}
\index{Testing!Regression}
\index{MARF!Testing}

If you want to test the newly built {\marf} before you deploy it,
you can run the regression tests. The regression
tests are a test suite to verify that {\marf}
runs on your platform in the way the developers expected it to.
For now to do so the option is to run manually tests located
in the \api{marf.junit} package as well as executing
all the \file{Test*} applications described later on in
the Applications chapter. A \api{Regression} testing application
to do the comprehensive testing and grouping the other test
applications as well as the JUnit tests. This application is
still under development as of 0.3.0.5.

\section{Uninstall}
\index{Uninstall}
\index{Installation!Removal}

Simply remove the pertinent \file{marf*.jar} file(s).

\section{Cleaning}

After the installation you can make room by removing the built
files from the source tree with the command \tool{make clean}.

% EOF

	\chapter{MARF Architecture}
\index{MARF!Architecture}

$Revision: 1.29 $

Before we begin, you should understand the basic
{\marf} system  architecture. Understanding how the
parts of {\marf} interact will make the follow up sections
somewhat clearer. This document presents architecture
of the {\marf} system, including the layout of the physical
directory structure, and Java packages.

Let's take a look at the general {\marf} structure in \xf{fig:arch}.
The \api{MARF} class is the central ``server'' and configuration ``placeholder'', which contains the major methods --
the core pipeline -- a typical pattern recognition process.
The figure presents basic abstract modules of the architecture.
When a developer needs to add or use a module, they derive
from the generic ones.

\begin{figure}
	\centering
	\includegraphics[angle=90,totalheight=\textheight]{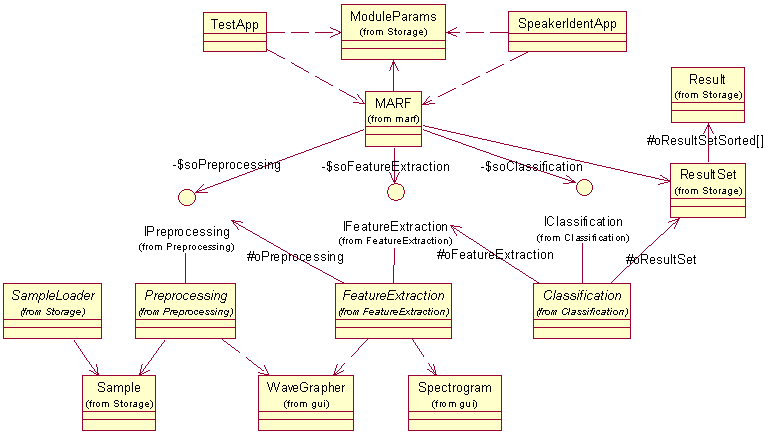}
	\caption{Overall Architecture}
	\label{fig:arch}
\end{figure}

The core pipeline\index{MARF!Core Pipeline} sequence diagram from an application
up until the very end result is presented in \xf{fig:pipeline}. It includes all major
participants as well as basic operations. The participants are the
modules responsible for a typical general pattern recognition pipeline.
A conceptual data-flow diagram of the pipeline is in \xf{fig:pipeline-flow}.
The grey areas indicate stub modules that are yet to be implemented.

\begin{figure}
	\centering
	\includegraphics[angle=90,totalheight=660pt,width=550pt]{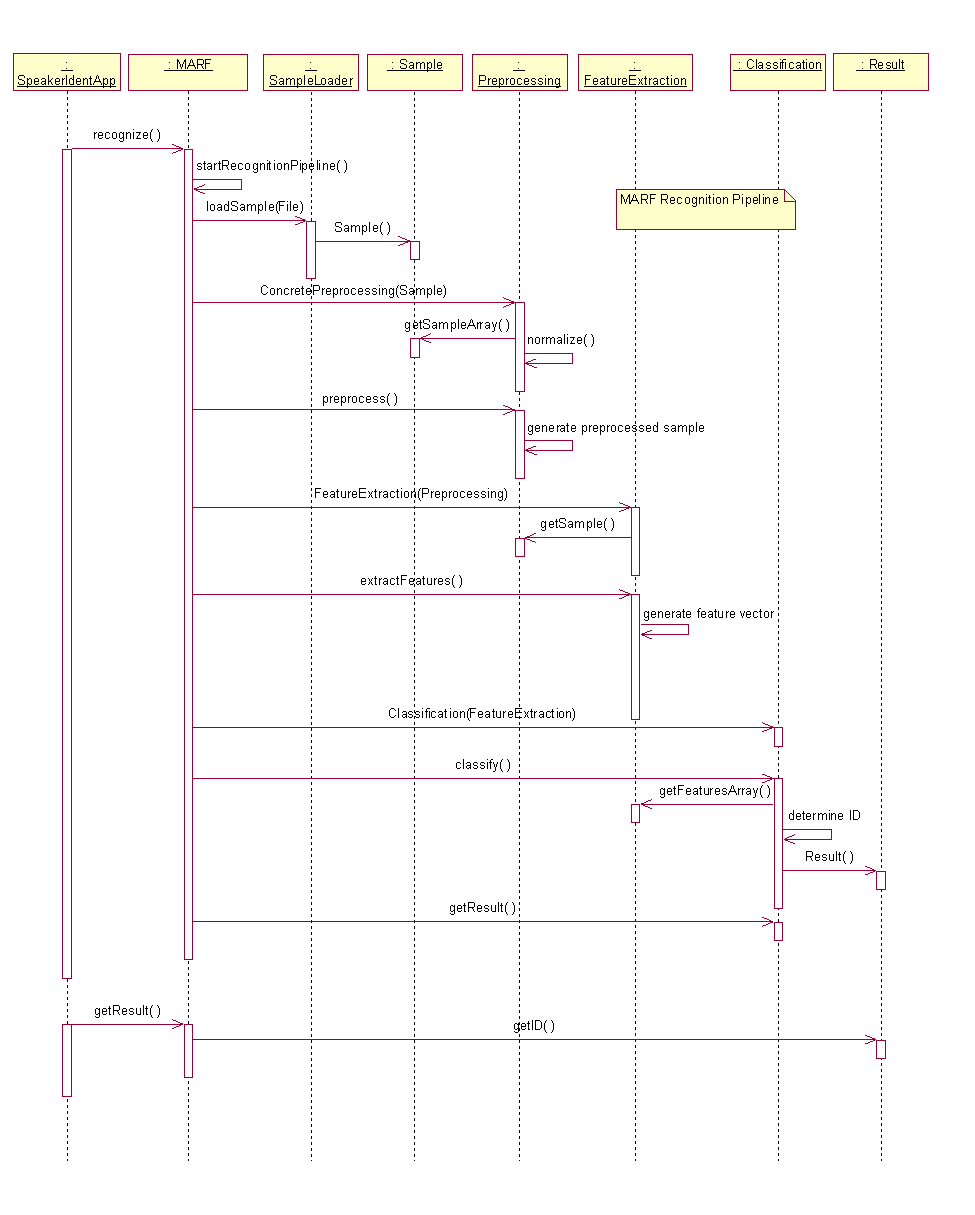}
	\caption{The Core Pipeline\index{MARF!Core Pipeline} Sequence Diagram}
	\label{fig:pipeline}
\end{figure}

\begin{figure}
	\centering
	\includegraphics[width=\textwidth]{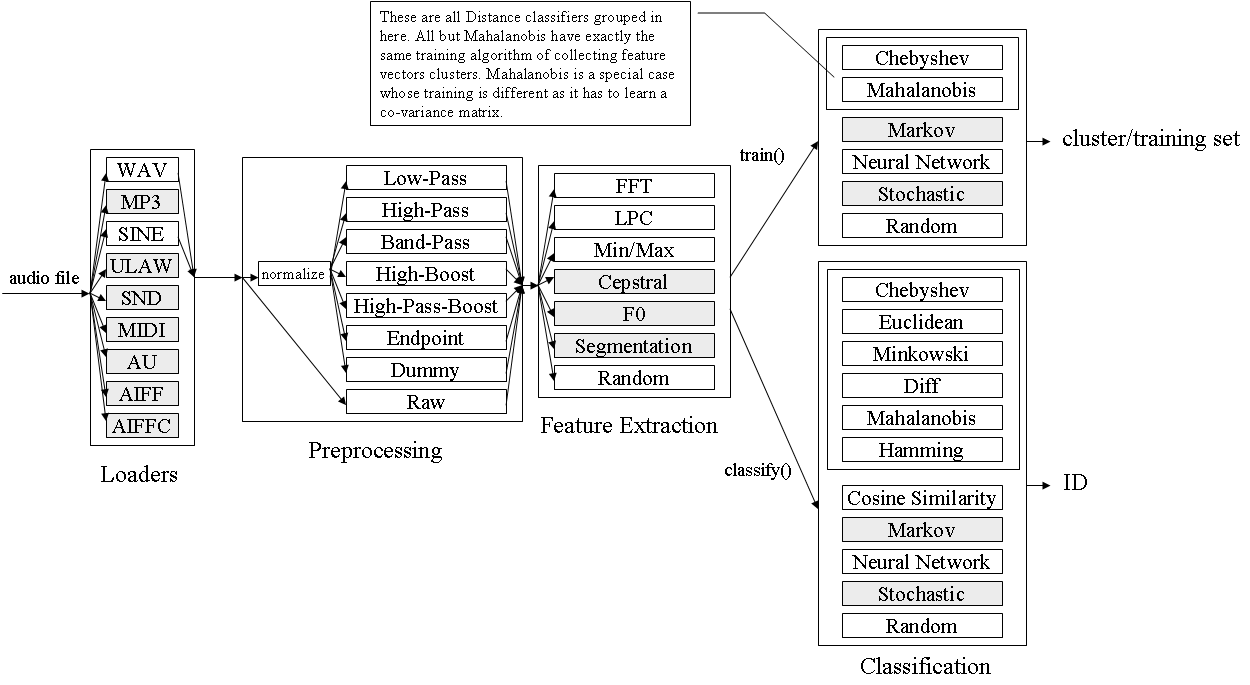}
	\caption{The Core Pipeline\index{MARF!Core Pipeline} Data Flow}
	\label{fig:pipeline-flow}
\end{figure}

Consequently, the framework has the mentioned basic modules,
as well as some additional entities to manage storage
and serialization of the input/output data.

\section{Application Point of View}
\index{MARF!Application Point of View}

An application, using the framework, has to choose
the concrete configuration and submodules for preprocessing,
feature extraction, and classification stages. There is an API the application
may use defined by each module or it can use them through the {\marf}.

There are two phases in {\marf}'s usage by an application:

\begin{itemize}
	\item Training, i.e. \api{train()}
	\item Recognition, i.e. \api{recognize()}
\end{itemize}

Training is performed on a virgin {\marf} installation to get
some training data in. Recognition is an actual identification process of a sample
against previously stored patterns during training.

\section{Packages and Physical Layout}
\index{MARF!Packages}

\begin{figure}
	\centering
	\includegraphics{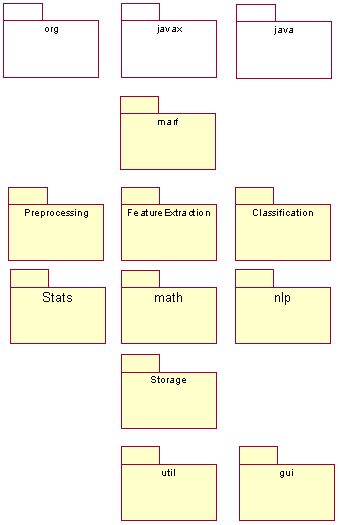}
	\caption{MARF Java Packages}
	\label{fig:packages}
\end{figure}

The Java package structure is in \xf{fig:packages}.
The following is the basic structure of {\marf}:

\vspace{15pt}
\hrule
\begin{verbatim}
marf.*

MARF.java - The MARF Server
            Supports Training and Recognition mode
            and keeps all the configuration settings.

marf.Preprocessing.* - The Preprocessing Package

/marf/Preprocessing/

   Preprocessing.java - Abstract Preprocessing Module, has to be subclassed
   PreprocessingExceotion.java

   /Endpoint/*.java   - Endpoint Filter as implementation of Preprocessing
   /Dummy/
       Dummy.java  - Normalization only
       Raw.java    - no preprocessing
   /FFTFilter/
       FFTFilter.java
       LowPassFilter.java
       HighPassFilter.java
       BandpassFilter.java - Band-pass Filter as implementation of Preprocessing
       HighFrequencyBoost.java

marf.FeatureExtraction.* - The Feature Extraction Package

/marf/FeatureExtraction/
    FeatureExtraction.java
    /FFT/FFT.java        - FFT implementation of Preprocessing
    /LPC/LPC.java        - LPC implementation of Preprocessing
    /MinMaxAmplitudes/MinMaxAmplitudes.java
    /Cepstral/*.java
    /Segmentation/*.java
    /F0/*.java

marf.Classification.* - The Classification Package

/marf/Classification/
    Classification.java
    ClassificationException.java
    /NeuralNetwork/
        NeuralNetwork.java
        Neuron.java
        Layer.java
    /Stochastic/*.java
    /Markov/*.java
    /Distance/
       Distance.java
       EuclideanDistance.java
       ChebyshevDistance.java
       MinkowskiDistance.java
       MahalonobisDistance.java
       DiffDistance.java

marf.Storage.* - The Physical Storage Management Interface

/marf/Storage/
    Sample.java
    ModuleParams.java
    TrainingSet.java
    FeatureSet.java
    Cluster.java
    Result.java
    ResultSet.java
    IStorageManager.java - Interface to be implemented by the above modules
    StorageManager.java  - The most common implementation of IStorageManager
    ISampleLoader.java   - All loaders implement this
    SampleLoader.java    - Should know how to load different sample format
    /Loaders/*.*         - WAV, MP3, ULAW, etc.
    IDatabase.java
    Database.java

marf.Stats.* - The Statistics Package meant to collect various types of stats.

/marf/Stats/
    StatsCollector.java - Time took, noise removed, patterns stored, modules available, etc.
    Ngram.java
    Observation.java
    ProbabilityTable.java
    StatisticalEstimators
    StatisticalObject.java
    WordStats.java
    /StatisticalEstimators/
        GLI.java
        KatzBackoff.java
        MLE.java
        SLI.java
        StatisticalEstimator.java
        /Smoothing/
            AddDelta.java
            AddOne.java
            GoodTuring.java
            Smoothing.java
            WittenBell.java

marf.gui.* - GUI to the graphs and configuration

/marf/gui/
    Spectrogram.java
    SpectrogramPanel.java
    WaveGrapher.java
    WaveGrapherPanel.java
    /util/
       BorderPanel.java
       ColoredStatusPanel.java
       SmartSizablePanel.java

marf.nlp.* - most of the NLP-related modules

/marf/nlp/
    Collocations/
    Parsing/
    Stemming/
    util/

marf.math.* - math-related algorithms are here

/marf/math/
    Algorithms.java
    MathException.java
    Matrix.java
    Vector.java

marf.util.* - important utility modules

/marf/util/
    Arrays.java
    BaseThread.java
    ByteUtils.java
    Debug.java
    ExpandedThreadGroup.java
    FreeVector.java
    InvalidSampleFormatException.java
    Logger.java
    MARFException.java
    Matrix.java
    NotImplementedException.java
    OptionProcessor.java
    SortComparator.java
    /comparators/
        FrequencyComparator.java
        RankComparator.java
        ResultComparator.java
\end{verbatim}
\hrule
\vspace{15pt}

\section{Current Limitations}
\index{Limiations}
\index{MARF!Current Limitaions}

Our current pipeline is maybe somewhat too rigid.
That is, there's no way to specify more than one preprocessing
to process the same sample in one pass (as of
0.3.0.2 the preprocessing modules can be chained, however, e.g.
one filter followed by another in a preprocessing pipeline).

Also, the pipeline often assumes that the whole sample is loaded before doing
anything with it, instead of sending parts of the sample a bit at a time.
Perhaps this simplifies things, but it won't allow us to deal with large
samples at the moment. However, it's not a problem for our framework
and the application because our samples are small enough and memory is cheap. Additionally,
we have streaming support already in the \api{WAVLoader} and some modules support it, but
the final conversion to streaming did not happen yet.

{\marf} provides only limited support for inter-module dependency. It is possible
to pass module-specific arguments, but problems like
number of parameters mismatch between feature extraction and classification,
and so on are not tracked.
There is also one instance of \api{ModuleParams} that exists in \api{MARF}
for now limiting combination of non-default feature extraction modules.

% EOF

	\chapter{Methodology}
\index{Methodology}

$Revision: 1.11 $

This section presents what methods and algorithms were implemented and used
in this project. We overview storage issues first, then preprocessing
methods followed by feature extraction, and ended by classification.

\section{Storage}

$Revision: 1.19 $

\begin{figure}
	\centering
	\includegraphics[angle=90,totalheight=660pt,width=550pt]{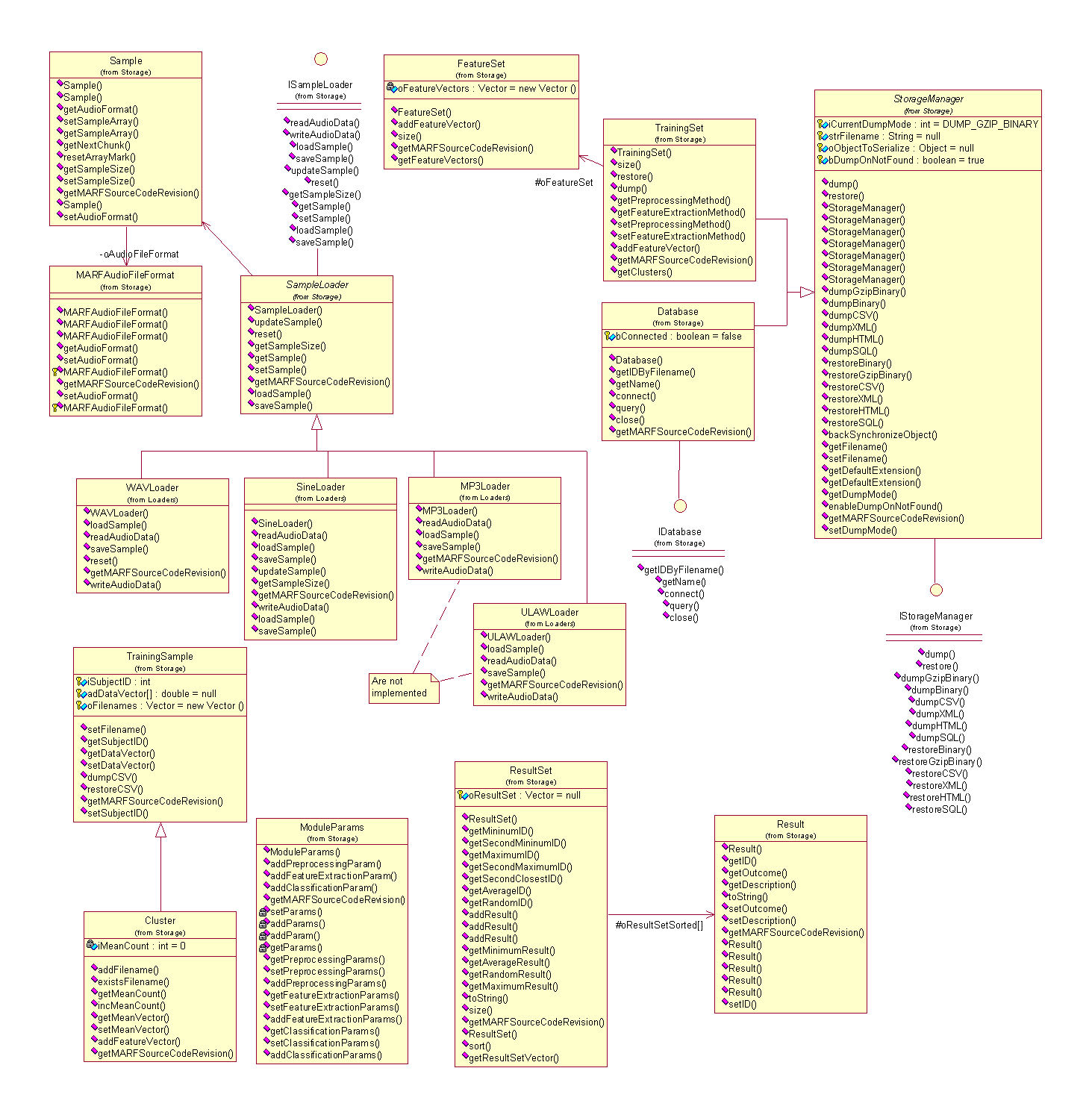}
	\caption{Storage}
	\label{fig:storage}
\end{figure}

Figure \ref{fig:storage}
presents basic \api{Storage} modules and their API.

\subsection{Speaker Database}

We store specific speakers in a comma-separated (CSV) file, \verb+speakers.txt+
within the application.
It has the following format:

\verb+<id:int>,<name:string>,<training-samples:list>,<testing-samples:list>+

Sample lists are defined as follows:

\verb+<*-sample-list> := filename1.wav|filename2.wav|...+

\subsection{Storing Features, Training, and Classification Data}

We defined a standard \verb+StorageManager+ interface for the modules to use.
That's part of the \verb+StorageManager+ interface which each module will override
because each a module has to know how to serialize itself, but the applications
using {\marf} should not care.
Thus, this \verb+StorageManager+ is a base class with abstract methods \verb+dump()+ and
\verb+restore()+. These methods would generalize the
model's serialization, in the sense that they are somehow ``read'' and ``written''.

We have to store data we used for training for later use in
the classification process. For this we pass FFT (\xf{sect:fft}) and LPC (\xf{sect:lpc})
feature vectors through the \verb+TrainingSet+/\verb+TrainingSample+ class pair,
which, as a result, store mean vectors (clusters) for our training models.

In the Neural Network we use XML. The only reason XML and text files
have been suggested is to allow us to easily modify values in
a text editor and verify the data visually.

In the Neural Network classification, we are using one net for
all the speakers. We had thought that one net for each speaker
would be ideal, but then we'll lose too much discrimination
power. But doing this means that the net will need complete
re-training for each new training utterance (or group thereof).

We have a training/testing script that lists the location of all the wave files to
be trained along with the identification of the speaker - \verb+testing.sh+.

%All resulting vectors (and associated speakers) are appended to the \verb+training.set+ file.
%Then the model is re-trained on whatever data needed and new models are dumped.

%In the stochastic models, if we had them, the complete set of
%utterances will be needed for the speaker to which the new
%utterance(s) are being added, for mean and variance calculations. This
%implies that all data needs storing.

\subsection{Assorted Training Notes}

$Revision: 1.6 $

For training we are using sets of feature vectors created by \api{FFT} or
\api{LPC} and passing them to the \api{NeuralNetwork} in the form of a 
cluster, or rather probabilistic ``cluster''. With this, there are some
issues:

\begin{enumerate}
\item
Mapping. We have to have a record of speakers and IDs for these speakers.
This can be the same for \api{NeuralNetwork} and \api{Stochastic} methods so long as the \api{NeuralNetwork}
will return the proper number and the ``clusters'' in the \api{Stochastic} module have
proper labeling.

\item
Feature vector generation. I think the best way to do this is for each
application using {\marf} to specify a feature file or directory which will
contain lines/files with the following info:

\verb+[a1, a2, ... , an] : <speaker id>+

Retraining for a new speaker would involve two phases 1) appending the
features to the file/dir, then 2) re-training the models. The
\api{Classification} modules will be aware of the scheme and re-train on all data
required.
\end{enumerate}

\subsubsection{Clusters Defined}

Given a set of feature vectors in $n$-dimensional space, if we want these to
represent $m$ ``items'' (in our case, speakers), we can make groupings of
these vectors with a center point $c_{i}$ (ie: $m$ center points which will
then represent the ``items''). Suen discussed an iterative algorithm to find
the optimal groupings (or clusters).
Anyway, I don't believe that Suen's clustering stuff is at all useful, as
we will know, through info from training, which speaker is associated with
the feature vector and can create the ``clusters'' with that information.

So for \api{NeuralNetwork}: no clusters, just regular training.
So for \api{Stochastic}: Clusters (kind of). I believe that we need to represent
a Gaussian curve with a mean vector and a co-variance matrix. This will be
created from the set of feature vectors for the speaker. But again, we know
who it is so the Optimal Clustering business is useless.

% EOF

\subsection{File Location}

We decided to keep all the data and intermediate files in the same
directory or subdirectories of that of the application.

\begin{itemize}

\item \verb+marf.Storage.TrainingSet.*+ - represent training sets
	(global clusters) used in training with different preprocessing and feature extraction
	methods; they can either be gzipped binary (.bin) or CSV text (.csv).

\item \verb+speakers.txt.stats+ - binary statistics file.

\item \verb+marf.Classification.NeuralNetwork.*.xml+ - XML file representing a trained
	Neural Net for all the speakers in the database.

\item \verb+training-samples/+ - directory with WAV files for training.

\item \verb+testing-samples/+ - directory with WAV files for testing.

\end{itemize}

\subsection{Sample and Feature Sizes}

Wave files are read and outputted as an array of
data points that represents the waveform of the signal.

Different methods will have different feature vector sizes.
It depends on what kind of precision one desires.
In the case of FFT, a 1024 FFT will result in 512 features,
being an array of ``doubles'' corresponding to the frequency range.

\cite{shaughnessy2000} said about using 3 ms for phoneme analysis and
something like one second for general voice analysis.  At 8 kHz, 1024 samples
represents 128 ms, this might be a good compromise.

\subsection{Parameter Passing}

A generic \verb+ModuleParams+ container class has been created to for an application
to be able to pass module-specific parameters when
specifying model files, training data,
amount of LPC coefficients, FFT window size, logging/stats files, etc.

\subsection{Result}

When classification is over, its result should be stored somehow
for further retrieval by the application. We have defined
the \verb+Result+ object to carry out this task. It contains
ID of the subject identified as well as some additional statistics
(such as second closest speaker and distances from other speakers, etc.)

\subsection{Sample Format}

$Revision: 1.20 $

The sample format used for our samples was the following:

\begin{itemize}
	\item Audio Format: PCM signed (WAV)
	\item Sample Rate: 8000 Hz
	\item Audio Sample Size: 16 bit
	\item Channels: 1 (mono)
	\item Duration: from about 7 to 20 seconds
\end{itemize}

All training and testing samples were recorded through an external sound recording
program (MS Sound Recorder) using a standard microphone. Each sample was
saved as a WAV file with the above properties and stored in the appropriate folders
where they would be loaded from within the main application. The PCM audio format
(which stands for Pulse Code Modulation) refers to the digital encoding of the audio
sample contained in the file and is the format used for WAV files. In a PCM
encoding, an analog signal is represented as a sequence of amplitude values. The
range of the amplitude value is given by the audio sample size which represents the
number of bits that a PCM value consists of. In our case, the audio sample size is
16-bit which means that that a PCM value can range from 0 to 65536. Since we are
using PCM-signed format, this gives an amplitude range between $-32768$ and $32768$.
That is, the amplitude values of each recorded sample can vary within this range.
Also, this sample size gives a greater range and thus provides better accuracy in
representing an audio signal then using a sample size of 8-bit which limited to a
range of $(-128, 128)$. Therefore, a 16-bit audio sample size was used for our
experiments in order to provide the best possible results. The sampling rate refers
to the number of amplitude values taken per second during audio digitization.
According to the Nyquist theorem, this rate must be at least twice the maximum rate
(frequency) of the analog signal that we wish to digitize (\cite{jervis}). Otherwise, the signal
cannot be properly regenerated in digitized form. Since we are using an 8 kHz
sampling rate, this means that actual analog frequency of each sample is limited to
4 kHz. However, this limitation does not pose a hindrance since the difference in
sound quality is negligible (\cite{shaughnessy2000}). The number of channels
refers to the output of the sound (1 for mono and 2 for stereo -- left and right
speakers). For our experiment, a single channel format was used to avoid complexity
during the sample loading process.

\subsection{Sample Loading Process}

To read audio information from a saved voice sample, a special sample-loading
component had to be implemented in order to load a sample into an internal data
structure for further processing. For this, certain sound libraries (\verb+javax.sound.sampled+)
were provided from the Java programming language which enabled us to
stream the audio data from the sample file. However once the data was captured, it
had to be converted into readable amplitude values since the library routines only
provide PCM values of the sample. This required the need to implement special
routines to convert raw PCM values to actual amplitude values (see \verb+SampleLoader+ class in
the \verb+Storage+ package).

\clearpage
The following pseudo-code represents the algorithm used to convert the PCM values
into real amplitude values (\cite{javasun}):

\vspace{15pt}
\hrule
\begin{verbatim}
function readAmplitudeValues(Double Array : audioData)
{
    Integer: MSB, LSB,
             index = 0;

    Byte Array: AudioBuffer[audioData.length * 2];

    read audio data from Audio stream into AudioBuffer;

    while(not reached the end of stream OR index not equal to audioData.length)
    {
        if(Audio data representation is BigEndian)
        {
            // First byte is MSB (high order)
            MSB = audioBuffer[2 * index];
            // Second byte is LSB (low order)
            LSB = audioBuffer[2 * index + 1];
        }
        else
        {
            // Vice-versa...
            LSB = audioBuffer[2 * index];
            MSB = audioBuffer[2 * index + 1];
        }

        // Merge high-order and low-order byte to form a 16-bit double value.
        // Values are divided by maximum range
        audioData[index] = (merge of MSB and LSB) / 32768;
    }
}
\end{verbatim}
\hrule
\vspace{15pt}

This function reads PCM values from a sample stream into a byte array that has
twice the length of \verb+audioData+; the array which will hold the converted amplitude
values (since sample size is 16-bit). Once the PCM values are read into \verb+audioBuffer+,
the high and low order bytes that make up the amplitude value are extracted
according to the type of representation defined in the sample's audio format.
If the data representation is {\it big endian}, the high order byte of each PCM value is
located at every even-numbered position in \verb+audioBuffer+. That is, the high order
byte of the first PCM value is found at position 0, 2 for the second value, 4 for
the third and so forth. Similarly, the low order byte of each PCM value is located
at every odd-numbered position (1, 3, 5, etc.). In other words, if the data
representation is {\it big endian}, the bytes of each PCM code are read from left to
right in the \verb+audioBuffer+. If the data representation is not {\it big endian}, then high
and low order bytes are inversed. That is, the high order byte for the first PCM
value in the array will be at position 1 and the low order byte will be at
position 0 (read right to left). Once the high and low order bytes are properly
extracted, the two bytes can be merged to form a 16-bit double value. This value is
then scaled down (divide by 32768) to represent an amplitude within a unit range $(-1, 1)$.
The resulting value is stored into the \verb+audioData+ array, which will be passed to
the calling routine once all the available audio data is entered into the array. An
additional routine was also required to write audio data from an array into wave
file. This routine involved the inverse of reading audio data from a sample file
stream. More specifically, the amplitude values inside an array are converted back
to PCM codes and are stored inside an array of bytes (used to create new audio
stream). The following illustrates how this works:

\vspace{15pt}
\hrule
\begin{verbatim}
public void writePCMValues(Double Array: audioData)
{
    Integer: word  = 0,
             index = 0;

    Byte Array: audioBytes[(number of ampl. values in audioData) * 2];

    while(index not equal to (number of ampl. values in audioData * 2))
    {
        word = (audioData[index] * 32768);
        extract high order byte and place it in appropriate position in audioBytes;
        extract low order byte and place it in appropriate position in audioBytes;
    }

    create new audio stream from audioBytes;
}
\end{verbatim}
\hrule
\vspace{15pt}

% EOF

\section{Assorted File Format Notes}

$Revision: 1.4 $

	We decided to stick to Mono-8000Hz-16bit WAV files.  8-bit might be
	okay too, but we could retain more precision with 16-bit files.  8000Hz is
	supposed to be all you need to contain all frequencies of the vocal
	spectrum
	(according to Nyquist anyways...).  If we use 44.1 kHz we'll just be wasting
	space and computation time.

	There are also MP3 and ULAW and other file format loaders stubs which are unimplemented
	as of this version of MARF.

Also: I was just thinking I think I may have made a bit of a mistake
downsampling to 8000Hz... I was saying that the voice ranges to about 8000Hz so
that's all we should need for the samples, but then I realized that if you have
an 8000Hz sample, it actually only represents 4000Hz, which would account for
the difference I noticed.. but maybe we should be using 16KHz samples.  On the
other hand, even at 4KHz the voice is still perfectly distinguishable...

I tried the WaveLoader with one of the samples provided by Stephen
(jimmy1.wav naturally!) and got some nice results! I graphed the PCM
obtained from the getaudioData() function and noticed quite a difference
from the PCM graph obtained with my ``test.wav". With ``test.wav", I was
getting unexpected results as the graph (``rawpcm.xls") didn't resemble any
wave form. This lead me to believe that I needed to convert the data on
order to represent it in wave form (done in the ``getWaveform()" function).
But after having tested the routine with ``jimmy1.wav", I got a beautiful
wave-like graph with just the PCM data which makes more sense since PCM
represents amplitude values! The reason for this is that my ``test.wav"
sample was actually 8-bit mono (less info.) rather than 16-bit mono as with
Stephen's samples. So basically, we don't need to do any ``conversion" if we
use 16-bit mono samples and we can scrap the ``getWaveform()" function.
I will come up with a ``Wave" class sometime this week which will take care
of loading wave files and windowing audio data. Also, we should only read
audio data that has actual sound, meaning that any silence (say -10 < db <
10) should be discarded from the sample when extracting audio data.
Just thinking out loud!

I agree here.  I was thinking perhaps the threshold could be determined from
the ``silence" sample.

% EOF

% EOF

\clearpage

\section{Preprocessing}
\label{sect:preprocessing}
\index{Preprocessing}
\index{Methodology!Preprocessing}

\noindent
\rule{7.0in}{.013in}

$Revision: 1.21 $

This section outlines the preprocessing mechanisms considered
and implemented in {\marf}. We present you with the API and structure in \xf{fig:preprocessing}, along with
the description of the methods.

\begin{figure}
	\centering
	\includegraphics[angle=90,height=660pt]{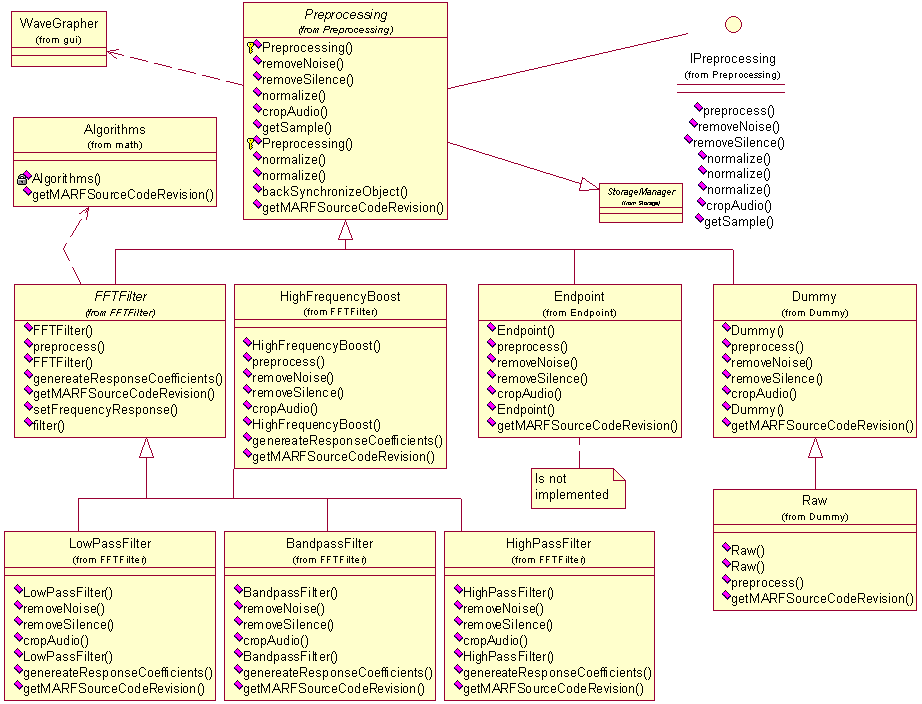}
	\caption{Preprocessing}
	\label{fig:preprocessing}
\end{figure}

\subsection{``Raw Meat''}
\index{Preprocessing!Raw}

$Revision: 1.4 $

\subsubsection{Description}

This is a basic ``pass-everything-through'' method that doesn't
do actually any preprocessing. Originally developed within the
framework, it was meant to be a base line method, but it gives
better top results out of many configurations. This method
is however not ``fair'' to others as it doesn't do any normalization
so the samples compared on the original data coming in. Likewise
silence and noise removal are not done in here.

\subsubsection{Implementation Summary}

\begin{itemize}
\item Implementation: \api{marf.Preprocessing.Dummy.Raw}
\item Depends on: \api{marf.Preprocessing.Dummy.Dummy}
\item Used by: \api{test}, \api{marf.MARF}, \api{SpeakerIdentApp}
\end{itemize}

% EOF

\subsection{Normalization}

Since not all voices will be recorded at exactly the same level, it is important to normalize
the amplitude of each sample in order to ensure that features will be comparable.  Audio
normalization is analogous to image normalization.  Since all samples are to be loaded as
floating point values in the range $[-1.0, 1.0]$, it should be ensured that every sample actually
does cover this entire range.

\begin{figure}
	\centering
	\includegraphics[width=400pt]{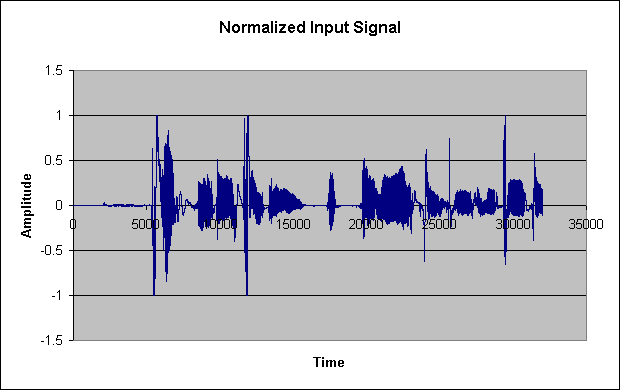}
	\caption{Normalization of aihua5.wav from the testing set.}
	\label{fig:prep-norm}
\end{figure}

The procedure is relatively simple: find the maximum amplitude in the sample, and then scale
the sample by dividing each point by this maximum. Figure \ref{fig:prep-norm} illustrates
normalized input wave signal.

\subsection{Noise Removal}

Any vocal sample taken in a less-than-perfect (which is always the case) environment will experience a certain
amount of room noise.  Since background noise exhibits a certain frequency characteristic,
if the noise is loud enough it may inhibit good recognition of a voice when the voice is
later tested in a different environment.  Therefore, it is necessary to remove as much
environmental interference as possible.

To remove room noise, it is first necessary to get a sample of the room noise by itself.
This sample, usually at least 30 seconds long, should provide the general frequency
characteristics of the noise when subjected to FFT analysis.  Using a technique similar
to overlap-add FFT filtering, room noise can then be removed from the vocal sample by
simply subtracting the noise's frequency characteristics from the vocal sample in question.

That is, if $S(x)$ is the sample, $N(x)$ is the noise, and $V(x)$ is the voice, all in the
frequency domain, then

\begin{center}$S(x) = N(x) + V(x)$\end{center}

Therefore, it should be possible to isolate the voice:

\begin{center}$V(x) = S(x) - N(x)$\end{center}

Unfortunately, time has not permitted us to implement this in practice yet.

\subsection{Silence Removal}

Silence removal got implemented in 0.3.0.6 in MARF's
\api{Preprocessing.removeSilence()} methods (and derivatives, except \api{Raw}).
For better results, \api{removeSilence()} should be executed after
normalization, where the default threshold of 1\% (0.01) works well
for the most cases.

The silence removal is performed in time domain where the amplitudes
below the threshold are discarded from the sample. This also makes
the sample smaller and less resemblent to other samples improving
overall recognition performance.

The actual threshold can be a settable parameter through
\api{ModuleParams}, which is a third parameter according
to the preprocessing parameter protocol.

% EOF

\subsection{Endpointing}
\index{Preprocessing!Endpointing}
\index{Endpointing}
\index{Methodology!Endpointing}

The Endpointing algorithm got implemented in {\marf} as follows.
By the {\em end-points} we mean the local minimums and maximums
in the amplitude changes. A variation of that is whether to
consider the sample edges and continuous data points (of the same
value) as end-points. In {\marf}, all these four cases
are considered as end-points by default with an option to
enable or disable the latter two cases via setters or the \api{ModuleParams}
facility. The endpointing algorithm is implemented in \api{Endpoint}
of the \api{marf.Preprocessing.Endpoint} package and appeared in 0.3.0.5.

\subsection{FFT Filter}\label{sect:fft-filter}

The FFT filter is used to modify the frequency domain of the input sample in
order to better measure the distinct frequencies we are interested in.
Two filters are useful to speech analysis: high frequency boost, and low-pass
filter (yet we provided more of them, to toy around).

Speech tends to fall off at a rate of 6 dB per octave, and therefore the high
frequencies can be boosted to introduce more precision in their analysis.
Speech, after all, is still characteristic of the speaker at high frequencies,
even though they have a lower amplitude.  Ideally this boost should be performed
via compression, which automatically boosts the quieter sounds while maintaining
the amplitude of the louder sounds.  However, we have simply done this using a
positive value for the filter's frequency response.
The low-pass filter (\xs{sect:low-pass}) is used as a simplified noise reducer, simply cutting off
all frequencies above a certain point.  The human voice does not generate sounds
all the way up to 4000 Hz, which is the maximum frequency of our test samples,
and therefore since this range will only be filled with noise, it may be better
just to cut it out.

Essentially the FFT filter is an implementation of the Overlap-Add method of FIR
filter design \cite{dspdimension}.  The process is a simple way to perform fast convolution, by
converting the input to the frequency domain, manipulating the frequencies
according to the desired frequency response, and then using an Inverse-FFT to
convert back to the time domain. \xf{fig:fft-filter} demonstrates
the normalized incoming wave form translated into the frequency domain.

\begin{figure}
	\centering
	\includegraphics[width=400pt]{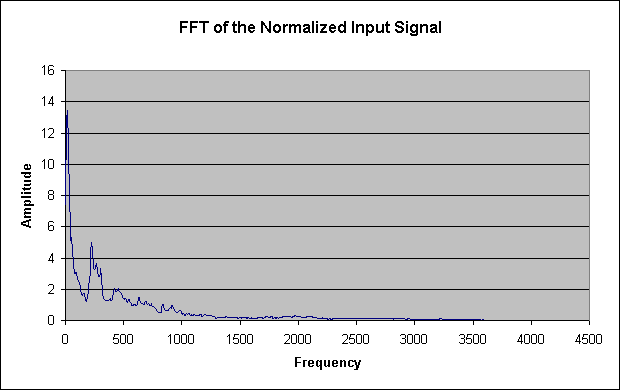}
	\caption{FFT of normalized aihua5.wav from the testing set.}
	\label{fig:fft-filter}
\end{figure}

The code applies the square root of the hamming window to the input windows
(which are overlapped by half-windows), applies the FFT, multiplies the results
by the desired frequency response, applies the Inverse-FFT, and applies the
square root of the hamming window again, to produce an undistorted output.

Another similar filter could be used for noise reduction, subtracting the
noise characteristics from the frequency response instead of multiplying,
thereby remove the room noise from the input sample.

\subsection{Low-Pass Filter}\label{sect:low-pass}

The low-pass filter has been realized on top of the FFT Filter,
by setting up frequency response to zero for frequencies
past certain threshold chosen heuristically
based on the window size where to cut off. We filtered out
all the frequencies past 2853 Hz.

\xf{fig:low-pass} presents an FFT graph of a
low-pass filtered signal.

\begin{figure}
	\centering
	\includegraphics[width=400pt]{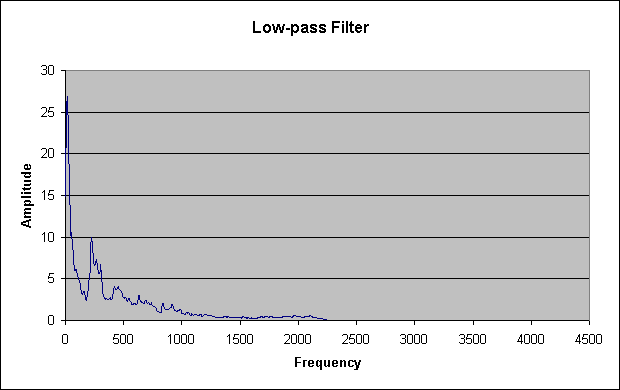}
	\caption{Low-pass filter applied to aihua5.wav.}
	\label{fig:low-pass}
\end{figure}

\subsection{High-Pass Filter}
\index{High-Pass Filter}
\index{Filters!High-Pass Filter}

$Revision: 1.7 $

As the low-pass filter, the high-pass filter (e.g. is in \xf{fig:high-pass})
has been realized on top of the FFT Filter,
in fact, it is the opposite to low-pass filter, and filters out
frequencies before 2853~Hz.
The implementation of the high-pass filter can be found in
\api{marf.Preprocessing.FFTFilter.HighPassFilter}.

\begin{figure}
	\centering
	\includegraphics[width=400pt]{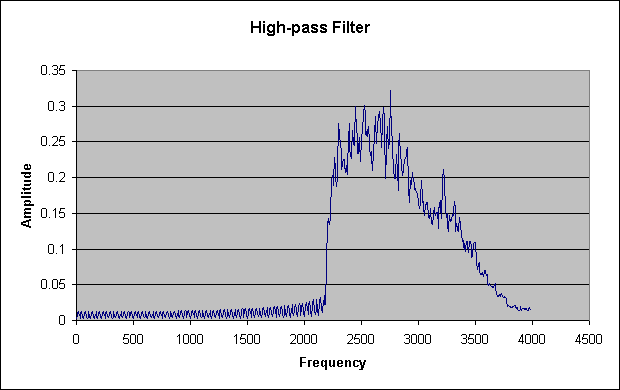}
	\caption{High-pass filter applied to aihua5.wav.}
	\label{fig:high-pass}
\end{figure}

% EOF

\subsection{Band-Pass Filter}

Band-pass filter in {\marf} is yet another instance of an FFT Filter (\xs{sect:fft-filter}),
with the default settings of the band of frequencies of $[1000, 2853]$ Hz. See \xf{fig:band-pass}.

\begin{figure}
	\centering
	\includegraphics[width=400pt]{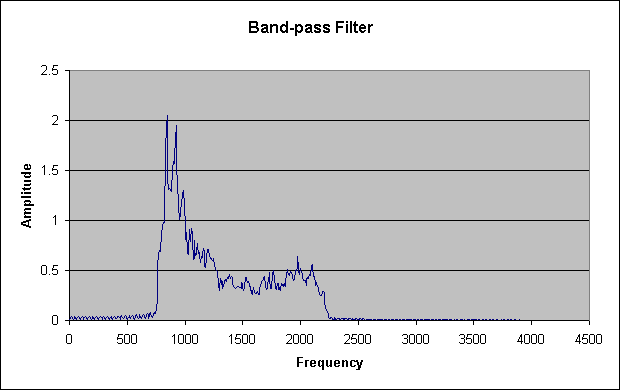}
	\caption{Band-pass filter applied to aihua5.wav.}
	\label{fig:band-pass}
\end{figure}

\subsection{High Frequency Boost}
\index{High Frequency Boost}
\index{Filters!High Frequency Boost}

$Revision: 1.11 $

This filter was also implemented on top of the FFT filter to boost the high-end
frequencies. The frequencies boosted after approx. 1000~Hz by a factor
of $5\pi$, heuristically determined, and then re-normalized. See \xf{fig:high-boost}.
The implementation of the high-frequency boost preprocessor can be found in
\api{marf.Preprocessing.FFTFilter.HighFrequencyBoost}.

\begin{figure}
	\centering
	\includegraphics[width=400pt]{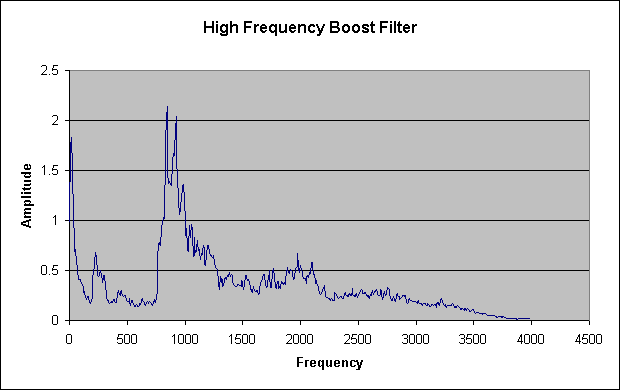}
	\caption{High frequency boost filter applied to aihua5.wav.}
	\label{fig:high-boost}
\end{figure}

% EOF

\subsection{High-Pass High Frequency Boost Filter}
\index{High-Pass High Frequency Boost Filter}
\index{Filters!High-Pass High Frequency Boost Filter}

$Revision: 1.2 $

For experimentation we said what would be very useful to do is
to test a high-pass filter along with high-frequency boost.
While there is no immediate class that does this, \api{MARF}
now chains the former and the latter via the new addition
to the preprocessing framework (in 0.3.0-devel-20050606)
where a constructor of one preprocessing module takes up another allowing
a preprocessing pipeline on its own. The results of this
experiment can be found in the Consolidated Results section.
While they did not yield a better recognition performance,
it was a good try to see. More tweaking and trying is
required to make a final decision on this approach as there
is an issue with the re-normalization of the entire input
instead of just the boosted part.

% EOF

% EOF

\clearpage

\section{Feature Extraction}
\index{Feature Extraction}
\index{Methodology!Feature Extraction}

$Revision: 1.14 $

This section outlines some concrete implementations of feature extraction methods of the {\marf} project.
First we present you with the API and structure, followed
by the description of the methods. The class diagram of this
module set is in \xf{fig:feat}.

\begin{figure}
	\centering
	\includegraphics[angle=90,height=660pt]{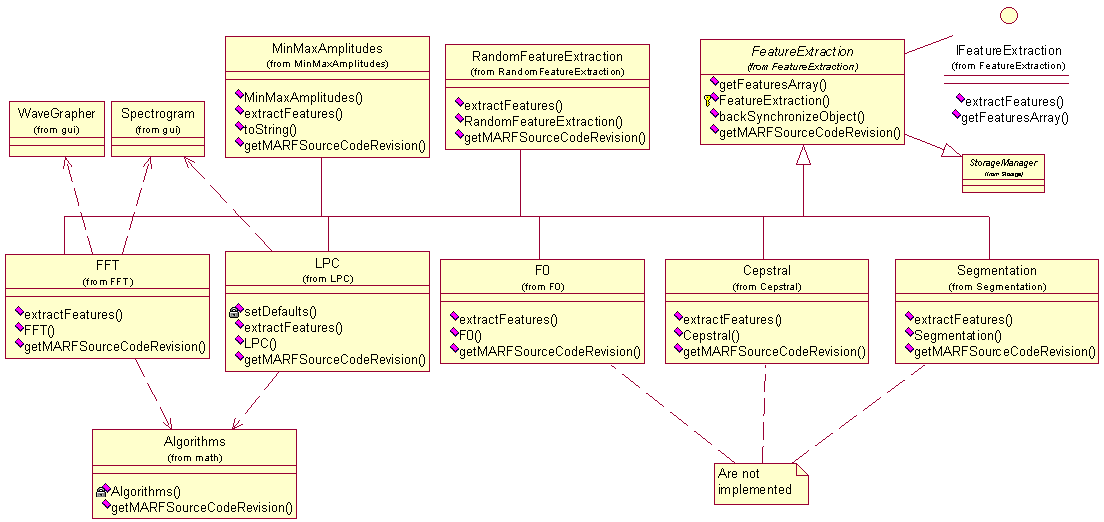}
	\caption{Feature Extraction Class Diagram}
	\label{fig:feat}
\end{figure}

\subsection{Hamming Window}
\index{Hamming Window}

$Revision: 1.13 $

\subsubsection{Implementation}
\index{Hamming Window!Implementation}

The Hamming Window implementation in {\marf} is in the
\api{marf.math.Algortithms.Hamming} class as of
version 0.3.0-devel-20050606 (a.k.a 0.3.0.2).

\subsubsection{Theory}
\index{Hamming Window!Theory}

In many DSP techniques, it is necessary to consider a smaller portion of the
entire speech sample rather than attempting to process the entire sample at
once.  The technique of cutting a sample into smaller pieces to be considered
individually is called ``windowing''.  The simplest kind of window to use is
the ``rectangle'', which is simply an unmodified cut from the larger sample.

$$
r(t) =
\left\{
{
	\begin{array}{ll}
		1 & \mbox{ for } (0 \le t \le N-1) \\
		0 & \mbox{ otherwise }
	\end{array}
}
\right.
$$

Unfortunately, rectangular windows can introduce errors, because near the edges
of the window there will potentially be a sudden drop from a high amplitude
to nothing, which can produce false ``pops'' and ``clicks'' in the analysis.

A better way to window the sample is to slowly fade out toward the edges, by
multiplying the points in the window by a ``window function''.  If we take
successive windows side by side, with the edges faded out, we will distort
our analysis because the sample has been modified by the window function.
To avoid this, it is necessary to overlap the windows so that all points in
the sample will be considered equally.  Ideally, to avoid all distortion, the
overlapped window functions should add up to a constant.  This is exactly what
the Hamming window does.  It is defined as:

$$ x = 0.54 - 0.46 \cdot \cos\left(\frac{2 \pi n}{l-1}\right) $$

\noindent
where $x$ is the new sample amplitude, $n$ is the index into the window, and $l$ is
the total length of the window.

% EOF

\subsection{Fast Fourier Transform (FFT)}\label{sect:fft}

The Fast Fourier Transform (FFT) algorithm is used both for feature extraction and as the basis for the
filter algorithm used in preprocessing.  Although a complete discussion of the
FFT algorithm is beyond the scope of this document, a short description of the
implementation will be provided here.

Essentially the FFT is an optimized version of the Discrete Fourier Transform.
It takes a window of size $2^{k}$ and returns a complex array of coefficients
for the corresponding frequency curve.  For feature extraction, only the
magnitudes of the complex values are used, while the FFT filter operates
directly on the complex results.

The implementation involves two steps: First, shuffling the input positions by a
binary reversion process, and then combining the results via a ``butterfly''
decimation in time to produce the final frequency coefficients.
The first step corresponds to breaking down the time-domain sample of size $n$
into $n$ frequency-domain samples of size 1.  The second step re-combines the $n$
samples of size 1 into 1 n-sized frequency-domain sample.

The code used in {\marf} has been translated from the C code provided in the book,
``Numeric Recipes in C'', \cite{numericalrecipes}.

\subsubsection{FFT Feature Extraction}

The frequency-domain view of a window of a time-domain sample gives us the
frequency characteristics of that window.  In feature identification, the
frequency characteristics of a voice can be considered as a list of ``features''
for that voice.  If we combine all windows of a vocal sample by
taking the average between them, we can get the average frequency
characteristics of the sample.  Subsequently, if we average the frequency
characteristics for samples from the same speaker, we are essentially finding
the center of the cluster for the speaker's samples.  Once all speakers have
their cluster centers recorded in the training set, the speaker of an
input sample should be identifiable by comparing its frequency analysis with
each cluster center by some classification method.

Since we are dealing with speech, greater accuracy should be attainable by
comparing corresponding phonemes with each other.  That is, ``th'' in ``the''
should bear greater similarity to ``th'' in ``this'' than will ``the" and ``this'' when
compared as a whole.

The only characteristic of the FFT to worry about is the window used as input.
Using a normal rectangular window can result in glitches in the frequency
analysis because a sudden cutoff of a high frequency may distort the results.
Therefore it is necessary to apply a Hamming window to the input sample, and
to overlap the windows by half.  Since the Hamming window adds up to a constant
when overlapped, no distortion is introduced.

When comparing phonemes, a window size of about 2 or 3 ms is appropriate, but
when comparing whole words, a window size of about 20 ms is more likely to be
useful.  A larger window size produces a higher resolution in the frequency
analysis.

\subsection{Linear Predictive Coding (LPC)}\label{sect:lpc}\index{LPC}

This section presents implementation of the LPC Classification module.

One method of feature extraction used in the {\marf} project was Linear
Predictive Coding (LPC) analysis. It evaluates windowed sections of
input speech waveforms and determines a set of coefficients
approximating the amplitude vs. frequency function. This approximation
aims to replicate the results of the Fast Fourier Transform yet only
store a limited amount of information: that which is most valuable to
the analysis of speech.

\subsubsection{Theory}

The LPC method is based on the formation of a spectral shaping filter,
$H(z)$, that, when applied to a input excitation source, $U(z)$, yields a
speech sample similar to the initial signal. The excitation source,
$U(z)$, is assumed to be a flat spectrum leaving all the useful
information in $H(z)$. The model of shaping filter used in most LPC
implementation is called an ``all-pole'' model, and is as follows:

$$ H(z) = \frac{G}{\left(1 - \displaystyle\sum_{k=1}^{p}(a_{k} z^{-k})\right)} $$

Where $p$ is the number of poles used. A pole is a root of the
denominator in the Laplace transform of the input-to-output
representation of the speech signal.

The coefficients $a_{k}$ are the final representation if the speech
waveform. To obtain these coefficients, the least-square
autocorrelation method was used. This method requires the use of the
autocorrelation of a signal defined as:

$$ R(k) = \displaystyle\sum_{m=k}^{n-1}(x(n) \cdot x(n-k)) $$

where $x(n)$ is the windowed input signal.

In the LPC analysis, the error in the approximation is used to derive
the algorithm. The error at time n can be expressed in the following
manner: $ e(n) = s(n) - \displaystyle\sum_{k=1}^{p}\left(a_{k} \cdot s(n-k)\right) $. Thusly,
the complete squared error of the spectral shaping filter $H(z)$ is:

$$ E = \displaystyle\sum_{n=-\infty}^{\infty}\left(x(n) - \displaystyle\sum_{k=1}^{p}(a_{k} \cdot x(n-k))\right) $$

To minimize the error, the partial derivative $\frac{{\delta}E}{{\delta}a_{k}}$ is
taken for each $k=1..p$, which yields $p$ linear equations of the form:

$$ \displaystyle\sum_{n=-\infty}^{\infty}(x(n-i) \cdot x(n)) = \displaystyle\sum_{k=1}^{p}(a_{k} \cdot \displaystyle\sum_{n=-\infty}^{\infty}(x(n-i) \cdot x(n-k)) $$

For $i=1..p$. Which, using the autocorrelation function, is:

$$ \displaystyle\sum_{k=1}^{p}(a_{k} \cdot R(i-k)) = R(i) $$

Solving these as a set of linear equations and observing that the
matrix of autocorrelation values is a Toeplitz matrix yields the
following recursive algorithm for determining the LPC coefficients:

$$ k_{m} = \frac{\left(R(m) - \displaystyle\sum_{k=1}^{m-1}\left(a_{m-1}(k)R(m-k)\right)\right)}{E_{m-1}} $$

$$ a_{m}(m) = k_{m} $$

$$ a_{m}(k) = a_{m-1}(k) - k_{m} \cdot a_{m}(m-k) \mbox{ for } 1 \le k \le m-1\mbox{,} $$

$$ E_{m} = (1 - k_{m}^2) \cdot E_{m-1} $$.

This is the algorithm implemented in the {\marf} LPC module.

\subsubsection{Usage for Feature Extraction}

The LPC coefficients were evaluated at each windowed iteration,
yielding a vector of coefficient of size $p$. These coefficients were
averaged across the whole signal to give a mean coefficient vector
representing the utterance. Thus a $p$ sized vector was used for
training and testing. The value of $p$ chosen was based on tests given
speed vs. accuracy. A $p$ value of around 20 was observed to be accurate and
computationally feasible.

\subsection{F0: The Fundamental Frequency}
\label{sect:f0}
\index{F0}
\index{Feature Extraction!F0}

$Revision: 1.6 $

[WORK ON THIS SECTION IS IN PROGRESS AS WE PROCEED WITH F0 IMPLEMENTATION IN {\marf}]

F0, the fundamental frequency, or ``pitch''.

Ian: ``The text (\cite{shaughnessy2000}) doesn't go into too much detail but gives a few techniques. Most
seem to involve another preprocessing to remove high frequencies and then
some estimation and postprocessing correction. Another, more detailed
source may be needed.''

Serguei: ``One of the prerequisites we already have: the low-pass filter that
does remove the high frequencies.''

% EOF

\subsection{Min/Max Amplitudes}
\label{sect:minmax}
\index{Min/Max Amplitudes}
\index{Feature Extraction!Min/Max Amplitudes}

$Revision: 1.4 $

\subsubsection{Description}

The Min/Max Amplitudes extraction simply involves
picking up $X$ maximums and $N$ minimums out of the
sample as features. If the length of the sample
is less than $X+N$, the difference is filled in
with the middle element of the sample.

TODO: This feature extraction does not perform very well yet
in any configuration because of the simplistic implementation:
the sample amplitudes are sorted and $N$ minimums and $X$ maximums
are picked up from both ends of the array. As the samples are
usually large, the values in each group are really close if
not identical making it hard for any of the classifiers
to properly discriminate the subjects. The future improvements
here will include attempts to pick up values in $N$ and $X$
distinct enough to be features and for the samples smaller than the $X+N$ sum,
use increments of the difference of smallest maximum and
largest minimum divided among missing elements in the middle
instead one the same value filling that space in.

% EOF

\subsection{Feature Extraction Aggregation}
\label{sect:aggregator}
\index{Feature Extraction Aggragation}
\index{Feature Extraction!Feature Extraction Aggragation}

$Revision: 1.2 $

\subsubsection{Description}

This method appeared in {\marf} as of 0.3.0.5.
This class by itself does not do any feature extraction, but
instead allows concatenation of the results of several actual feature
extractors to be combined in a single result. This should give the
classification modules more discriminatory power (e.g. when combining
the results of FFT and F0 together).
\api{FeatureExtractionAggregator} itself still implements
the \api{FeatureExtraction} API in order to be used in the main
pipeline of \api{MARF}.

The aggregator expects \api{ModuleParams} to be set to the
enumeration constants of a module to be invoked followed by that module's
enclosed instance \api{ModuleParams}. As of this implementation,
that enclosed instance of \api{ModuleParams} isn't really used, so
the {\bf main limitation} of the aggregator is that all the aggregated
feature extractors act with their default settings. This will happen until
the pipeline is re-designed a bit to include this capability.

The aggregator clones the incoming preprocessed sample for each
feature extractor and runs each module in a separate thread. A the
end, the results of each tread are collected in the same order as
specified by the initial \api{ModuleParams} and returned
as a concatenated feature vector. Some meta-information is available
if needed.

\subsubsection{Implementation}

Class: \api{marf.FeatureExtraction.FeatureExtractionAggregator}.

% EOF

\subsection{Random Feature Extraction}

By default given a window of size 256 samples, it picks
at random a number from a Gaussian distribution, and
multiplies by the incoming sample frequencies.
This all adds up and we have a feature vector at the end.
This should be the bottom line performance of all feature
extraction methods. It can also be used as a relatively fast
testing module.

% EOF

\clearpage

%\input{training}
%\clearpage

\section{Classification}
\index{Classification}

\noindent
\rule{7.0in}{.013in}

$Revision: 1.13 $

This section outlines classification methods of the {\marf} project.
First, we present you with the API and overall structure, followed
by the description of the methods. Overall structure of the modules
is in \xf{fig:classification}.

\begin{figure}
	\centering
	\includegraphics[angle=90,height=660pt]{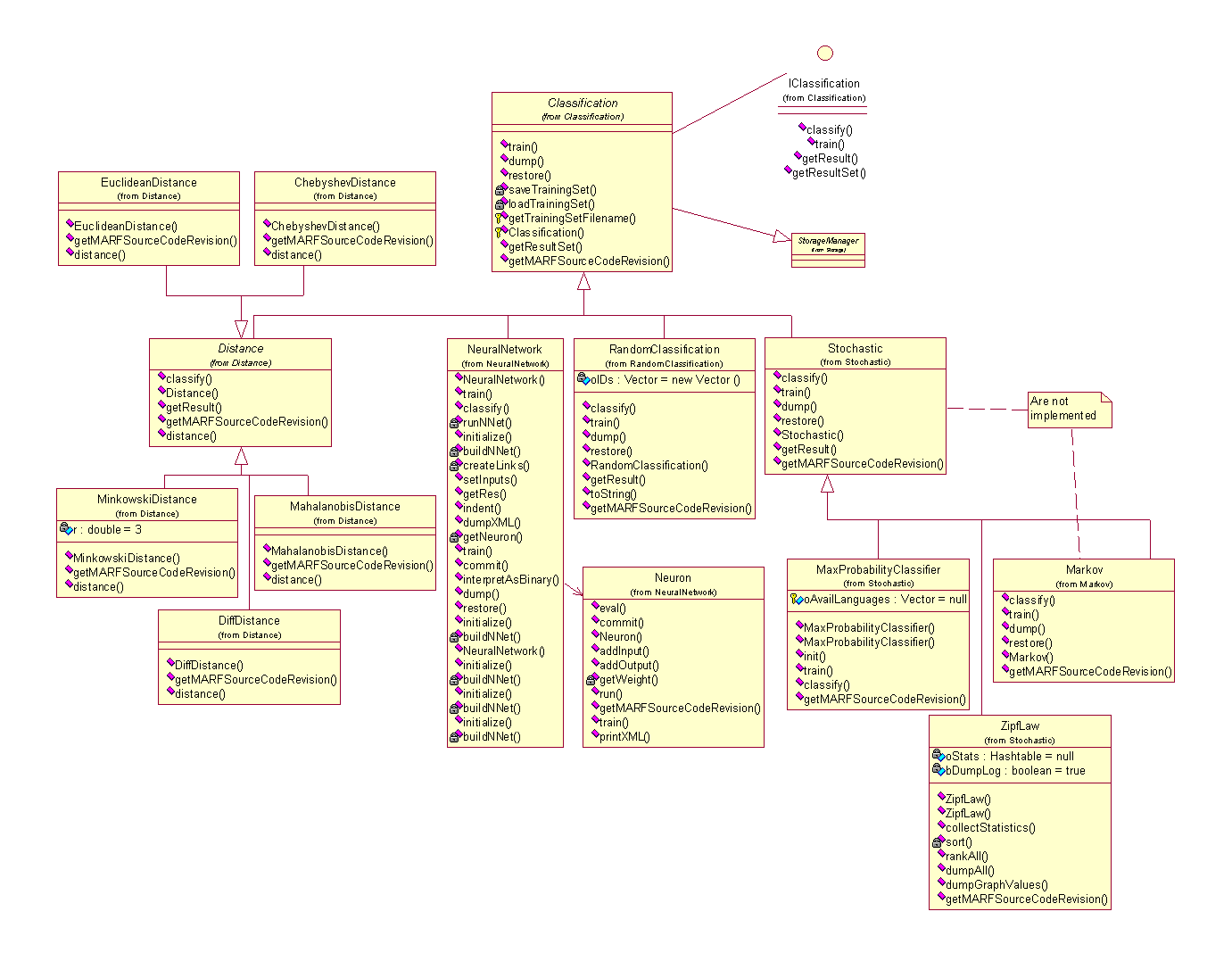}
	\caption{Classification}
	\label{fig:classification}
\end{figure}

\subsection{Chebyshev Distance}

Chebyshev distance is used along with other distance classifiers
for comparison. Chebyshev distance is also known as
a city-block or Manhattan distance. Here's its mathematical representation:

$$ d(x,y) = \displaystyle\sum_{k=1}^{n}(|x_{k}-y_{k}|) $$

\noindent
where $x$ and $y$ are feature vectors of the same length $n$.

\subsection{Euclidean Distance}

The Euclidean Distance classifier uses an Euclidean distance equation to find the
distance between two feature vectors.

If $A=(x_{1},x_{2})$ and $B=(y_{1},y_{2})$ are two 2-dimensional vectors, then the distance between
$A$ and $B$ can be defined as the square root of the sum of the squares of their
differences:

$$ d(x,y) = \sqrt{(x_{1}-y_{1})^{2} + {(x_{2}-y_{2})}^{2}} $$

This equation can be generalized to n-dimensional vectors by simply adding terms
under the square root.

$$ d(x,y) = \sqrt{(x_{n}-y_{n})^{2} + {(x_{n-1}-y_{n-1})}^{2} + ... + {(x_{1}-y_{1})}^{2}} $$

or

$$ d(x,y) = \sqrt{\displaystyle\sum_{k=1}^{n}(x_{k}-y_{k})^{2}} $$

or

$$ d(x,y) = \sqrt{(x-y)^{T}(x-y)} $$

A cluster is chosen based on smallest distance to the feature vector in question.

\subsection{Minkowski Distance}

Minkowski distance measurement is a generalization of both Euclidean
and Chebyshev distances.

$$ d(x,y) = \left(\displaystyle\sum_{k=1}^{n}(|x_{k}-y_{k}|)^{r}\right)^\frac{1}{r} $$

\noindent
where $r$ is a Minkowski factor. When $r=1$, it becomes Chebyshev distance,
and when $r=2$, it is the Euclidean one. $x$ and $y$ are feature vectors of the same length $n$.

\subsection{Mahalanobis Distance}
\label{sect:mahalanobis}
\index{Distance!Mahalanobis}
\index{Classification!Mahalanobis Distance}

$Revision: 1.7 $

\subsubsection{Summary}

\begin{itemize}
\item Implementation: \api{marf.Classification.Distance.MahalanobisDistance}
\item Depends on: \api{marf.Classification.Distance.Distance}
\item Used by: \api{test}, \api{marf.MARF}, \api{SpeakerIdentApp}
\end{itemize}

\subsubsection{Theory}

This distance classification is meant to be able to detect features
that tend to vary together in the same cluster if linear
transformations are applied to them, so it becomes invariant
from these transformations unlike all the other, previously seen
distance classifiers.

$$ d(x,y) = \sqrt{(x-y) C^{-1} (x-y)^{T}} $$

\noindent
where $x$ and $y$ are feature vectors of the same length $n$, and
$C$ is a covariance matrix, learnt during training for co-related
features.

In this release, namely 0.3.0-devel,
the covariance matrix being an identity matrix, $ C = I $,
making Mahalanobis distance be the same as the Euclidean one.
We need to complete the learning of the covariance matrix
to complete this classifier.

% EOF

\subsection{Diff Distance}
\label{sect:diff-distance}
\index{Distance!Diff}
\index{Classification!Diff Distance}

$Revision: 1.2 $

\subsubsection{Summary}

\begin{itemize}
\item Implementation: \api{marf.Classification.Distance.DiffDistance}
\item Depends on: \api{marf.Classification.Distance.Distance}
\item Used by: \api{test}, \api{marf.MARF}, \api{SpeakerIdentApp}
\end{itemize}

\subsubsection{Theory}

When Serguei Mokhov invented this classifier in May 2005, the original idea
was based on the way the \tool{diff} UNIX utility works. Later, for
performance enhancements it was modified. The essence of the diff distance is to
count how one input vector is different from the other in terms of
elements correspondence. If the Chebyshev distance between the two
corresponding elements is greater than some error $e$, then this
distance is accounted for plus some additional distance penalty $p$
is added. Both factors $e$ and $p$ can vary depending on desired
configuration. If the two elements are equal or pretty close
(the difference is less than $e$) then a small ``bonus'' of $e$ is subtracted
from the distance.

$$ d(x,y) = \sum_{i}{|x_i-y_i| + p, \mathit{if} |x_i-y_i| > e, \mathit{or} (-e)} $$

\noindent
where $x$ and $y$ are feature vectors of the same length.

% EOF

\subsection{Artificial Neural Network}
\label{sect:nnet}
\index{Artificial Neural Network}
\index{Neural Network}
\index{Methodology!Artificial Neural Network}
\index{Methodology!Neural Network}
\index{Algorithm!Artificial Neural Network}
\index{Algorithm!Neural Network}

This section presents implementation of the Neural Network Classification module.

One method of classification used is an Artificial Neural
Network. Such a network is meant to represent the neuronal
organization in organisms. Its use as a classification method lies is
in the training of the network to output a certain value given a
particular input \cite{artificialintelligence}.

\subsubsection{Theory}

A neuron consists of a set of inputs with associated weights, a
threshold, an activation function ($f(x)$) and an output value. The
output value will propagate to further neurons (as input values) in
the case where the neuron is not part of the ``output'' layer of the
network. The relation of the inputs to the activation function is as
follows:

$output \longleftarrow f(in)$

where $in = \displaystyle\sum_{i=0}^{n}(w_{i} \cdot a_{i}) - t$, ``vector'' $a$ is the
input activations, ``vector'' $w$ is the associated weights and $t$ is the
threshold of the network. The following activation function was used:

$sigmoid(x; c) = \frac{1}{(1 + e^{-cx})}$

where $c$ is a constant. The advantage of this function is that it is
differentiable over the region $(-\infty,+\infty)$ and has derivative:

$\frac{d(sigmoid(x; c))}{dx} = c \cdot sigmoid(x; c) \cdot (1 - sigmoid(x; c))$

The structure of the network used was a Feed-Forward Neural
Network. This implies that the neurons are organized in sets,
representing layers, and that a neuron in layer $j$, has inputs from
layer $j-1$ and output to layer $j+1$ only. This structure facilitates the
evaluation and the training of a network. For instance, in the
evaluation of a network on an input vector $I$, the output of neuron in
the first layer is calculated, followed by the second layer, and so
on.

\subsubsection{Training}

Training in a Feed-Forward Neural Network is done through the an
algorithm called Back-Propagation Learning. It is based on the error
of the final result of the network. The error the propagated backward
throughout the network, based on the amount the neuron contributed to
the error. It is defined as follows:

$w_{i,j} \longleftarrow \beta w_{i,j} + \alpha \cdot a_{j} \cdot \Delta_{i}$

where

$\Delta_{i} = Err_{i} \cdot \frac{df}{dx(in_{i})}$  for neuron $i$ in the output layer

and

$\Delta_{i} = \frac{df}{dt(in_{i})} \cdot \displaystyle\sum_{j=0}^{n}(\Delta_{j})$ for neurons in other layers

The parameters $\alpha$ and $\beta$ are used to avoid local minima in
the training optimization process. They weight the combination of the
old weight with the addition of the new change. Usual values for these
are determined experimentally.

The Back-Propagation training method was used in conjunction with
epoch training. Given a set of training input vectors $Tr$, the
Back-Propagation training is done on each run. However, the new weight
vectors for each neuron, ``vector'' $w'$, are stored and not used. After
all the inputs in $Tr$ have been trained, the new weights are committed
and a set of test input vectors $Te$, are run, and a mean error is
calculated. This mean error determines whether to continue epoch
training or not.

\subsubsection{Usage as a Classifier}

As a classifier, a Neural Network is used to map feature vectors to
speaker identifiers. The neurons in the input layer correspond to each
feature in the feature vector. The output of the network is the binary
interpretation of the output layer. Therefore the Neural Network has
an input layer of size $m$, where $m$ is the size of all feature vectors
and the output layer has size $\lceil(\log_{2}(n))\rceil$, where $n$ is the maximum
speaker identifier.

A network of this structure is trained with the set of input vectors
corresponding to the set of training samples for each speaker. The
network is epoch trained to optimize the results. This fully trained
network is then used for classification in the recognition process.

% EOF

\subsection{Random Classification}

That might sound strange, but we have a random classifier in {\marf}.
This is more or less testing module just to quickly test
the PR pipeline.  It picks an ID in the pseudo-random manner from the list of
trained IDs of subjects to classification.  It also serves as
a bottom-line of performance (i.e. recognition rate) for all
the other, slightly more sophisticated classification methods meaning
performance of the aforementioned methods must be better than
that of the Random; otherwise, there is a problem.

% EOF

\clearpage

% EOF

	%
% $Id: statistic-processing.tex,v 1.2 2006/02/26 23:38:46 mokhov Exp $
%

\chapter{Statistics Processing}
\label{chapt:statistic-processing}
\index{Statistics}
\index{Statistics!Processing}

$Revision: 1.2 $

This chapter describes what kind of statistics
processing is done in {\marf} in terms of statistics collection,
estimators, and smoothing and their use.

{\todo}

\section{Statistics Collection}
\index{Statistics!Collection}

{\todo}

\section{Statistical Estimators and Smoothing}
\index{Statistics!Estimators and Smoothing}

{\todo}

% EOF

	\chapter{Natural Language Processing (NLP)}
\index{Natural Language Processing}
\index{NLP}

$Revision: 1.5 $

This chapter will describe the NLP facilities now
present in {\marf}. This includes:

\begin{enumerate}
\item Probabilistic Parsing of English
\item Stemming
\item Collocations
\item Related N-gram Models, Zipf's Law and Maximum Probability Classifiers, Statistical Estimators and Smoothing.
\end{enumerate}

Please for now refer the related applications in the
Applications chapter in \xc{chapt:apps}.

{\todo}

\section{Zipf's Law}
\index{Zipf's Law}

See \xs{sect:zipf-law-app}.

{\todo}

\section{Statistical N-gram Models}
\index{Statistical N-gram Models}

See \xs{sect:lang-ident-app}.

{\todo}

\section{Probabilistic Parsing}
\index{Probabilistic Parsing}

See \xs{sect:probabilistic-parsing-app}.

{\todo}

\section{Collocations}
\index{Collocations}
{\todo}

\section{Stemming}
\index{Stemming}
{\todo}

% EOF

	\chapter{GUI}
\index{GUI}
\index{MARF!GUI}

$Revision: 1.6 $

[UPDATE ME]

\begin{figure}
	\centering
	\includegraphics[width=\textwidth]{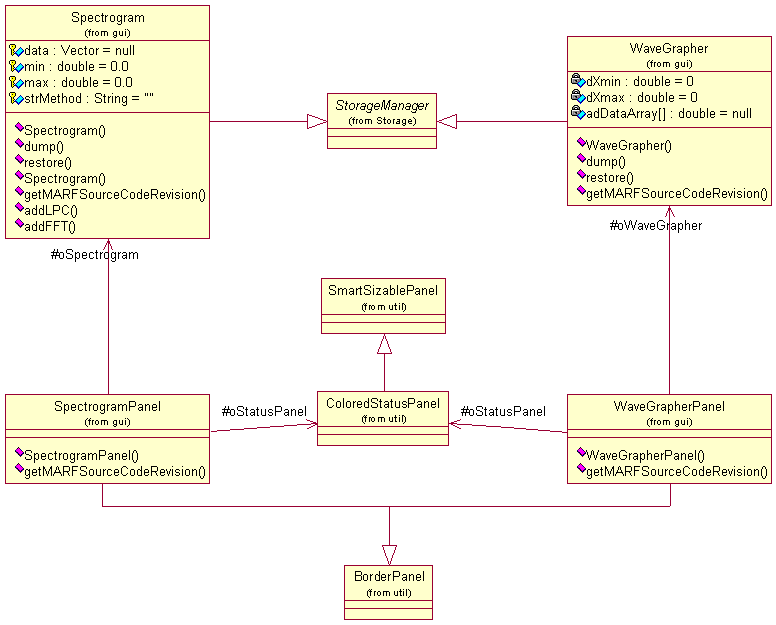}
	\caption{GUI Package}
	\label{fig:gui}
\end{figure}

Even though this section is entitled as GUI, we don't
really have too much GUI yet (it's planned though,
see TODO, \ref{appx:todo}). We do have a couple
of things under the \api{marf.gui} package, which we do
occasionally use and eventually they will expand to be a real
GUI classes. This tiny package is in Figure \ref{fig:gui}.

{\todo}

\section{Spectrogram}

$Revision: 1.4 $

Sometimes it is useful to visualize the data we are
playing with. One of the typical thing when dealing with
sounds, specifically voice, people are interested in
spectrograms of frequency distributions. The \api{Spectrogram}
class was designed to handle that and produce spectrograms
from both FFT and LPC algorithms and simply draw them. We did not
manage to make it a true GUI component yet, but instead we made
it to dump the spectrograms into PPM-format image files to be
looked at using some graphical package. Two examples of
such spectrograms are in the Appendix \ref{appx:spectra}.

We are just taking all the \verb+Output[]+ for the spectrogram. It's
supposed to be only half (\cite{shaughnessy2000}).
We took a hamming window of the waveform at $1/2$ intervals of 128
samples (ie: 8 kHz, 16 ms). By half intervals we mean that the second half
of the window was the first half of the next. O'Shaughnessy in \cite{shaughnessy2000} says this is a
good way to use the window. Thus, any streaming of waveform must consider
this.

What we did for both the FFT spectrogram and LPC determination was to
multiply the signal by the window and do a \api{doFFT()} or a \api{doLPC()} coefficient
determination on the resulting array of $N$ windowed samples. This gave us
an approximation of the stable signal at $s(i \cdot N/2)$.
Or course, we will have to experiment with windows and see which one is better, but
there may be no definitive best.

% EOF

\section{Wave Grapher}

$Revision: 1.4 $

\api{WaveGrapher} is another class designed, as the name
suggests, to draw the wave form of the incoming/preprocessed
signal. Well, it doesn't actually draw a thing, but dumps
the sample points into a tab-delimited text file to be loaded
into some plotting software, such as \tool{gnuplot} or \tool{Excel}. We also
use it to produce graphs of the signal in the frequency domain
instead of time domain. Examples of the graphs of data obtained
via this class are in the Preprocessing Section (\ref{sect:preprocessing}).

% EOF

% EOF

	\chapter{Sample Data and Experimentation}
\index{Sample Data}
\index{Experiments}

$Revision: 1.21 $

\section{Sample Data}
\index{Sample Data}

$Revision: 1.17 $

We have both female and male speakers, with age
ranging from a college student to a University professor.
The table \ref{tab:samples} has a list of people who have contributed their
voice samples for our project (with the first four being ourselves).
We want to thank them once again for helping us out.

\begin{table}
\begin{minipage}[b]{\textwidth}
\centering
\begin{tabular}{|c|c|c|c|} \hline
ID          & Name                & Training Samples & Testing Samples\\ \hline\hline
1           & Serge               & 14               & 1 \\
2           & Ian                 & 14               & 1 \\
3           & Steve               & 12               & 3 \\
4           & Jimmy               & 14               & 1 \\
5           & Dr. C.Y. Suen       & 2                & 1 \\
6           & Margarita Mokhova   & 14               & 1 \\
7           & Alexei Mokhov       & 14               & 1 \\
8           & Alexandr Mokhov     & 14               & 1 \\
9           & Graham Sinclair     & 12               & 2 \\
10          & Jihed Halimi        & 2                & 1 \\
11          & Madhumita Banerjee  & 3                & 1 \\
13          & Irina Dymova        & 3                & 1 \\
14          & Aihua Wu            & 14               & 1 \\
15          & Nick                & 9                & 1 \\
16          & Michelle Khalife    & 14               & 1 \\
17          & Shabana             & 7                & 1 \\
18          & Van Halen           & 8                & 1 \\
19          & RHCP                & 8                & 1 \\
20          & Talal Al-Khoury     & 14               & 1 \\
21          & Ke Gong             & 14               & 1 \\
22          & Emily Wu Rong       & 14               & 1 \\
23          & Emily Ying Lu       & 14               & 1 \\
24          & Shaozhen Fang       & 14               & 1 \\
25          & Chunlei He          & 14               & 1 \\
26          & Shuxin Fan          & 15               & 1 \\
27          & Shivani Bhat        & 14               & 1 \\
28          & Marinela Meladinova & 14               & 1 \\
29          & Fei Fang            & 14               & 2 \\ \hline\hline
{\bf Total} & {\bf 27}            & {\bf 319}        & {\bf 32} \\ \hline
\end{tabular}
\end{minipage}
\caption{Speakers contributed their voice samples.}
\label{tab:samples}
\end{table}

% EOF

\section{Comparison Setup}

The main idea was to compare combinations (in {\marf}: {\it configurations})
of different methods and variations within them in terms of recognition
rate performance. That means that having several preprocessing modules, several feature
extraction modules, and several classification modules, we can (and did)
try all their possible combinations.

That includes:

\begin{enumerate}
	\item Preprocessing: No-filtering, normalization, low-pass, high-pass,
	      band-pass, and high-frequency boost, high-pass and boost filters,
	      and endpointing.
	\item Feature Extraction: FFT/LPC/Min-Max/Random algorithms comparison.
	\item Classification: Distance classifiers, such as Chebyshev, Euclidean,
	      Minkowski, Mahalanobis, and Diff distances, as well as Neural Network and Random
	      classification.
\end{enumerate}

For this purpose we have written a \api{SpeakerIdentApp}, a command-line application
(so far, but GUI is planned) for TI speaker identification. We ran it for every possible configuration
with the following shell script, namely \tool{testing.sh}:

\vspace{15pt}
\hrule
{\scriptsize \begin{verbatim}
#!/bin/tcsh -f

#
# Batch Processing of Training/Testing Samples
# NOTE: Make take quite some time to execute
#
# Copyright (C) 2002 - 2009 The MARF Research and Development Group
#
# $Header: /cvsroot/marf/apps/SpeakerIdentApp/testing.sh,v 1.48 2009/02/22 02:11:41 mokhov Exp $
#

#
# Set environment variables, if needed
#

setenv CLASSPATH .:marf.jar
setenv EXTDIRS

#
# Set flags to use in the batch execution
#

set java = 'java -ea -verify -Xmx512m'
#set debug = '-debug'
set debug = ''
set graph = ''
#set graph = '-graph'
#set spectrogram = '-spectrogram'
set spectrogram = ''

if($1 == '--reset') then
    echo "Resetting Stats..."
    $java SpeakerIdentApp --reset
    exit 0
endif

if($1 == '--retrain') then
    echo "Training..."

    # Always reset stats before retraining the whole thing
    $java SpeakerIdentApp --reset

    foreach preprep ("" "-silence" "-noise" "-silence -noise")
        foreach prep (-norm -boost -low -high -band -bandstop -highpassboost -raw -endp -lowcfe -highcfe -bandcfe -bandstopcfe)
            foreach feat (-fft -lpc -randfe -minmax -aggr)

                # Here we specify which classification modules to use for
                # training.
                #
                # NOTE: for most distance classifiers it's not important
                # which exactly it is, because the one of the generic Distance is used.
                # Exception from this rule is the Mahalanobis Distance, which needs
                # to learn its Covariance Matrix.

                foreach class (-cheb -mah -randcl -nn)
                    echo "Config: $preprep $prep $feat $class $spectrogram $graph $debug"
                    date

                    # XXX: We cannot cope gracefully right now with these combinations in the
                    # typical PC/JVM set up --- too many links in the fully-connected NNet,
                    # so can run out of memory quite often; hence, skip them for now.
                    if("$class" == "-nn" && ("$feat" == "-fft" || "$feat" == "-randfe" || "$feat" == "-aggr")) then
                        echo "skipping..."
                        continue
                    endif

                    time $java SpeakerIdentApp --train training-samples $preprep $prep $feat $class $spectrogram $graph $debug
                end

            end
        end
    end

endif

echo "Testing..."

foreach preprep ("" "-silence" "-noise" "-silence -noise")
    foreach prep (-norm -boost -low -high -band -bandstop -highpassboost -raw -endp -lowcfe -highcfe -bandcfe -bandstopcfe)
        foreach feat (-fft -lpc -randfe -minmax -aggr)
            foreach class (-eucl -cheb -mink -mah -diff -hamming -cos -randcl -nn)
                echo "=-=-=-=-=-=-=-=-=-=-=-=-=-=-=-=-=-=-=-=-=-=-="
                echo "Config: $preprep $prep $feat $class $spectrogram $graph $debug"
                date
                echo "============================================="

                # XXX: We cannot cope gracefully right now with these combinations in the
                # typical PC/JVM set up --- too many links in the fully-connected NNet,
                # so can run out of memory quite often; hence, skip them for now.
                if("$class" == "-nn" && ("$feat" == "-fft" || "$feat" == "-randfe" || "$feat" == "-aggr")) then
                    echo "skipping..."
                    continue
                endif

                time $java SpeakerIdentApp --batch-ident testing-samples $preprep $prep $feat $class $spectrogram $graph $debug

                echo "---------------------------------------------"
            end
        end
    end
end

echo "Stats:"

$java SpeakerIdentApp --stats | tee stats.txt
$java SpeakerIdentApp --best-score | tee best-score.tex
date | tee stats-date.tex

echo "Testing Done"

exit 0

# EOF
\end{verbatim}
}
\hrule
\vspace{15pt}

The above script is for Linux/UNIX environments. To run a similar script
from Windows, use \tool{testing.bat} for classification and the \tool{retrain}
shortcut for re-training and classification. These have been completed
during the development of the 0.3.0 series.

See the results section (\ref{sect:results}) for results analysis.

\section{What Else Could/Should/Will Be Done}

There is a lot more that we realistically could do, but due to lack of time, these things
are not in yet. If you would like to contribute, let us know, meanwhile we'll keep working
at our speed.

\subsection{Combination of Feature Extraction Methods}

For example, assuming we use a combination of LPC coefficients and F0
estimation, we could compare the results of different combinations of
these, and discuss them later. Same with the
Neural Nets (modifying number of layers and number or neurons, etc.).

We could also do a 1024 FFT analysis and compare it against a 128 FFT
analysis.  (That is, the size of the resulting feature vector would be 512 or 64 respectively).
With LPC, one can specify the number of coefficients you want, the more you
have the more precise the analysis will be.

\subsection{Entire Recognition Path}

The \api{LPC} module is used to generate a mean vector of LPC coefficients for
the utterance. \api{F0} is used to find the average fundamental frequency of the
utterance. The results are concatenated to form the output vector, in a
particular order. The classifier would take into account the weighting of
the features: Neural Network would do so implicitly if it benefits the speaker
matching, and stochastic can be modified to give more weight to the F0 or
vice versa, depending on what we see best (i.e.: the covariance matrix in the
Mahalanobis distance (\ref{sect:mahalanobis})).

\subsection{More Methods}

Things like $F_0$, stochastic, and some other methods have not made to this release.
More detailed on this aspect, please refer to the TODO list in the Appendix.

% EOF

	\chapter{Experimentation Results}
\label{sect:results}
\index{Results}

$Revision: 1.23 $

\section{Notes}

Before we get to numbers, few notes and observations first:

\begin{enumerate}

\item By increasing the number of samples our results got
      better; with few exceptions, however. This can be explained by
      the diversity of the recording equipment, a lot less than uniform
      number of samples per speaker, and absence of noise
      removal. All the samples were recorded in not the same environments.
      The results then start averaging after awhile.

\item Another observation we made from our output, is that
      when the speaker is guessed incorrectly, quite often the second
      guess is correct, so we included this in our results as if we were
      ``guessing'' right from the second attempt.

\item FUN. Interesting to note, that we also tried to take some
      samples of music bands, and feed it to our application
      along with the speakers, and application's performance didn't suffer,
      yet even improved because the samples were treated in
      the same manner. The groups were not mentioned in the table,
      so we name them here: Van Halen [8:1] and Red Hot Chili Peppers [10:1] (where numbers
      represent [training:testing] samples used).

\end{enumerate}

\clearpage

\section{\api{SpeakerIdentApp}'s Options}

Configuration parameters were extracted from the command line,
which \api{SpeakerIdentApp} can be invoked with. They mean the following:

\vspace{15pt}
\hrule
\scriptsize
\begin{verbatim}
Usage:
  java SpeakerIdentApp --train <samples-dir> [options]        -- train mode
                       --single-train <sample> [options]      -- add a single sample to the training set
                       --ident <sample> [options]             -- identification mode
                       --batch-ident <samples-dir> [options]  -- batch identification mode
                       --gui                                  -- use GUI as a user interface
                       --stats                                -- display stats
                       --best-score                           -- display best classification result
                       --reset                                -- reset stats
                       --version                              -- display version info
                       --help | -h                            -- display this help and exit

Options (one or more of the following):

Preprocessing:

  -silence      - remove silence (can be combined with any below)
  -noise        - remove noise (can be combined with any below)
  -raw          - no preprocessing
  -norm         - use just normalization, no filtering
  -low          - use low-pass FFT filter
  -high         - use high-pass FFT filter
  -boost        - use high-frequency-boost FFT preprocessor
  -band         - use band-pass FFT filter
  -endp         - use endpointing
  -lowcfe       - use low-pass CFE filter
  -highcfe      - use high-pass CFE filter
  -bandcfe      - use band-pass CFE filter
  -bandstopcfe  - use band-stop CFE filter

Feature Extraction:

  -lpc          - use LPC
  -fft          - use FFT
  -minmax       - use Min/Max Amplitudes
  -randfe       - use random feature extraction
  -aggr         - use aggregated FFT+LPC feature extraction

Classification:

  -nn           - use Neural Network
  -cheb         - use Chebyshev Distance
  -eucl         - use Euclidean Distance
  -mink         - use Minkowski Distance
  -diff         - use Diff-Distance
  -randcl       - use random classification

Misc:

  -debug        - include verbose debug output
  -spectrogram  - dump spectrogram image after feature extraction
  -graph        - dump wave graph before preprocessing and after feature extraction
  <integer>     - expected speaker ID


\end{verbatim}

\normalsize
\hrule
\vspace{15pt}

\clearpage

% NOTE: stats-date.tex, best-score.tex, and stats.tex are generated by
%       the SpeakerIdentApp after running testing.sh

\section{Consolidated Results}

Our ultimate results \footnote{authoritative as of Tue Jan 24 06:23:21 EST 2006
} for all configurations we can have
and samples we've got are below.
Looks like our best results are with
``-endp -lpc -cheb'',
``-raw -aggr -eucl'',
``-norm -aggr -diff'',
``-norm -aggr -cheb'',
``-raw -aggr -mah'',
``-raw -fft -mah'',
``-raw -fft -eucl'', and
``-norm -fft -diff''
with the top result being around \bestscore{82.76
} and the second-best is
around \bestscore{86.21
} (see \xt{tab:results1}).

\begin{table}
\begin{minipage}[b]{\textwidth}
\centering
\begin{tabular}{|c|c|l|c|c|r|} \hline
Guess & Run \# & Configuration & GOOD & BAD & Recognition Rate,\%\\ \hline\hline
1st & 1 & -endp -lpc -cheb  & 24 & 5 & 82.76\\ \hline
1st & 2 & -raw -aggr -eucl  & 22 & 7 & 75.86\\ \hline
1st & 3 & -norm -aggr -diff  & 22 & 7 & 75.86\\ \hline
1st & 4 & -norm -aggr -cheb  & 22 & 7 & 75.86\\ \hline
1st & 5 & -raw -aggr -mah  & 22 & 7 & 75.86\\ \hline
1st & 6 & -raw -fft -mah  & 22 & 7 & 75.86\\ \hline
1st & 7 & -raw -fft -eucl  & 22 & 7 & 75.86\\ \hline
1st & 8 & -norm -fft -diff  & 22 & 7 & 75.86\\ \hline
1st & 9 & -norm -fft -cheb  & 21 & 8 & 72.41\\ \hline
1st & 10 & -raw -aggr -cheb  & 21 & 8 & 72.41\\ \hline
1st & 11 & -endp -lpc -mah  & 21 & 8 & 72.41\\ \hline
1st & 12 & -endp -lpc -eucl  & 21 & 8 & 72.41\\ \hline
1st & 13 & -raw -fft -mink  & 21 & 8 & 72.41\\ \hline
1st & 14 & -norm -fft -mah  & 21 & 8 & 72.41\\ \hline
1st & 15 & -norm -fft -eucl  & 21 & 8 & 72.41\\ \hline
1st & 16 & -norm -aggr -eucl  & 20 & 9 & 68.97\\ \hline
1st & 17 & -low -aggr -diff  & 20 & 9 & 68.97\\ \hline
1st & 18 & -norm -aggr -mah  & 20 & 9 & 68.97\\ \hline
1st & 19 & -raw -fft -cheb  & 20 & 9 & 68.97\\ \hline
1st & 20 & -raw -aggr -mink  & 20 & 9 & 68.97\\ \hline
1st & 21 & -norm -aggr -mink  & 19 & 10 & 65.52\\ \hline
1st & 22 & -low -fft -cheb  & 19 & 10 & 65.52\\ \hline
1st & 23 & -raw -lpc -mink  & 19 & 10 & 65.52\\ \hline
1st & 24 & -raw -lpc -diff  & 19 & 10 & 65.52\\ \hline
1st & 25 & -raw -lpc -eucl  & 19 & 10 & 65.52\\ \hline
1st & 26 & -raw -lpc -mah  & 19 & 10 & 65.52\\ \hline
1st & 27 & -raw -lpc -cheb  & 19 & 10 & 65.52\\ \hline
1st & 28 & -low -aggr -eucl  & 19 & 10 & 65.52\\ \hline
1st & 29 & -norm -lpc -mah  & 19 & 10 & 65.52\\ \hline
\end{tabular}
\end{minipage}
\caption{Consolidated results, Part 1.}
\label{tab:results1}
\end{table}

\begin{table}
\begin{minipage}[b]{\textwidth}
\centering
\begin{tabular}{|c|c|l|c|c|r|} \hline
Guess & Run \# & Configuration & GOOD & BAD & Recognition Rate,\%\\ \hline\hline
1st & 30 & -norm -lpc -mink  & 19 & 10 & 65.52\\ \hline
1st & 31 & -norm -lpc -diff  & 19 & 10 & 65.52\\ \hline
1st & 32 & -norm -lpc -eucl  & 19 & 10 & 65.52\\ \hline
1st & 33 & -low -aggr -cheb  & 19 & 10 & 65.52\\ \hline
1st & 34 & -norm -lpc -cheb  & 19 & 10 & 65.52\\ \hline
1st & 35 & -low -aggr -mah  & 19 & 10 & 65.52\\ \hline
1st & 36 & -low -fft -mah  & 19 & 10 & 65.52\\ \hline
1st & 37 & -norm -fft -mink  & 19 & 10 & 65.52\\ \hline
1st & 38 & -low -fft -diff  & 19 & 10 & 65.52\\ \hline
1st & 39 & -low -fft -eucl  & 19 & 10 & 65.52\\ \hline
1st & 40 & -raw -aggr -diff  & 19 & 10 & 65.52\\ \hline
1st & 41 & -high -aggr -mink  & 18 & 11 & 62.07\\ \hline
1st & 42 & -high -aggr -eucl  & 18 & 11 & 62.07\\ \hline
1st & 43 & -endp -lpc -mink  & 18 & 11 & 62.07\\ \hline
1st & 44 & -high -aggr -mah  & 18 & 11 & 62.07\\ \hline
1st & 45 & -raw -fft -diff  & 18 & 11 & 62.07\\ \hline
1st & 46 & -high -aggr -cheb  & 17 & 12 & 58.62\\ \hline
1st & 47 & -low -aggr -mink  & 17 & 12 & 58.62\\ \hline
1st & 48 & -low -fft -mink  & 17 & 12 & 58.62\\ \hline
1st & 49 & -high -fft -cheb  & 16 & 13 & 55.17\\ \hline
1st & 50 & -high -fft -mah  & 16 & 13 & 55.17\\ \hline
1st & 51 & -low -lpc -cheb  & 16 & 13 & 55.17\\ \hline
1st & 52 & -high -fft -mink  & 16 & 13 & 55.17\\ \hline
1st & 53 & -high -fft -eucl  & 16 & 13 & 55.17\\ \hline
1st & 54 & -low -lpc -eucl  & 15 & 14 & 51.72\\ \hline
1st & 55 & -low -lpc -mah  & 15 & 14 & 51.72\\ \hline
1st & 56 & -low -lpc -mink  & 14 & 15 & 48.28\\ \hline
1st & 57 & -low -lpc -diff  & 14 & 15 & 48.28\\ \hline
1st & 58 & -high -lpc -cheb  & 14 & 15 & 48.28\\ \hline
1st & 59 & -raw -lpc -nn  & 14 & 15 & 48.28\\ \hline
\end{tabular}
\end{minipage}
\caption{Consolidated results, Part 2.}
\label{tab:results2}
\end{table}

\begin{table}
\begin{minipage}[b]{\textwidth}
\centering
\begin{tabular}{|c|c|l|c|c|r|} \hline
Guess & Run \# & Configuration & GOOD & BAD & Recognition Rate,\%\\ \hline\hline
1st & 60 & -band -aggr -diff  & 13 & 16 & 44.83\\ \hline
1st & 61 & -norm -lpc -nn  & 13 & 16 & 44.83\\ \hline
1st & 62 & -band -fft -diff  & 13 & 16 & 44.83\\ \hline
1st & 63 & -high -lpc -eucl  & 12 & 17 & 41.38\\ \hline
1st & 64 & -high -aggr -diff  & 12 & 17 & 41.38\\ \hline
1st & 65 & -endp -fft -diff  & 12 & 17 & 41.38\\ \hline
1st & 66 & -endp -fft -eucl  & 12 & 17 & 41.38\\ \hline
1st & 67 & -band -lpc -mink  & 12 & 17 & 41.38\\ \hline
1st & 68 & -band -lpc -mah  & 12 & 17 & 41.38\\ \hline
1st & 69 & -band -lpc -eucl  & 12 & 17 & 41.38\\ \hline
1st & 70 & -endp -fft -cheb  & 12 & 17 & 41.38\\ \hline
1st & 71 & -band -lpc -cheb  & 12 & 17 & 41.38\\ \hline
1st & 72 & -endp -fft -mah  & 12 & 17 & 41.38\\ \hline
1st & 73 & -high -lpc -mah  & 12 & 17 & 41.38\\ \hline
1st & 74 & -endp -aggr -diff  & 12 & 17 & 41.38\\ \hline
1st & 75 & -endp -aggr -eucl  & 12 & 17 & 41.38\\ \hline
1st & 76 & -endp -aggr -mah  & 12 & 17 & 41.38\\ \hline
1st & 77 & -endp -aggr -cheb  & 12 & 17 & 41.38\\ \hline
1st & 78 & -high -lpc -diff  & 11 & 18 & 37.93\\ \hline
1st & 79 & -band -aggr -eucl  & 11 & 18 & 37.93\\ \hline
1st & 80 & -endp -fft -mink  & 11 & 18 & 37.93\\ \hline
1st & 81 & -band -aggr -cheb  & 11 & 18 & 37.93\\ \hline
1st & 82 & -band -lpc -diff  & 11 & 18 & 37.93\\ \hline
1st & 83 & -band -aggr -mah  & 11 & 18 & 37.93\\ \hline
1st & 84 & -band -fft -mah  & 11 & 18 & 37.93\\ \hline
1st & 85 & -band -fft -eucl  & 11 & 18 & 37.93\\ \hline
1st & 86 & -endp -aggr -mink  & 11 & 18 & 37.93\\ \hline
1st & 87 & -high -fft -diff  & 11 & 18 & 37.93\\ \hline
1st & 88 & -high -lpc -mink  & 10 & 19 & 34.48\\ \hline
1st & 89 & -raw -minmax -mink  & 10 & 19 & 34.48\\ \hline
\end{tabular}
\end{minipage}
\caption{Consolidated results, Part 3.}
\label{tab:results3}
\end{table}

\begin{table}
\begin{minipage}[b]{\textwidth}
\centering
\begin{tabular}{|c|c|l|c|c|r|} \hline
Guess & Run \# & Configuration & GOOD & BAD & Recognition Rate,\%\\ \hline\hline
1st & 90 & -raw -minmax -eucl  & 10 & 19 & 34.48\\ \hline
1st & 91 & -band -aggr -mink  & 10 & 19 & 34.48\\ \hline
1st & 92 & -raw -minmax -mah  & 10 & 19 & 34.48\\ \hline
1st & 93 & -endp -minmax -eucl  & 10 & 19 & 34.48\\ \hline
1st & 94 & -endp -minmax -cheb  & 10 & 19 & 34.48\\ \hline
1st & 95 & -endp -minmax -mah  & 10 & 19 & 34.48\\ \hline
1st & 96 & -band -fft -cheb  & 10 & 19 & 34.48\\ \hline
1st & 97 & -raw -minmax -cheb  & 9 & 20 & 31.03\\ \hline
1st & 98 & -endp -lpc -nn  & 9 & 20 & 31.03\\ \hline
1st & 99 & -endp -lpc -diff  & 9 & 20 & 31.03\\ \hline
1st & 100 & -endp -minmax -mink  & 8 & 21 & 27.59\\ \hline
1st & 101 & -endp -randfe -diff  & 7 & 22 & 24.14\\ \hline
1st & 102 & -endp -randfe -cheb  & 7 & 22 & 24.14\\ \hline
1st & 103 & -endp -minmax -diff  & 7 & 22 & 24.14\\ \hline
1st & 104 & -endp -minmax -nn  & 7 & 22 & 24.14\\ \hline
1st & 105 & -band -fft -mink  & 7 & 22 & 24.14\\ \hline
1st & 106 & -norm -randfe -eucl  & 6 & 23 & 20.69\\ \hline
1st & 107 & -raw -minmax -diff  & 6 & 23 & 20.69\\ \hline
1st & 108 & -endp -randfe -eucl  & 6 & 23 & 20.69\\ \hline
1st & 109 & -endp -randfe -mah  & 6 & 23 & 20.69\\ \hline
1st & 110 & -low -randfe -mink  & 6 & 23 & 20.69\\ \hline
1st & 111 & -norm -randfe -mah  & 6 & 23 & 20.69\\ \hline
1st & 112 & -norm -minmax -eucl  & 6 & 23 & 20.69\\ \hline
1st & 113 & -norm -minmax -cheb  & 6 & 23 & 20.69\\ \hline
1st & 114 & -low -minmax -mink  & 6 & 23 & 20.69\\ \hline
1st & 115 & -norm -minmax -mah  & 6 & 23 & 20.69\\ \hline
1st & 116 & -norm -randfe -mink  & 6 & 23 & 20.69\\ \hline
1st & 117 & -endp -randfe -mink  & 5 & 24 & 17.24\\ \hline
1st & 118 & -low -minmax -mah  & 5 & 24 & 17.24\\ \hline
1st & 119 & -low -randfe -eucl  & 5 & 24 & 17.24\\ \hline
\end{tabular}
\end{minipage}
\caption{Consolidated results, Part 4.}
\label{tab:results4}
\end{table}

\begin{table}
\begin{minipage}[b]{\textwidth}
\centering
\begin{tabular}{|c|c|l|c|c|r|} \hline
Guess & Run \# & Configuration & GOOD & BAD & Recognition Rate,\%\\ \hline\hline
1st & 120 & -high -minmax -mah  & 5 & 24 & 17.24\\ \hline
1st & 121 & -raw -randfe -mink  & 5 & 24 & 17.24\\ \hline
1st & 122 & -low -randfe -mah  & 5 & 24 & 17.24\\ \hline
1st & 123 & -low -minmax -diff  & 5 & 24 & 17.24\\ \hline
1st & 124 & -high -minmax -diff  & 5 & 24 & 17.24\\ \hline
1st & 125 & -high -minmax -eucl  & 5 & 24 & 17.24\\ \hline
1st & 126 & -low -minmax -eucl  & 5 & 24 & 17.24\\ \hline
1st & 127 & -low -minmax -cheb  & 5 & 24 & 17.24\\ \hline
1st & 128 & -norm -randfe -diff  & 4 & 25 & 13.79\\ \hline
1st & 129 & -norm -randfe -cheb  & 4 & 25 & 13.79\\ \hline
1st & 130 & -high -randfe -mink  & 4 & 25 & 13.79\\ \hline
1st & 131 & -low -randfe -diff  & 4 & 25 & 13.79\\ \hline
1st & 132 & -high -randfe -diff  & 4 & 25 & 13.79\\ \hline
1st & 133 & -high -randfe -eucl  & 4 & 25 & 13.79\\ \hline
1st & 134 & -high -randfe -cheb  & 4 & 25 & 13.79\\ \hline
1st & 135 & -low -randfe -cheb  & 4 & 25 & 13.79\\ \hline
1st & 136 & -norm -minmax -mink  & 4 & 25 & 13.79\\ \hline
1st & 137 & -raw -randfe -eucl  & 4 & 25 & 13.79\\ \hline
1st & 138 & -band -lpc -nn  & 4 & 25 & 13.79\\ \hline
1st & 139 & -raw -randfe -mah  & 4 & 25 & 13.79\\ \hline
1st & 140 & -high -minmax -mink  & 4 & 25 & 13.79\\ \hline
1st & 141 & -raw -minmax -nn  & 4 & 25 & 13.79\\ \hline
1st & 142 & -high -randfe -mah  & 4 & 25 & 13.79\\ \hline
1st & 143 & -high -minmax -cheb  & 4 & 25 & 13.79\\ \hline
1st & 144 & -high -minmax -nn  & 4 & 25 & 13.79\\ \hline
1st & 145 & -endp -randfe -randcl  & 4 & 25 & 13.79\\ \hline
1st & 146 & -band -minmax -mink  & 3 & 26 & 10.34\\ \hline
1st & 147 & -band -minmax -diff  & 3 & 26 & 10.34\\ \hline
1st & 148 & -band -minmax -eucl  & 3 & 26 & 10.34\\ \hline
1st & 149 & -band -minmax -mah  & 3 & 26 & 10.34\\ \hline
\end{tabular}
\end{minipage}
\caption{Consolidated results, Part 5.}
\label{tab:results5}
\end{table}

\begin{table}
\begin{minipage}[b]{\textwidth}
\centering
\begin{tabular}{|c|c|l|c|c|r|} \hline
Guess & Run \# & Configuration & GOOD & BAD & Recognition Rate,\%\\ \hline\hline
1st & 150 & -raw -minmax -randcl  & 3 & 26 & 10.34\\ \hline
1st & 151 & -band -minmax -cheb  & 3 & 26 & 10.34\\ \hline
1st & 152 & -low -lpc -nn  & 3 & 26 & 10.34\\ \hline
1st & 153 & -raw -randfe -diff  & 3 & 26 & 10.34\\ \hline
1st & 154 & -norm -minmax -diff  & 3 & 26 & 10.34\\ \hline
1st & 155 & -boost -lpc -randcl  & 3 & 26 & 10.34\\ \hline
1st & 156 & -raw -randfe -cheb  & 3 & 26 & 10.34\\ \hline
1st & 157 & -boost -minmax -nn  & 3 & 26 & 10.34\\ \hline
1st & 158 & -highpassboost -lpc -nn  & 3 & 26 & 10.34\\ \hline
1st & 159 & -norm -minmax -nn  & 3 & 26 & 10.34\\ \hline
1st & 160 & -highpassboost -minmax -nn  & 3 & 26 & 10.34\\ \hline
1st & 161 & -boost -minmax -randcl  & 3 & 26 & 10.34\\ \hline
1st & 162 & -boost -lpc -nn  & 3 & 26 & 10.34\\ \hline
1st & 163 & -raw -aggr -randcl  & 2 & 27 & 6.90\\ \hline
1st & 164 & -band -randfe -mah  & 2 & 27 & 6.90\\ \hline
1st & 165 & -highpassboost -lpc -randcl  & 2 & 27 & 6.90\\ \hline
1st & 166 & -band -randfe -diff  & 2 & 27 & 6.90\\ \hline
1st & 167 & -band -randfe -eucl  & 2 & 27 & 6.90\\ \hline
1st & 168 & -low -minmax -nn  & 2 & 27 & 6.90\\ \hline
1st & 169 & -boost -randfe -randcl  & 2 & 27 & 6.90\\ \hline
1st & 170 & -band -randfe -cheb  & 2 & 27 & 6.90\\ \hline
1st & 171 & -raw -lpc -randcl  & 2 & 27 & 6.90\\ \hline
1st & 172 & -highpassboost -aggr -randcl  & 2 & 27 & 6.90\\ \hline
1st & 173 & -boost -fft -randcl  & 2 & 27 & 6.90\\ \hline
1st & 174 & -highpassboost -minmax -diff  & 1 & 28 & 3.45\\ \hline
1st & 175 & -boost -randfe -eucl  & 1 & 28 & 3.45\\ \hline
1st & 176 & -highpassboost -minmax -eucl  & 1 & 28 & 3.45\\ \hline
1st & 177 & -boost -lpc -mink  & 1 & 28 & 3.45\\ \hline
1st & 178 & -boost -lpc -diff  & 1 & 28 & 3.45\\ \hline
1st & 179 & -boost -fft -mah  & 1 & 28 & 3.45\\ \hline
\end{tabular}
\end{minipage}
\caption{Consolidated results, Part 6.}
\label{tab:results6}
\end{table}

\begin{table}
\begin{minipage}[b]{\textwidth}
\centering
\begin{tabular}{|c|c|l|c|c|r|} \hline
Guess & Run \# & Configuration & GOOD & BAD & Recognition Rate,\%\\ \hline\hline
1st & 180 & -boost -lpc -eucl  & 1 & 28 & 3.45\\ \hline
1st & 181 & -low -fft -randcl  & 1 & 28 & 3.45\\ \hline
1st & 182 & -low -minmax -randcl  & 1 & 28 & 3.45\\ \hline
1st & 183 & -boost -minmax -mah  & 1 & 28 & 3.45\\ \hline
1st & 184 & -highpassboost -minmax -mah  & 1 & 28 & 3.45\\ \hline
1st & 185 & -boost -randfe -cheb  & 1 & 28 & 3.45\\ \hline
1st & 186 & -high -randfe -randcl  & 1 & 28 & 3.45\\ \hline
1st & 187 & -highpassboost -minmax -cheb  & 1 & 28 & 3.45\\ \hline
1st & 188 & -highpassboost -fft -mah  & 1 & 28 & 3.45\\ \hline
1st & 189 & -boost -lpc -cheb  & 1 & 28 & 3.45\\ \hline
1st & 190 & -boost -aggr -mah  & 1 & 28 & 3.45\\ \hline
1st & 191 & -highpassboost -lpc -mink  & 1 & 28 & 3.45\\ \hline
1st & 192 & -highpassboost -lpc -diff  & 1 & 28 & 3.45\\ \hline
1st & 193 & -endp -lpc -randcl  & 1 & 28 & 3.45\\ \hline
1st & 194 & -highpassboost -lpc -eucl  & 1 & 28 & 3.45\\ \hline
1st & 195 & -high -minmax -randcl  & 1 & 28 & 3.45\\ \hline
1st & 196 & -highpassboost -lpc -cheb  & 1 & 28 & 3.45\\ \hline
1st & 197 & -norm -fft -randcl  & 1 & 28 & 3.45\\ \hline
1st & 198 & -band -aggr -randcl  & 1 & 28 & 3.45\\ \hline
1st & 199 & -low -randfe -randcl  & 1 & 28 & 3.45\\ \hline
1st & 200 & -boost -aggr -mink  & 1 & 28 & 3.45\\ \hline
1st & 201 & -boost -aggr -diff  & 1 & 28 & 3.45\\ \hline
1st & 202 & -endp -aggr -randcl  & 1 & 28 & 3.45\\ \hline
1st & 203 & -boost -aggr -eucl  & 1 & 28 & 3.45\\ \hline
1st & 204 & -boost -fft -mink  & 1 & 28 & 3.45\\ \hline
1st & 205 & -boost -randfe -mah  & 1 & 28 & 3.45\\ \hline
1st & 206 & -boost -fft -diff  & 1 & 28 & 3.45\\ \hline
1st & 207 & -boost -fft -eucl  & 1 & 28 & 3.45\\ \hline
1st & 208 & -highpassboost -randfe -mink  & 1 & 28 & 3.45\\ \hline
1st & 209 & -highpassboost -randfe -diff  & 1 & 28 & 3.45\\ \hline
\end{tabular}
\end{minipage}
\caption{Consolidated results, Part 7.}
\label{tab:results7}
\end{table}

\begin{table}
\begin{minipage}[b]{\textwidth}
\centering
\begin{tabular}{|c|c|l|c|c|r|} \hline
Guess & Run \# & Configuration & GOOD & BAD & Recognition Rate,\%\\ \hline\hline
1st & 210 & -boost -minmax -mink  & 1 & 28 & 3.45\\ \hline
1st & 211 & -boost -minmax -diff  & 1 & 28 & 3.45\\ \hline
1st & 212 & -highpassboost -randfe -eucl  & 1 & 28 & 3.45\\ \hline
1st & 213 & -boost -minmax -eucl  & 1 & 28 & 3.45\\ \hline
1st & 214 & -low -aggr -randcl  & 1 & 28 & 3.45\\ \hline
1st & 215 & -band -fft -randcl  & 1 & 28 & 3.45\\ \hline
1st & 216 & -boost -aggr -cheb  & 1 & 28 & 3.45\\ \hline
1st & 217 & -band -randfe -randcl  & 1 & 28 & 3.45\\ \hline
1st & 218 & -boost -fft -cheb  & 1 & 28 & 3.45\\ \hline
1st & 219 & -highpassboost -aggr -mink  & 1 & 28 & 3.45\\ \hline
1st & 220 & -highpassboost -aggr -diff  & 1 & 28 & 3.45\\ \hline
1st & 221 & -highpassboost -fft -mink  & 1 & 28 & 3.45\\ \hline
1st & 222 & -endp -minmax -randcl  & 1 & 28 & 3.45\\ \hline
1st & 223 & -highpassboost -fft -diff  & 1 & 28 & 3.45\\ \hline
1st & 224 & -highpassboost -aggr -eucl  & 1 & 28 & 3.45\\ \hline
1st & 225 & -highpassboost -randfe -cheb  & 1 & 28 & 3.45\\ \hline
1st & 226 & -high -lpc -nn  & 1 & 28 & 3.45\\ \hline
1st & 227 & -boost -minmax -cheb  & 1 & 28 & 3.45\\ \hline
1st & 228 & -highpassboost -fft -eucl  & 1 & 28 & 3.45\\ \hline
1st & 229 & -boost -lpc -mah  & 1 & 28 & 3.45\\ \hline
1st & 230 & -norm -randfe -randcl  & 1 & 28 & 3.45\\ \hline
1st & 231 & -highpassboost -aggr -cheb  & 1 & 28 & 3.45\\ \hline
1st & 232 & -highpassboost -fft -cheb  & 1 & 28 & 3.45\\ \hline
1st & 233 & -band -minmax -randcl  & 1 & 28 & 3.45\\ \hline
1st & 234 & -boost -aggr -randcl  & 1 & 28 & 3.45\\ \hline
1st & 235 & -highpassboost -lpc -mah  & 1 & 28 & 3.45\\ \hline
1st & 236 & -highpassboost -aggr -mah  & 1 & 28 & 3.45\\ \hline
1st & 237 & -high -lpc -randcl  & 1 & 28 & 3.45\\ \hline
1st & 238 & -highpassboost -randfe -mah  & 1 & 28 & 3.45\\ \hline
1st & 239 & -boost -randfe -mink  & 1 & 28 & 3.45\\ \hline
\end{tabular}
\end{minipage}
\caption{Consolidated results, Part 8.}
\label{tab:results8}
\end{table}

\begin{table}
\begin{minipage}[b]{\textwidth}
\centering
\begin{tabular}{|c|c|l|c|c|r|} \hline
Guess & Run \# & Configuration & GOOD & BAD & Recognition Rate,\%\\ \hline\hline
1st & 240 & -boost -randfe -diff  & 1 & 28 & 3.45\\ \hline
1st & 241 & -highpassboost -minmax -mink  & 1 & 28 & 3.45\\ \hline
1st & 242 & -raw -randfe -randcl  & 0 & 29 & 0.00\\ \hline
1st & 243 & -highpassboost -fft -randcl  & 0 & 29 & 0.00\\ \hline
1st & 244 & -band -lpc -randcl  & 0 & 29 & 0.00\\ \hline
1st & 245 & -endp -fft -randcl  & 0 & 29 & 0.00\\ \hline
1st & 246 & -raw -fft -randcl  & 0 & 29 & 0.00\\ \hline
1st & 247 & -norm -lpc -randcl  & 0 & 29 & 0.00\\ \hline
1st & 248 & -highpassboost -randfe -randcl  & 0 & 29 & 0.00\\ \hline
1st & 249 & -high -aggr -randcl  & 0 & 29 & 0.00\\ \hline
1st & 250 & -band -randfe -mink  & 0 & 29 & 0.00\\ \hline
1st & 251 & -low -lpc -randcl  & 0 & 29 & 0.00\\ \hline
1st & 252 & -highpassboost -minmax -randcl  & 0 & 29 & 0.00\\ \hline
1st & 253 & -norm -aggr -randcl  & 0 & 29 & 0.00\\ \hline
1st & 254 & -high -fft -randcl  & 0 & 29 & 0.00\\ \hline
1st & 255 & -band -minmax -nn  & 0 & 29 & 0.00\\ \hline
1st & 256 & -norm -minmax -randcl  & 0 & 29 & 0.00\\ \hline
2nd & 1 & -endp -lpc -cheb  & 24 & 5 & 82.76\\ \hline
2nd & 2 & -raw -aggr -eucl  & 24 & 5 & 82.76\\ \hline
2nd & 3 & -norm -aggr -diff  & 24 & 5 & 82.76\\ \hline
2nd & 4 & -norm -aggr -cheb  & 24 & 5 & 82.76\\ \hline
2nd & 5 & -raw -aggr -mah  & 24 & 5 & 82.76\\ \hline
2nd & 6 & -raw -fft -mah  & 24 & 5 & 82.76\\ \hline
2nd & 7 & -raw -fft -eucl  & 24 & 5 & 82.76\\ \hline
2nd & 8 & -norm -fft -diff  & 24 & 5 & 82.76\\ \hline
2nd & 9 & -norm -fft -cheb  & 24 & 5 & 82.76\\ \hline
2nd & 10 & -raw -aggr -cheb  & 22 & 7 & 75.86\\ \hline
2nd & 11 & -endp -lpc -mah  & 22 & 7 & 75.86\\ \hline
2nd & 12 & -endp -lpc -eucl  & 22 & 7 & 75.86\\ \hline
2nd & 13 & -raw -fft -mink  & 25 & 4 & 86.21\\ \hline
\end{tabular}
\end{minipage}
\caption{Consolidated results, Part 9.}
\label{tab:results9}
\end{table}

\begin{table}
\begin{minipage}[b]{\textwidth}
\centering
\begin{tabular}{|c|c|l|c|c|r|} \hline
Guess & Run \# & Configuration & GOOD & BAD & Recognition Rate,\%\\ \hline\hline
2nd & 14 & -norm -fft -mah  & 24 & 5 & 82.76\\ \hline
2nd & 15 & -norm -fft -eucl  & 24 & 5 & 82.76\\ \hline
2nd & 16 & -norm -aggr -eucl  & 24 & 5 & 82.76\\ \hline
2nd & 17 & -low -aggr -diff  & 21 & 8 & 72.41\\ \hline
2nd & 18 & -norm -aggr -mah  & 24 & 5 & 82.76\\ \hline
2nd & 19 & -raw -fft -cheb  & 22 & 7 & 75.86\\ \hline
2nd & 20 & -raw -aggr -mink  & 25 & 4 & 86.21\\ \hline
2nd & 21 & -norm -aggr -mink  & 24 & 5 & 82.76\\ \hline
2nd & 22 & -low -fft -cheb  & 22 & 7 & 75.86\\ \hline
2nd & 23 & -raw -lpc -mink  & 23 & 6 & 79.31\\ \hline
2nd & 24 & -raw -lpc -diff  & 23 & 6 & 79.31\\ \hline
2nd & 25 & -raw -lpc -eucl  & 23 & 6 & 79.31\\ \hline
2nd & 26 & -raw -lpc -mah  & 23 & 6 & 79.31\\ \hline
2nd & 27 & -raw -lpc -cheb  & 23 & 6 & 79.31\\ \hline
2nd & 28 & -low -aggr -eucl  & 23 & 6 & 79.31\\ \hline
2nd & 29 & -norm -lpc -mah  & 23 & 6 & 79.31\\ \hline
2nd & 30 & -norm -lpc -mink  & 23 & 6 & 79.31\\ \hline
2nd & 31 & -norm -lpc -diff  & 23 & 6 & 79.31\\ \hline
2nd & 32 & -norm -lpc -eucl  & 23 & 6 & 79.31\\ \hline
2nd & 33 & -low -aggr -cheb  & 22 & 7 & 75.86\\ \hline
2nd & 34 & -norm -lpc -cheb  & 23 & 6 & 79.31\\ \hline
2nd & 35 & -low -aggr -mah  & 23 & 6 & 79.31\\ \hline
2nd & 36 & -low -fft -mah  & 23 & 6 & 79.31\\ \hline
2nd & 37 & -norm -fft -mink  & 24 & 5 & 82.76\\ \hline
2nd & 38 & -low -fft -diff  & 21 & 8 & 72.41\\ \hline
2nd & 39 & -low -fft -eucl  & 23 & 6 & 79.31\\ \hline
2nd & 40 & -raw -aggr -diff  & 22 & 7 & 75.86\\ \hline
2nd & 41 & -high -aggr -mink  & 21 & 8 & 72.41\\ \hline
2nd & 42 & -high -aggr -eucl  & 21 & 8 & 72.41\\ \hline
2nd & 43 & -endp -lpc -mink  & 20 & 9 & 68.97\\ \hline
\end{tabular}
\end{minipage}
\caption{Consolidated results, Part 10.}
\label{tab:results10}
\end{table}

\begin{table}
\begin{minipage}[b]{\textwidth}
\centering
\begin{tabular}{|c|c|l|c|c|r|} \hline
Guess & Run \# & Configuration & GOOD & BAD & Recognition Rate,\%\\ \hline\hline
2nd & 44 & -high -aggr -mah  & 21 & 8 & 72.41\\ \hline
2nd & 45 & -raw -fft -diff  & 22 & 7 & 75.86\\ \hline
2nd & 46 & -high -aggr -cheb  & 20 & 9 & 68.97\\ \hline
2nd & 47 & -low -aggr -mink  & 24 & 5 & 82.76\\ \hline
2nd & 48 & -low -fft -mink  & 24 & 5 & 82.76\\ \hline
2nd & 49 & -high -fft -cheb  & 20 & 9 & 68.97\\ \hline
2nd & 50 & -high -fft -mah  & 21 & 8 & 72.41\\ \hline
2nd & 51 & -low -lpc -cheb  & 22 & 7 & 75.86\\ \hline
2nd & 52 & -high -fft -mink  & 19 & 10 & 65.52\\ \hline
2nd & 53 & -high -fft -eucl  & 21 & 8 & 72.41\\ \hline
2nd & 54 & -low -lpc -eucl  & 20 & 9 & 68.97\\ \hline
2nd & 55 & -low -lpc -mah  & 20 & 9 & 68.97\\ \hline
2nd & 56 & -low -lpc -mink  & 18 & 11 & 62.07\\ \hline
2nd & 57 & -low -lpc -diff  & 21 & 8 & 72.41\\ \hline
2nd & 58 & -high -lpc -cheb  & 19 & 10 & 65.52\\ \hline
2nd & 59 & -raw -lpc -nn  & 14 & 15 & 48.28\\ \hline
2nd & 60 & -band -aggr -diff  & 14 & 15 & 48.28\\ \hline
2nd & 61 & -norm -lpc -nn  & 15 & 14 & 51.72\\ \hline
2nd & 62 & -band -fft -diff  & 14 & 15 & 48.28\\ \hline
2nd & 63 & -high -lpc -eucl  & 17 & 12 & 58.62\\ \hline
2nd & 64 & -high -aggr -diff  & 18 & 11 & 62.07\\ \hline
2nd & 65 & -endp -fft -diff  & 17 & 12 & 58.62\\ \hline
2nd & 66 & -endp -fft -eucl  & 16 & 13 & 55.17\\ \hline
2nd & 67 & -band -lpc -mink  & 16 & 13 & 55.17\\ \hline
2nd & 68 & -band -lpc -mah  & 16 & 13 & 55.17\\ \hline
2nd & 69 & -band -lpc -eucl  & 16 & 13 & 55.17\\ \hline
2nd & 70 & -endp -fft -cheb  & 17 & 12 & 58.62\\ \hline
2nd & 71 & -band -lpc -cheb  & 17 & 12 & 58.62\\ \hline
2nd & 72 & -endp -fft -mah  & 16 & 13 & 55.17\\ \hline
2nd & 73 & -high -lpc -mah  & 17 & 12 & 58.62\\ \hline
\end{tabular}
\end{minipage}
\caption{Consolidated results, Part 11.}
\label{tab:results11}
\end{table}

\begin{table}
\begin{minipage}[b]{\textwidth}
\centering
\begin{tabular}{|c|c|l|c|c|r|} \hline
Guess & Run \# & Configuration & GOOD & BAD & Recognition Rate,\%\\ \hline\hline
2nd & 74 & -endp -aggr -diff  & 18 & 11 & 62.07\\ \hline
2nd & 75 & -endp -aggr -eucl  & 17 & 12 & 58.62\\ \hline
2nd & 76 & -endp -aggr -mah  & 17 & 12 & 58.62\\ \hline
2nd & 77 & -endp -aggr -cheb  & 18 & 11 & 62.07\\ \hline
2nd & 78 & -high -lpc -diff  & 18 & 11 & 62.07\\ \hline
2nd & 79 & -band -aggr -eucl  & 15 & 14 & 51.72\\ \hline
2nd & 80 & -endp -fft -mink  & 14 & 15 & 48.28\\ \hline
2nd & 81 & -band -aggr -cheb  & 14 & 15 & 48.28\\ \hline
2nd & 82 & -band -lpc -diff  & 17 & 12 & 58.62\\ \hline
2nd & 83 & -band -aggr -mah  & 15 & 14 & 51.72\\ \hline
2nd & 84 & -band -fft -mah  & 15 & 14 & 51.72\\ \hline
2nd & 85 & -band -fft -eucl  & 15 & 14 & 51.72\\ \hline
2nd & 86 & -endp -aggr -mink  & 14 & 15 & 48.28\\ \hline
2nd & 87 & -high -fft -diff  & 18 & 11 & 62.07\\ \hline
2nd & 88 & -high -lpc -mink  & 14 & 15 & 48.28\\ \hline
2nd & 89 & -raw -minmax -mink  & 12 & 17 & 41.38\\ \hline
2nd & 90 & -raw -minmax -eucl  & 12 & 17 & 41.38\\ \hline
2nd & 91 & -band -aggr -mink  & 13 & 16 & 44.83\\ \hline
2nd & 92 & -raw -minmax -mah  & 12 & 17 & 41.38\\ \hline
2nd & 93 & -endp -minmax -eucl  & 12 & 17 & 41.38\\ \hline
2nd & 94 & -endp -minmax -cheb  & 12 & 17 & 41.38\\ \hline
2nd & 95 & -endp -minmax -mah  & 12 & 17 & 41.38\\ \hline
2nd & 96 & -band -fft -cheb  & 14 & 15 & 48.28\\ \hline
2nd & 97 & -raw -minmax -cheb  & 11 & 18 & 37.93\\ \hline
2nd & 98 & -endp -lpc -nn  & 12 & 17 & 41.38\\ \hline
2nd & 99 & -endp -lpc -diff  & 19 & 10 & 65.52\\ \hline
2nd & 100 & -endp -minmax -mink  & 12 & 17 & 41.38\\ \hline
2nd & 101 & -endp -randfe -diff  & 8 & 21 & 27.59\\ \hline
2nd & 102 & -endp -randfe -cheb  & 8 & 21 & 27.59\\ \hline
2nd & 103 & -endp -minmax -diff  & 12 & 17 & 41.38\\ \hline
\end{tabular}
\end{minipage}
\caption{Consolidated results, Part 12.}
\label{tab:results12}
\end{table}

\begin{table}
\begin{minipage}[b]{\textwidth}
\centering
\begin{tabular}{|c|c|l|c|c|r|} \hline
Guess & Run \# & Configuration & GOOD & BAD & Recognition Rate,\%\\ \hline\hline
2nd & 104 & -endp -minmax -nn  & 7 & 22 & 24.14\\ \hline
2nd & 105 & -band -fft -mink  & 13 & 16 & 44.83\\ \hline
2nd & 106 & -norm -randfe -eucl  & 10 & 19 & 34.48\\ \hline
2nd & 107 & -raw -minmax -diff  & 10 & 19 & 34.48\\ \hline
2nd & 108 & -endp -randfe -eucl  & 9 & 20 & 31.03\\ \hline
2nd & 109 & -endp -randfe -mah  & 9 & 20 & 31.03\\ \hline
2nd & 110 & -low -randfe -mink  & 9 & 20 & 31.03\\ \hline
2nd & 111 & -norm -randfe -mah  & 10 & 19 & 34.48\\ \hline
2nd & 112 & -norm -minmax -eucl  & 9 & 20 & 31.03\\ \hline
2nd & 113 & -norm -minmax -cheb  & 10 & 19 & 34.48\\ \hline
2nd & 114 & -low -minmax -mink  & 8 & 21 & 27.59\\ \hline
2nd & 115 & -norm -minmax -mah  & 9 & 20 & 31.03\\ \hline
2nd & 116 & -norm -randfe -mink  & 10 & 19 & 34.48\\ \hline
2nd & 117 & -endp -randfe -mink  & 8 & 21 & 27.59\\ \hline
2nd & 118 & -low -minmax -mah  & 6 & 23 & 20.69\\ \hline
2nd & 119 & -low -randfe -eucl  & 9 & 20 & 31.03\\ \hline
2nd & 120 & -high -minmax -mah  & 6 & 23 & 20.69\\ \hline
2nd & 121 & -raw -randfe -mink  & 9 & 20 & 31.03\\ \hline
2nd & 122 & -low -randfe -mah  & 9 & 20 & 31.03\\ \hline
2nd & 123 & -low -minmax -diff  & 6 & 23 & 20.69\\ \hline
2nd & 124 & -high -minmax -diff  & 6 & 23 & 20.69\\ \hline
2nd & 125 & -high -minmax -eucl  & 6 & 23 & 20.69\\ \hline
2nd & 126 & -low -minmax -eucl  & 6 & 23 & 20.69\\ \hline
2nd & 127 & -low -minmax -cheb  & 6 & 23 & 20.69\\ \hline
2nd & 128 & -norm -randfe -diff  & 10 & 19 & 34.48\\ \hline
2nd & 129 & -norm -randfe -cheb  & 10 & 19 & 34.48\\ \hline
2nd & 130 & -high -randfe -mink  & 5 & 24 & 17.24\\ \hline
2nd & 131 & -low -randfe -diff  & 9 & 20 & 31.03\\ \hline
2nd & 132 & -high -randfe -diff  & 7 & 22 & 24.14\\ \hline
2nd & 133 & -high -randfe -eucl  & 6 & 23 & 20.69\\ \hline
\end{tabular}
\end{minipage}
\caption{Consolidated results, Part 13.}
\label{tab:results13}
\end{table}

\begin{table}
\begin{minipage}[b]{\textwidth}
\centering
\begin{tabular}{|c|c|l|c|c|r|} \hline
Guess & Run \# & Configuration & GOOD & BAD & Recognition Rate,\%\\ \hline\hline
2nd & 134 & -high -randfe -cheb  & 7 & 22 & 24.14\\ \hline
2nd & 135 & -low -randfe -cheb  & 9 & 20 & 31.03\\ \hline
2nd & 136 & -norm -minmax -mink  & 10 & 19 & 34.48\\ \hline
2nd & 137 & -raw -randfe -eucl  & 8 & 21 & 27.59\\ \hline
2nd & 138 & -band -lpc -nn  & 4 & 25 & 13.79\\ \hline
2nd & 139 & -raw -randfe -mah  & 8 & 21 & 27.59\\ \hline
2nd & 140 & -high -minmax -mink  & 6 & 23 & 20.69\\ \hline
2nd & 141 & -raw -minmax -nn  & 4 & 25 & 13.79\\ \hline
2nd & 142 & -high -randfe -mah  & 6 & 23 & 20.69\\ \hline
2nd & 143 & -high -minmax -cheb  & 5 & 24 & 17.24\\ \hline
2nd & 144 & -high -minmax -nn  & 4 & 25 & 13.79\\ \hline
2nd & 145 & -endp -randfe -randcl  & 6 & 23 & 20.69\\ \hline
2nd & 146 & -band -minmax -mink  & 5 & 24 & 17.24\\ \hline
2nd & 147 & -band -minmax -diff  & 5 & 24 & 17.24\\ \hline
2nd & 148 & -band -minmax -eucl  & 5 & 24 & 17.24\\ \hline
2nd & 149 & -band -minmax -mah  & 5 & 24 & 17.24\\ \hline
2nd & 150 & -raw -minmax -randcl  & 3 & 26 & 10.34\\ \hline
2nd & 151 & -band -minmax -cheb  & 4 & 25 & 13.79\\ \hline
2nd & 152 & -low -lpc -nn  & 4 & 25 & 13.79\\ \hline
2nd & 153 & -raw -randfe -diff  & 6 & 23 & 20.69\\ \hline
2nd & 154 & -norm -minmax -diff  & 9 & 20 & 31.03\\ \hline
2nd & 155 & -boost -lpc -randcl  & 3 & 26 & 10.34\\ \hline
2nd & 156 & -raw -randfe -cheb  & 6 & 23 & 20.69\\ \hline
2nd & 157 & -boost -minmax -nn  & 4 & 25 & 13.79\\ \hline
2nd & 158 & -highpassboost -lpc -nn  & 4 & 25 & 13.79\\ \hline
2nd & 159 & -norm -minmax -nn  & 3 & 26 & 10.34\\ \hline
2nd & 160 & -highpassboost -minmax -nn  & 4 & 25 & 13.79\\ \hline
2nd & 161 & -boost -minmax -randcl  & 5 & 24 & 17.24\\ \hline
2nd & 162 & -boost -lpc -nn  & 4 & 25 & 13.79\\ \hline
2nd & 163 & -raw -aggr -randcl  & 4 & 25 & 13.79\\ \hline
\end{tabular}
\end{minipage}
\caption{Consolidated results, Part 14.}
\label{tab:results14}
\end{table}

\begin{table}
\begin{minipage}[b]{\textwidth}
\centering
\begin{tabular}{|c|c|l|c|c|r|} \hline
Guess & Run \# & Configuration & GOOD & BAD & Recognition Rate,\%\\ \hline\hline
2nd & 164 & -band -randfe -mah  & 5 & 24 & 17.24\\ \hline
2nd & 165 & -highpassboost -lpc -randcl  & 2 & 27 & 6.90\\ \hline
2nd & 166 & -band -randfe -diff  & 4 & 25 & 13.79\\ \hline
2nd & 167 & -band -randfe -eucl  & 5 & 24 & 17.24\\ \hline
2nd & 168 & -low -minmax -nn  & 3 & 26 & 10.34\\ \hline
2nd & 169 & -boost -randfe -randcl  & 3 & 26 & 10.34\\ \hline
2nd & 170 & -band -randfe -cheb  & 4 & 25 & 13.79\\ \hline
2nd & 171 & -raw -lpc -randcl  & 3 & 26 & 10.34\\ \hline
2nd & 172 & -highpassboost -aggr -randcl  & 2 & 27 & 6.90\\ \hline
2nd & 173 & -boost -fft -randcl  & 3 & 26 & 10.34\\ \hline
2nd & 174 & -highpassboost -minmax -diff  & 2 & 27 & 6.90\\ \hline
2nd & 175 & -boost -randfe -eucl  & 2 & 27 & 6.90\\ \hline
2nd & 176 & -highpassboost -minmax -eucl  & 2 & 27 & 6.90\\ \hline
2nd & 177 & -boost -lpc -mink  & 2 & 27 & 6.90\\ \hline
2nd & 178 & -boost -lpc -diff  & 2 & 27 & 6.90\\ \hline
2nd & 179 & -boost -fft -mah  & 2 & 27 & 6.90\\ \hline
2nd & 180 & -boost -lpc -eucl  & 2 & 27 & 6.90\\ \hline
2nd & 181 & -low -fft -randcl  & 3 & 26 & 10.34\\ \hline
2nd & 182 & -low -minmax -randcl  & 2 & 27 & 6.90\\ \hline
2nd & 183 & -boost -minmax -mah  & 2 & 27 & 6.90\\ \hline
2nd & 184 & -highpassboost -minmax -mah  & 2 & 27 & 6.90\\ \hline
2nd & 185 & -boost -randfe -cheb  & 2 & 27 & 6.90\\ \hline
2nd & 186 & -high -randfe -randcl  & 2 & 27 & 6.90\\ \hline
2nd & 187 & -highpassboost -minmax -cheb  & 2 & 27 & 6.90\\ \hline
2nd & 188 & -highpassboost -fft -mah  & 2 & 27 & 6.90\\ \hline
2nd & 189 & -boost -lpc -cheb  & 2 & 27 & 6.90\\ \hline
2nd & 190 & -boost -aggr -mah  & 2 & 27 & 6.90\\ \hline
2nd & 191 & -highpassboost -lpc -mink  & 2 & 27 & 6.90\\ \hline
2nd & 192 & -highpassboost -lpc -diff  & 2 & 27 & 6.90\\ \hline
2nd & 193 & -endp -lpc -randcl  & 1 & 28 & 3.45\\ \hline
\end{tabular}
\end{minipage}
\caption{Consolidated results, Part 15.}
\label{tab:results15}
\end{table}

\begin{table}
\begin{minipage}[b]{\textwidth}
\centering
\begin{tabular}{|c|c|l|c|c|r|} \hline
Guess & Run \# & Configuration & GOOD & BAD & Recognition Rate,\%\\ \hline\hline
2nd & 194 & -highpassboost -lpc -eucl  & 2 & 27 & 6.90\\ \hline
2nd & 195 & -high -minmax -randcl  & 1 & 28 & 3.45\\ \hline
2nd & 196 & -highpassboost -lpc -cheb  & 2 & 27 & 6.90\\ \hline
2nd & 197 & -norm -fft -randcl  & 3 & 26 & 10.34\\ \hline
2nd & 198 & -band -aggr -randcl  & 1 & 28 & 3.45\\ \hline
2nd & 199 & -low -randfe -randcl  & 1 & 28 & 3.45\\ \hline
2nd & 200 & -boost -aggr -mink  & 2 & 27 & 6.90\\ \hline
2nd & 201 & -boost -aggr -diff  & 2 & 27 & 6.90\\ \hline
2nd & 202 & -endp -aggr -randcl  & 3 & 26 & 10.34\\ \hline
2nd & 203 & -boost -aggr -eucl  & 2 & 27 & 6.90\\ \hline
2nd & 204 & -boost -fft -mink  & 2 & 27 & 6.90\\ \hline
2nd & 205 & -boost -randfe -mah  & 2 & 27 & 6.90\\ \hline
2nd & 206 & -boost -fft -diff  & 2 & 27 & 6.90\\ \hline
2nd & 207 & -boost -fft -eucl  & 2 & 27 & 6.90\\ \hline
2nd & 208 & -highpassboost -randfe -mink  & 2 & 27 & 6.90\\ \hline
2nd & 209 & -highpassboost -randfe -diff  & 2 & 27 & 6.90\\ \hline
2nd & 210 & -boost -minmax -mink  & 2 & 27 & 6.90\\ \hline
2nd & 211 & -boost -minmax -diff  & 2 & 27 & 6.90\\ \hline
2nd & 212 & -highpassboost -randfe -eucl  & 2 & 27 & 6.90\\ \hline
2nd & 213 & -boost -minmax -eucl  & 2 & 27 & 6.90\\ \hline
2nd & 214 & -low -aggr -randcl  & 2 & 27 & 6.90\\ \hline
2nd & 215 & -band -fft -randcl  & 2 & 27 & 6.90\\ \hline
2nd & 216 & -boost -aggr -cheb  & 2 & 27 & 6.90\\ \hline
2nd & 217 & -band -randfe -randcl  & 2 & 27 & 6.90\\ \hline
2nd & 218 & -boost -fft -cheb  & 2 & 27 & 6.90\\ \hline
2nd & 219 & -highpassboost -aggr -mink  & 2 & 27 & 6.90\\ \hline
2nd & 220 & -highpassboost -aggr -diff  & 2 & 27 & 6.90\\ \hline
2nd & 221 & -highpassboost -fft -mink  & 2 & 27 & 6.90\\ \hline
2nd & 222 & -endp -minmax -randcl  & 2 & 27 & 6.90\\ \hline
2nd & 223 & -highpassboost -fft -diff  & 2 & 27 & 6.90\\ \hline
\end{tabular}
\end{minipage}
\caption{Consolidated results, Part 16.}
\label{tab:results16}
\end{table}

\begin{table}
\begin{minipage}[b]{\textwidth}
\centering
\begin{tabular}{|c|c|l|c|c|r|} \hline
Guess & Run \# & Configuration & GOOD & BAD & Recognition Rate,\%\\ \hline\hline
2nd & 224 & -highpassboost -aggr -eucl  & 2 & 27 & 6.90\\ \hline
2nd & 225 & -highpassboost -randfe -cheb  & 2 & 27 & 6.90\\ \hline
2nd & 226 & -high -lpc -nn  & 2 & 27 & 6.90\\ \hline
2nd & 227 & -boost -minmax -cheb  & 2 & 27 & 6.90\\ \hline
2nd & 228 & -highpassboost -fft -eucl  & 2 & 27 & 6.90\\ \hline
2nd & 229 & -boost -lpc -mah  & 2 & 27 & 6.90\\ \hline
2nd & 230 & -norm -randfe -randcl  & 1 & 28 & 3.45\\ \hline
2nd & 231 & -highpassboost -aggr -cheb  & 2 & 27 & 6.90\\ \hline
2nd & 232 & -highpassboost -fft -cheb  & 2 & 27 & 6.90\\ \hline
2nd & 233 & -band -minmax -randcl  & 2 & 27 & 6.90\\ \hline
2nd & 234 & -boost -aggr -randcl  & 2 & 27 & 6.90\\ \hline
2nd & 235 & -highpassboost -lpc -mah  & 2 & 27 & 6.90\\ \hline
2nd & 236 & -highpassboost -aggr -mah  & 2 & 27 & 6.90\\ \hline
2nd & 237 & -high -lpc -randcl  & 1 & 28 & 3.45\\ \hline
2nd & 238 & -highpassboost -randfe -mah  & 2 & 27 & 6.90\\ \hline
2nd & 239 & -boost -randfe -mink  & 2 & 27 & 6.90\\ \hline
2nd & 240 & -boost -randfe -diff  & 2 & 27 & 6.90\\ \hline
2nd & 241 & -highpassboost -minmax -mink  & 2 & 27 & 6.90\\ \hline
2nd & 242 & -raw -randfe -randcl  & 0 & 29 & 0.00\\ \hline
2nd & 243 & -highpassboost -fft -randcl  & 1 & 28 & 3.45\\ \hline
2nd & 244 & -band -lpc -randcl  & 0 & 29 & 0.00\\ \hline
2nd & 245 & -endp -fft -randcl  & 0 & 29 & 0.00\\ \hline
2nd & 246 & -raw -fft -randcl  & 2 & 27 & 6.90\\ \hline
2nd & 247 & -norm -lpc -randcl  & 1 & 28 & 3.45\\ \hline
2nd & 248 & -highpassboost -randfe -randcl  & 1 & 28 & 3.45\\ \hline
2nd & 249 & -high -aggr -randcl  & 0 & 29 & 0.00\\ \hline
2nd & 250 & -band -randfe -mink  & 1 & 28 & 3.45\\ \hline
2nd & 251 & -low -lpc -randcl  & 2 & 27 & 6.90\\ \hline
2nd & 252 & -highpassboost -minmax -randcl  & 2 & 27 & 6.90\\ \hline
2nd & 253 & -norm -aggr -randcl  & 1 & 28 & 3.45\\ \hline
\end{tabular}
\end{minipage}
\caption{Consolidated results, Part 17.}
\label{tab:results17}
\end{table}

\begin{table}
\begin{minipage}[b]{\textwidth}
\centering
\begin{tabular}{|c|c|l|c|c|r|} \hline
Guess & Run \# & Configuration & GOOD & BAD & Recognition Rate,\%\\ \hline\hline
2nd & 254 & -high -fft -randcl  & 3 & 26 & 10.34\\ \hline
2nd & 255 & -band -minmax -nn  & 1 & 28 & 3.45\\ \hline
2nd & 256 & -norm -minmax -randcl  & 1 & 28 & 3.45\\ \hline
\end{tabular}
\end{minipage}
\caption{Consolidated results, Part 18.}
\label{tab:results18}
\end{table}

% EOF

	\chapter{Applications}
\label{chapt:apps}
\index{Applications}
\index{MARF!Applications}

$Revision: 1.18 $

This chapter describes the applications that employ {\marf}
for one purpose or another. Most of them are our own applications
that demonstrate how to use various {\marf}'s features. Others
are either research or internal projects of their own. If you
use {\marf} or its derivative in your application and would like
to be listed here, please let us know at \url{marf-devel@sf.lists.net}.

%
% Internal Research Applications
%

\section{MARF Research Applications}
\index{MARF!Research Applications}

%
% SpeakerIdentApp
%

\subsection{SpeakerIdentApp - Text-Independent Speaker Identification Application}
\index{Applications!SpeakerIdentApp}
\index{Research Applications!SpeakerIdentApp}

\api{SpeakerIdentApp} is an application for text-independent
speaker identification that exercises the most of the {\marf}'s
features.
\api{SpeakerIdentApp} is broadly presented through
the rest of this manual.

%
% ZipfLawApp
%

\subsection{Zipf's Law Application}
\label{sect:zipf-law-app}
\index{Zipf's Law}
\index{Zipf's Law!Application}
\index{Applications!Zipf's Law}

$Revision: 1.16 $

Originally written on February 7, 2003.

\paragraph{The Program}

%is Java has built-in \api{StreamTokenizer}; you just have
%to instruct it how you want tokens to look like, and get
%them from any input stream, one-by-one. Java also provides
%us with built-in types and data-structures to manage
%collections (build, sort, store/retrieve) efficiently.

\paragraph{Data Structures}

The main statistics ``place-holder'' is a \api{oStats} \api{Hashtable}, nicely
provided by {\java}, which hashes by words as keys to other
data structures, called \api{WordStats}. The \api{WordStats} data structure
contains word's spelling, called lexeme (borrowed this term
from compiler design), word's frequency (initially $0$), and
word's rank (initially $-1$, i.e. unset). \api{WordStats} provides a typical
variety of methods to access and alter any of those three values.
There is an array, called \api{oSortedStatRefs}, which will eventually
hold references to \api{WordStats} objects in the \api{oStats} hashtable in
the sorted by the frequency.

Hashtable entry might look like this in pseudo-notation:

\verb+{ <word> } -> [ WordStats{ <word>, <f>, <r> } ]+

\subsubsection{Mini User Manual}

\paragraph{System Requirements}

The program was mostly developed under Linux, so there's
a \file{Makefile} and a testing shell script to simplify some routine tasks.
For JVM, any JDK 1.4.* and above will do. \tool{bash} would be nice
to have to be able to run the batch script. Since the application
itself is written in {\java}, it's not bound to specific architecture,
thus may be compiled and run without the makefiles and scripts
on virtually any operating system.

\paragraph{How To Run It}

There are at least three ways how to run the program, listed in order of complexity
(in terms of number of learning and typing involved).
In order for the below to work you'd need some corpora in
the same directory as the application.

\paragraph{Using the \tool{testing.sh} Script}

The script is written using \tool{bash} syntax; hence, \tool{bash} should be
present. The script ensures
that the program was compiled first, by invoking \tool{make}, then
in a for loop feeds all the \file{*.txt} files (presumably our corpora) along with the \file{ZipfLaw.java}
program itself to the executable, one by one, and redirects its standard output
to the files as \file{$<$corpus-name$>$[$<$options>$]$.log} in the current directory. These files will contain the
grouping of the 100 most frequent words every 1000 words as well as frequency-of-frequency counts at the end.

\noindent
Type:

\verb+./testing.sh+

\noindent
to run the batch through with the default settings.

\verb+./testing.sh <options>+

\noindent
to override the default settings with $<$options$>$. $<$options$>$ are the same
as that of the \api{ZipfLaw} application (see \ref{sect:zipf-law-app}).

\paragraph{Using \file{Makefile}}

If you want to build and to run the application just for individual,
pre-set corpora, use the \tool{make} utility provided with UNIXen.

Type (assuming GNU-style \tool{make}, or \tool{gmake}):

\noindent
\verb+make+

	to just compile the thing

\noindent
\verb+make <corpus-name> [ > <filename>.log ]+

	to (possibly compile if not done yet) and run it for the corpus
	\verb+<corpus-name>+ and to optionally redirect output to the file
	named \verb+<filename>.log+
	The \verb+<corpus-name>+ are described in the \file{Makefile} itself.

\noindent
\verb+make clean+

	to clean up the \file{*.class} and \file{*.log} files that happened to be
	generated and so on.

\noindent
\verb+make run+

	is actually equivalent to running \tool{testing.sh} with no options.

\noindent
\verb+make report+

	to produce a PDF file out of the {\LaTeX} source of the report.

\paragraph{Running The ZipfLaw Application}
\label{sect:zipf-law-app}

You can run the application itself without any wrapping scripts
and provide options to it. This is a command-line application,
so there is no GUI associated with it yet. To run the application
you have to compile it first. You can use either \tool{make} with no
arguments to compile or use the standard Java compiler.

\noindent
\verb+make+

or

\noindent
\verb+javac -cp marf.jar:. ZipfLaw.java+

\noindent
After having compiled the thing, you can run it with the JVM.
There is one required argument - either corpus file to analyze
or \verb+--help+. If it's a corpus, it
may be accompanied with one or more options overriding the default
settings. Here are the options as per the application's output:

\vspace{15pt}
\hrule
\begin{verbatim}
Usage:
    java ZipfLaw --help | -h | [ OPTIONS ] <corpus-file>
    java ZipfLaw --list [ OPTIONS ] <corpus-file>

Options (one or more of the following):
    --case  - make it case-sensitive
    --num   - parse numerical values
    --quote - consider quotes and count quoted strings as one token
    --eos   - make typical ends of sentences (<?>, <!>, <.>) significant
    --nolog - dump Zipf's law graph values as-is instead of log/log
    --list  - lists already pre-collected statistics for a given corpus

\end{verbatim}

\hrule
\vspace{15pt}

If the filename isn't specified, that will be stated and the usage
instructions above displayed.
The output filename generated from the input file name with the
options (if any) pre-pended and it ends with the extension of \file{.csv},
which can directly be opened by OpenOffice Calc or Microsoft Excel.

\subsubsection{Experiments}

Various experiments on diverse corpora were conducted to find
out whether Zipf's Law can possibly fail. For that purpose
I used the following corpora:

\begin{itemize}
\item a technical white paper of Dr. Probst (\cite{probst95}) of 20k in size, the filename is \file{multiprocessor.txt}
\item three ordinary corpora (non-technical literature)
	(\cite{greif, ulysses, speak}) -- \file{grfst10.txt}, 853k;
	\file{ulysses.txt}, 1.5M; and \file{hwswc10.txt}, 271k
\item my personal mailbox in UNIX format, raw as is, 5.5M
\item the source code of this application itself, \file{ZipfLaw.java}, 8.0k
\end{itemize}

\paragraph{Default Setup}

This is very simplistic approach in the application, where
everything but a letter (26 caps, and 26 lowercase) is a
blank, and as such is discarded. All the words were folded to the
lower case as well. This default setup can be overridden by
specifying the command-line options described above.

\paragraph{Extreme Setup}

One of the option combinations, that makes the program
case-sensitive, considers the numbers, and treats ``!'',
``.'', and ``?'' as special tokens.

\subsubsection{Results}

We failed to prove Zipf wrong. With {\bf any} of the corpora and the settings.
Further, the log/log graphs of the frequency and the rank
for the default and the extreme setup are provided. The graphs don't change very much
in shape. For other option combinations, the graphs are not provided
since they don't vary much.
It turned out to be the that capitalization, end of sentence symbols,
and numbers, if treated as tokens, don't make much of a difference,
as if they simply aren't there.

In the distribution archive, you will find the \file{*.log} and \file{*.csv} files
of the test runs, which contain the described statistics. You are welcome
to do \verb+make clean+ and re-run the tests on your own. NOTE, by default
the output goes to the standard output, so it's a good idea to redirect
it to a file especially if a corpus is a very large one.

\paragraph{Graphs, The Default Setup}

\begin{figure}[h!]
	\begin{center}
	\includegraphics[width=0.5\textwidth]{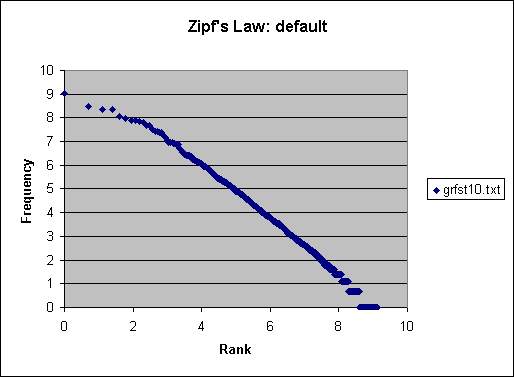}
	\caption{Zipf's Law for the ``Greifenstein'' corpus with the default setup.}
	\end{center}
\end{figure}

\begin{figure}
    \begin{center}
    \includegraphics[width=0.5\textwidth]{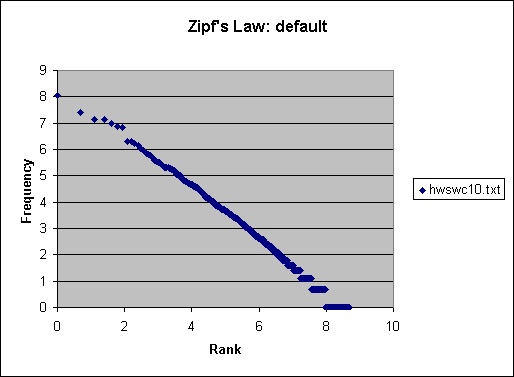}
    \caption{Zipf's Law for the ``How to Speak and Write Correctly'' corpus with the default setup.}
    \end{center}
\end{figure}

\begin{figure}
    \begin{center}
    \includegraphics[width=0.5\textwidth]{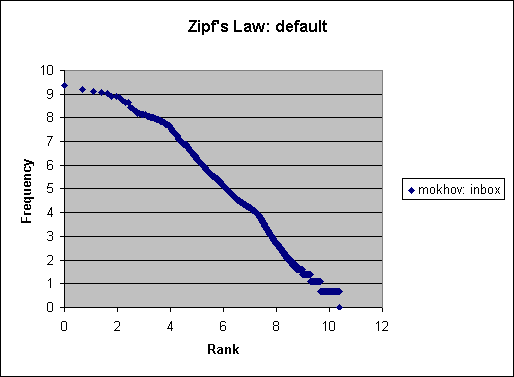}
    \caption{Zipf's Law for my 5.6 Mb INBOX with the default setup.}
    \end{center}
\end{figure}

\begin{figure}
    \begin{center}
    \includegraphics[width=0.5\textwidth]{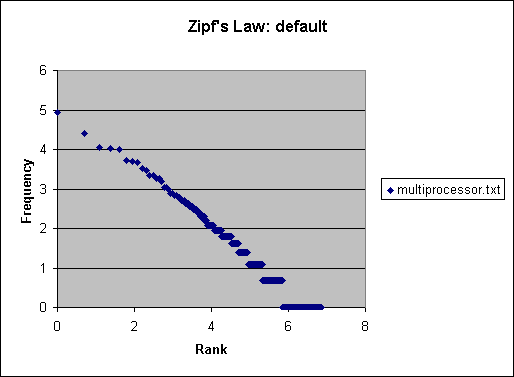}
    \caption{Zipf's Law for the white paper ``The United States Needs a Scalable Shared-Memory Multiprocessor, But Might Not Get One!'' with the default setup.}
    \end{center}
\end{figure}

\begin{figure}
    \begin{center}
    \includegraphics[width=0.5\textwidth]{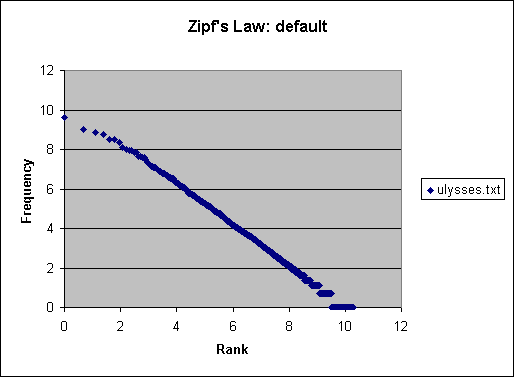}
    \caption{Zipf's Law for the ``Ulysses'' corpus with the default setup.}
    \end{center}
\end{figure}

\begin{figure}
    \begin{center}
    \includegraphics[width=0.5\textwidth]{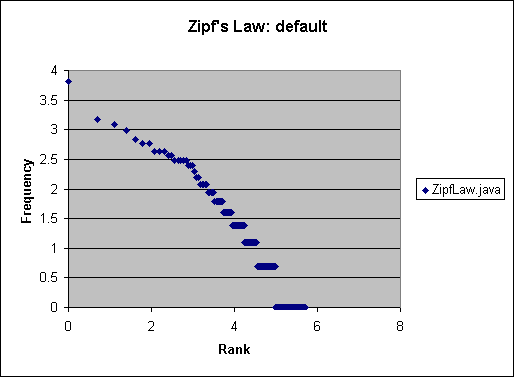}
    \caption{Zipf's Law for the ``ZipfLaw.java'' program itself with the default setup.}
    \end{center}
\end{figure}

\clearpage

\paragraph{Graphs, The Extreme Setup}

\begin{figure}[h!]
    \begin{center}
    \includegraphics[width=0.5\textwidth]{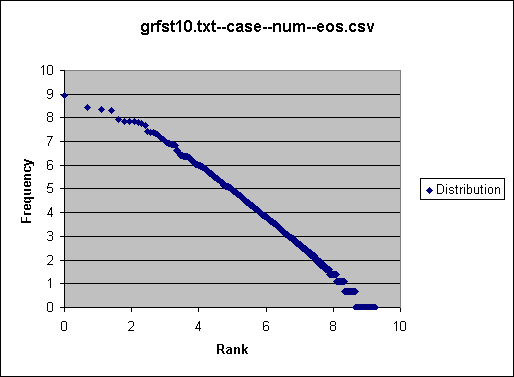}
    \caption{Zipf's Law for the ``Greifenstein'' corpus with the extreme setup.}
    \end{center}
\end{figure}

\begin{figure}
    \begin{center}
    \includegraphics[width=0.5\textwidth]{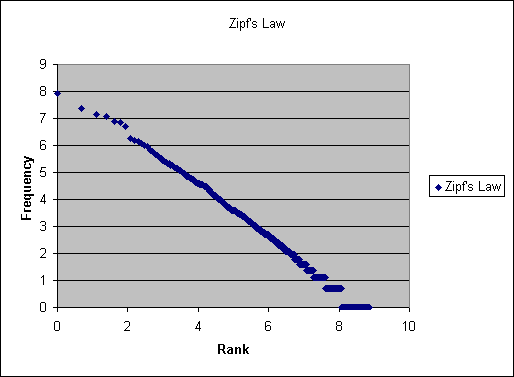}
    \caption{Zipf's Law for the ``How to Speak and Write Correctly'' corpus with the extreme setup.}
    \end{center}
\end{figure}

\begin{figure}
    \begin{center}
    \includegraphics[width=0.5\textwidth]{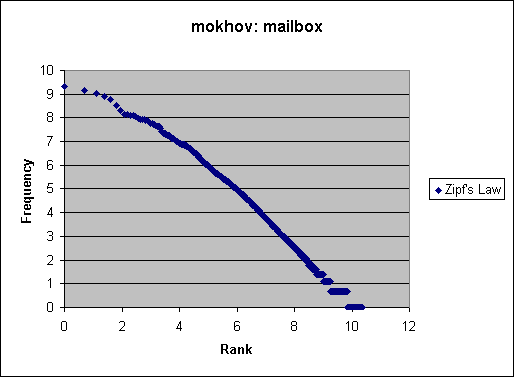}
    \caption{Zipf's Law for my 5.6 Mb INBOX  with the extreme setup.}
    \end{center}
\end{figure}

\begin{figure}
	\begin{center}
	\includegraphics[width=0.5\textwidth]{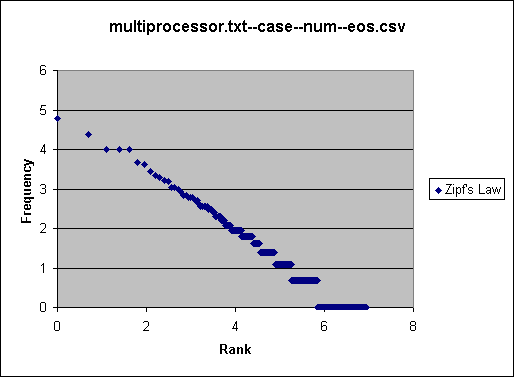}
	\caption{Zipf's Law for the white paper ``The United States Needs a Scalable Shared-Memory Multiprocessor, But Might Not Get One! with the extreme setup}
	\end{center}
\end{figure}

\begin{figure}
	\begin{center}
	\includegraphics[width=0.5\textwidth]{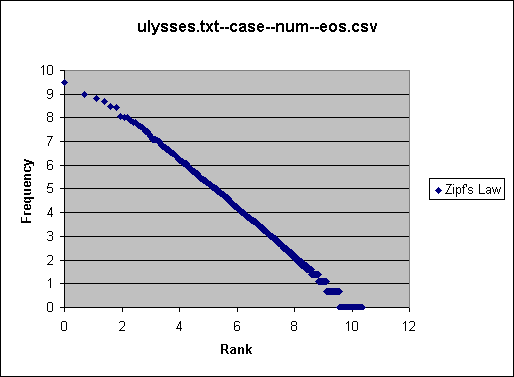}
	\caption{Zipf's Law for the ``Ulysses'' corpus with the extreme setup.}
	\end{center}
\end{figure}

\begin{figure}
	\begin{center}
	\includegraphics[width=0.5\textwidth]{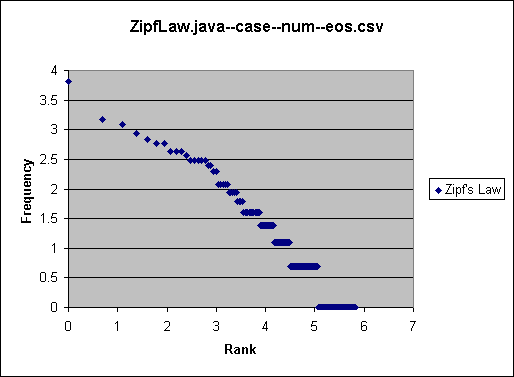}
	\caption{Zipf's Law for the ``ZipfLaw.java'' program itself with the extreme setup.}
	\end{center}
\end{figure}

\subsubsection{Conclusions}

Zipf's Law holds so far. However, more experimentation is required,
for example, other punctuation characters than
that ending a sentence were not considered. Then other languages than English is a good
area to explore as well.

% EOF

%
% LangIdentApp
%

\subsection{Language Identification Application}
\label{sect:lang-ident-app}
\index{Language Identification}
\index{Language Identification!Application}
\index{Applications!Language Identification}

$Revision: 1.28 $

Originally written on March 17, 2003.

\subsubsection{The Program}

The Mini-User Manual is provided in the \xs{sect:lang-ident-app-manual}. The source
is provided in the electronic form only from the release tarballs or the CVS repository.

NOTE: in the code there are a lot of files. Not all of them might
be of a great interest to you since
some of them are stubs right now and don't provide much functionality in them
or the functionality is not linked anyhow to the main application.
These files are:

\begin{verbatim}
./marf/nlp/Collocations:
    ChiSquareTest.java
    CollocationWindow.java
    TTest.java

./marf/Stats/StatisticalEstimators:
    GLI.java
    KatzBackoff.java
    SLI.java

./marf/util/comparators:
    RankComparator.java

./marf/nlp/Stemming:
    StemmingEN.java
    Stemming.java
\end{verbatim}

\subsubsection{Hypotheses}
\index{Language Identification!Hypotheses}

\paragraph{Identifying Non-Latin-Script-Based Languages}

The methodology, if implemented correctly, should work
for natural languages that use non-Latin scripts for
their writing. Of course, to test this, such scripts
will have to be romanized (either transcribed or transliterated
using Latin, more precisely, ASCII characters).

\paragraph{Identifying Programming Languages}
\index{Language Identification!Programming Languages}

I have thought of if the proposed methodology works well for natural
languages, it would probably work at least as good for the programming
languages.

\paragraph{Sophisticated Statistical Estimators Should Work Better}

``Good'' (or ``better'' if you will) statistical estimators presented in \xc{chapt:statitstic-processing}
should give better results (higher language recognition rate) over simpler ones.
By ``simpler'' we mean the MLE and Add-* family. By more sophisticated we mean Witten-Bell,
Good-Turing, and the combination of the Statistical Estimators.

\paragraph{Zipf's Law Can Also Be Effective in Language Identification}
\index{Language Identification!Zipf's Law}
\index{Zipf's Law!Language Identification}

Zipf's Law can be reasonably effective and very ``cheap'' in language identification
task. This one is yet to be verified.

\subsubsection{Initial Experiments Setup}

\paragraph{Languages}

Several natural and programming languages in were used in the experiments.

\paragraph*{Natural Lanugages (NLs)}

\begin{itemize}
\item English (en - ISO 2-letter code code \cite{iso-codes})
\item French (fr, desaccented)
\item Spanish (es, desaccented)
\item Italian (it)
\item Arabic (ar, transcribed in ASCII according to the qalam rules \cite{qalam})
\item Hebrew (he, transcribed in ASCII)
\item Bulgarian (bg, transcribed in ASCII)
\item Russian (ru, transcribed in ASCII)
\end{itemize}

\paragraph*{Programming Languages (PLs)}

\begin{itemize}
\item {\C}/{\cpp} (together, since {\C} is a proper subset of {\cpp})
\item {\java}
\item {\perl}
\end{itemize}

\paragraph*{Statistical Estimators Implemented}

\begin{itemize}
\item MLE in \api{MLE}
\item Add-One in \api{AddOne}
\item Add-Delta (ELE) in \api{AddDelta}
\item Witten-Bell in \api{WittenBell}
\item Good-Turing in \api{GoodTuring}
\end{itemize}

\paragraph*{N-gram Models}

\begin{itemize}
\item Unigram
\item Bigram
\item Trigram
\end{itemize}

\paragraph{Corpora}

\paragraph*{English}

The English language corpora is (not very surprisingly) was the biggest one.
To simplify the training process, we combined them all in one file
\file{en.txt}. It includes
\cite{probst95, ulysses, greif, speak, fannyhill, lysistrata, canterby, defoe, rousseau, tasso}.
Additionally, a few chapters of ``The Little Prince''.
Total size of the combined file is 7Mb.

\paragraph*{French}

For French we used few chapters of ``Le Petit Prince''.
The totals combined size is 12k, \file{fr.txt}.

\paragraph*{Spanish}

Like the French corpora, the Spanish one includes several
chapters of ``Le Petit Prince'' in Spanish (from the same source).
The total size is 12k, \file{es.txt}.

\paragraph*{Arabic}

We have compiled a few surah from transliterated Quran (\cite{quran})
as well as a couple of texts transcribed by Michelle Khalif\'e
from a proverb \cite{jeha} and a passage from a newspaper in Arabic \cite{arnews}.
Total size: 208k, \file{ar.txt}.

\paragraph*{Hebrew}

We used a few poems written by their author (\cite{hepoems}) in ASCII alphabet.
Total size is 37k, \file{he.txt}.

\paragraph*{Russian}

Latinized classics (one whole book) was used (see \cite{ohotnik}).
Total size is 877k, \file{ru.txt}.

\paragraph*{Bulgarian}

A few transcribed poems were used for training from \cite{bgpoems}.
Total size: 21k, \file{bg.txt}.

\paragraph*{Italian}

We used the ``Pinocchio'' book \cite{pinocchio} of the size of 245k, \file{it.txt}.

\paragraph*{C/C++}

Various \file{.c} and \file{.cpp} files were used from a variety of projects
and examples for the ``COMP444 - System Software Design'', ``COMP471 - Computer Graphics'',
``COMP361'', and ``COMP229 - System Software courses''. The total size is 137k, \file{cpp.txt}.

\paragraph*{Java}

As Java ``corpora'' we used the sources of this application at some point in the
development cycle and source files for the MARF project itself.
The total size is 260k, \file{java.txt}.

\paragraph*{Perl}

For {\perl}, we used many of Serguei's scripts written to help with marking of
assignments and accept electronic submissions as well as a couple tools
for CVS written in {\perl} from the Internet (Google keywords: \tool{cvs2cl.pl} and \tool{cvs2html.pl}).
Size: 299k, \file{perl.txt}.

%%%%%%%%%%%%%%%%%%%%%%%%%%%%%%%%%%%%%%%%%%%%%%%
%%%%%%%%%%%%%%%%%%%%%%%%%%%%%%%%%%%%%%%%%%%%%%%

\subsubsection{Methodology}

\paragraph{Add-Delta Family}

Add-Delta is defined as:

$P(\mbox{ngram}) = \frac{C_{\mbox{ngram}} + \delta}{N + \delta \cdot V}$

\noindent
where $V$ is a vocabulary size, $N$ is the number of n-grams that start with {\em ngram}.
By implementing this general Add-Delta smoothing we get MLE, ELE, and Add-One ``for free''
as special cases of Add-Delta:

\begin{itemize}
\item $\delta = 0$ is MLE
\item $\delta = 1$ is Add One
\item $\delta = 0.5$ is ELE
\end{itemize}

\paragraph{Witten-Bell and Good Turing}

These two estimators were implemented as given in hopes to get a better
recognition rate over the Add-Delta family.

\subsubsection{Difficulties Encountered}

During the experimentations we faced several problems, some of which
are worth mentioning.

\paragraph{``Bag-of-Languages'' and Alphabets}

From this point now on, by an {\em alphabet} in this document we mean something more than what people understand
by the language alphabet. In our case an alphabet may include
characters other than letters, such as numbers, punctuation, even
a blank sometimes, all being derived from a training corpus.

Initially, we attempted to treat programming languages as if they
were natural ones. That way, from the developer's standpoint we deal with
them all uniformly. This assumption could be viewed as cheating
to some extent however because programming languages have a lot larger
alphabets that are necessary lexical parts of the language in addition
to the statements written using ASCII letters. Therefore, this gives a lot of
discriminatory power as compared to the NLs when these characters are
inputted by an user. Treating PLs as using
only ASCII Latin base should lead to a lot
of confusion with English (and sometimes other NLs) because most of the keywords
are English words in addition to literal text strings and comments present within the code.

Among NLs that were transcribed or transliterated in Latin there are alphabetical differences.
For instance, in Arabic there are three h-like sounds that have no
English equivalent, so sometimes numbers 3, 7, and 5 are used for that purpose (or in more
standard LAiLA notation \cite{qalam} capitals are used instead. To be
more fair to others, we let numerals to be a part of the alphabet as well.
Analogous problem existed when using capitals for different sounds as opposed
to direct lowercase transliteration in Arabic making lowercasing
a not necessarily good idea.

Russian and Bulgarian (transcribed from Cyrillic scripts) use
(') and some other symbols (like ([) or (]) in Bulgarian) to represent
certain letters or sounds; hence, they always have to be a part of the
alphabet.
This has caused some problems again, and I thought of another separation that is
needed: Latin-based, Cyrillic-based, and Semitic-based languages,
and our ``bag-of-languages'' approach might no do so well. (We, however, just split
PLs, Latin-based and non-Latin as the end result.)

Even for Latin-based languages that can be a problem. For example, the
letter $k$ does not exist in Spanish or Italian (it may if referred to
the foreign words, such as ``kilogram'' or proper names but is not used otherwise).
So are $w$ and $x$ and maybe some others. The same
with French~--~the letters $k$ and $w$ are very rare (so they didn't happen to be in ``Le Petit
Prince'' corpus used, for example).

\paragraph{Alphabet Normalization}

We {\em do} get different alphabet sizes
of my corpora for a language. The alphabets are derived from the corpora
themselves, so depending on the size some characters that appear in
one corpora might not appear in another. Thus, when we perform classification
task for a given sentence, the models compared may be with differently-sized
alphabets thereby returning a probability of 0.0 for the n-gram's characters that
have not been present yet in a given trained model. This can
be viewed as a non-uniform smoothing of the models and implies necessity
of the normalization of the alphabets of all the language models after
accounting for n-grams has been made, and only then smooth.

Language model normalization in has not been implemented yet. Such normalization, however, will
provoke a problem of data sparseness similar to the one described below.
This presents a problem for smoothing techniques, because some counts
we get, $N$ of either n-grams or (n-1)-grams, will have values of 0.0 and
division by 0.0 will become a problem.

\paragraph{N-grams with $n > 2$}

The implemented maximum so far is $n = 3$, but it is a general problem for any $n$.
The problem stems from
the excessive data sparseness of the models for $n > 2$. Taking for
example MLE~---~it won't be able to cope with it properly without
special care because $N$ there is now a two-dimensional matrix,
which can easily be 0.0 in places and the division by zero is unavoidable.
Analogous problem exists in the Good-Turing smoothing.
To solve this we have said if $N = 0$ make $N=1$. Maybe by doing
so (as this is a quite a kludge) we have created more trouble,
but that was the ``design decision'' in the first implementation.

\subsubsection{Experiments}

\paragraph{Bag-of-Languages and the Language Split}

We came up with a few testing sentences/statements for all languages
(can be found in the \file{test.langs} file, see \xf{fig:lang-ident-sentences}).
Then, based on my random observations above we conducted more guided experiments.

\begin{figure}\small
\hrule\vskip4pt
\input{test.langs.tex}
\caption{Sample sentences in various languages used for testing. Was used in ``Bag-of-Languages'' experiments.}
\label{fig:lang-ident-sentences}
\vskip4pt\hrule
\end{figure}

\begin{figure}\small
\hrule\vskip4pt
\input{test.latin.langs.tex}
\caption{Subset of the sample sentences in languages with Latin base.}
\label{fig:lang-ident-latin}
\vskip4pt\hrule
\end{figure}

The sentences from the \xf{fig:lang-ident-sentences} were used as-is to pipe to
the program for classification for the bag-of-languages approach. This
file has been split into parts (see \xf{fig:lang-ident-latin}, \xf{fig:lang-ident-nonlatin},
and \xf{fig:lang-ident-pls}) to try out other approaches as well (see \xs{sect:lang-ident-training}).

\begin{figure}\small
\hrule\vskip4pt
\input{test.non-latin.langs.tex}
\caption{Subset of the sample sentences in languages with non-Latin base.}
\label{fig:lang-ident-nonlatin}
\vskip4pt\hrule
\end{figure}

\begin{figure}\small
\hrule\vskip4pt
\input{test.pls.langs.tex}
\caption{Subset of the sample statements in programming languages.}
\label{fig:lang-ident-pls}
\vskip4pt\hrule
\end{figure}

\paragraph{Tokenization}

We used two types of tokenizer, {\bf restricted} and {\bf unrestricted} to experiment with
the diverse alphabets.
The ``restricted tokenizer'' means lowercase-folded ASCII characters and numbers (more corresponding
to the original requirements). The ``unrestricted tokenizer''
means additional characters are allowed and it is case-sensitive.
In both tokenizers blanks are discarded.
An implementation of these tokenizer settings via command-line options is still a TODO,
so we were simply changing the code and recompiling. The code has an unrestricted tokenizer
(\file{NLPStreamTokenizer.java} under \file{marf/nlp/util}).

\subsubsection{Training}
\label{sect:lang-ident-training}

We trained language models to include the following:

\begin{itemize}
\item all the languages (both NLs and PLs) with the restricted tokenizer
\item all the languages with the unrestricted tokenizer
\item latin-based NLs (English, French, Spanish, and Italian) with the restricted tokenizer
\item non-latin-based romanized NLs (Arabic, Hebrew, Russian, and Bulgarian) with the unrestricted tokenizer
\item PLs ({\java}, {\C}/{\cpp}, {\perl}) with the unrestricted tokenizer.
\end{itemize}

%%%%%%%%%%%%%%%%%%%%%%%%%%%%%%%%%%%%%%%%%%%%%%%%%%%%%
%%%%%%%%%%%%%%%%%%%%%%%%%%%%%%%%%%%%%%%%%%%%%%%%%%%%%

\subsubsection{Results}

\paragraph{Brief Summary}

(Brief because there's more elaborate Conclusion section).

So far, the results are good in places, sometimes pitiful.
Trigrams alone generally were very poor and slow for us.
Unigrams and bigrams performed quite well, however.

More detailed results can be observed in the \xa{sect:lang-classification-results}.
Below are the numbers as a recognition rate of the
sentences presented earlier for every language model.
Note that numbers by themselves may not convey
enough information, one has to look at the detailed results further
to actually realize that the number of samples is debatable
to be good and so are the training corpora is not
uniform. One might also want to look which languages
get confused with each other.

\paragraph{Summary of Recognition Rates}

\vspace{15pt}
\hrule
\vspace{15pt}

Language Model:

\begin{itemize}
\item ``Bag-of-Languages''
\item Unrestricted tokenizer
\end{itemize}

\begin{verbatim}
                 unigram  bigram  trigram

MLE              54.17%   16.67%  16.67%
add-delta (ELE ) 58.33%   12.50%  16.67%
add-one          58.33%   12.50%  16.67%
Witten-Bell      16.67%   29.17%  16.67%
Good-Turing      16.67%   12.50%  16.67%
\end{verbatim}

\vspace{15pt}
\hrule
\vspace{15pt}

Language Model:

\begin{itemize}
\item NLs transcribed in ASCII (Arabic, Hebrew, Bulgarian, and Russian)
\item Unrestricted tokenizer
\end{itemize}

\begin{verbatim}
                 unigram  bigram  trigram

MLE              66.67%   33.33%  33.33%
add-delta (ELE)  77.78%   11.11%  55.56%
add-one          77.78%   11.11%  55.56%
Witten-Bell      55.56%   66.67%  33.33%
Good-Turing      55.56%   55.56%  55.56%
\end{verbatim}

\vspace{15pt}
\hrule
\vspace{15pt}

Language Model:

\begin{itemize}
\item PLs only ({\C}/{\cpp}, {\java}, and {\perl})
\item Unrestricted tokenizer
\end{itemize}

\begin{verbatim}
                 unigram  bigram  trigram

MLE              33.33%   33.33%  33.33%
add-delta (ELE)  33.33%   50.00%  33.33%
add-one          33.33%   50.00%  33.33%
Witten-Bell      33.33%   50.00%  33.33%
Good-Turing      33.33%   33.33%  33.33%
\end{verbatim}

\vspace{15pt}
\hrule
\vspace{15pt}

Language Model:

\begin{itemize}
\item Latin-based Langauges only (English, French, Spanish, and Italian)
\item Restricted tokenizer
\end{itemize}

\begin{verbatim}
                 unigram  bigram  trigram

MLE              77.78%   33.33%  44.44%
add-delta (ELE)  77.78%   44.44%  55.56%
add-one          77.78%   55.56%  55.56%
Witten-Bell      44.44%   44.44%  44.44%
Good-Turing      44.44%   44.44%  55.56%
\end{verbatim}

\vspace{15pt}
\hrule
\vspace{15pt}

Language Model:

\begin{itemize}
\item ``Bag-of-Languages''
\item Restricted tokenizer
\end{itemize}

\begin{verbatim}
                 unigram  bigram  trigram

MLE              62.50%   16.67%  16.67%
add-delta (ELE)  62.50%    4.17%  12.50%
add-one          62.50%    8.33%  12.50%
Witten-Bell      16.67%   33.33%  16.67%
Good-Turing      16.67%   20.83%  25.00%
\end{verbatim}

%\subsubsection{Sample Interactive Run}

%\scriptsize
%\input{lang-ident-app-sample-run}
%\normalsize

\subsubsection{Conclusions}

\paragraph{Concrete Points}

\begin{itemize}

\item
The best results we've got so far from simpler language models;
especially using languages with the Latin base only.

\item
The methodology did work for non-Latin languages as well. Not 100\% of the time,
but around 60\%, but this is a start.

\item
Witten-Bell and Good-Turing performed rather poorly in our tests in general.
We think we need a lot larger corpora to test out Witten-Bell and Good-Turing
smoothing more thoroughly. This can be confirmed by some of the results
where Good-Turing gave us all English and English is the biggest
corpus we've got.

\item
Identification of the Latin-based languages among themselves was the best one.
It worked OK in the bag-of-languages approach.

\item
The strict tokenizer and bag-of-languages were surprisingly good (or at least
better than we expected).

\item
Recognition of the programming languages according to the conducted
experiments can be qualified as ``so-so'' when PLs are compared to
each other only. They were recognized slightly better in the bag-of-languages
approach (due to the larger alphabets).

\end{itemize}

\paragraph{Generalities}

Some of the hypotheses didn't hold (``better techniques would do better'' and
``PLs can be identified as easily as NLs''),
some didn't have time allotted to them yet (``try out Zipf's Law for the purpose
of the language identification'').

In the next releases, we want to experiment with more things, for example,
cross-validation and held-out estimation as well as
linear interpolations and Katz Backoff.

\subsubsection{Mini User Manual}
\label{sect:lang-ident-app-manual}

\paragraph{System Requirements}

The application was primarily developed under Linux, so there's
a \file{Makefile} and a testing shell script to simplify some routine tasks.
For JVM, any JDK 1.4.* and above will do. \tool{tcsh} would be nice
to have to be able to run the batch script. Since the application
itself is written in {\java}, it's not bound to specific architecture,
thus may be compiled and run without the makefiles and scripts
on virtually any operating system.

\paragraph{How To Run It}

In order for the below to work you'd need some corpora in
the same directory as the application. (Check the reference
section for the corpora used in the experiments.)
There are thousands of ways how to run the program. Some of them are listed below.

\paragraph{Using the Shell Scripts}

There are two scripts out there -- \tool{training.sh} and \tool{testing.sh}.
The former is used to do batch training on all the languages and all the models,
the latter to perform batch-testing of the models. They hide the complexity
of typing many options to the users. If you are ever to use them, tweak
them appropriately for the specific languages and n-gram models if you don't
all all-or-nothing testing.

The scripts are written using the \tool{tcsh} syntax; hence, \tool{tcsh} should be
present. The scripts ensure
that the program was compiled first, by invoking \tool{make}, then
in several \verb+for()+ loops feed all the options and filenames to the application.

Type:

\noindent
\verb+./training.sh+

to train the models.

\noindent
\verb+./testing.sh+

to test the models. NOTE: to start testing right away, you need the \file{*.gzbin} files
(pre-trained models) which should be copied from a \file{training-*} directory
of your choice to the application's directory.

\paragraph{Running The LangIdentApp Application}
\label{sect:lang-ident-app}

You can run the application itself without any wrapping scripts
and provide options to it. This is a command-line application,
so there is no GUI associated with it yet (next release). To run the application
you have to compile it first. You can use either \tool{make} with no
arguments to compile or use a standard Java compiler.

\noindent
\tool{make}

or

\noindent
\verb+javac -cp marf.jar:. LangIdentApp.java+

After having compiled the application, you can run it with the JVM.
There are several required options:

\begin{itemize}
\item
\verb+-char+ makes sure we deal with characters instead of strings of characters as a part of an n-gram

\item
one of the statistical estimators (see below); if none present, it won't pick any default one

\item
language parameter (it may seem awkward to require it for identification, but this will fixed,
so for now use anything for it, like ``foo''). Thus, the ``language'' is a typically two-to-four letter
abbrieviation of the language you are trying to train on (w.g. ``en'', ``es'', ``java'', etc.).

\item
corpus - a path to the corpus file for training. For testing just use ``bar''.
\end{itemize}

If you want an interactive mode to enter sentences yourself, use the \verb+-interactive+
option. E.g.:

\verb+java -cp marf.jar:. LangIdentApp --ident -char -interactive -bigram -add-delta foo bar+

Here are the options as per the application's output:

\vspace{15pt}
\hrule
\begin{verbatim}

Language Identification Application, $Revision: 1.4 $, $Date: 2009/02/22 02:16:00 $
Serguei A. Mokhov, mokhov@cs.concordia.ca
March 2003 - 2009


Usage:
    java LangIdentApp --help | -h
    java LangIdentApp --version
    java LangIdentApp --train [ --debug ] [ OPTIONS ] <language> <corpus-file>
    java LangIdentApp --ident [ --debug ] [ OPTIONS ] foo <bar|corpus-file>

Options (one or more of the following):
    -interactive   interactive mode for classification instead of reading from a file
    -char          use characters as n-grams (should always be present for this app)

    -unigram       use UNIGRAM model
    -bigram        use BIGRAM model
    -trigram       use TRIGRAM model

    -mle           use MLE
    -add-one       use Add-One smoothing
    -add-delta     use Add-Delta (ELE, d=0.5) smoothing
    -witten-bell   use Witten-Bell smoothing
    -good-turing   use Good-Turing smoothing

\end{verbatim}

\hrule
\vspace{15pt}

If the filename isn't specified, that will be stated and the usage
instructions above displayed.

\subsubsection{List of Files}

\paragraph{Directories}

\begin{itemize}

\item
\file{marf/nlp} -- that's where most of the code is for for use by this application, the \api{marf.nlp} package.
As discussed at the beginning, it has all the possible implementation files, but some of them are just
unimplemented stubs.

\item
\file{index/} -- this directory contains indexing of file names of corpora per language
(see \xs{sect:corpora-per-language})

\item
\file{test/} -- this directory contains testing files with sentences in various languages for testing
(see \xs{sect:lang-test-sentences})

\item
\file{expected/} -- this directory contains expected output files for classification
(see \xs{sect:lang-expected-classification})

%\item
%\file{corpora/} -- this directory contains all the corpora I had collected

\item
\file{training-*/} -- these directories contain all pre-trained models by us
for the described experiments and will be supplied in the training-sets tarballs.

\end{itemize}

\paragraph{Corpora per Language}
\label{sect:corpora-per-language}

The below is the list of files of ``pointers''
to the training corpora %under the \verb+/corpora+ directory
for the corresponding languages.

\begin{verbatim}
ar.train.corpora
bg.train.corpora
cpp.train.corpora
en.train.corpora
es.train.corpora
fr.train.corpora
he.train.corpora
it.train.corpora
java.train.corpora
perl.train.corpora
ru.train.corpora
\end{verbatim}

\paragraph{Expected Results}
\label{sect:lang-expected-classification}

The below files present the ideal results of batch
identification that correspond the \file{test.*.langs} files below,
and can be compared to those produced by the \tool{testing.sh} script.

\begin{verbatim}
expected.langs
expected.latin.langs
expected.non-latin.langs
expected.pls.langs
\end{verbatim}

\paragraph{Application}

The application and its makefile.

\begin{verbatim}
LangIdentApp.java
Makefile
marf.jar
\end{verbatim}

\paragraph{Scripts}

The wrapper scripts for batch training and testing.

\begin{verbatim}
testing.sh
training.sh
\end{verbatim}

\paragraph{Test Sentences}
\label{sect:lang-test-sentences}

\begin{itemize}
\item
\file{test.langs} --- the bag-of-languages

\item
\file{test.latin.langs} --- English, French, Spanish, and Italian sentences

\item
\file{test.non-latin.langs} --- Arabic, Hebrew, Russian, and Bulgarian sentences

\item
\file{test.pls.langs} --- Programming Languages (a few statements in {\cpp}, {\java}, and {\perl})
\end{itemize}

%%%%%%%%%%%%%%%%%%%%%%%%%%%%%%%%%%%%%%%%%%%%%%%%%%%%
%%%%%%%%%%%%%%%%%%%%%%%%%%%%%%%%%%%%%%%%%%%%%%%%%%%%

\subsubsection{Classification Results}
\label{sect:lang-classification-results}

\paragraph{``Bag-of-Languages'', Unrestricted Tokenizer}

\tiny
\hrule\vskip4pt
\begin{verbatim}
UNIGRAM, ADD-DELTA                                                 UNIGRAM, ADD-ONE

Ideal                        Actual                       Match    Ideal                        Actual                       Match

Language identified: [en]    Language identified: [en]    1        Language identified: [en]    Language identified: [en]    1
Language identified: [en]    Language identified: [en]    2        Language identified: [en]    Language identified: [en]    2
Language identified: [es]    Language identified: [fr]             Language identified: [es]    Language identified: [fr]
Language identified: [en]    Language identified: [en]    3        Language identified: [en]    Language identified: [en]    3
Language identified: [ru]    Language identified: [ru]    4        Language identified: [ru]    Language identified: [ru]    4
Language identified: [fr]    Language identified: [fr]    5        Language identified: [fr]    Language identified: [fr]    5
Language identified: [ru]    Language identified: [ru]    6        Language identified: [ru]    Language identified: [ru]    6
Language identified: [es]    Language identified: [es]    7        Language identified: [es]    Language identified: [es]    7
Language identified: [he]    Language identified: [he]    8        Language identified: [he]    Language identified: [he]    8
Language identified: [it]    Language identified: [it]    9        Language identified: [it]    Language identified: [it]    9
Language identified: [ar]    Language identified: [ru]             Language identified: [ar]    Language identified: [ru]
Language identified: [bg]    Language identified: [bg]    10       Language identified: [bg]    Language identified: [bg]    10
Language identified: [java]  Language identified: [es]             Language identified: [java]  Language identified: [es]
Language identified: [cpp]   Language identified: [perl]           Language identified: [cpp]   Language identified: [perl]
Language identified: [cpp]   Language identified: [perl]           Language identified: [cpp]   Language identified: [perl]
Language identified: [he]    Language identified: [he]    11       Language identified: [he]    Language identified: [he]    11
Language identified: [java]  Language identified: [it]             Language identified: [java]  Language identified: [it]
Language identified: [perl]  Language identified: [cpp]            Language identified: [perl]  Language identified: [cpp]
Language identified: [ar]    Language identified: [ar]    12       Language identified: [ar]    Language identified: [ar]    12
Language identified: [perl]  Language identified: [es]             Language identified: [perl]  Language identified: [es]
Language identified: [ru]    Language identified: [ru]    13       Language identified: [ru]    Language identified: [ru]    13
Language identified: [fr]    Language identified: [it]             Language identified: [fr]    Language identified: [it]
Language identified: [ar]    Language identified: [bg]             Language identified: [ar]    Language identified: [bg]
Language identified: [en]    Language identified: [en]    14       Language identified: [en]    Language identified: [en]    14

Total                        24                           14       Total                        24                           14
%                            58.33                                 %                            58.33
\end{verbatim}
\vskip4pt\hrule

\clearpage

\tiny
\hrule\vskip4pt
\begin{verbatim}
UNIGRAM, GOOD-TURING                                               UNIGRAM, MLE

Ideal                        Actual                       Match    Ideal                        Actual                       Match

Language identified: [en]    Language identified: [en]    1        Language identified: [en]    Language identified: [en]    1
Language identified: [en]    Language identified: [en]    2        Language identified: [en]    Language identified: [en]    2
Language identified: [es]    Language identified: [en]             Language identified: [es]    Language identified: [fr]
Language identified: [en]    Language identified: [en]    3        Language identified: [en]    Language identified: [en]    3
Language identified: [ru]    Language identified: [en]             Language identified: [ru]    Language identified: [bg]
Language identified: [fr]    Language identified: [en]             Language identified: [fr]    Language identified: [fr]    4
Language identified: [ru]    Language identified: [en]             Language identified: [ru]    Language identified: [ru]    5
Language identified: [es]    Language identified: [en]             Language identified: [es]    Language identified: [es]    6
Language identified: [he]    Language identified: [en]             Language identified: [he]    Language identified: [he]    7
Language identified: [it]    Language identified: [en]             Language identified: [it]    Language identified: [it]    8
Language identified: [ar]    Language identified: [en]             Language identified: [ar]    Language identified: [ru]
Language identified: [bg]    Language identified: [en]             Language identified: [bg]    Language identified: [bg]    9
Language identified: [java]  Language identified: [en]             Language identified: [java]  Language identified: [es]
Language identified: [cpp]   Language identified: [en]             Language identified: [cpp]   Language identified: [perl]
Language identified: [cpp]   Language identified: [en]             Language identified: [cpp]   Language identified: [perl]
Language identified: [he]    Language identified: [en]             Language identified: [he]    Language identified: [he]    10
Language identified: [java]  Language identified: [en]             Language identified: [java]  Language identified: [it]
Language identified: [perl]  Language identified: [en]             Language identified: [perl]  Language identified: [cpp]
Language identified: [ar]    Language identified: [en]             Language identified: [ar]    Language identified: [ar]    11
Language identified: [perl]  Language identified: [en]             Language identified: [perl]  Language identified: [es]
Language identified: [ru]    Language identified: [en]             Language identified: [ru]    Language identified: [ru]    12
Language identified: [fr]    Language identified: [en]             Language identified: [fr]    Language identified: [it]
Language identified: [ar]    Language identified: [en]             Language identified: [ar]    Language identified: [bg]
Language identified: [en]    Language identified: [en]    4        Language identified: [en]    Language identified: [en]    13

Total                        24                           4        Total                        24                           13
%                            16.67                                 %                            54.17
\end{verbatim}
\vskip4pt\hrule

\tiny
\hrule\vskip4pt
\begin{verbatim}
UNIGRAM, WITTEN-BELL                                               BIGRAM, ADD-DELTA

Ideal                        Actual                       Match    Ideal                        Actual                       Match

Language identified: [en]    Language identified: [en]    1        Language identified: [en]    Language identified: [bg]
Language identified: [en]    Language identified: [en]    2        Language identified: [en]    Language identified: [fr]
Language identified: [es]    Language identified: [en]             Language identified: [es]    Language identified: [es]    1
Language identified: [en]    Language identified: [en]    3        Language identified: [en]    Language identified: [bg]
Language identified: [ru]    Language identified: [en]             Language identified: [ru]    Language identified: [bg]
Language identified: [fr]    Language identified: [en]             Language identified: [fr]    Language identified: [fr]    2
Language identified: [ru]    Language identified: [en]             Language identified: [ru]    Language identified: [bg]
Language identified: [es]    Language identified: [en]             Language identified: [es]    Language identified: [fr]
Language identified: [he]    Language identified: [en]             Language identified: [he]    Language identified: [bg]
Language identified: [it]    Language identified: [en]             Language identified: [it]    Language identified: [fr]
Language identified: [ar]    Language identified: [en]             Language identified: [ar]    Language identified: [fr]
Language identified: [bg]    Language identified: [en]             Language identified: [bg]    Language identified: [es]
Language identified: [java]  Language identified: [en]             Language identified: [java]  Language identified: [fr]
Language identified: [cpp]   Language identified: [en]             Language identified: [cpp]   Language identified: [perl]
Language identified: [cpp]   Language identified: [en]             Language identified: [cpp]   Language identified: [cpp]   3
Language identified: [he]    Language identified: [en]             Language identified: [he]    Language identified: [bg]
Language identified: [java]  Language identified: [en]             Language identified: [java]  Language identified: [fr]
Language identified: [perl]  Language identified: [en]             Language identified: [perl]  Language identified: [es]
Language identified: [ar]    Language identified: [en]             Language identified: [ar]    Language identified: [bg]
Language identified: [perl]  Language identified: [en]             Language identified: [perl]  Language identified: [fr]
Language identified: [ru]    Language identified: [en]             Language identified: [ru]    Language identified: [bg]
Language identified: [fr]    Language identified: [en]             Language identified: [fr]    Language identified: [es]
Language identified: [ar]    Language identified: [en]             Language identified: [ar]    Language identified: [es]
Language identified: [en]    Language identified: [en]    4        Language identified: [en]    Language identified: [bg]

Total                        24                           4        Total                        24                           3
%                            16.67                                 %                            12.50
\end{verbatim}
\vskip4pt\hrule

\clearpage

\tiny
\hrule\vskip4pt
\begin{verbatim}
BIGRAM,ADD-ONE                                                     BIGRAM, GOOD-TURING

Ideal                        Actual                       Match    Ideal                        Actual                       Match

Language identified: [en]    Language identified: [bg]             Language identified: [en]    Language identified: [en]    1
Language identified: [en]    Language identified: [fr]             Language identified: [en]    Language identified: [perl]
Language identified: [es]    Language identified: [es]    1        Language identified: [es]    Language identified: [perl]
Language identified: [en]    Language identified: [bg]             Language identified: [en]    Language identified: [perl]
Language identified: [ru]    Language identified: [bg]             Language identified: [ru]    Language identified: [perl]
Language identified: [fr]    Language identified: [fr]    2        Language identified: [fr]    Language identified: [perl]
Language identified: [ru]    Language identified: [bg]             Language identified: [ru]    Language identified: [perl]
Language identified: [es]    Language identified: [fr]             Language identified: [es]    Language identified: [perl]
Language identified: [he]    Language identified: [bg]             Language identified: [he]    Language identified: [perl]
Language identified: [it]    Language identified: [fr]             Language identified: [it]    Language identified: [perl]
Language identified: [ar]    Language identified: [fr]             Language identified: [ar]    Language identified: [perl]
Language identified: [bg]    Language identified: [es]             Language identified: [bg]    Language identified: [perl]
Language identified: [java]  Language identified: [fr]             Language identified: [java]  Language identified: [perl]
Language identified: [cpp]   Language identified: [perl]           Language identified: [cpp]   Language identified: [perl]
Language identified: [cpp]   Language identified: [cpp]   3        Language identified: [cpp]   Language identified: [perl]
Language identified: [he]    Language identified: [bg]             Language identified: [he]    Language identified: [perl]
Language identified: [java]  Language identified: [fr]             Language identified: [java]  Language identified: [perl]
Language identified: [perl]  Language identified: [es]             Language identified: [perl]  Language identified: [perl]  2
Language identified: [ar]    Language identified: [bg]             Language identified: [ar]    Language identified: [perl]
Language identified: [perl]  Language identified: [fr]             Language identified: [perl]  Language identified: [perl]  3
Language identified: [ru]    Language identified: [bg]             Language identified: [ru]    Language identified: [perl]
Language identified: [fr]    Language identified: [es]             Language identified: [fr]    Language identified: [perl]
Language identified: [ar]    Language identified: [es]             Language identified: [ar]    Language identified: [perl]
Language identified: [en]    Language identified: [bg]             Language identified: [en]    Language identified: [perl]

Total                        24                           3        Total                        24                           3
%                            12.50                                 %                            12.50
\end{verbatim}
\vskip4pt\hrule

\tiny
\hrule\vskip4pt
\begin{verbatim}
BIGRAM, MLE                                                        BIGRAM, WITTEN-BELL

Ideal                        Actual                       Match    Ideal                        Actual                       Match

Language identified: [en]    Language identified: [en]    1        Language identified: [en]    Language identified: [java]
Language identified: [en]    Language identified: [en]    2        Language identified: [en]    Language identified: [java]
Language identified: [es]    Language identified: [en]             Language identified: [es]    Language identified: [es]    1
Language identified: [en]    Language identified: [en]    3        Language identified: [en]    Language identified: [java]
Language identified: [ru]    Language identified: [en]             Language identified: [ru]    Language identified: [java]
Language identified: [fr]    Language identified: [en]             Language identified: [fr]    Language identified: [es]
Language identified: [ru]    Language identified: [en]             Language identified: [ru]    Language identified: [java]
Language identified: [es]    Language identified: [en]             Language identified: [es]    Language identified: [ar]
Language identified: [he]    Language identified: [en]             Language identified: [he]    Language identified: [ar]
Language identified: [it]    Language identified: [en]             Language identified: [it]    Language identified: [it]    2
Language identified: [ar]    Language identified: [en]             Language identified: [ar]    Language identified: [ar]    3
Language identified: [bg]    Language identified: [en]             Language identified: [bg]    Language identified: [java]
Language identified: [java]  Language identified: [en]             Language identified: [java]  Language identified: [es]
Language identified: [cpp]   Language identified: [en]             Language identified: [cpp]   Language identified: [java]
Language identified: [cpp]   Language identified: [en]             Language identified: [cpp]   Language identified: [cpp]   4
Language identified: [he]    Language identified: [en]             Language identified: [he]    Language identified: [ru]
Language identified: [java]  Language identified: [en]             Language identified: [java]  Language identified: [java]  5
Language identified: [perl]  Language identified: [en]             Language identified: [perl]  Language identified: [ru]
Language identified: [ar]    Language identified: [en]             Language identified: [ar]    Language identified: [ar]    6
Language identified: [perl]  Language identified: [en]             Language identified: [perl]  Language identified: [java]
Language identified: [ru]    Language identified: [en]             Language identified: [ru]    Language identified: [ar]
Language identified: [fr]    Language identified: [en]             Language identified: [fr]    Language identified: [java]
Language identified: [ar]    Language identified: [en]             Language identified: [ar]    Language identified: [ar]    7
Language identified: [en]    Language identified: [en]    4        Language identified: [en]    Language identified: [java]

Total                        24                           4        Total                        24                           7
%                            16.67                                 %                            29.17
\end{verbatim}
\vskip4pt\hrule

\clearpage

\tiny
\hrule\vskip4pt
\begin{verbatim}
TRIGRAM, ADD-DELTA                                                 TRIGRAM, ADD-ONE

Ideal                        Actual                       Match    Ideal                        Actual                       Match

Language identified: [en]    Language identified: [he]             Language identified: [en]    Language identified: [he]
Language identified: [en]    Language identified: [fr]             Language identified: [en]    Language identified: [fr]
Language identified: [es]    Language identified: [fr]             Language identified: [es]    Language identified: [fr]
Language identified: [en]    Language identified: [he]             Language identified: [en]    Language identified: [he]
Language identified: [ru]    Language identified: [he]             Language identified: [ru]    Language identified: [he]
Language identified: [fr]    Language identified: [fr]    1        Language identified: [fr]    Language identified: [fr]    1
Language identified: [ru]    Language identified: [he]             Language identified: [ru]    Language identified: [he]
Language identified: [es]    Language identified: [fr]             Language identified: [es]    Language identified: [fr]
Language identified: [he]    Language identified: [ar]             Language identified: [he]    Language identified: [ar]
Language identified: [it]    Language identified: [fr]             Language identified: [it]    Language identified: [fr]
Language identified: [ar]    Language identified: [fr]             Language identified: [ar]    Language identified: [fr]
Language identified: [bg]    Language identified: [fr]             Language identified: [bg]    Language identified: [fr]
Language identified: [java]  Language identified: [fr]             Language identified: [java]  Language identified: [fr]
Language identified: [cpp]   Language identified: [java]           Language identified: [cpp]   Language identified: [java]
Language identified: [cpp]   Language identified: [java]           Language identified: [cpp]   Language identified: [java]
Language identified: [he]    Language identified: [he]    2        Language identified: [he]    Language identified: [he]    2
Language identified: [java]  Language identified: [fr]             Language identified: [java]  Language identified: [fr]
Language identified: [perl]  Language identified: [es]             Language identified: [perl]  Language identified: [es]
Language identified: [ar]    Language identified: [ar]    3        Language identified: [ar]    Language identified: [ar]    3
Language identified: [perl]  Language identified: [fr]             Language identified: [perl]  Language identified: [fr]
Language identified: [ru]    Language identified: [ar]             Language identified: [ru]    Language identified: [ar]
Language identified: [fr]    Language identified: [fr]    4        Language identified: [fr]    Language identified: [fr]    4
Language identified: [ar]    Language identified: [fr]             Language identified: [ar]    Language identified: [fr]
Language identified: [en]    Language identified: [he]             Language identified: [en]    Language identified: [he]

Total                        24                           4        Total                        24                           4
%                            16.67                                 %                            16.67
\end{verbatim}
\vskip4pt\hrule

\tiny
\hrule\vskip4pt
\begin{verbatim}
TRIGRAM, GOOD-TURING                                               TRIGRAM, MLE

Ideal                        Actual                       Match    Ideal                        Actual                       Match

Language identified: [en]    Language identified: [he]             Language identified: [en]    Language identified: [en]    1
Language identified: [en]    Language identified: [he]             Language identified: [en]    Language identified: [en]    2
Language identified: [es]    Language identified: [ar]             Language identified: [es]    Language identified: [en]
Language identified: [en]    Language identified: [he]             Language identified: [en]    Language identified: [en]    3
Language identified: [ru]    Language identified: [he]             Language identified: [ru]    Language identified: [en]
Language identified: [fr]    Language identified: [ar]             Language identified: [fr]    Language identified: [en]
Language identified: [ru]    Language identified: [he]             Language identified: [ru]    Language identified: [en]
Language identified: [es]    Language identified: [ar]             Language identified: [es]    Language identified: [en]
Language identified: [he]    Language identified: [ar]             Language identified: [he]    Language identified: [en]
Language identified: [it]    Language identified: [he]             Language identified: [it]    Language identified: [en]
Language identified: [ar]    Language identified: [ar]    1        Language identified: [ar]    Language identified: [en]
Language identified: [bg]    Language identified: [es]             Language identified: [bg]    Language identified: [en]
Language identified: [java]  Language identified: [ru]             Language identified: [java]  Language identified: [en]
Language identified: [cpp]   Language identified: [ar]             Language identified: [cpp]   Language identified: [en]
Language identified: [cpp]   Language identified: [ar]             Language identified: [cpp]   Language identified: [en]
Language identified: [he]    Language identified: [he]    2        Language identified: [he]    Language identified: [en]
Language identified: [java]  Language identified: [he]             Language identified: [java]  Language identified: [en]
Language identified: [perl]  Language identified: [ar]             Language identified: [perl]  Language identified: [en]
Language identified: [ar]    Language identified: [ar]    3        Language identified: [ar]    Language identified: [en]
Language identified: [perl]  Language identified: [ar]             Language identified: [perl]  Language identified: [en]
Language identified: [ru]    Language identified: [ar]             Language identified: [ru]    Language identified: [en]
Language identified: [fr]    Language identified: [he]             Language identified: [fr]    Language identified: [en]
Language identified: [ar]    Language identified: [ar]    4        Language identified: [ar]    Language identified: [en]
Language identified: [en]    Language identified: [he]             Language identified: [en]    Language identified: [en]    4

Total                        24                           4        Total                        24                           4
%                            16.67                                 %                            16.67
\end{verbatim}
\vskip4pt\hrule

\clearpage

\tiny
\hrule\vskip4pt
\begin{verbatim}
TRIGRAM, WITTEN-BELL

Ideal                        Actual                       Match

Language identified: [en]    Language identified: [en]    1
Language identified: [en]    Language identified: [en]    2
Language identified: [es]    Language identified: [en]
Language identified: [en]    Language identified: [en]    3
Language identified: [ru]    Language identified: [en]
Language identified: [fr]    Language identified: [en]
Language identified: [ru]    Language identified: [en]
Language identified: [es]    Language identified: [en]
Language identified: [he]    Language identified: [en]
Language identified: [it]    Language identified: [en]
Language identified: [ar]    Language identified: [en]
Language identified: [bg]    Language identified: [en]
Language identified: [java]  Language identified: [en]
Language identified: [cpp]   Language identified: [en]
Language identified: [cpp]   Language identified: [en]
Language identified: [he]    Language identified: [en]
Language identified: [java]  Language identified: [en]
Language identified: [perl]  Language identified: [en]
Language identified: [ar]    Language identified: [en]
Language identified: [perl]  Language identified: [en]
Language identified: [ru]    Language identified: [en]
Language identified: [fr]    Language identified: [en]
Language identified: [ar]    Language identified: [en]
Language identified: [ar]    Language identified: [en]    4

Total                        24                           4
%                            16.67
\end{verbatim}
\vskip4pt\hrule

\clearpage

\paragraph{Non-Latin-Based Languages, Unrestricted Tokenizer}

\tiny
\hrule\vskip4pt
\begin{verbatim}
UNIGRAM, ADD-DELTA                                                 UNIGRAM, ADD-ONE

Ideal                        Actual                       Match    Ideal                        Actual                       Match

Language identified: [ru]    Language identified: [ru]    1        Language identified: [ru]    Language identified: [ru]    1
Language identified: [ru]    Language identified: [ru]    2        Language identified: [ru]    Language identified: [ru]    2
Language identified: [he]    Language identified: [he]    3        Language identified: [he]    Language identified: [he]    3
Language identified: [ar]    Language identified: [ru]             Language identified: [ar]    Language identified: [ru]
Language identified: [bg]    Language identified: [bg]    4        Language identified: [bg]    Language identified: [bg]    4
Language identified: [he]    Language identified: [he]    5        Language identified: [he]    Language identified: [he]    5
Language identified: [ar]    Language identified: [ar]    6        Language identified: [ar]    Language identified: [ar]    6
Language identified: [ru]    Language identified: [ru]    7        Language identified: [ru]    Language identified: [ru]    7
Language identified: [ar]    Language identified: [bg]             Language identified: [ar]    Language identified: [bg]

Total                        9                            7        Total                        9                            7
%                            77.78                                 %                            77.78
\end{verbatim}
\vskip4pt\hrule

\tiny
\hrule\vskip4pt
\begin{verbatim}
UNIGRAM, GOOD-TURING                                               UNIGRAM, MLE

Ideal                        Actual                       Match    Ideal                        Actual                       Match

Language identified: [ru]    Language identified: [ru]    1        Language identified: [ru]    Language identified: [bg]
Language identified: [ru]    Language identified: [ru]    2        Language identified: [ru]    Language identified: [ru]    1
Language identified: [he]    Language identified: [ru]             Language identified: [he]    Language identified: [he]    2
Language identified: [ar]    Language identified: [ru]             Language identified: [ar]    Language identified: [ru]
Language identified: [bg]    Language identified: [bg]    3        Language identified: [bg]    Language identified: [bg]    3
Language identified: [he]    Language identified: [ru]             Language identified: [he]    Language identified: [he]    4
Language identified: [ar]    Language identified: [ar]    4        Language identified: [ar]    Language identified: [ar]    5
Language identified: [ru]    Language identified: [ru]    5        Language identified: [ru]    Language identified: [ru]    6
Language identified: [ar]    Language identified: [ru]             Language identified: [ar]    Language identified: [bg]

Total                        9                            5        Total                        9                            6
%                            55.56                                 %                            66.67
\end{verbatim}
\vskip4pt\hrule

\tiny
\hrule\vskip4pt
\begin{verbatim}
UNIGRAM, WITTEN-BELL                                               BIGRAM, ADD-DELTA

Ideal                        Actual                       Match    Ideal                        Actual                       Match

Language identified: [ru]    Language identified: [ru]    1        Language identified: [ru]    Language identified: [bg]
Language identified: [ru]    Language identified: [ru]    2        Language identified: [ru]    Language identified: [bg]
Language identified: [he]    Language identified: [ru]             Language identified: [he]    Language identified: [bg]
Language identified: [ar]    Language identified: [ru]             Language identified: [ar]    Language identified: [bg]
Language identified: [bg]    Language identified: [bg]    3        Language identified: [bg]    Language identified: [bg]    1
Language identified: [he]    Language identified: [ru]             Language identified: [he]    Language identified: [bg]
Language identified: [ar]    Language identified: [ar]    4        Language identified: [ar]    Language identified: [bg]
Language identified: [ru]    Language identified: [ru]    5        Language identified: [ru]    Language identified: [bg]
Language identified: [ar]    Language identified: [ru]             Language identified: [ar]    Language identified: [bg]

Total                        9                            5        Total                        9                            1
%                            55.56                                 %                            11.11
\end{verbatim}
\vskip4pt\hrule

\tiny
\hrule\vskip4pt
\begin{verbatim}
BIGRAM, ADD-ONE                                                    BIGRAM, GOOD-TURING

Ideal                        Actual                       Match    Ideal                        Actual                       Match

Language identified: [ru]    Language identified: [bg]             Language identified: [ru]    Language identified: [ru]    1
Language identified: [ru]    Language identified: [bg]             Language identified: [ru]    Language identified: [ru]    2
Language identified: [he]    Language identified: [bg]             Language identified: [he]    Language identified: [ar]
Language identified: [ar]    Language identified: [bg]             Language identified: [ar]    Language identified: [ar]
Language identified: [bg]    Language identified: [bg]    1        Language identified: [bg]    Language identified: [bg]    3
Language identified: [he]    Language identified: [bg]             Language identified: [he]    Language identified: [ru]
Language identified: [ar]    Language identified: [bg]             Language identified: [ar]    Language identified: [ar]    4
Language identified: [ru]    Language identified: [bg]             Language identified: [ru]    Language identified: [ru]    5
Language identified: [ar]    Language identified: [bg]             Language identified: [ar]    Language identified: [ru]

Total                        9                            1        Total                        9                            5
%                            11.11                                 %                            55.56
\end{verbatim}
\vskip4pt\hrule

\clearpage

\tiny
\hrule\vskip4pt
\begin{verbatim}
BIGRAM, MLE                                                        BIGRAM, WITTEN-BELL

Ideal                        Actual                       Match    Ideal                        Actual                       Match

Language identified: [ru]    Language identified: [ru]    1        Language identified: [ru]    Language identified: [ru]    1
Language identified: [ru]    Language identified: [ru]    2        Language identified: [ru]    Language identified: [ru]    2
Language identified: [he]    Language identified: [ru]             Language identified: [he]    Language identified: [ar]
Language identified: [ar]    Language identified: [ru]             Language identified: [ar]    Language identified: [ar]    3
Language identified: [bg]    Language identified: [ru]             Language identified: [bg]    Language identified: [bg]    4
Language identified: [he]    Language identified: [ru]             Language identified: [he]    Language identified: [ru]
Language identified: [ar]    Language identified: [ru]             Language identified: [ar]    Language identified: [ar]    5
Language identified: [ru]    Language identified: [ru]    3        Language identified: [ru]    Language identified: [ar]
Language identified: [ar]    Language identified: [ru]             Language identified: [ar]    Language identified: [ar]    6

Total                        9                            3        Total                        9                            6
%                            33.33                                 %                            66.67
\end{verbatim}
\vskip4pt\hrule

\tiny
\hrule\vskip4pt
\begin{verbatim}
TRIGRAM, ADD-DELTA                                                 TRIGRAM, ADD-ONE

Ideal                        Actual                       Match    Ideal                        Actual                       Match

Language identified: [ru]    Language identified: [he]             Language identified: [ru]    Language identified: [he]
Language identified: [ru]    Language identified: [he]             Language identified: [ru]    Language identified: [he]
Language identified: [he]    Language identified: [ar]             Language identified: [he]    Language identified: [ar]
Language identified: [ar]    Language identified: [ar]    1        Language identified: [ar]    Language identified: [ar]    1
Language identified: [bg]    Language identified: [bg]    2        Language identified: [bg]    Language identified: [bg]    2
Language identified: [he]    Language identified: [he]    3        Language identified: [he]    Language identified: [he]    3
Language identified: [ar]    Language identified: [ar]    4        Language identified: [ar]    Language identified: [ar]    4
Language identified: [ru]    Language identified: [ar]             Language identified: [ru]    Language identified: [ar]
Language identified: [ar]    Language identified: [ar]    5        Language identified: [ar]    Language identified: [ar]    5

Total                        9                            5        Total                        9                            5
%                            55.56                                 %                            55.56
\end{verbatim}
\vskip4pt\hrule

\tiny
\hrule\vskip4pt
\begin{verbatim}
TRIGRAM, GOOD-TURING                                               TRIGRAM, MLE

Ideal                        Actual                       Match    Ideal                        Actual                       Match

Language identified: [ru]    Language identified: [he]             Language identified: [ru]    Language identified: [ru]    1
Language identified: [ru]    Language identified: [he]             Language identified: [ru]    Language identified: [ru]    2
Language identified: [he]    Language identified: [ar]             Language identified: [he]    Language identified: [ru]
Language identified: [ar]    Language identified: [ar]    1        Language identified: [ar]    Language identified: [ru]
Language identified: [bg]    Language identified: [bg]    2        Language identified: [bg]    Language identified: [ru]
Language identified: [he]    Language identified: [he]    3        Language identified: [he]    Language identified: [ru]
Language identified: [ar]    Language identified: [ar]    4        Language identified: [ar]    Language identified: [ru]
Language identified: [ru]    Language identified: [ar]             Language identified: [ru]    Language identified: [ru]    3
Language identified: [ar]    Language identified: [ar]    5        Language identified: [ar]    Language identified: [ru]

Total                        9                            5        Total                        9                            3
%                            55.56                                 %                            33.33
\end{verbatim}
\vskip4pt\hrule

\tiny
\hrule\vskip4pt
\begin{verbatim}
TRIGRAM, WITTEN-BELL

Ideal                        Actual                       Match

Language identified: [ru]    Language identified: [ru]    1
Language identified: [ru]    Language identified: [ru]    2
Language identified: [he]    Language identified: [ru]
Language identified: [ar]    Language identified: [ru]
Language identified: [bg]    Language identified: [ru]
Language identified: [he]    Language identified: [ru]
Language identified: [ar]    Language identified: [ru]
Language identified: [ru]    Language identified: [ru]    3
Language identified: [ar]    Language identified: [ru]

Total                        9                            3
%                            33.33
\end{verbatim}
\vskip4pt\hrule

\clearpage

\paragraph{Programming Languages, Unrestricted Tokenizer}

\tiny
\hrule\vskip4pt
\begin{verbatim}
UNIGRAM,ADD-DELTA                UNIGRAM,ADD-ONE

Ideal                          Actual                         Match  Ideal                          Actual                         Match

Language identified: [java]    Language identified: [java]    1      Language identified: [java]    Language identified: [java]    1
Language identified: [cpp]     Language identified: [perl]           Language identified: [cpp]     Language identified: [perl]
Language identified: [cpp]     Language identified: [perl]           Language identified: [cpp]     Language identified: [perl]
Language identified: [java]    Language identified: [java]    2      Language identified: [java]    Language identified: [java]    2
Language identified: [perl]    Language identified: [cpp]            Language identified: [perl]    Language identified: [cpp]
Language identified: [perl]    Language identified: [cpp]            Language identified: [perl]    Language identified: [cpp]

Total    6    2        Total    6    2
%    33.33            %    33.33
\end{verbatim}
\vskip4pt\hrule

\tiny
\hrule\vskip4pt
\begin{verbatim}
UNIGRAM,GOOD-TURING                UNIGRAM,MLE

Ideal                          Actual                         Match  Ideal                          Actual                         Match

Language identified: [java]    Language identified: [perl]           Language identified: [java]    Language identified: [java]    1
Language identified: [cpp]     Language identified: [perl]           Language identified: [cpp]     Language identified: [perl]
Language identified: [cpp]     Language identified: [perl]           Language identified: [cpp]     Language identified: [perl]
Language identified: [java]    Language identified: [java]    1      Language identified: [java]    Language identified: [java]    2
Language identified: [perl]    Language identified: [java]           Language identified: [perl]    Language identified: [cpp]
Language identified: [perl]    Language identified: [perl]    2      Language identified: [perl]    Language identified: [cpp]

Total    6    2        Total    6    2
%    33.33            %    33.33
\end{verbatim}
\vskip4pt\hrule

\tiny
\hrule\vskip4pt
\begin{verbatim}
UNIGRAM,WITTEN-BELL                                                  BIGRAM,ADD-DELTA

Ideal                          Actual                         Match  Ideal                          Actual                         Match

Language identified: [java]    Language identified: [perl]           Language identified: [java]    Language identified: [perl]
Language identified: [cpp]     Language identified: [perl]           Language identified: [cpp]     Language identified: [perl]
Language identified: [cpp]     Language identified: [perl]           Language identified: [cpp]     Language identified: [cpp]     1
Language identified: [java]    Language identified: [java]    1      Language identified: [java]    Language identified: [perl]
Language identified: [perl]    Language identified: [java]           Language identified: [perl]    Language identified: [perl]    2
Language identified: [perl]    Language identified: [perl]    2      Language identified: [perl]    Language identified: [perl]    3

Total                          6                              2      Total                          6                              3
%                              33.33                                 %                              50.00
\end{verbatim}
\vskip4pt\hrule

\tiny
\hrule\vskip4pt
\begin{verbatim}
BIGRAM,ADD-ONE                                                       BIGRAM,GOOD-TURING

Ideal                          Actual                         Match  Ideal                          Actual                         Match

Language identified: [java]    Language identified: [perl]           Language identified: [java]    Language identified: [perl]
Language identified: [cpp]     Language identified: [perl]           Language identified: [cpp]     Language identified: [perl]
Language identified: [cpp]     Language identified: [cpp]     1      Language identified: [cpp]     Language identified: [perl]
Language identified: [java]    Language identified: [perl]           Language identified: [java]    Language identified: [perl]
Language identified: [perl]    Language identified: [perl]    2      Language identified: [perl]    Language identified: [perl]    1
Language identified: [perl]    Language identified: [perl]    3      Language identified: [perl]    Language identified: [perl]    2

Total                          6                              3      Total                          6                              2
%                              50.00                                 %                              33.33
\end{verbatim}
\vskip4pt\hrule

\clearpage

\tiny
\hrule\vskip4pt
\begin{verbatim}
BIGRAM,MLE                                                           BIGRAM,WITTEN-BELL

Ideal                          Actual                         Match  Ideal                          Actual                         Match

Language identified: [java]    Language identified: [java]    1      Language identified: [java]    Language identified: [java]    1
Language identified: [cpp]     Language identified: [java]           Language identified: [cpp]     Language identified: [java]
Language identified: [cpp]     Language identified: [java]           Language identified: [cpp]     Language identified: [cpp]     2
Language identified: [java]    Language identified: [java]    2      Language identified: [java]    Language identified: [java]    3
Language identified: [perl]    Language identified: [java]           Language identified: [perl]    Language identified: [java]
Language identified: [perl]    Language identified: [java]           Language identified: [perl]    Language identified: [java]

Total                          6                              2      Total                          6                              3
%                              33.33                                 %                              50.00
\end{verbatim}
\vskip4pt\hrule

\tiny
\hrule\vskip4pt
\begin{verbatim}
TRIGRAM,ADD-DELTA                                                    TRIGRAM,ADD-ONE

Ideal                          Actual                         Match  Ideal                          Actual                         Match

Language identified: [java]    Language identified: [java]    1      Language identified: [java]    Language identified: [java]    1
Language identified: [cpp]     Language identified: [java]           Language identified: [cpp]     Language identified: [java]
Language identified: [cpp]     Language identified: [java]           Language identified: [cpp]     Language identified: [java]
Language identified: [java]    Language identified: [java]    2      Language identified: [java]    Language identified: [java]    2
Language identified: [perl]    Language identified: [java]           Language identified: [perl]    Language identified: [java]
Language identified: [perl]    Language identified: [java]           Language identified: [perl]    Language identified: [java]

Total                          6                              2      Total                          6                              2
%                              33.33                                 %                              33.33
\end{verbatim}
\vskip4pt\hrule

\tiny
\hrule\vskip4pt
\begin{verbatim}
TRIGRAM,GOOD-TURING                                                  TRIGRAM,MLE

Ideal                          Actual                         Match  Ideal                          Actual                         Match

Language identified: [java]    Language identified: [perl]           Language identified: [java]    Language identified: [java]    1
Language identified: [cpp]     Language identified: [perl]           Language identified: [cpp]     Language identified: [java]
Language identified: [cpp]     Language identified: [perl]           Language identified: [cpp]     Language identified: [java]
Language identified: [java]    Language identified: [perl]           Language identified: [java]    Language identified: [java]    2
Language identified: [perl]    Language identified: [perl]    1      Language identified: [perl]    Language identified: [java]
Language identified: [perl]    Language identified: [perl]    2      Language identified: [perl]    Language identified: [java]

Total                          6                              2      Total                          6                              2
%                              33.33                                 %                              33.33
\end{verbatim}
\vskip4pt\hrule

\tiny
\hrule\vskip4pt
\begin{verbatim}
TRIGRAM,WITTEN-BELL

Ideal                          Actual                         Match

Language identified: [java]    Language identified: [java]    1
Language identified: [cpp]     Language identified: [java]
Language identified: [cpp]     Language identified: [java]
Language identified: [java]    Language identified: [java]    2
Language identified: [perl]    Language identified: [java]
Language identified: [perl]    Language identified: [java]

Total                          6                              2
%                              33.33
\end{verbatim}
\vskip4pt\hrule

\clearpage

%%%%%%%%%%%%%%%%%%%%%%%%%%%%%%%%%%%%%%%%%%%%%%%%%%%%%%%%%%%%%%%%%%%%%%%%%%%%%%%%%

\paragraph{Latin-Based Languages, Restricted Tokenizer}

\tiny
\hrule\vskip4pt
\begin{verbatim}
unigram,add-delta                                                unigram,add-one

Ideal                        Actual                       Match  Ideal                        Actual                       Match

Language identified: [en]    Language identified: [en]    1      Language identified: [en]    Language identified: [en]    1
Language identified: [en]    Language identified: [en]    2      Language identified: [en]    Language identified: [en]    2
Language identified: [es]    Language identified: [fr]           Language identified: [es]    Language identified: [fr]
Language identified: [en]    Language identified: [en]    3      Language identified: [en]    Language identified: [en]    3
Language identified: [fr]    Language identified: [fr]    4      Language identified: [fr]    Language identified: [fr]    4
Language identified: [es]    Language identified: [es]    5      Language identified: [es]    Language identified: [es]    5
Language identified: [it]    Language identified: [it]    6      Language identified: [it]    Language identified: [it]    6
Language identified: [fr]    Language identified: [it]           Language identified: [fr]    Language identified: [it]
Language identified: [en]    Language identified: [en]    7      Language identified: [en]    Language identified: [en]    7

Total                        9                            7      Total                        9                            7
%                            77.78                               %                            77.78

\tiny
\hrule\vskip4pt
\begin{verbatim}
unigram,good-turing                                              unigram,mle

Ideal                        Actual                       Match  Ideal                        Actual                       Match

Language identified: [en]    Language identified: [en]    1      Language identified: [en]    Language identified: [en]    1
Language identified: [en]    Language identified: [en]    2      Language identified: [en]    Language identified: [en]    2
Language identified: [es]    Language identified: [en]           Language identified: [es]    Language identified: [fr]
Language identified: [en]    Language identified: [en]    3      Language identified: [en]    Language identified: [en]    3
Language identified: [fr]    Language identified: [en]           Language identified: [fr]    Language identified: [fr]    4
Language identified: [es]    Language identified: [en]           Language identified: [es]    Language identified: [es]    5
Language identified: [it]    Language identified: [en]           Language identified: [it]    Language identified: [it]    6
Language identified: [fr]    Language identified: [en]           Language identified: [fr]    Language identified: [it]
Language identified: [en]    Language identified: [en]    4      Language identified: [en]    Language identified: [en]    7

Total                        9                            4      Total                        9                            7
%                            44.44                               %                            77.78
\end{verbatim}
\vskip4pt\hrule

\tiny
\hrule\vskip4pt
\begin{verbatim}
unigram,witten-bell                                              bigram,add-one

Ideal                        Actual                       Match  Ideal                        Actual                       Match

Language identified: [en]    Language identified: [en]    1      Language identified: [en]    Language identified: [en]    1
Language identified: [en]    Language identified: [en]    2      Language identified: [en]    Language identified: [es]
Language identified: [es]    Language identified: [en]           Language identified: [es]    Language identified: [es]    2
Language identified: [en]    Language identified: [en]    3      Language identified: [en]    Language identified: [en]    3
Language identified: [fr]    Language identified: [en]           Language identified: [fr]    Language identified: [es]
Language identified: [es]    Language identified: [en]           Language identified: [es]    Language identified: [fr]
Language identified: [it]    Language identified: [en]           Language identified: [it]    Language identified: [es]
Language identified: [fr]    Language identified: [en]           Language identified: [fr]    Language identified: [es]
Language identified: [en]    Language identified: [en]    4      Language identified: [en]    Language identified: [en]    4

Total                        9                            4      Total                        9                            4
%                            44.44                               %                            44.44
\end{verbatim}
\vskip4pt\hrule

\clearpage

\tiny
\hrule\vskip4pt
\begin{verbatim}
bigram,add-one                                                   bigram,good-turing

Ideal                        Actual                       Match  Ideal                        Actual                       Match

Language identified: [en]    Language identified: [en]    1      Language identified: [en]    Language identified: [en]    1
Language identified: [en]    Language identified: [es]           Language identified: [en]    Language identified: [en]    2
Language identified: [es]    Language identified: [es]    2      Language identified: [es]    Language identified: [en]
Language identified: [en]    Language identified: [en]    3      Language identified: [en]    Language identified: [en]    3
Language identified: [fr]    Language identified: [es]           Language identified: [fr]    Language identified: [en]
Language identified: [es]    Language identified: [es]    4      Language identified: [es]    Language identified: [fr]
Language identified: [it]    Language identified: [es]           Language identified: [it]    Language identified: [en]
Language identified: [fr]    Language identified: [es]           Language identified: [fr]    Language identified: [en]
Language identified: [en]    Language identified: [en]    5      Language identified: [en]    Language identified: [en]    4

Total                        9                            5      Total                        9                            4
%                            55.56                               %                            44.44
\end{verbatim}
\vskip4pt\hrule

\tiny
\hrule\vskip4pt
\begin{verbatim}
bigram,mle                                                       bigram,witten-belll

Ideal                        Actual                       Match  Ideal                        Actual                       Match

Language identified: [en]    Language identified: [en]    1      Language identified: [en]    Language identified: [en]    1
Language identified: [en]    Language identified: [en]    2      Language identified: [en]    Language identified: [en]    2
Language identified: [es]    Language identified: [en]           Language identified: [es]    Language identified: [en]
Language identified: [en]    Language identified: [en]    3      Language identified: [en]    Language identified: [en]    3
Language identified: [fr]    Language identified: [en]           Language identified: [fr]    Language identified: [en]
Language identified: [es]    Language identified: [en]           Language identified: [es]    Language identified: [es]
Language identified: [it]    Language identified: [en]           Language identified: [it]    Language identified: [en]
Language identified: [fr]    Language identified: [en]           Language identified: [fr]    Language identified: [en]
Language identified: [en]    Language identified: [en]    4      Language identified: [en]    Language identified: [en]    4

Total                        9                            4      Total                        9                            4
%                            44.44                               %                            44.44
\end{verbatim}
\vskip4pt\hrule

\tiny
\hrule\vskip4pt
\begin{verbatim}
trigram,add-delta                                                trigram,add-one

Ideal                        Actual                       Match  Ideal                        Actual                       Match

Language identified: [en]    Language identified: [en]    1      Language identified: [en]    Language identified: [en]    1
Language identified: [en]    Language identified: [es]           Language identified: [en]    Language identified: [es]
Language identified: [es]    Language identified: [es]    2      Language identified: [es]    Language identified: [es]    2
Language identified: [en]    Language identified: [en]    3      Language identified: [en]    Language identified: [en]    3
Language identified: [fr]    Language identified: [es]           Language identified: [fr]    Language identified: [es]
Language identified: [es]    Language identified: [es]    4      Language identified: [es]    Language identified: [es]    4
Language identified: [it]    Language identified: [es]           Language identified: [it]    Language identified: [es]
Language identified: [fr]    Language identified: [es]           Language identified: [fr]    Language identified: [es]
Language identified: [en]    Language identified: [en]    5      Language identified: [en]    Language identified: [en]    5

Total                        9                            5      Total                        9                            5
%                            55.56                               %                            55.56
\end{verbatim}
\vskip4pt\hrule

\clearpage

\tiny
\hrule\vskip4pt
\begin{verbatim}
trigram,good-turing                                              trigram,mle

Ideal                        Actual                       Match  Ideal                        Actual                       Match

Language identified: [en]    Language identified: [en]    1      Language identified: [en]    Language identified: [en]    1
Language identified: [en]    Language identified: [fr]           Language identified: [en]    Language identified: [en]    2
Language identified: [es]    Language identified: [fr]           Language identified: [es]    Language identified: [en]
Language identified: [en]    Language identified: [en]    2      Language identified: [en]    Language identified: [en]    3
Language identified: [fr]    Language identified: [en]           Language identified: [fr]    Language identified: [en]
Language identified: [es]    Language identified: [fr]           Language identified: [es]    Language identified: [en]
Language identified: [it]    Language identified: [it]    3      Language identified: [it]    Language identified: [en]
Language identified: [fr]    Language identified: [fr]    4      Language identified: [fr]    Language identified: [en]
Language identified: [en]    Language identified: [en]    5      Language identified: [en]    Language identified: [en]    4

Total                        9                            5      Total                        9                            4
%                            55.56                               %                            44.44
\end{verbatim}
\vskip4pt\hrule

\tiny
\hrule\vskip4pt
\begin{verbatim}
trigram,witten-bell

Ideal                        Actual                       Match

Language identified: [en]    Language identified: [en]    1
Language identified: [en]    Language identified: [en]    2
Language identified: [es]    Language identified: [en]
Language identified: [en]    Language identified: [en]    3
Language identified: [fr]    Language identified: [en]
Language identified: [es]    Language identified: [en]
Language identified: [it]    Language identified: [en]
Language identified: [fr]    Language identified: [en]
Language identified: [en]    Language identified: [en]    4

Total                        9                            4
%                            44.44
\end{verbatim}
\vskip4pt\hrule

\clearpage

%%%%%%%%%%%%%%%%%%%%%%%%%%%%%%%%%%%%%%%%%%%%%%%%%%%%%%%%%%%%%%%%%%%%%%%%

\paragraph{``Bag-of-Languages'', Restricted Tokenizer}

\tiny
\hrule\vskip4pt
\begin{verbatim}
unigram, add-delta                                                unigram, add-one

Ideal                        Actual                       Match   Ideal                        Actual                       Match

Language identified: [en]    Language identified: [en]    1       Language identified: [en]    Language identified: [en]    1
Language identified: [en]    Language identified: [en]    2       Language identified: [en]    Language identified: [en]    2
Language identified: [es]    Language identified: [fr]            Language identified: [es]    Language identified: [fr]
Language identified: [en]    Language identified: [en]    3       Language identified: [en]    Language identified: [en]    3
Language identified: [ru]    Language identified: [ru]    4       Language identified: [ru]    Language identified: [ru]    4
Language identified: [fr]    Language identified: [fr]    5       Language identified: [fr]    Language identified: [fr]    5
Language identified: [ru]    Language identified: [ru]    6       Language identified: [ru]    Language identified: [ru]    6
Language identified: [es]    Language identified: [es]    7       Language identified: [es]    Language identified: [es]    7
Language identified: [he]    Language identified: [he]    8       Language identified: [he]    Language identified: [he]    8
Language identified: [it]    Language identified: [it]    9       Language identified: [it]    Language identified: [it]    9
Language identified: [ar]    Language identified: [ru]            Language identified: [ar]    Language identified: [ru]
Language identified: [bg]    Language identified: [bg]    10      Language identified: [bg]    Language identified: [bg]    10
Language identified: [java]  Language identified: [java]  11      Language identified: [java]  Language identified: [java]    11
Language identified: [cpp]   Language identified: [en]            Language identified: [cpp]   Language identified: [en]
Language identified: [cpp]   Language identified: [it]            Language identified: [cpp]   Language identified: [it]
Language identified: [he]    Language identified: [he]    12      Language identified: [he]    Language identified: [he]    12
Language identified: [java]  Language identified: [it]            Language identified: [java]  Language identified: [it]
Language identified: [perl]  Language identified: [cpp]           Language identified: [perl]  Language identified: [cpp]
Language identified: [ar]    Language identified: [ar]    13      Language identified: [ar]    Language identified: [ar]    13
Language identified: [perl]  Language identified: [perl]  14      Language identified: [perl]  Language identified: [perl]    14
Language identified: [ru]    Language identified: [he]            Language identified: [ru]    Language identified: [he]
Language identified: [fr]    Language identified: [it]            Language identified: [fr]    Language identified: [it]
Language identified: [ar]    Language identified: [bg]            Language identified: [ar]    Language identified: [bg]
Language identified: [en]    Language identified: [en]    15      Language identified: [en]    Language identified: [en]    15

Total                        24                           15      Total                        24                           15
%                            62.50                                %                            62.50
\end{verbatim}
\vskip4pt\hrule

\clearpage

\tiny
\hrule\vskip4pt
\begin{verbatim}
unigram, good-turing                                              unigram, mle

Ideal                        Actual                       Match   Ideal                        Actual                       Match

Language identified: [en]    Language identified: [en]    1       Language identified: [en]    Language identified: [en]    1
Language identified: [en]    Language identified: [en]    2       Language identified: [en]    Language identified: [en]    2
Language identified: [es]    Language identified: [en]            Language identified: [es]    Language identified: [fr]
Language identified: [en]    Language identified: [en]    3       Language identified: [en]    Language identified: [en]    3
Language identified: [ru]    Language identified: [en]            Language identified: [ru]    Language identified: [ru]    4
Language identified: [fr]    Language identified: [en]            Language identified: [fr]    Language identified: [fr]    5
Language identified: [ru]    Language identified: [en]            Language identified: [ru]    Language identified: [ru]    6
Language identified: [es]    Language identified: [en]            Language identified: [es]    Language identified: [es]    7
Language identified: [he]    Language identified: [en]            Language identified: [he]    Language identified: [he]    8
Language identified: [it]    Language identified: [en]            Language identified: [it]    Language identified: [it]    9
Language identified: [ar]    Language identified: [en]            Language identified: [ar]    Language identified: [ru]
Language identified: [bg]    Language identified: [en]            Language identified: [bg]    Language identified: [bg]    10
Language identified: [java]  Language identified: [en]            Language identified: [java]  Language identified: [java]    11
Language identified: [cpp]   Language identified: [en]            Language identified: [cpp]   Language identified: [en]
Language identified: [cpp]   Language identified: [en]            Language identified: [cpp]   Language identified: [it]
Language identified: [he]    Language identified: [en]            Language identified: [he]    Language identified: [he]    12
Language identified: [java]  Language identified: [en]            Language identified: [java]  Language identified: [it]
Language identified: [perl]  Language identified: [en]            Language identified: [perl]  Language identified: [cpp]
Language identified: [ar]    Language identified: [en]            Language identified: [ar]    Language identified: [ar]    13
Language identified: [perl]  Language identified: [en]            Language identified: [perl]  Language identified: [perl]    14
Language identified: [ru]    Language identified: [en]            Language identified: [ru]    Language identified: [he]
Language identified: [fr]    Language identified: [en]            Language identified: [fr]    Language identified: [it]
Language identified: [ar]    Language identified: [en]            Language identified: [ar]    Language identified: [bg]
Language identified: [en]    Language identified: [en]    4       Language identified: [en]    Language identified: [en]    15

Total                        24                           4       Total                        24                           15
%                            16.67                                %                            62.50
\end{verbatim}
\vskip4pt\hrule

\tiny
\hrule\vskip4pt
\begin{verbatim}
unigram, witten-bell                                              bigram,add-delta

Ideal                        Actual                       Match   Ideal                        Actual                       Match

Language identified: [en]    Language identified: [en]    1       Language identified: [en]    Language identified: [bg]
Language identified: [en]    Language identified: [en]    2       Language identified: [en]    Language identified: [es]
Language identified: [es]    Language identified: [en]            Language identified: [es]    Language identified: [es]    1
Language identified: [en]    Language identified: [en]    3       Language identified: [en]    Language identified: [he]
Language identified: [ru]    Language identified: [en]            Language identified: [ru]    Language identified: [bg]
Language identified: [fr]    Language identified: [en]            Language identified: [fr]    Language identified: [es]
Language identified: [ru]    Language identified: [en]            Language identified: [ru]    Language identified: [bg]
Language identified: [es]    Language identified: [en]            Language identified: [es]    Language identified: [bg]
Language identified: [he]    Language identified: [en]            Language identified: [he]    Language identified: [bg]
Language identified: [it]    Language identified: [en]            Language identified: [it]    Language identified: [es]
Language identified: [ar]    Language identified: [en]            Language identified: [ar]    Language identified: [fr]
Language identified: [bg]    Language identified: [en]            Language identified: [bg]    Language identified: [es]
Language identified: [java]  Language identified: [en]            Language identified: [java]  Language identified: [es]
Language identified: [cpp]   Language identified: [en]            Language identified: [cpp]   Language identified: [es]
Language identified: [cpp]   Language identified: [en]            Language identified: [cpp]   Language identified: [perl]
Language identified: [he]    Language identified: [en]            Language identified: [he]    Language identified: [bg]
Language identified: [java]  Language identified: [en]            Language identified: [java]  Language identified: [es]
Language identified: [perl]  Language identified: [en]            Language identified: [perl]  Language identified: [es]
Language identified: [ar]    Language identified: [en]            Language identified: [ar]    Language identified: [bg]
Language identified: [perl]  Language identified: [en]            Language identified: [perl]  Language identified: [es]
Language identified: [ru]    Language identified: [en]            Language identified: [ru]    Language identified: [bg]
Language identified: [fr]    Language identified: [en]            Language identified: [fr]    Language identified: [es]
Language identified: [ar]    Language identified: [en]            Language identified: [ar]    Language identified: [bg]
Language identified: [en]    Language identified: [en]    4       Language identified: [en]    Language identified: [bg]

Total                        24                           4       Total                        24                           1
%                            16.67                                %                            4.17
\end{verbatim}
\vskip4pt\hrule

\clearpage

\tiny
\hrule\vskip4pt
\begin{verbatim}
bigram,add-one                                                    bigram, good-turing

Ideal                        Actual                       Match   Ideal                        Actual                       Match

Language identified: [en]    Language identified: [bg]            Language identified: [en]    Language identified: [en]    1
Language identified: [en]    Language identified: [es]            Language identified: [en]    Language identified: [en]    2
Language identified: [es]    Language identified: [es]    1       Language identified: [es]    Language identified: [en]
Language identified: [en]    Language identified: [he]            Language identified: [en]    Language identified: [java]
Language identified: [ru]    Language identified: [bg]            Language identified: [ru]    Language identified: [he]
Language identified: [fr]    Language identified: [es]            Language identified: [fr]    Language identified: [en]
Language identified: [ru]    Language identified: [bg]            Language identified: [ru]    Language identified: [ru]    3
Language identified: [es]    Language identified: [es]    2       Language identified: [es]    Language identified: [ar]
Language identified: [he]    Language identified: [bg]            Language identified: [he]    Language identified: [en]
Language identified: [it]    Language identified: [es]            Language identified: [it]    Language identified: [ru]
Language identified: [ar]    Language identified: [fr]            Language identified: [ar]    Language identified: [ru]
Language identified: [bg]    Language identified: [es]            Language identified: [bg]    Language identified: [fr]
Language identified: [java]  Language identified: [es]            Language identified: [java]  Language identified: [java]    4
Language identified: [cpp]   Language identified: [es]            Language identified: [cpp]   Language identified: [fr]
Language identified: [cpp]   Language identified: [perl]          Language identified: [cpp]   Language identified: [en]
Language identified: [he]    Language identified: [bg]            Language identified: [he]    Language identified: [en]
Language identified: [java]  Language identified: [es]            Language identified: [java]  Language identified: [ru]
Language identified: [perl]  Language identified: [es]            Language identified: [perl]  Language identified: [en]
Language identified: [ar]    Language identified: [bg]            Language identified: [ar]    Language identified: [java]
Language identified: [perl]  Language identified: [es]            Language identified: [perl]  Language identified: [perl]    5
Language identified: [ru]    Language identified: [bg]            Language identified: [ru]    Language identified: [en]
Language identified: [fr]    Language identified: [es]            Language identified: [fr]    Language identified: [en]
Language identified: [ar]    Language identified: [fr]            Language identified: [ar]    Language identified: [java]
Language identified: [en]    Language identified: [bg]            Language identified: [en]    Language identified: [java]

Total                        24                           2       Total                        24                           5
%                            8.33                                 %                            20.83
\end{verbatim}
\vskip4pt\hrule

\tiny
\hrule\vskip4pt
\begin{verbatim}
bigram, mle                                                       bigram, witten-bell

Ideal                        Actual                       Match   Ideal                        Actual                       Match

Language identified: [en]    Language identified: [en]    1       Language identified: [en]    Language identified: [en]    1
Language identified: [en]    Language identified: [en]    2       Language identified: [en]    Language identified: [en]    2
Language identified: [es]    Language identified: [en]            Language identified: [es]    Language identified: [en]
Language identified: [en]    Language identified: [en]    3       Language identified: [en]    Language identified: [perl]
Language identified: [ru]    Language identified: [en]            Language identified: [ru]    Language identified: [perl]
Language identified: [fr]    Language identified: [en]            Language identified: [fr]    Language identified: [perl]
Language identified: [ru]    Language identified: [en]            Language identified: [ru]    Language identified: [ru]    3
Language identified: [es]    Language identified: [en]            Language identified: [es]    Language identified: [ar]
Language identified: [he]    Language identified: [en]            Language identified: [he]    Language identified: [en]
Language identified: [it]    Language identified: [en]            Language identified: [it]    Language identified: [ru]
Language identified: [ar]    Language identified: [en]            Language identified: [ar]    Language identified: [ar]    4
Language identified: [bg]    Language identified: [en]            Language identified: [bg]    Language identified: [cpp]
Language identified: [java]  Language identified: [en]            Language identified: [java]  Language identified: [es]
Language identified: [cpp]   Language identified: [en]            Language identified: [cpp]   Language identified: [perl]
Language identified: [cpp]   Language identified: [en]            Language identified: [cpp]   Language identified: [en]
Language identified: [he]    Language identified: [en]            Language identified: [he]    Language identified: [he]    5
Language identified: [java]  Language identified: [en]            Language identified: [java]  Language identified: [ru]
Language identified: [perl]  Language identified: [en]            Language identified: [perl]  Language identified: [en]
Language identified: [ar]    Language identified: [en]            Language identified: [ar]    Language identified: [ar]    6
Language identified: [perl]  Language identified: [en]            Language identified: [perl]  Language identified: [perl]    7
Language identified: [ru]    Language identified: [en]            Language identified: [ru]    Language identified: [en]
Language identified: [fr]    Language identified: [en]            Language identified: [fr]    Language identified: [en]
Language identified: [ar]    Language identified: [en]            Language identified: [ar]    Language identified: [en]
Language identified: [en]    Language identified: [en]    4       Language identified: [en]    Language identified: [en]    8

Total                        24                           4       Total                        24                           8
%                            16.67                                %                            33.33
\end{verbatim}
\vskip4pt\hrule

\clearpage

\tiny
\hrule\vskip4pt
\begin{verbatim}
trigram, add-delta                                                trigram, add-one

Ideal                        Actual                       Match   Ideal                        Actual                       Match

Language identified: [en]    Language identified: [bg]            Language identified: [en]    Language identified: [bg]
Language identified: [en]    Language identified: [es]            Language identified: [en]    Language identified: [es]
Language identified: [es]    Language identified: [es]    1       Language identified: [es]    Language identified: [es]    1
Language identified: [en]    Language identified: [bg]            Language identified: [en]    Language identified: [bg]
Language identified: [ru]    Language identified: [bg]            Language identified: [ru]    Language identified: [bg]
Language identified: [fr]    Language identified: [es]            Language identified: [fr]    Language identified: [es]
Language identified: [ru]    Language identified: [bg]            Language identified: [ru]    Language identified: [bg]
Language identified: [es]    Language identified: [es]    2       Language identified: [es]    Language identified: [es]    2
Language identified: [he]    Language identified: [ar]            Language identified: [he]    Language identified: [ar]
Language identified: [it]    Language identified: [es]            Language identified: [it]    Language identified: [es]
Language identified: [ar]    Language identified: [es]            Language identified: [ar]    Language identified: [es]
Language identified: [bg]    Language identified: [es]            Language identified: [bg]    Language identified: [es]
Language identified: [java]  Language identified: [es]            Language identified: [java]  Language identified: [es]
Language identified: [cpp]   Language identified: [es]            Language identified: [cpp]   Language identified: [es]
Language identified: [cpp]   Language identified: [es]            Language identified: [cpp]   Language identified: [es]
Language identified: [he]    Language identified: [ru]            Language identified: [he]    Language identified: [ru]
Language identified: [java]  Language identified: [es]            Language identified: [java]  Language identified: [es]
Language identified: [perl]  Language identified: [es]            Language identified: [perl]  Language identified: [es]
Language identified: [ar]    Language identified: [ar]    3       Language identified: [ar]    Language identified: [ar]    3
Language identified: [perl]  Language identified: [es]            Language identified: [perl]  Language identified: [es]
Language identified: [ru]    Language identified: [ar]            Language identified: [ru]    Language identified: [ar]
Language identified: [fr]    Language identified: [es]            Language identified: [fr]    Language identified: [es]
Language identified: [ar]    Language identified: [es]            Language identified: [ar]    Language identified: [es]
Language identified: [en]    Language identified: [bg]            Language identified: [en]    Language identified: [bg]

Total                        24                           3       Total                        24                           3
%                            12.50                                %                            12.50
\end{verbatim}
\vskip4pt\hrule

\tiny
\hrule\vskip4pt
\begin{verbatim}
trigram,good-turing                                               trigram,mle

Ideal                        Actual                       Match   Ideal                        Actual                       Match

Language identified: [en]    Language identified: [en]    1       Language identified: [en]    Language identified: [en]    1
Language identified: [en]    Language identified: [fr]            Language identified: [en]    Language identified: [en]    2
Language identified: [es]    Language identified: [fr]            Language identified: [es]    Language identified: [en]
Language identified: [en]    Language identified: [en]    2       Language identified: [en]    Language identified: [en]    3
Language identified: [ru]    Language identified: [en]            Language identified: [ru]    Language identified: [en]
Language identified: [fr]    Language identified: [en]            Language identified: [fr]    Language identified: [en]
Language identified: [ru]    Language identified: [en]            Language identified: [ru]    Language identified: [en]
Language identified: [es]    Language identified: [fr]            Language identified: [es]    Language identified: [en]
Language identified: [he]    Language identified: [ar]            Language identified: [he]    Language identified: [en]
Language identified: [it]    Language identified: [it]    3       Language identified: [it]    Language identified: [en]
Language identified: [ar]    Language identified: [fr]            Language identified: [ar]    Language identified: [en]
Language identified: [bg]    Language identified: [fr]            Language identified: [bg]    Language identified: [en]
Language identified: [java]  Language identified: [fr]            Language identified: [java]  Language identified: [en]
Language identified: [cpp]   Language identified: [fr]            Language identified: [cpp]   Language identified: [en]
Language identified: [cpp]   Language identified: [fr]            Language identified: [cpp]   Language identified: [en]
Language identified: [he]    Language identified: [en]            Language identified: [he]    Language identified: [en]
Language identified: [java]  Language identified: [fr]            Language identified: [java]  Language identified: [en]
Language identified: [perl]  Language identified: [fr]            Language identified: [perl]  Language identified: [en]
Language identified: [ar]    Language identified: [ar]    4       Language identified: [ar]    Language identified: [en]
Language identified: [perl]  Language identified: [fr]            Language identified: [perl]  Language identified: [en]
Language identified: [ru]    Language identified: [ar]            Language identified: [ru]    Language identified: [en]
Language identified: [fr]    Language identified: [fr]    5       Language identified: [fr]    Language identified: [en]
Language identified: [ar]    Language identified: [fr]            Language identified: [ar]    Language identified: [en]
Language identified: [en]    Language identified: [en]    6       Language identified: [en]    Language identified: [en]    4

Total                        24                           6       Total                        24                           4
%                            25.00                                %                            16.67
\end{verbatim}
\vskip4pt\hrule

\clearpage

\tiny
\hrule\vskip4pt
\begin{verbatim}
trigram,witten-bell

Ideal                        Actual                       Match

Language identified: [en]    Language identified: [en]    1
Language identified: [en]    Language identified: [en]    2
Language identified: [es]    Language identified: [en]
Language identified: [en]    Language identified: [en]    3
Language identified: [ru]    Language identified: [en]
Language identified: [fr]    Language identified: [en]
Language identified: [ru]    Language identified: [en]
Language identified: [es]    Language identified: [en]
Language identified: [he]    Language identified: [en]
Language identified: [it]    Language identified: [en]
Language identified: [ar]    Language identified: [en]
Language identified: [bg]    Language identified: [en]
Language identified: [java]  Language identified: [en]
Language identified: [cpp]   Language identified: [en]
Language identified: [cpp]   Language identified: [en]
Language identified: [he]    Language identified: [en]
Language identified: [java]  Language identified: [en]
Language identified: [perl]  Language identified: [en]
Language identified: [ar]    Language identified: [en]
Language identified: [perl]  Language identified: [en]
Language identified: [ru]    Language identified: [en]
Language identified: [fr]    Language identified: [en]
Language identified: [ar]    Language identified: [en]
Language identified: [en]    Language identified: [en]    4

Total                        24                           4
%                            16.67
\end{verbatim}
\vskip4pt\hrule

\normalsize

% EOF

%
% ProbabilisticParsingApp
%

%
% $Id: probabilistic-parsing-app.tex,v 1.30 2006/09/22 20:31:52 mokhov Exp $
%

\subsection{CYK Probabilistic Parsing with \api{ProbabilisticParsingApp}}
\label{sect:probabilistic-parsing-app}
\index{CYK Probabilistic Parsing}
\index{CYK Probabilistic Parsing!Application}
\index{Applications!CYK Probabilistic Parsing}

$Revision: 1.30 $

Originally written on April 12, 2003.

\subsubsection{Introduction}

This section describes the implementation of the CYK probabilistic parsing
algorithm \cite{cyk} implemented in {\java} and discusses the experiments, grammars used,
the results, and difficulties encountered.

\subsubsection{The Program}

\paragraph{Manual and the Source Code}

Mini User Manual along with instructions on how to run the application are provided
in the \xs{sect:prob-manual}. The source code is provided in the electronic form only with few extracts
in the presented in the document.

\paragraph{Grammar File Format}

Serguei has developed a ``grammar compiler'' for the Compiler Design course, and
we have adapted it to accept probabilistic grammars. The grammar compiler reads a source
grammar text file and compiles it (some rudimentary lexical and semantic checks are in place).
As a result, a successfully ``compiled'' \api{Grammar} object
has a set of terminals, non-terminals, and rules stored in a binary file.
Parsers re-load this compiled grammar and do the main parsing of what they are supposed to parse.

\begin{figure}
\hrule\vskip4pt
\begin{verbatim}
<LHS> ::= PROBABILITY RHS %EOL

#  single-line comment; shell style

// single-line comment; C++ style

/*
 * multi-line comment; C style
 */
\end{verbatim}
\caption{Grammar File Format}
\label{fig:grammar-format}
\vskip4pt\hrule
\end{figure}

The grammar file for the grammar file has the format presented in \xf{fig:grammar-format}. Description
of the elements is below. The example of the grammar rules is in \xf{fig:grammar-example}.
{\bf Whenever one changes the grammar, it has to be recompiled to take effect.}

\begin{itemize}

\item
    \verb+<LHS>+ is a single non-terminal on the left-hand side of the rule.

\item
    \verb+::=+ is a rule
    operator separating LHS and RHS.

\item
    \verb+PROBABILITY+ is a floating-point number indicating rule's
    probability.

\item
    \verb+RHS+ for this particular assignment has to be in CNF, i.e. either \verb+<B> <C>+
    or \verb+terminal+ with \verb+<A>+ and \verb+<B>+ being non-terminals.

\item
    All non-terminals have to be enclosed within the angle brackets \verb+<+ and \verb+>+.

\item
    All grammar rules have to terminated by \verb+%EOL+ that acts a semicolon in {\C}/{\cpp} or
    {\java}. It indicates where to stop processing current rule and look for the next
    (in case a rule spans several text lines).

\item
    Amount of white space between grammar elements doesn't matter much.

\item
    The grammar file has also a notion of comments. The grammar compiler accepts
    shell-like single line comments when lines start with \verb+#+ as well as {\C} or {\cpp} comments
    like \verb+//+ and \verb+/* */+ with the same effect as that of {\C} and {\cpp}.

\end{itemize}

\begin{figure}
\hrule\vskip4pt
\begin{verbatim}
/*
 * 80% of sentences are noun phrases
 * followed by verb phrases.
 */
<S> ::= 0.8 <NP> <VP> %EOL

// A very rare verb
<V> ::= 0.0001 disambiguate %EOL

# EOF
\end{verbatim}
\caption{Grammar File Example}
\label{fig:grammar-example}
\vskip4pt\hrule
\end{figure}

\paragraph{Data Structures}

The main storage data structure is an instance of the \api{Grammar} class that holds
vectors of
terminals (see file \file{marf/nlp/Parsing/GrammarCompiler/Terminal.java}),
non-terminals (see file \file{marf/nlp/Parsing/GrammarCompiler/NonTerminal.java}),
and rules (see file \\\file{marf/nlp/Parsing/GrammarCompiler/ProbabilisticRule.java}).

While the grammar is being parsed and compiled there are also
various grammar tokens and their types involved.
Since they are
not very much relevant to the subject of this application we won't talk about them
(examine the contents of the \file{marf/nlp/Parsing/} and \file{marf/nlp/Parsing/GrammarCompiler/}
directories if you care).

The CYK algorithm's data structures, the $\pi$ and \verb+back+ arrays,
are represented as 3-dimensional array of doubles and vectors of back-indices
respectively: \verb+double[][][] oParseMatrix+ and \verb+Vector[][][] aoBack+
in \file{marf/nlp/Parsing/ProbabilisticParser.java}. There is also a vector of
words of the incoming sentence to be parsed, \verb+Vector oWords+.

%%%%%%%%%%%%%%%%%%%%%%%%%%%%%%%%%%%%%%%%%%%%%%%
%%%%%%%%%%%%%%%%%%%%%%%%%%%%%%%%%%%%%%%%%%%%%%%

\subsubsection{Methodology}

We have experimented (not to full extent yet) with three grammars:
{\em Basic}, {\em Extended}, and {\em Realistic}. Description of the grammars and how
they were obtained is presented below. The set of testing sentences was initially
based on the given (Basic) grammar to see whether the CYK algorithm indeed
parses all syntactically valid sentences and rejects the rest. Then the
sentence set was augmented from various sources (e.g. the paper Serguei presented
and on top of his head). Finally, the original requirements were attempted to be used
as a source of grammar.

\paragraph{Basic Grammar}

The basic grammar given in the requirements has been
used at first to develop and debug the application.
The basic grammar is in \xf{fig:basic-grammar}.

\begin{figure}
\hrule\vskip4pt
\tiny
\input{grammar.original.tex}
\normalsize
\caption{Original Basic Grammar}
\label{fig:basic-grammar}
\vskip4pt\hrule
\end{figure}

\paragraph{Extended Grammar}

The basic grammar was extended with few rules
from \cite{jurafsky}, p. 449.
The probability scores are {\bf artificial} and
adapted from the basic grammar and the book's grammar, and
recomputed {\em approximately}\footnote{Preserving proportional relevance from all sources
and making sure they all add up to 100\%} by hand.
The extended grammar is in \xf{fig:extended-grammar}.
Both basic and extended grammars used the same set of
testing sentences presented in \xf{fig:test-sentences}.
There is a number of sentences for which we have never come up
with rules as initially intended, so there are no parses
for the them in the output can be seen yet, a TODO.

\begin{figure}
\hrule\vskip4pt
\tiny
\input{grammar.extended.tex}
\normalsize
\caption{Extended Grammar}
\label{fig:extended-grammar}
\vskip4pt\hrule
\end{figure}

\begin{figure}
\hrule\vskip4pt
\input{test-sentences.tex}
\caption{Input sentences for the Basic and Extended Grammars}
\label{fig:test-sentences}
\vskip4pt\hrule
\end{figure}

\paragraph{Realistic Grammar}\label{sect:realistic-grammar}

Without having 100\% completed the extended grammar, we jumped
to develop something more ``realistic'', the Realistic Grammar.
Since the two previous grammars are quite artificial, to test
out it on some more ``realistic'' data we came up with the
best grammar we could
from the sentences of the requirements (other than
the sample grammar; the actual requirements were used).
The sentences, some were expanded, are in \xf{fig:asmt-sentences},
and the grammar itself is in \xf{fig:asmt-grammar}. Many of the rules
of the form of \verb+A->BC+ still may not have truly correct probabilities
corresponding to the paper due to the above two reasons.

\begin{figure}
\hrule\vskip4pt
\input{asmt-sentences.tex}
\caption{Requirements' sentences used as a parsed ``corpus''}
\label{fig:asmt-sentences}
\vskip4pt\hrule
\end{figure}

\begin{figure}
\vskip4pt\hrule\vskip4pt
\tiny
\input{grammar.asmt.tex}
\normalsize
\caption{Realistic Grammar}
\label{fig:asmt-grammar}
\vskip4pt\hrule
\end{figure}

\paragraph{Grammar Restrictions}

All incoming sentences, though case-sensitive, are required to be
in the lower-case unless a given word is a proper noun or the
pronoun {\it I}. The current system doesn't deal with punctuation
either, so complex sentences that use commas, semicolons, and
other punctuation characters may not (and appear not to) be
parsed correctly (such punctuation is simply ignored).

However, all the above restrictions can be solved just at the
grammar level without touching a single line of code,
but they were not dealt with yet in the first prototype.

\paragraph{Difficulties Encountered}
\label{sect:prob-app-troubles}

\paragraph*{Learning Probabilistic Grammars}

The major problem of this type of grammars is to learn them.
This requires at least having a some decent POS tagger
(e.g. in \cite{brill}) and decent knowledge of the English grammar
and then sitting down and computing the probabilities
of the rules manually. This is a very time-consuming
and unjustified enormous effort for documents of relatively
small size (let alone medium-size or million-word corpora).
Hence, there is a need for automatic tools and/or pre-existing
(and freely available!) treebanks to learn the grammars from.
For example, we have spent two days developing grammar for a one-sheet
document (see \xs{sect:realistic-grammar}) and the end result is that
we can only parse 3 (three) sentences so far with it.

\paragraph*{Chomsky Normal Form}

Another problem is the conversion of existing grammars
or maintaining the current grammar to make sure it is in the CNF
as the CYK algorithm \cite{cyk} requires. This is a problem because
for a human maintainer the number of rules grows, so it becomes
less obvious what a initial rule was like, and there's always
a possibility to create more undesired parses that way. We had
to do that for the Extended and Realistic Grammars that were
based on the grammar rules from the book \cite{jurafsky}.

To illustrate the problem, the rules similar to the
below have to be converted into CNF:

\begin{verbatim}
<A> -> <B>
<A> -> t <C>
<A> -> <B> <C> <D>
<A> -> <B> <C> t <D> <E>
\end{verbatim}

The conversion implies creating new non-terminals augmenting
the number of rules, which may be hard to trace later on when
there are many.

\begin{verbatim}
<A>   -> <B> <A>
<A>   -> <T> <C>
<A>   -> <BC> <D>
<A>   -> <BCT> <DE>
<T>   -> t
<BC>  -> <B> <C>
<DE>  -> <D> <DE>
<BCT> -> <BC> <T>
\end{verbatim}

The rule-set has been doubled in the above example. That creates
a problem of distributing the probability to the new rules as
well as the problem below.

\paragraph*{Exponential Growth of Storage Requirements and Run Time}

While playing with the Basic and Extended grammars, we didn't pay
much attention to the run-time aspect of the algorithm (even though
the number of the nested for-loops looked alerting) because
it was a second or less for the sentence set in \xf{fig:test-sentences}.
It became a problem, however, when we came up with the Realistic
grammar. The run-time for this grammar has jumped to 16 (sixteen !!!)
{\bf minutes} (!!!) in average for the sentence set in \xf{fig:asmt-sentences}.
This was rather discouraging (unless the goal is not the speed but
the most correct result at the hopefully near end). The problem stems
from the algorithm's complexity and huge data sparseness for large grammars.
One of the major reasons of the data sparseness problem is the CNF requirement as
described above: the number of non-terminals grows rapidly largely
increasing one of the dimensions of our $\pi$ and \verb+back+ arrays causing
number of iteration of the parsing algorithm to increase exponentially (or it is
at least cubic). And a lot of time is spent to find out that there's no
a parse for a sentence (given all the words of the sentence in the grammar).

\paragraph*{Data Sparseness and Smoothing}

The bigger grammar is the more severe the data sparseness is in our arrays
in this algorithm causing the above problem of the run-time and storage
requirements. Smoothing techniques we previously implemented in \api{LangIdentApp}
can be applied to at least get rid of zero-probabilities in the $\pi$ array, but we believe
the smoothing might hurt rather than improve the parsing performance;
yet we haven't tested this claim out yet (a TODO). A better way could be smooth
the grammar rules' probabilities.

%%%%%%%%%%%%%%%%%%%%%%%%%%%%%%%%%%%%%%%%%%%%%%%%%%%%%
%%%%%%%%%%%%%%%%%%%%%%%%%%%%%%%%%%%%%%%%%%%%%%%%%%%%%

\subsubsection{Results and Conclusions}

\paragraph{Generalities}

The algorithm does indeed seem to work and accept
syntactically-valid sentences while rejecting the ungrammatical
ones. Though to exhaust all the possible grammatical
and ungrammatical sentences in testing would require a lot more time.
There is also some disappointment
with respect to the running time increase when
the grammar grows and the other problems mentioned before.

To overcome the run-time problem, we'd have to use some
more sophisticated data structures than a plain 3D array
of doubles, but this is like fighting with the disease, instead
of its cause. The CYK algorithm has to be optimized
or even re-born to allow more than just CNF grammars
and be faster at the same time.

The restrictions on the sentences mentioned earlier can all
be overcome by just only tweaking the grammar (but increasing
the run-time along the way).

While most of the results of my test runs are in the further sections,
below we present one interactive session sample as well
as our favorite parse.

\paragraph{Sample Interactive Run}

\scriptsize
\begin{verbatim}
junebug.mokhov [a3] % java ProbabilisticParsingApp --parse

Probabilistic Parsing
Serguei A. Mokhov, mokhov@cs
April 2003



Entering interactive mode... Type \q to exit.
sentence> my rabbit has a white smile
my rabbit has a white smile

Parse for the sentence [ my rabbit has a white smile ] is below:

SYNOPSIS:

<NONTERMINAL> (PROBABILITY) [ SPAN: words of span ]

<S> (0.0020480000000000008) [ 0-5: my rabbit has a white smile ]
        <NP> (0.04000000000000001) [ 0-1: my rabbit ]
                <DET> (0.1) [ 0-0: my ]
                <NOMINAL> (0.4) [ 1-1: rabbit ]
        <VP> (0.06400000000000002) [ 2-5: has a white smile ]
                <V> (0.8) [ 2-2: has ]
                <NP> (0.08000000000000002) [ 3-5: a white smile ]
                        <DET> (0.4) [ 3-3: a ]
                        <NOMINAL> (0.2) [ 4-5: white smile ]
                                <ADJ> (1.0) [ 4-4: white ]
                                <NOMINAL> (0.2) [ 5-5: smile ]


sentence> my rabbit has a smile
my rabbit has a smile

Parse for the sentence [ my rabbit has a smile ] is below:

SYNOPSIS:

<NONTERMINAL> (PROBABILITY) [ SPAN: words of span ]

<S> (0.0020480000000000008) [ 0-4: my rabbit has a smile ]
        <NP> (0.04000000000000001) [ 0-1: my rabbit ]
                <DET> (0.1) [ 0-0: my ]
                <NOMINAL> (0.4) [ 1-1: rabbit ]
        <VP> (0.06400000000000002) [ 2-4: has a smile ]
                <V> (0.8) [ 2-2: has ]
                <NP> (0.08000000000000002) [ 3-4: a smile ]
                        <DET> (0.4) [ 3-3: a ]
                        <NOMINAL> (0.2) [ 4-4: smile ]


sentence> my rabbit has a telephone
my rabbit has a telephone

There's no parse for [ my rabbit has a telephone ]

sentence> the cat eats the rabbit
the cat eats the rabbit

Parse for the sentence [ the cat eats the rabbit ] is below:

SYNOPSIS:

<NONTERMINAL> (PROBABILITY) [ SPAN: words of span ]

<S> (0.006400000000000002) [ 0-4: the cat eats the rabbit ]
        <NP> (0.2) [ 0-1: the cat ]
                <DET> (0.5) [ 0-0: the ]
                <NOMINAL> (0.4) [ 1-1: cat ]
        <VP> (0.04000000000000001) [ 2-4: eats the rabbit ]
                <V> (0.2) [ 2-2: eats ]
                <NP> (0.2) [ 3-4: the rabbit ]
                        <DET> (0.5) [ 3-3: the ]
                        <NOMINAL> (0.4) [ 4-4: rabbit ]


sentence> a white cat has a white smile
a white cat has a white smile

Parse for the sentence [ a white cat has a white smile ] is below:

SYNOPSIS:

<NONTERMINAL> (PROBABILITY) [ SPAN: words of span ]

<S> (0.008192000000000003) [ 0-6: a white cat has a white smile ]
        <NP> (0.16000000000000003) [ 0-2: a white cat ]
                <DET> (0.4) [ 0-0: a ]
                <NOMINAL> (0.4) [ 1-2: white cat ]
                        <ADJ> (1.0) [ 1-1: white ]
                        <NOMINAL> (0.4) [ 2-2: cat ]
        <VP> (0.06400000000000002) [ 3-6: has a white smile ]
                <V> (0.8) [ 3-3: has ]
                <NP> (0.08000000000000002) [ 4-6: a white smile ]
                        <DET> (0.4) [ 4-4: a ]
                        <NOMINAL> (0.2) [ 5-6: white smile ]
                                <ADJ> (1.0) [ 5-5: white ]
                                <NOMINAL> (0.2) [ 6-6: smile ]


sentence> cat white my has rabbit
cat white my has rabbit

There's no parse for [ cat white my has rabbit ]

sentence> smile rabbit eats my cat
smile rabbit eats my cat

There's no parse for [ smile rabbit eats my cat ]

sentence> cat eats rabbit
cat eats rabbit

There's no parse for [ cat eats rabbit ]

sentence> the cat eats the rabbit
the cat eats the rabbit

Parse for the sentence [ the cat eats the rabbit ] is below:

SYNOPSIS:

<NONTERMINAL> (PROBABILITY) [ SPAN: words of span ]

<S> (0.006400000000000002) [ 0-4: the cat eats the rabbit ]
        <NP> (0.2) [ 0-1: the cat ]
                <DET> (0.5) [ 0-0: the ]
                <NOMINAL> (0.4) [ 1-1: cat ]
        <VP> (0.04000000000000001) [ 2-4: eats the rabbit ]
                <V> (0.2) [ 2-2: eats ]
                <NP> (0.2) [ 3-4: the rabbit ]
                        <DET> (0.5) [ 3-3: the ]
                        <NOMINAL> (0.4) [ 4-4: rabbit ]


sentence> \q
junebug.mokhov [a3] %
\end{verbatim}

\normalsize

\paragraph{Favorite Parse}

As of this release, we were able to only parse the three sentences
from the ``requirements corpus'' and below is the favorite:

\begin{verbatim}
<S> (7.381879524149979E-10) [ 0-7: describe your grammar and how you developed it ]
    <S> (4.11319751814313E-4) [ 0-2: describe your grammar ]
        <V> (0.206897) [ 0-0: describe ]
        <NOMINAL> (0.039760823193600005) [ 1-2: your grammar ]
            <ADJ> (0.439024) [ 1-1: your ]
            <NOMINAL> (0.113208) [ 2-2: grammar ]
    <ConjS> (1.7946815078995936E-5) [ 3-7: and how you developed it ]
        <Conj> (0.92) [ 3-3: and ]
        <S> (1.9507407694560798E-5) [ 4-7: how you developed it ]
            <WhNP> (0.094285785) [ 4-5: how you ]
                <WhWord> (0.5) [ 4-4: how ]
                <Pronoun> (0.571429) [ 5-5: you ]
            <VP> (0.002068965931032) [ 6-7: developed it ]
                <V> (0.0344828) [ 6-6: developed ]
                <Pronoun> (0.333333) [ 7-7: it ]
\end{verbatim}

\subsubsection{Mini User Manual}
\label{sect:prob-manual}

\paragraph{System Requirements}

The program was mostly developed under Linux, so there's
a \file{Makefile} and a testing shell script to simplify some routine tasks.
For JVM, any JDK 1.4.* and above will do. \tool{bash} would be nice
to have to be able to run the batch script. Since the application
itself is written in {\java}, it's not bound to a specific architecture,
thus may be compiled and run without the makefiles and scripts
on virtually any operating system.

\paragraph{How To Run It}

There are thousands of ways how to run the program. Some of them are listed below.

\paragraph*{Using the Shell Script}

There is a script out there -- \tool{testing.sh}. It simply does compilation
and batch processing for all the three grammars and two sets of test
sentences in one run.
The script is written using \tool{bash} syntax; hence, \tool{bash} should be
present.

Type:

\noindent
\verb+./testing.sh+

or

\noindent
\verb+time ./testing.sh+

to execute  the batch (and time it in the second case). And example
of what one can get is below. NOTE: processing the first grammar, in
\file{grammar.asmt.txt}, may take awhile (below it took us 16 minutes),
so be aware of that fact (see the reasons in \xs{sect:prob-app-troubles}).

E.g.:

\begin{verbatim}
junebug.mokhov [a3] % time ./testing.sh
Making sure java files get compiled...
javac -g ProbabilisticParsingApp.java
Testing...
Compiling grammar: grammar.asmt.txt
Parsing...991.318u 1.511s 16:35.55 99.7%        0+0k 0+0io 5638pf+0w
Done
Look for parsing results in grammar.asmt.txt-parse.log

Compiling grammar: grammar.extended.txt
Parsing...1.591u 0.062s 0:01.82 90.6%   0+0k 0+0io 5637pf+0w
Done
Look for parsing results in grammar.extended.txt-parse.log

Compiling grammar: grammar.original.txt
Parsing...0.455u 0.066s 0:00.72 70.8%   0+0k 0+0io 5599pf+0w
Done
Look for parsing results in grammar.original.txt-parse.log

Testing done.
995.675u 1.906s 16:41.68 99.5%  0+0k 0+0io 42211pf+0w
junebug.mokhov [a3] %
\end{verbatim}

\paragraph*{Running ProbabilisitcParsingApp}
\label{sect:prob-parsing-app}

You can run the application itself without any wrapping scripts
and provide options to it. This is a command-line application,
so there is no GUI associated with it yet. To run the application
you have to compile it first. You can use either \tool{make} with no
arguments to compile or use a standard Java compiler.

Type:

\noindent
\verb+make+

or

\noindent
\verb+javac -cp marf.jar:. ProbabilisitcParsingApp.java+

After having compiled the application, you can run it with a JVM.
There are mutually-exclusive required options:

\begin{itemize}

\item
\verb+--train <grammar-file>+ -- to compile a grammar from the specified text file.
This is the first thing you need to do before trying to parse any sentences. Once compiled,
you don't need to recompile it each time you run the parser unless you made a fresh
copy of the application or made changes to the grammar file or plan to use a grammar
from another file.

\item
\verb+--parse+ -- to actually run the parser in the interactive mode (or batch mode, just use
input file redirection with your test sentences). To run the parser successfully there should
be compiled grammar first (see \verb+--train+).

\end{itemize}

E.g.:

To compile the Basic Grammar:

\verb+java -cp marf.jar:. ProbabilisitcParsingApp.java --train grammars/grammar.original.txt+

To batch process the sentence set from a file:

\verb+java -cp marf.jar:. ProbabilisitcParsingApp.java --parse < data/test-sentences.txt+

To run the application interactively:

\verb+java -cp marf.jar:. ProbabilisitcParsingApp.java --parse+

\noindent
Complete usage information:

\vspace{15pt}
\hrule
\begin{verbatim}

Probabilistic Parsing
Serguei A. Mokhov, mokhov@cs.concordia.ca
April 2003 - 2009


Usage:
    java ProbabilisticParsingApp --help | -h
        : to display this help and exit

    java ProbabilisticParsingApp --version
        : to display version and exit

    java ProbabilisticParsingApp --train [ OPTIONS ] <grammar-file>
        : to compile grammar from the <grammar-file>

    java ProbabilisticParsingApp --parse [ OPTIONS ]
        : to parse sentences from standard input

Where options are of the following:

    --debug  - enable debugging (more verbose output)
    -case    - make it case-sensitive
    -num     - parse numerical values
    -quote   - consider quotes and count quoted strings as one token
    -eos     - make typical ends of sentences (<?>, <!>, <.>) significant


\end{verbatim}

\hrule
\vspace{15pt}

\paragraph{List of Files of Interest}

\paragraph*{Directories}

\begin{itemize}

\item
\file{marf/nlp/} -- that's where most of the code is for this application is, the \api{marf.nlp} package.

\item
\file{marf/nlp/Parsing/} -- that's where most of the Parsing code is for Probabilistic Grammars and
Grammars in general

\item
\file{marf/nlp/Parsing/GrammarCompiler/} -- that's where the Grammar Compiler modules are

\end{itemize}

\paragraph*{The Application}

The application, its makefile, and
the wrapper script for batch training and testing.

\begin{verbatim}
ProbabilisticParsingApp.java
Makefile
testing.sh
marf.jar
\end{verbatim}

\paragraph*{Grammars}

\noindent
\file{grammars/grammar.original.txt} -- The Basic Grammar

\noindent
\file{grammars/grammar.extended.txt} -- The Extended Grammar

\noindent
\file{grammars/grammar.asmt.txt} -- The Realistic Grammar

\paragraph*{Test Sentences}

\noindent
\file{data/test-sentences.txt} -- the sample sentences from all over the place.

\noindent
\file{data/asmt-sentences.txt} -- the sentences derived from the requirements sheet.

%%%%%%%%%%%%%%%%%%%%%%%%%%%%%%%%%%%%%%%%%%%%%%%%%%%%
%%%%%%%%%%%%%%%%%%%%%%%%%%%%%%%%%%%%%%%%%%%%%%%%%%%%

\subsubsection{Results}
\label{sect:prob-parse-results}

\paragraph{Basic Grammar}

\tiny
\hrule\vskip4pt
\begin{verbatim}

Probabilistic Parsing
Serguei A. Mokhov, mokhov@cs
April 2003



Entering interactive mode... Type \q to exit.
sentence> my rabbit has a white smile

Parse for the sentence [ my rabbit has a white smile ] is below:

SYNOPSIS:

<NONTERMINAL> (PROBABILITY) [ SPAN: words of span ]

<S> (0.0020480000000000008) [ 0-5: my rabbit has a white smile ]
    <NP> (0.04000000000000001) [ 0-1: my rabbit ]
        <DET> (0.1) [ 0-0: my ]
        <NOMINAL> (0.4) [ 1-1: rabbit ]
    <VP> (0.06400000000000002) [ 2-5: has a white smile ]
        <V> (0.8) [ 2-2: has ]
        <NP> (0.08000000000000002) [ 3-5: a white smile ]
            <DET> (0.4) [ 3-3: a ]
            <NOMINAL> (0.2) [ 4-5: white smile ]
                <ADJ> (1.0) [ 4-4: white ]
                <NOMINAL> (0.2) [ 5-5: smile ]


sentence> my rabbit has a smile

Parse for the sentence [ my rabbit has a smile ] is below:

SYNOPSIS:

<NONTERMINAL> (PROBABILITY) [ SPAN: words of span ]

<S> (0.0020480000000000008) [ 0-4: my rabbit has a smile ]
    <NP> (0.04000000000000001) [ 0-1: my rabbit ]
        <DET> (0.1) [ 0-0: my ]
        <NOMINAL> (0.4) [ 1-1: rabbit ]
    <VP> (0.06400000000000002) [ 2-4: has a smile ]
        <V> (0.8) [ 2-2: has ]
        <NP> (0.08000000000000002) [ 3-4: a smile ]
            <DET> (0.4) [ 3-3: a ]
            <NOMINAL> (0.2) [ 4-4: smile ]


sentence> my rabbit has a telephone

There's no parse for [ my rabbit has a telephone ]

sentence> rabbit my a white has smile

There's no parse for [ rabbit my a white has smile ]

sentence> a slim blue refrigerator jumped gracefully out of the bottle

There's no parse for [ a slim blue refrigerator jumped gracefully out of the bottle ]

sentence> my smile has a rabbit

Parse for the sentence [ my smile has a rabbit ] is below:

SYNOPSIS:

<NONTERMINAL> (PROBABILITY) [ SPAN: words of span ]

<S> (0.0020480000000000008) [ 0-4: my smile has a rabbit ]
    <NP> (0.020000000000000004) [ 0-1: my smile ]
        <DET> (0.1) [ 0-0: my ]
        <NOMINAL> (0.2) [ 1-1: smile ]
    <VP> (0.12800000000000003) [ 2-4: has a rabbit ]
        <V> (0.8) [ 2-2: has ]
        <NP> (0.16000000000000003) [ 3-4: a rabbit ]
            <DET> (0.4) [ 3-3: a ]
            <NOMINAL> (0.4) [ 4-4: rabbit ]


sentence> the cat eats my white rabbit

Parse for the sentence [ the cat eats my white rabbit ] is below:

SYNOPSIS:

<NONTERMINAL> (PROBABILITY) [ SPAN: words of span ]

<S> (0.0012800000000000005) [ 0-5: the cat eats my white rabbit ]
    <NP> (0.2) [ 0-1: the cat ]
        <DET> (0.5) [ 0-0: the ]
        <NOMINAL> (0.4) [ 1-1: cat ]
    <VP> (0.008000000000000002) [ 2-5: eats my white rabbit ]
        <V> (0.2) [ 2-2: eats ]
        <NP> (0.04000000000000001) [ 3-5: my white rabbit ]
            <DET> (0.1) [ 3-3: my ]
            <NOMINAL> (0.4) [ 4-5: white rabbit ]
                <ADJ> (1.0) [ 4-4: white ]
                <NOMINAL> (0.4) [ 5-5: rabbit ]


sentence> a white smile eats the cat

Parse for the sentence [ a white smile eats the cat ] is below:

SYNOPSIS:

<NONTERMINAL> (PROBABILITY) [ SPAN: words of span ]

<S> (0.0025600000000000015) [ 0-5: a white smile eats the cat ]
    <NP> (0.08000000000000002) [ 0-2: a white smile ]
        <DET> (0.4) [ 0-0: a ]
        <NOMINAL> (0.2) [ 1-2: white smile ]
            <ADJ> (1.0) [ 1-1: white ]
            <NOMINAL> (0.2) [ 2-2: smile ]
    <VP> (0.04000000000000001) [ 3-5: eats the cat ]
        <V> (0.2) [ 3-3: eats ]
        <NP> (0.2) [ 4-5: the cat ]
            <DET> (0.5) [ 4-4: the ]
            <NOMINAL> (0.4) [ 5-5: cat ]


sentence> my cat has a white rabbit

Parse for the sentence [ my cat has a white rabbit ] is below:

SYNOPSIS:

<NONTERMINAL> (PROBABILITY) [ SPAN: words of span ]

<S> (0.0040960000000000015) [ 0-5: my cat has a white rabbit ]
    <NP> (0.04000000000000001) [ 0-1: my cat ]
        <DET> (0.1) [ 0-0: my ]
        <NOMINAL> (0.4) [ 1-1: cat ]
    <VP> (0.12800000000000003) [ 2-5: has a white rabbit ]
        <V> (0.8) [ 2-2: has ]
        <NP> (0.16000000000000003) [ 3-5: a white rabbit ]
            <DET> (0.4) [ 3-3: a ]
            <NOMINAL> (0.4) [ 4-5: white rabbit ]
                <ADJ> (1.0) [ 4-4: white ]
                <NOMINAL> (0.4) [ 5-5: rabbit ]


sentence> my cat has white rabbit

There's no parse for [ my cat has white rabbit ]

sentence> cat has a white rabbit

There's no parse for [ cat has a white rabbit ]

sentence> smile has my cat

There's no parse for [ smile has my cat ]

sentence> can you book TWA flights

There's no parse for [ can you book TWA flights ]

sentence> the lion jumped through the hoop

There's no parse for [ the lion jumped through the hoop ]

sentence> the trainer jumped the lion through the hoop

There's no parse for [ the trainer jumped the lion through the hoop ]

sentence> the butter melted in the pan

There's no parse for [ the butter melted in the pan ]

sentence> the cook melted the butter in the pan

There's no parse for [ the cook melted the butter in the pan ]

sentence> the rich love their money

There's no parse for [ the rich love their money ]

sentence> the rich love sometimes too

There's no parse for [ the rich love sometimes too ]

sentence> the contractor built the houses last summer

There's no parse for [ the contractor built the houses last summer ]

sentence> the contractor built last summer

There's no parse for [ the contractor built last summer ]

sentence>
\end{verbatim}
\vskip4pt\hrule

\paragraph{Extended Grammar}

\tiny
\hrule\vskip4pt
\begin{verbatim}

Probabilistic Parsing
Serguei A. Mokhov, mokhov@cs
April 2003



Entering interactive mode... Type \q to exit.
sentence> my rabbit has a white smile

Parse for the sentence [ my rabbit has a white smile ] is below:

SYNOPSIS:

<NONTERMINAL> (PROBABILITY) [ SPAN: words of span ]

<S> (1.915200000000001E-6) [ 0-5: my rabbit has a white smile ]
    <NP> (0.0020000000000000005) [ 0-1: my rabbit ]
        <DET> (0.05) [ 0-0: my ]
        <NOMINAL> (0.2) [ 1-1: rabbit ]
    <VP> (0.002736000000000001) [ 2-5: has a white smile ]
        <V> (0.5) [ 2-2: has ]
        <NP> (0.005760000000000002) [ 3-5: a white smile ]
            <DET> (0.4) [ 3-3: a ]
            <NOMINAL> (0.07200000000000002) [ 4-5: white smile ]
                <ADJ> (0.8) [ 4-4: white ]
                <NOMINAL> (0.1) [ 5-5: smile ]


sentence> my rabbit has a smile

Parse for the sentence [ my rabbit has a smile ] is below:

SYNOPSIS:

<NONTERMINAL> (PROBABILITY) [ SPAN: words of span ]

<S> (2.6600000000000012E-6) [ 0-4: my rabbit has a smile ]
    <NP> (0.0020000000000000005) [ 0-1: my rabbit ]
        <DET> (0.05) [ 0-0: my ]
        <NOMINAL> (0.2) [ 1-1: rabbit ]
    <VP> (0.003800000000000001) [ 2-4: has a smile ]
        <V> (0.5) [ 2-2: has ]
        <NP> (0.008000000000000002) [ 3-4: a smile ]
            <DET> (0.4) [ 3-3: a ]
            <NOMINAL> (0.1) [ 4-4: smile ]


sentence> my rabbit has a telephone

There's no parse for [ my rabbit has a telephone ]

sentence> rabbit my a white has smile

There's no parse for [ rabbit my a white has smile ]

sentence> a slim blue refrigerator jumped gracefully out of the bottle

There's no parse for [ a slim blue refrigerator jumped gracefully out of the bottle ]

sentence> my smile has a rabbit

Parse for the sentence [ my smile has a rabbit ] is below:

SYNOPSIS:

<NONTERMINAL> (PROBABILITY) [ SPAN: words of span ]

<S> (2.6600000000000012E-6) [ 0-4: my smile has a rabbit ]
    <NP> (0.0010000000000000002) [ 0-1: my smile ]
        <DET> (0.05) [ 0-0: my ]
        <NOMINAL> (0.1) [ 1-1: smile ]
    <VP> (0.007600000000000002) [ 2-4: has a rabbit ]
        <V> (0.5) [ 2-2: has ]
        <NP> (0.016000000000000004) [ 3-4: a rabbit ]
            <DET> (0.4) [ 3-3: a ]
            <NOMINAL> (0.2) [ 4-4: rabbit ]


sentence> the cat eats my white rabbit

Parse for the sentence [ the cat eats my white rabbit ] is below:

SYNOPSIS:

<NONTERMINAL> (PROBABILITY) [ SPAN: words of span ]

<S> (9.576000000000004E-7) [ 0-5: the cat eats my white rabbit ]
    <NP> (0.020000000000000004) [ 0-1: the cat ]
        <DET> (0.5) [ 0-0: the ]
        <NOMINAL> (0.2) [ 1-1: cat ]
    <VP> (1.3680000000000005E-4) [ 2-5: eats my white rabbit ]
        <V> (0.1) [ 2-2: eats ]
        <NP> (0.0014400000000000005) [ 3-5: my white rabbit ]
            <DET> (0.05) [ 3-3: my ]
            <NOMINAL> (0.14400000000000004) [ 4-5: white rabbit ]
                <ADJ> (0.8) [ 4-4: white ]
                <NOMINAL> (0.2) [ 5-5: rabbit ]


sentence> a white smile eats the cat

Parse for the sentence [ a white smile eats the cat ] is below:

SYNOPSIS:

<NONTERMINAL> (PROBABILITY) [ SPAN: words of span ]

<S> (3.8304000000000015E-6) [ 0-5: a white smile eats the cat ]
    <NP> (0.005760000000000002) [ 0-2: a white smile ]
        <DET> (0.4) [ 0-0: a ]
        <NOMINAL> (0.07200000000000002) [ 1-2: white smile ]
            <ADJ> (0.8) [ 1-1: white ]
            <NOMINAL> (0.1) [ 2-2: smile ]
    <VP> (0.0019000000000000004) [ 3-5: eats the cat ]
        <V> (0.1) [ 3-3: eats ]
        <NP> (0.020000000000000004) [ 4-5: the cat ]
            <DET> (0.5) [ 4-4: the ]
            <NOMINAL> (0.2) [ 5-5: cat ]


sentence> my cat has a white rabbit

Parse for the sentence [ my cat has a white rabbit ] is below:

SYNOPSIS:

<NONTERMINAL> (PROBABILITY) [ SPAN: words of span ]

<S> (3.830400000000002E-6) [ 0-5: my cat has a white rabbit ]
    <NP> (0.0020000000000000005) [ 0-1: my cat ]
        <DET> (0.05) [ 0-0: my ]
        <NOMINAL> (0.2) [ 1-1: cat ]
    <VP> (0.005472000000000002) [ 2-5: has a white rabbit ]
        <V> (0.5) [ 2-2: has ]
        <NP> (0.011520000000000004) [ 3-5: a white rabbit ]
            <DET> (0.4) [ 3-3: a ]
            <NOMINAL> (0.14400000000000004) [ 4-5: white rabbit ]
                <ADJ> (0.8) [ 4-4: white ]
                <NOMINAL> (0.2) [ 5-5: rabbit ]


sentence> my cat has white rabbit

There's no parse for [ my cat has white rabbit ]

sentence> cat has a white rabbit

Parse for the sentence [ cat has a white rabbit ] is below:

SYNOPSIS:

<NONTERMINAL> (PROBABILITY) [ SPAN: words of span ]

<S> (3.283200000000001E-5) [ 0-4: cat has a white rabbit ]
    <NOMINAL> (0.2) [ 0-0: cat ]
    <VP> (0.005472000000000002) [ 1-4: has a white rabbit ]
        <V> (0.5) [ 1-1: has ]
        <NP> (0.011520000000000004) [ 2-4: a white rabbit ]
            <DET> (0.4) [ 2-2: a ]
            <NOMINAL> (0.14400000000000004) [ 3-4: white rabbit ]
                <ADJ> (0.8) [ 3-3: white ]
                <NOMINAL> (0.2) [ 4-4: rabbit ]


sentence> smile has my cat

Parse for the sentence [ smile has my cat ] is below:

SYNOPSIS:

<NONTERMINAL> (PROBABILITY) [ SPAN: words of span ]

<S> (2.8500000000000007E-6) [ 0-3: smile has my cat ]
    <NOMINAL> (0.1) [ 0-0: smile ]
    <VP> (9.500000000000002E-4) [ 1-3: has my cat ]
        <V> (0.5) [ 1-1: has ]
        <NP> (0.0020000000000000005) [ 2-3: my cat ]
            <DET> (0.05) [ 2-2: my ]
            <NOMINAL> (0.2) [ 3-3: cat ]


sentence> can you book TWA flights

Parse for the sentence [ can you book TWA flights ] is below:

SYNOPSIS:

<NONTERMINAL> (PROBABILITY) [ SPAN: words of span ]

<S> (3.192E-5) [ 0-4: can you book TWA flights ]
    <AuxNP> (0.09600000000000002) [ 0-1: can you ]
        <Aux> (0.4) [ 0-0: can ]
        <Pronoun> (0.4) [ 1-1: you ]
    <VP> (0.0033249999999999994) [ 2-4: book TWA flights ]
        <V> (0.1) [ 2-2: book ]
        <NP> (0.034999999999999996) [ 3-4: TWA flights ]
            <ProperNoun> (0.4) [ 3-3: TWA ]
            <NOMINAL> (0.25) [ 4-4: flights ]


sentence> the lion jumped through the hoop

There's no parse for [ the lion jumped through the hoop ]

sentence> the trainer jumped the lion through the hoop

There's no parse for [ the trainer jumped the lion through the hoop ]

sentence> the butter melted in the pan

There's no parse for [ the butter melted in the pan ]

sentence> the cook melted the butter in the pan

There's no parse for [ the cook melted the butter in the pan ]

sentence> the rich love their money

There's no parse for [ the rich love their money ]

sentence> the rich love sometimes too

There's no parse for [ the rich love sometimes too ]

sentence> the contractor built the houses last summer

There's no parse for [ the contractor built the houses last summer ]

sentence> the contractor built last summer

There's no parse for [ the contractor built last summer ]

sentence>
\end{verbatim}
\vskip4pt\hrule

\paragraph{Realistic Grammar}

\tiny
\hrule\vskip4pt
\begin{verbatim}

Probabilistic Parsing
Serguei A. Mokhov, mokhov@cs
April 2003



Entering interactive mode... Type \q to exit.
sentence> you should submit a paper listing and report and an electronic version of your code

There's no parse for [ you should submit a paper listing and report and an electronic version of your code ]

sentence> implement the CYK algorithm to find the best parse for a given sentence

There's no parse for [ implement the CYK algorithm to find the best parse for a given sentence ]

sentence> your program should take as input a probabilistic grammar and a sentence and display the best parse tree along with its probability

There's no parse for [ your program should take as input a probabilistic grammar and a sentence and display the best parse tree along with its probability ]

sentence> you are not required to use a specific programming language

There's no parse for [ you are not required to use a specific programming language ]

sentence> Perl, CPP, C or Java are appropriate for this task
WARNING: Non-word token encountered: Token[','], line 1
WARNING: Non-word token encountered: Token[','], line 1

There's no parse for [ Perl, CPP, C or Java are appropriate for this task ]

sentence> as long as you explain it, your grammar can be in any format you wish
WARNING: Non-word token encountered: Token[','], line 1

There's no parse for [ as long as you explain it, your grammar can be in any format you wish ]

sentence> your grammar can be in any format you wish as long as you explain it

There's no parse for [ your grammar can be in any format you wish as long as you explain it ]

sentence> experiment with your grammar

There's no parse for [ experiment with your grammar ]

sentence> is it restrictive enough

There's no parse for [ is it restrictive enough ]

sentence> does it refuse ungrammatical sentences

Parse for the sentence [ does it refuse ungrammatical sentences ] is below:

SYNOPSIS:

<NONTERMINAL> (PROBABILITY) [ SPAN: words of span ]

<S> (1.294887213288665E-8) [ 0-4: does it refuse ungrammatical sentences ]
    <AuxNP> (0.01999998) [ 0-1: does it ]
        <Aux> (0.1) [ 0-0: does ]
        <Pronoun> (0.333333) [ 1-1: it ]
    <VP> (6.474442540885865E-6) [ 2-4: refuse ungrammatical sentences ]
        <V> (0.0344828) [ 2-2: refuse ]
        <NOMINAL> (0.001104462402208) [ 3-4: ungrammatical sentences ]
            <ADJ> (0.0243902) [ 3-3: ungrammatical ]
            <NOMINAL> (0.0566038) [ 4-4: sentences ]


sentence> does it cover all grammatical sentences

Parse for the sentence [ does it cover all grammatical sentences ] is below:

SYNOPSIS:

<NONTERMINAL> (PROBABILITY) [ SPAN: words of span ]

<S> (1.2948872132886648E-9) [ 0-5: does it cover all grammatical sentences ]
    <AuxNP> (0.01999998) [ 0-1: does it ]
        <Aux> (0.1) [ 0-0: does ]
        <Pronoun> (0.333333) [ 1-1: it ]
    <VP> (6.474442540885865E-7) [ 2-5: cover all grammatical sentences ]
        <V> (0.0344828) [ 2-2: cover ]
        <NP> (5.52231201104E-5) [ 3-5: all grammatical sentences ]
            <PreDet> (0.5) [ 3-3: all ]
            <NOMINAL> (0.001104462402208) [ 4-5: grammatical sentences ]
                <ADJ> (0.0243902) [ 4-4: grammatical ]
                <NOMINAL> (0.0566038) [ 5-5: sentences ]


sentence> write a report to describe your code and your experimentation

There's no parse for [ write a report to describe your code and your experimentation ]

sentence> your report must describe the program

There's no parse for [ your report must describe the program ]

sentence> your report must describe the experiments

There's no parse for [ your report must describe the experiments ]

sentence> describe your code itself

There's no parse for [ describe your code itself ]

sentence> indicate the instructions necessary to run your code

There's no parse for [ indicate the instructions necessary to run your code ]

sentence> describe your choice of test sentences

There's no parse for [ describe your choice of test sentences ]

sentence> describe your grammar and how you developed it

Parse for the sentence [ describe your grammar and how you developed it ] is below:

SYNOPSIS:

<NONTERMINAL> (PROBABILITY) [ SPAN: words of span ]

<S> (7.381879524149979E-10) [ 0-7: describe your grammar and how you developed it ]
    <S> (4.11319751814313E-4) [ 0-2: describe your grammar ]
        <V> (0.206897) [ 0-0: describe ]
        <NOMINAL> (0.039760823193600005) [ 1-2: your grammar ]
            <ADJ> (0.439024) [ 1-1: your ]
            <NOMINAL> (0.113208) [ 2-2: grammar ]
    <ConjS> (1.7946815078995936E-5) [ 3-7: and how you developed it ]
        <Conj> (0.92) [ 3-3: and ]
        <S> (1.9507407694560798E-5) [ 4-7: how you developed it ]
            <WhNP> (0.094285785) [ 4-5: how you ]
                <WhWord> (0.5) [ 4-4: how ]
                <Pronoun> (0.571429) [ 5-5: you ]
            <VP> (0.002068965931032) [ 6-7: developed it ]
                <V> (0.0344828) [ 6-6: developed ]
                <Pronoun> (0.333333) [ 7-7: it ]


sentence> what problems do you see with your current implementation

There's no parse for [ what problems do you see with your current implementation ]

sentence> what problems do you see with your current grammar

There's no parse for [ what problems do you see with your current grammar ]

sentence> how would you improve it

There's no parse for [ how would you improve it ]

sentence> both the code and the report must be submitted electronically and in paper

There's no parse for [ both the code and the report must be submitted electronically and in paper ]

sentence> in class, you must submit a listing of your program and results and the report
WARNING: Non-word token encountered: Token[','], line 1

There's no parse for [ in class, you must submit a listing of your program and results and the report ]

sentence> through the electronic submission form you must submit the code of your program and results and an electronic version of your report

There's no parse for [ through the electronic submission form you must submit the code of your program and results and an electronic version of your report ]

sentence>
\end{verbatim}
\vskip4pt\hrule

\normalsize

\subsubsection{Code}
\label{sect:prob-parse-code}

The entire source code is provided in the electronic form, but
here we provide the two methods extracted from
\file{marf/nlp/Parsing/ProbabilisticParser.java} that are actual
implementation of the CYK algorithm and the \verb+build_tree()+ function,
named \api{parse()} and \file{dumpParseTree()} respectively.
The code below has experienced minor clean-ups in the report
version with most of the debug and error handling information removed. For the
unaltered code please refer the above mentioned file itself.

\paragraph{\texttt{ProbabilisticParser.parse() -- CYK}}

\tiny
\begin{verbatim}
public boolean parse()
throws SyntaxError
{
    // Restore grammar from the disk
    restore();

    // Split the string into words

    this.oWords = new Vector();

    int iTokenType;

    while((iTokenType = this.oStreamTokenizer.nextToken()) != StreamTokenizer.TT_EOF)
    {
        switch(iTokenType)
        {
            case StreamTokenizer.TT_WORD:
            {
                this.oWords.add(new String(this.oStreamTokenizer.sval));
                break;
            }

            case StreamTokenizer.TT_NUMBER:
            default:
            {
                System.err.println("WARNING: Non-word token encountered: " + this.oStreamTokenizer);
                break;
            }
        }
    }

    // CYK
    Vector oNonTerminals = this.oGrammar.getNonTerminalList();

    this.adParseMatrix = new double[this.oWords.size()][this.oWords.size()][oNonTerminals.size()];
    this.aoBack        = new Vector[this.oWords.size()][this.oWords.size()][oNonTerminals.size()];

    // Base case

    for(int i = 0; i < this.oWords.size(); i++)
    {
        String strTerminal = this.oWords.elementAt(i).toString();

        /*
         * Fail-fast: if the terminal is not in the grammar (no rule 'A->wd' found ),
         * there is no point to compute the parse
         */
        if(this.oGrammar.containsTerminal(strTerminal) == -1)
        {
            return false;
        }

        for(int A = 0; A < oNonTerminals.size(); A++)
        {
            ProbabilisticRule oRule = (ProbabilisticRule)this.oGrammar.getRule(strTerminal, A);

            // Current pair: 'A->wd' does not form a Rule
            if(oRule == null)
            {
                continue;
            }

            this.adParseMatrix[i][i][A] = oRule.getProbability();
        }
    }

    /*
     * Recursive case
     * ('recursive' as authors call it, but it's implemented iteratively
     * and me being just a copy-cat here)
     */
    for(int iSpan = 2; iSpan <= this.oWords.size(); iSpan++)
    {
        for(int iBegin = 0; iBegin < this.oWords.size() - iSpan + 1; iBegin++)
        {
            int iEnd = iBegin + iSpan - 1;

            // For every split m of the incoming sentence ...
            for(int m = iBegin; m <= iEnd - 1; m++)
            {
                // Check how the split divides B and C in A->BC
                for(int iA = 0; iA < oNonTerminals.size(); iA++)
                {
                    for(int iB = 0; iB < oNonTerminals.size(); iB++)
                    {
                        for(int iC = 0; iC < oNonTerminals.size(); iC++)
                        {
                            ProbabilisticRule oRule = (ProbabilisticRule)this.oGrammar.getRule(iA, iB, iC);

                            if(oRule == null)
                            {
                                continue;
                            }

                            double dProb =
                                this.adParseMatrix[iBegin][m][iB] *
                                this.adParseMatrix[m + 1][iEnd][iC] *
                                oRule.getProbability();

                            if(dProb > this.adParseMatrix[iBegin][iEnd][iA])
                            {
                                this.adParseMatrix[iBegin][iEnd][iA] = dProb;

                                Vector oBackIndices = new Vector();

                                oBackIndices.add(new Integer(m));
                                oBackIndices.add(new Integer(iB));
                                oBackIndices.add(new Integer(iC));

                                this.aoBack[iBegin][iEnd][iA] = oBackIndices;
                            }
                        } // for C
                    } // for B
                } // for A
            } // split
        }
    } // "recursive" case

    // No parse
    if(this.adParseMatrix[0][this.oWords.size() - 1][0] == 0)
    {
        return false;
    }

    return true;
}
\end{verbatim}
\normalsize

\paragraph{\texttt{ProbabilisticParser.dmpParseTree() -- build\_tree()}}

\tiny
\begin{verbatim}
public void dumpParseTree(int piLevel, int i, int j, int piA)
{
    NonTerminal oLHS = (NonTerminal)this.oGrammar.getNonTerminalList().elementAt(piA);

    indent(piLevel);

    // Termination case

    if(this.aoBack[i][j][piA] == null)
    {
        return;
    }

    // Recursive case

    int m = ((Integer)this.aoBack[i][j][piA].elementAt(0)).intValue();
    int iB = ((Integer)this.aoBack[i][j][piA].elementAt(1)).intValue();
    int iC = ((Integer)this.aoBack[i][j][piA].elementAt(2)).intValue();

    dumpParseTree(piLevel + 1, i, m, iB);
    dumpParseTree(piLevel + 1,  m + 1, j, iC);
}
\end{verbatim}
\normalsize

% EOF

%
% Internal Testing Applications
%

\section{MARF Testing Applications}
\index{MARF!Testing Applications}

%
% TestFilters
%

\subsection{TestFilters}
\index{Applications!TestFilters}
\index{Testing Applications!TestFilters}

$Revision: 1.7 $

\api{TestFilters} is one of the testing applications in {\marf}. It tests how four types of
FFT-filter-based preprocessors work with simulated or real wave type sound samples.
From user's point of view, \api{TestFilters} provides with usage, preprocessors, and loaders
command-line options.
By entering \texttt{--help}, or \texttt{-h}, or
when there are no arguments, the usage information will be displayed.
It explains the arguments used in \api{TestFilters}. The first argument is the type of
preprocessor/filter. These are high-pass filter (\texttt{--high}), low-pass filter (\texttt{--low});
band-pass filter (\texttt{--band}); high frequency boost preprocessor (\texttt{--boost})
to be chosen. Next argument is the type of loader to use to load initial testing sample.
\api{TestFilters} uses two types of
loaders, \api{SineLoader} and \api{WAVLoader}.
Users should enter either \texttt{--sine} or \texttt{--wave} as the second argument
to specify the desired loader. The argument \texttt{--sine} uses \api{SineLoader}
that will generate a plain sine wave to be fed to the selected preprocessor.
While the \texttt{--wave} argument uses \api{WAVLoader} to load a
real wave sound sample. In the latter case, users need to input the third argument --
sample file in the WAV format to feed to
\api{WAVLoader}. After selecting all necessary arguments, user can run and get the output of
\api{TestFilters} within seconds.

\noindent
Complete usage information:

\vspace{15pt}
\hrule
\begin{verbatim}

Usage:

    java TestFilters PREPROCESSOR LOADER [ SAMPLEFILE ]

    java TestFilters --help | -h
        displays this help and exits

    java TestFilters --version
        displays application and MARF versions

Where PREPROCESSOR is one of the following:
    --high           use high-pass filter
    --band           use band-pass filter
    --low            use low-pass filter
    --boost          use high-frequency-boost preprocessor
    --highpassboost  use high-pass-and-boost preprocessor

Where LOADER is one of the following:
    --sine    use SineLoader (generated sine wave)
    --wave    use WAVLoader; requires a SAMPLEFILE argument

\end{verbatim}

\hrule
\vspace{15pt}

The application is made to exercise the following {\marf} modules:

The main are the FFT-based filters in the
\api{marf.Preprocessing.FFTFilter.*} package.

\begin{enumerate}
\item
\api{BandpassFilter}
\item
\api{HighFrequencyBoost}
\item
\api{HighPassFilter}
\item
\api{LowPassFilter}
\end{enumerate}

Additionally, some other units were employed in this application:

\begin{enumerate}
\item
\api{marf.MARF}
\item
\api{marf.util.OptionProcessor}
\item
\api{marf.Storage.Sample}
\item
\api{marf.Storage.SampleLoader}
\item
\api{marf.Storage.WAVLoader}
\item
\api{marf.Preprocessing.Preprocessing}
\end{enumerate}

The main \api{MARF} module enumerates
these preprocessing modules as \api{HIGH\_FREQUENCY\_BOOST\_FFT\_FILTER}, \api{BANDPASS\_FFT\_FILTER}, \api{LOW\_PASS\_FFT\_FILTER}, and
\api{HIGH\_PASS\_FFT\_FILTER}, and incoming sample file format as \api{WAV}, \api{SINE}. \api{OptionProcessor} helps maintaining and
validating command-line options. \api{Sample} maintains incoming and processed sample data. \api{SampleLoader} provides sample loading
interface for all the MARF loaders. It must be overridden by a concrete sample loader such as \api{SineLoader} or \api{WAVLoader}.
\api{Preprocessing} does
general preprocessing such as \api{preprocess()} (overridden by the filters),
\api{normalize()}, \api{removeSilence()} and \api{removeNoise()} out of which
for this application the former three are used.
In the end, above modules work together to test the work of the filters and produce
the output to STDOUT.
The output of \api{TestFilters} is the filtered data from the original signal fed to each
of the preprocessors. It provides both users and programmers internal
information of the effect of MARF preprocessors so they can be compared with the
expected output in the \file{expected} folder to detect any errors if the underlying
algorithm has been changed.

%
% TestLPC
%

\subsection{TestLPC}
\index{Applications!TestLPC}
\index{Testing Applications!TestLPC}

$Revision: 1.5 $

The \api{TestLPC} application targets
the LPC unit testing as a preprocessing module.
Through a number of options it also allows choosing
between two implemented loaders -- \api{WAVLoader}
and \api{SineLoader}. To faciliate option processing
\api{marf.util.OptionProcessor} used that provides
an ability of maintaining and validating valid/active
option sets. The application also utilizes the \api{Dummy}
preprocessing module to perform the normalization of
incoming sample.

The application supports the following set of options:

\begin{itemize}
\item
\texttt{--help} or \texttt{-h} cause the application to
display the usage information and exit. The usage information
is also displayed if no option was supplied.

\item
\texttt{--sine} forces the use of \api{SineLoader} for
sample data generation. The output of this option is also
saved under \file{expected/sine.out} for regression testing.

\item
\texttt{--wave} forces the application to use the \api{WAVLoader}.
This option requires a mandatory filename argument of a wave file
to run the LPC algorithm against.
\end{itemize}

\noindent
Complete usage information:

\vspace{15pt}
\hrule
\begin{verbatim}

Usage:
    java TestLPC --help | -h
        displays this help and exits

    java TestLPC --version
        displays application and MARF versions

    java TestLPC --sine
        runs the LPC algorithm on a generated sine wave

    java TestLPC --wave <wave-file>
        runs the LPC algorithm against specified voice sample

\end{verbatim}

\hrule
\vspace{15pt}

% EOF

%
% TestFFT
%

\subsection{TestFFT}
\index{Applications!TestFFT}
\index{Testing Applications!TestFFT}

$Revision: 1.6 $

\api{TestFFT} is one of the testing applications in {\marf}.
It aims to test how the FFT (Fast Fourier Transform) algorithm works in
\marf{FeatureExtraction} by loading simulated or real wave sound samples.

From user's point of view, \api{TestFFT} provides with usage and loaders
command-line options. By entering \texttt{--help}, or \texttt{-h}, or
even no arguments, the usage information will be displayed. It explains
the arguments used in \api{TestFFT}. Another argument is the type of
loader. \api{TestFFT} uses two types of loaders, \api{Sineloader} and \api{WAVLoader}.
Users can enter \texttt{--sine} or \texttt{--wave} as the second argument.
The argument \texttt{--sine} uses \api{SineLoader} that will generate a
plain sine wave to be fed to a generic preprocessor.
If the argument \texttt{--wave} is specified, the application uses \api{WAVLoader}
to load a wave sample from file. In the latter case, users need to supply
one more argument -- sample file in the format of *.wav to be loaded by \api{WAVLoader}.
After selecting all necessary arguments, user can run and get the output of \api{TestFFT} within seconds.

\noindent
Complete usage information:

\vspace{15pt}
\hrule
\begin{verbatim}

Usage:
    java TestFFT --help | -h
        displays this help and exits

    java TestFFT --version
        displays application and MARF versions

    java TestFFT --sine
        runs the FFT algorithm on a generated sine wave

    java TestFFT --wave <wave-file>
        runs the FFT algorithm against specified voice sample

\end{verbatim}

\hrule
\vspace{15pt}

The application is made to exercise the {\marf}'s
FFT-based feature extraction algorithm located in the \api{marf.FeatureExtraction.FFT} package.
Additionally, the following {\marf} modules are utilized:

\begin{enumerate}
\item
\api{marf.MARF}

\item
\api{marf.util.OptionProcessor}

\item
\api{marf.Storage.Sample}

\item
\api{marf.Storage.SampleLoader}

\item
\api{marf.Storage.WAVLoader}

\item
\api{marf.Preprocessing.Dummy}
\end{enumerate}

The main \api{MARF} module enumerates incoming sample file format as
\api{WAV}, \api{SINE}. \api{OptionProcessor} helps maintain and validate
command-line options. \api{Sample} maintains and processed incoming
sample data. \api{SampleLoader} provides sample loading interface for
all the {\marf} loaders. It must be overridden by a concrete sample loader
such as \api{SineLoader} or \api{WAVLoader}. \api{Preprocessing} does general
preprocessing such with \api{preprocess()} (overridden by \api{Dummy}), \api{normalize()},
\api{removeSilence()} and \api{removeNoise()} out of which for this application the
former three are used. Then, processed data will be serve for an parameter of \api{FeatureExtraction.FFT.FFT}
to extract sample's features. In the end, above modules work together to test the work of the FFT
algorithm and produce the output to STDOUT.

As we know, the output of \api{TestFFT} extracts the features data by FFT feature extraction algorithm.
It gives both users and programmers direct information of the effect of MARF feature extraction,
and can be compared with the expected output in the \file{expected} folder to detect any errors
if the underlying algorithm has been changed.

% EOF

%
% TestWaveLoader
%

\subsection{TestWaveLoader}
\index{Applications!TestWaveLoader}
\index{Testing Applications!TestWaveLoader}

$Revision: 1.5 $

\api{TestWaveLoader} is one of the testing applications in {\marf}. It tests functionality of
the \api{WAVLoader} of {\marf}.

From user's point of view, \api{TestWaveLoader} provides with usage, input wave sample, output wave
sample, and output textual file command-line options.
By entering \texttt{--help}, or \texttt{-h}, or
when there are no arguments, the usage information will be displayed.
It explains the arguments used in \api{TestWaveLoader}. The first argument is an input wave sample file whose name is a mandatory argument.
The second and the third arguments are the output wave sample file and output textual sample file of the loaded input sample. The names of the output
files are optional arguments. If user does not provide any or both of the last two arguments, the output files will be saved in the files provided by 
\api{TestWaveLoader}.

\noindent
Complete usage information:

\vspace{15pt}
\hrule
\begin{verbatim}

Usage:
  java TestWaveLoader --help | -h

     displays usage information and exits

  java TestWaveLoader --version

     displays application and MARF versions

  java TestWaveLoader <input-wave-sample> [ <output-wave-sample> [ <output-txt-sample> ] ]

     loads a wave sample from the <input-wave-sample> file and
     stores the output wave sample in <output-wave-sample> and
     its textual equivalent is stored in <output-txt-sample>. If
     the second argument if omitted, the output file name is
     assumed to be "out.<input-wave-sample>". Likewise, if
     the third argument if omitted, the output file name is
     assumed to be "<output-wave-sample>.txt".

\end{verbatim}

\hrule
\vspace{15pt}

The application is made to exercise the following {\marf} modules.
The main module is the \api{WAVLoader} in the \api{marf.Storage.Loaders} package.
the \api{Sample} and the \api{SampleLoader} modules in the \api{marf.Storage} help \api{WAVLoader} prepare loading wave files.
\api{Sample} maintains and processes incoming sample data. \api{SampleLoader} provides sample loading interface for
all the {\marf} loaders. It must be overridden by a concrete sample loader. 
For loading wave samples, \api{SampleLoader} needs \api{WAVLoader} implementation.
Three modules work together to load in and write back a wave sample whose name was
provided by users in the first argument, to save the loaded sample into a newly-named wave file,
to save loaded input data into a data file referenced by \texttt{oDatFile}, 
and to output sample's duration and size to STDOUT.

As we know, the output of \api{TestWaveLoader} saves the loaded wave sample in a newly named output wave file.
Its output also saves the input file data into a textual file.
\api{TestWaveLoader} gives both users and programmers direct information of the results of MARF wave loader.
The output sample can be compared with the expected output in the \file{expected} folder to detect any errors.

% EOF

%
% TestLoaders
%

\subsection{TestLoaders}
\index{Applications!TestLoaders}
\index{Testing Applications!TestLoaders}

$Revision: 1.4 $

\api{TestLoaders} is another testing applications of {\marf}.
It generalizes the testing machinery of \api{TestWaveLoader} for 
all possible loaders we may have. For now, it tests functionality of
the \api{WAVLoader} and \api{SineLoader} of {\marf}, the only two
implemented loaders; the others give non-implemented exceptions instead. 

From user's point of view, \api{TestLoaders} provides with usage, loader types, input sample, output
sample, and output textual file command-line options.
By entering \texttt{--help}, or \texttt{-h}, or when there are no arguments, the usage information will be displayed.
Entering \texttt{--version}, the \api{TestLoaders}' version and {\marf}'s version is displayed.
The usage info explains the arguments used in \api{TestLoaders}.
The first argument is a type of loader that is a mandatory argument.
The second argument is input file name that is mandatory for all loaders except for \api{SineLoader}.
The third and forth arguments are the output wave sample file and output textual sample file names of the loaded 
input sample. The names of the output files are optional arguments. If user does not provide any or both of
the last two arguments, the output files will be saved in the file names derived from the original.

\noindent
Complete usage information:

\vspace{15pt}
\hrule
\begin{verbatim}

Usage:

  java TestLoaders --help | -h

     displays usage information and exits

  java TestLoaders --version

     displays TestLoaders version and MARF's version and exits

  java TestLoaders <LOADER> [ <input-sound-sample> ] [ <output-sound-sample> [ <output-txt-sample> ] ]

     selects a loader from various of loaders according to the LOADER option
     and then loads a corresponding type of sample from
     the <input-sound-sample> file and stores the output sound sample in
     <output-sound-sample> and its textual equivalent is stored in <output-txt-sample>.
     If the second argument if omitted, the output file name is
     assumed to be "out.<input-sound-sample>". Likewise,
     if the third argument omitted, the output file name is
     assumed to be "<output-sound-sample>.txt".

  Where LOADER is one of the following:

     --sine    use SineLoader (generated sine wave)
     --wave    use WAVLoader; requires a <input-sound-sample> argument
     --mp3     use MP3Loader; requires a <input-sound-sample> argument
     --ulaw    use ULAWLoader; requires a <input-sound-sample> argument
     --aiff    use AIFFLoader; requires a <input-sound-sample> argument
     --aiffc   use AIFFCLoader; requires a <input-sound-sample> argument
     --au      use AULoader; requires a <input-sound-sample> argument
     --snd     use SNDLoader; requires a <input-sound-sample> argument
     --midi    use MIDILoader; requires a <input-sound-sample> argument

\end{verbatim}

\hrule
\vspace{15pt}

The application is made to exercise the following {\marf} modules.
The main modules for testing are in the \api{marf.Storage.Loaders} package.
The \api{OptionProcessor} module in the \api {marf.util} helps handling the different loader types according
to the users input argument. The \api{Sample} and the \api{SampleLoader} modules in the \api{marf.Storage} help \api{WAVLoader} 
prepare loading input files. \api{Sample} maintains and processes incoming sample data. \api{SampleLoader}
provides sample loading interface for all the {\marf} loaders. It must be overridden by a concrete sample loader.
The modules work together to load in and write back a sample,
and save the loaded sample into a file, to save loaded input data into a data file referenced by \texttt{oDatFile}, 
and to output sample's duration and size to STDOUT. While for loading sine samples, it needs \api{SineLoader} implementation,
and instead of saving data file, it saves a csv file referenced by \texttt{oDatFile}.

The output of \api{TestLoaders} saves the loaded wave sample in a newly named output wave file.
Its output also saves the input file data into a textual file.
\api{TestLoaders} gives both users and programmers direct information of the results of MARF loaders.
Input sample can be compared with the expected output in the \file{expected} folder to detect any errors.

% EOF

%
% MathTestApp
%

\subsection{MathTestApp}
\index{Applications!MathTestApp}
\index{Testing Applications!MathTestApp}

$Revision: 1.3 $

The \api{MathTestApp} application targets
testing of the math-related implementations in the
\api{marf.math} package. As of this writing, the
application mainy exercises the \api{Matrix} class
as this is the one used by the \api{MahalanobisDistance}
classifier. It also does the necessary testing
of the \api{Vector} class as well.

The application supports the following set of options:

\begin{itemize}
\item
\texttt{--help} or \texttt{-h} cause the application to
display the usage information and exit.

\item
\texttt{--version} displays the application's and the
unrelying {\marf}'s versions and exits.
\end{itemize}

\noindent
Complete usage information:

\vspace{15pt}
\hrule
\begin{verbatim}

Usage:

    java MathTestApp
        runs the tests

    java MathTestApp --help | -h
        displays this help and exits

    java MathTestApp --version
        displays application and MARF versions

    java MathTestApp --debug
        run tests with the debug mode enabled

\end{verbatim}

\hrule
\vspace{15pt}

Before running tests, the application validates the
MARF version, and then produces the output to STDOUT of
various linear matrix operations, such as inversion,
identity, multiplication, transposal, scaling, rotation,
translation, and shear. The stored
output in the \file{expected} folder, \file{math.out},
contains expected output the authors believe is correct
and which is used in the regression testing.

% EOF

%
% TestPlugin
%

\subsection{TestPlugin}
\index{Applications!TestPlugin}
\index{Testing Applications!TestPlugin}

$Revision: 1.2 $

The \api{TestPlugin} testing application in {\marf}
tests and serves as an example of how to write
\api{SampleLoader}, \api{Preprocessing}, \api{FeatureExtraction}, and
\api{Classification} plugins to be used by the main
\api{MARF}'s pipeline.

\noindent
Complete usage information:

\vspace{15pt}
\hrule

\hrule
\vspace{15pt}

The application is made to exercise the {\marf}'s
interfaces \api{ISampleLoader}, \api{IPreprocessing}, \\\api{IFeatureExtraction}, and
\api{IClassification} along with the \api{MARF} itself. These interfaces
are implemented in the variety of ways. The sample loader is nearly identical
to the functionality of \api{WAVLoader} except that it yanks out checks
for sample format making it ``unrestricted'' (for illustratory purposes).
Random preprocessing is used to multiply the incoming amplitudes by
a pseudo-random Gausian distribution. The feature extraction is then
performed by convering the incoming variable-length input into chunks of
fixed size elements of which are added pairwise. Finally, the classification
does summing, logical AND, OR, and XOR ``hashing'' and applying a modulus
of the number of speakers we have. The actual samples are either the real
wave samples or a generated sine wave used.

% EOF

%
% TestPlugin
%

\subsection{TestNN}
\index{Applications!TestNN}
\index{Testing Applications!TestNN}

$Revision: 1.1 $

\api{TestNN} is one a testing applications in {\marf}
that aims to test how the neural network algorithm described in \xs{sect:nnet} works in
\api{Classification}.

From user's point of view, \api{TestNN} provides with usage and loaders
command-line options. By entering \texttt{--help} or \texttt{-h}
the usage information will be displayed.

\noindent
Complete usage information:

\vspace{15pt}
\hrule
\begin{verbatim}
Usage:
    java TestNN [ OPTIONS ] <sample-xml-file>
    java TestNN --help | -h | --version
    java -jar TestNN.jar [ OPTIONS ] <sample-xml-file>

    java -jar TestNN.jar --help | -h | --version

Options (one or more of the following):
    --debug       - enable debugging
    -dtd          - validate the DTD of the corresponding XML document
    -error=VALUE  - specify minimum desired error
    -epochs=VALUE - specify numer of epochs for training
    -train=NUMBER - number of training interations
    -test=NUMBER  - number of testing iterations

\end{verbatim}

\hrule
\vspace{15pt}

The application is made to exercise the {\marf}'s implementation of the
artificial neural network algorithm located in the \api{marf.Classification.NeuralNetwork} package.
Additionally, the following {\marf} modules are utilized:

\begin{enumerate}
\item
\api{marf.util.OptionProcessor}

\item
\api{marf.util.Arrays}

\item
\api{marf.util.Debug}
\end{enumerate}

% EOF

%
% External Applications
%

\section{External Applications}
\index{MARF!External Applications}

%
% GIPSY
%

\subsection{GIPSY}
\index{Applications!GIPSY}
\index{External Applications!GIPSY}

{\gipsy} stands for General Intensional Programming System \cite{gipsy}, which is
a distributed research, development, and investigation platform about intensional and hybrid
programming languages. {\gipsy} is a world on its own developed at Concordia University and
for more information on the system please see \cite{gipsy} and \cite{mokhovmcthesis05}.
{\gipsy} makes use of a variety of {\marf}'s utility and storage modules, as of this
writing this includes:

\begin{itemize}
\item
	\api{BaseThread} for most threading part of the {\gipsy} for multithreaded
	compilation and execution.
\item
	\api{ExpandedThreadGroup} to group a number of compiler or executor threads
	and have a group control over them.
\item
	\api{Debug} for, well, debugging
\item
	\api{Logger} to log compiler and executor activities
\item
	\api{StorageManager} for binary serialization
\item
	\api{OptionProcessor} for option processing in five core {\gipsy} utilities
\item
	\api{Arrays} for common arrays functionality
\end{itemize}

There is a provision to also use {\marf}'s NLP modules.

%
% ENCSAssetsDesktop
%

\subsection{ENCSAssetsDesktop}
\label{sect:encs-assets-desktop}
\index{Applications!ENCSAssetsDesktop}
\index{External Applications!ENCSAssetsDesktop}

This is a private little application developed for the Engineering and Computer Science
faculty of Concordia University's Faculty Information Systems team. The application
synchronize inventory data between a Palm device database and a relational database
powering an online inventory application. This application primarily exercises:

\begin{itemize}
\item
    \api{BaseThread} for most threading part of the {\gipsy} for multithreaded
    compilation and execution.
\item
    \api{ExpandedThreadGroup} to group a number of compiler or executor threads
    and have a group control over them.
\item
    \api{Arrays} for common arrays functionality
\item
	\api{ByteUtils} for any-type-to-byte-array-and-back conversion.
\end{itemize}

%
% ENCSAssets
%

\subsection{ENCSAssets}
\index{Applications!ENCSAssets}
\index{External Applications!ENCSAssets}

This is a web frontend to an inventory database developed by the same team as
in \xs{sect:encs-assets-desktop}. It primarily exercises the threading
and arrays parts of {\marf}, namely \api{BaseThread} and \api{Arrays}.

%
% ENCSAssetsCron
%

\subsection{ENCSAssetsCron}
\index{Applications!ENCSAssetsCron}
\index{External Applications!ENCSAssetsCron}

This is a cron-oriented frontend to an inventory database developed by the same team as
in \xs{sect:encs-assets-desktop}. It primarily exercises the \api{OptionProcessor}
of {\marf}.

% EOF

	\chapter{Conclusions}

$Revision: 1.13 $

\section{Review of Results}

Our best configuration yielded \bestscore{82.76
} correctness
of our work when identifying subjects. Having a total
of 29 testing samples (including the two music bands) that means
24 subjects identified correctly out of 29.

The main reasons the recognition rate could be low
is due to non-uniform sample taking, lack of preprocessing
techniques optimizations and parameter tweaking, lack of noise removal, lack or incomplete
sophisticated classification modules (e.g. Stochastic models),
and lack the samples themselves to train and test on.
You are welcome to contribute in any of these areas to make the framework better.

Even though for commercial and University-level research
standards \bestscore{82.76
} recognition rate is considered to be
low as opposed to a required minimum of 95\%-97\% and above,
we think it is still reasonably well off
to demonstrate the framework operation.

\section{Acknowledgments}

We would like to thank Dr. Suen and Mr. Sadri
for the course and help provided.
This {\LaTeX} documentation was possible due to an
excellent introduction of it by Dr. Peter Grogono
in \cite{grogono2001}.

% EOF

	\addcontentsline{toc}{chapter}{Bibliography}

% Look for report.bib file if you seek bibliography entries.
\bibliography{report}
\bibliographystyle{alpha}

% EOF

	\appendix

\chapter{Spectrogram Examples}
\label{appx:spectra}

$Revision: 1.14 $

As produced by the \api{Spectrogram} class.

\begin{figure}[h]
	\centering
	\includegraphics[width=500pt]{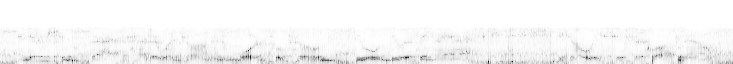}
	\caption{LPC spectrogram obtained for ian15.wav}
	\includegraphics[width=500pt]{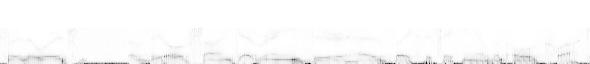}
	\caption{LPC spectrogram obtained for \file{graham13.wav}}
\end{figure}

%%%%%%%%%%%%%%%%%%%%%%%%%%%%%%%%%%%%%%%%%%%%%%%%%%

\chapter{MARF Source Code}

You can download
the code from \verb+<http://marf.sourceforge.net>+, specifically:

\begin{itemize}
	\item The latest unstable version:\\\verb+<http://marf.sourceforge.net/marf.tar.gz>+
	\item Browse code and revision history online:\\\verb+<http://cvs.sourceforge.net/cgi-bin/viewcvs.cgi/marf/>+
\end{itemize}

API documentation in the HTML format can be found in the documentation
distribution, or for the latest version please consult:
\verb+<http://marf.sourceforge.net/api/>+. If you want to participate
in development, there is a developers version of the API:
\verb+<http://marf.sourceforge.net/api-dev/>+, which includes
all the private constructs into the docs as well.

The current development version can also be retrieved via \tool{CVS}. The
process outlined in \xa{appx:cvs}.

\clearpage

%%%%%%%%%%%%%%%%%%%%%%%%%%%%%%%%%%%%%%%%%%%%%%%%%%

\chapter{The CVS Repository}
\label{appx:cvs}

The {\marf} source code is stored and managed using the
CVS code management system at SourceForge.
Anonymous CVS
is available to pull the CVS code tree from the
{\marf} package to your local machine.

\section{Getting The Source Via Anonymous CVS}

If you would like to keep up with the current sources on a regular
basis, you can fetch them from SourceForge's CVS server
and then use CVS to
retrieve updates from time to time.

\begin{itemize}
\item
     You will need a local copy of CVS
     (Concurrent Version Control System), which you can get from
     \verb+<http://www.cvshome.org/>+ or
     any GNU software archive site. There is also WinCVS and
     CVS mode built in JBulider if you plan to use these products
     on Win32 platforms.

\item
     Do an initial login to the CVS server:

\begin{verbatim}
cvs -d:pserver:anonymous@marf.cvs.sourceforge.net:/cvsroot/marf login
\end{verbatim}

     You will be prompted for a password; just press ENTER.
     You should only need to do this once, since the password will be
     saved in \verb+.cvspass+ in your home directory.

\item
     Fetch the {\marf} sources:

\begin{verbatim}
cvs -z3 -d:pserver:anonymous@marf.cvs.sourceforge.net:/cvsroot/marf co -P marf
\end{verbatim}

     which installs the {\marf} sources into a
     subdirectory \verb+marf+
     of the directory you are currently in.

     If you'd like to download sample applications which use {\marf}:

\begin{verbatim}
cvs -z3 -d:pserver:anonymous@marf.cvs.sourceforge.net:/cvsroot/marf co -P apps
\end{verbatim}

\item
     Whenever you want to update to the latest CVS sources,
     \verb+cd+ into
     the \verb+marf+ or \verb+apps+ subdirectories, and issue

\begin{verbatim}
cvs -z3 update -d -P
\end{verbatim}

     This will fetch only the changes since the last time you updated.

\item
     You can save yourself some typing by making a file \verb+.cvsrc+
     in your home directory that contains

\begin{verbatim}
cvs -z3
update -d -P
\end{verbatim}

     This supplies the \verb+-z3+ option to all \verb+cvs+ commands, and the
     \verb+-d+ and \verb+-P+ options to \verb+cvs update+.  Then you just have
     to say

\begin{verbatim}
cvs update
\end{verbatim}

     to update your files.

\end{itemize}

%%%%%%%%%%%%%%%%%%%%%%%%%%%%%%%%%%%%%%%%%%%%%%%%%%

\chapter{\api{SpeakerIdentApp} and \api{SpeakersIdentDb} Source Code}

\section{\texttt{SpeakerIdentApp.java}}

\vspace{15pt}
\hrule
{\scriptsize \begin{verbatim}
import java.io.File;
import java.util.Calendar;
import java.util.Date;

import marf.MARF;
import marf.Classification.Distance.HammingDistance;
import marf.Storage.ModuleParams;
import marf.Storage.TrainingSet;
import marf.util.Debug;
import marf.util.MARFException;
import marf.util.OptionProcessor;


/**
 * <p>Identifies a speaker independently of text, based on the MARF framework,
 * <a href="http://marf.sf.net">http://marf.sf.net</a>.
 * </p>
 *
 * $Id: SpeakerIdentApp.java,v 1.70 2009/03/03 12:06:54 mokhov Exp $
 *
 * @author Serguei Mokhov
 * @author Stephen Sinclair
 * @author Ian Clement
 * @author Dimitrios Nicolacopoulos
 * @author The MARF Research and Development Group
 *
 * @version 0.3.0, $Revision: 1.70 $
 * @since 0.0.1
 */
public class SpeakerIdentApp
{
    /*
     * ----------------
     * Apps. Versioning
     * ----------------
     */

    /**
     * Current major version of the application.
     */
    public static final int MAJOR_VERSION = 0;

    /**
     * Current minor version of the application.
     */
    public static final int MINOR_VERSION = 3;

    /**
     * Current revision of the application.
     */
    public static final int REVISION      = 0;

    
    /*
     * ----------------------------------
     * Defaults
     * ----------------------------------
     */

    /**
     * Default file name extension for sample files; case-insensitive.
     * @since 0.3.0.6
     */
    public static final String DEFAULT_SAMPLE_FILE_EXTENSION = "wav";
    //public static final String DEFAULT_SAMPLE_FILE_EXTENSION = "txt";

    /*
     * ----------------------------------
     * Major and Misc Options Enumeration
     * ----------------------------------
     */

    /**
     * Numeric equivalent of the option <code>--train</code>.
     * @since 0.3.0.5
     */
    public static final int OPT_TRAIN = 0;

    /**
     * Numeric equivalent of the option <code>--ident</code>.
     * @since 0.3.0.5
     */
    public static final int OPT_IDENT = 1;

    /**
     * Numeric equivalent of the option <code>--stats</code>.
     * @since 0.3.0.5
     */
    public static final int OPT_STATS = 2;

    /**
     * Numeric equivalent of the option <code>--reset</code>.
     * @since 0.3.0.5
     */
    public static final int OPT_RESET = 3;

    /**
     * Numeric equivalent of the option <code>--version</code>.
     * @since 0.3.0.5
     */
    public static final int OPT_VERSION = 4;

    /**
     * Numeric equivalent of the option <code>--help</code>.
     * @since 0.3.0.5
     */
    public static final int OPT_HELP_LONG = 5;

    /**
     * Numeric equivalent of the option <code>-h</code>.
     * @since 0.3.0.5
     */
    public static final int OPT_HELP_SHORT = 6;

    /**
     * Numeric equivalent of the option <code>-debug</code>.
     * @since 0.3.0.5
     */
    public static final int OPT_DEBUG = 7;

    /**
     * Numeric equivalent of the option <code>-graph</code>.
     * @since 0.3.0.5
     */
    public static final int OPT_GRAPH = 8;

    /**
     * Numeric equivalent of the option <code>-spectrogram</code>.
     * @since 0.3.0.5
     */
    public static final int OPT_SPECTROGRAM = 9;

    /**
     * Numeric equivalent of the option <code>&lt;speaker ID&gt;</code>.
     * @since 0.3.0.5
     */
    public static final int OPT_EXPECTED_SPEAKER_ID = 10;

    /**
     * Numeric equivalent of the option <code>--batch-ident</code>.
     * @since 0.3.0.5
     */
    public static final int OPT_BATCH_IDENT = 11;

    /**
     * Numeric equivalent of the option <code>--single-train</code>.
     * @since 0.3.0.5
     */
    public static final int OPT_SINGLE_TRAIN = 12;

    /**
     * Numeric equivalent of the option <code>&lt;sample-file-or-directory-name&gt;</code>.
     * @since 0.3.0.5
     */
    public static final int OPT_DIR_OR_FILE = 13;

    /**
     * Numeric equivalent of the option <code>--best-score</code>.
     * @since 0.3.0.5
     */
    public static final int OPT_BEST_SCORE = 14;
    
    /**
     * Numeric equivalent of the option <code>--gui</code> to have a GUI.
     * @since 0.3.0.6
     */
    public static final int OPT_GUI = 15;

    /**
     * Numeric equivalent of the option <code>-noise</code> to enable noise removal.
     * @since 0.3.0.6
     */
    public static final int OPT_NOISE_REMOVAL = 16;

    /**
     * Numeric equivalent of the option <code>-silence</code> to enable silence removal.
     * @since 0.3.0.6
     */
    public static final int OPT_SILENCE_REMOVAL = 17;


    /*
     * ---------------------
     * State Data Structures
     * ---------------------
     */

    /**
     * Instance of the database of speakers to track stats per
     * configuration for all speakers.
     * @since 0.3.0.5
     */
    protected static SpeakersIdentDb soDBPerConfig = new SpeakersIdentDb("speakers.txt");

    /**
     * Instance of the database of speakers to track stats per
     * speaker for all configuration.
     * @since 0.3.0.6
     */
    protected static SpeakersIdentDb soDBPerSpeaker = new SpeakersIdentDb("speakers.txt");

//    protected static SpeakersIdentDb soDBPerConfig = new SpeakersIdentDb("languages.txt");
//    protected static SpeakersIdentDb soDBPerSpeaker = new SpeakersIdentDb("languages.txt");

    /**
     * Instance of the option processing utility.
     * @since 0.3.0.5
     */
    protected static OptionProcessor soGetOpt = new OptionProcessor();

    /*
     * -----------------
     * Static State Init
     * -----------------
     */

    static
    {
        // Main options
        soGetOpt.addValidOption(OPT_TRAIN, "--train");
        soGetOpt.addValidOption(OPT_SINGLE_TRAIN, "--single-train");
        soGetOpt.addValidOption(OPT_IDENT, "--ident");
        soGetOpt.addValidOption(OPT_BATCH_IDENT, "--batch-ident");
        soGetOpt.addValidOption(OPT_STATS, "--stats");
        soGetOpt.addValidOption(OPT_BEST_SCORE, "--best-score");
        soGetOpt.addValidOption(OPT_VERSION, "--version");
        soGetOpt.addValidOption(OPT_RESET, "--reset");
        soGetOpt.addValidOption(OPT_HELP_LONG, "--help");
        soGetOpt.addValidOption(OPT_HELP_SHORT, "-h");
        soGetOpt.addValidOption(OPT_GUI, "--gui");

        // Sample Loading
        soGetOpt.addValidOption(MARF.WAV, "-wav");
        soGetOpt.addValidOption(MARF.TEXT, "-text");

        // Preprocessing
        soGetOpt.addValidOption(OPT_NOISE_REMOVAL, "-noise");
        soGetOpt.addValidOption(OPT_SILENCE_REMOVAL, "-silence");

        soGetOpt.addValidOption(MARF.DUMMY, "-norm");
        soGetOpt.addValidOption(MARF.HIGH_FREQUENCY_BOOST_FFT_FILTER, "-boost");
        soGetOpt.addValidOption(MARF.HIGH_PASS_FFT_FILTER, "-high");
        soGetOpt.addValidOption(MARF.LOW_PASS_FFT_FILTER, "-low");
        soGetOpt.addValidOption(MARF.BANDPASS_FFT_FILTER, "-band");
        soGetOpt.addValidOption(MARF.BAND_STOP_FFT_FILTER, "-bandstop");
        soGetOpt.addValidOption(MARF.HIGH_PASS_BOOST_FILTER, "-highpassboost");
        soGetOpt.addValidOption(MARF.RAW, "-raw");
        soGetOpt.addValidOption(MARF.ENDPOINT, "-endp");
        soGetOpt.addValidOption(MARF.HIGH_PASS_CFE_FILTER, "-highcfe");
        soGetOpt.addValidOption(MARF.LOW_PASS_CFE_FILTER, "-lowcfe");
        soGetOpt.addValidOption(MARF.BAND_PASS_CFE_FILTER, "-bandcfe");
        soGetOpt.addValidOption(MARF.BAND_STOP_CFE_FILTER, "-bandstopcfe");
        
        // Feature extraction
        soGetOpt.addValidOption(MARF.FFT, "-fft");
        soGetOpt.addValidOption(MARF.LPC, "-lpc");
        soGetOpt.addValidOption(MARF.RANDOM_FEATURE_EXTRACTION, "-randfe");
        soGetOpt.addValidOption(MARF.MIN_MAX_AMPLITUDES, "-minmax");
        soGetOpt.addValidOption(MARF.FEATURE_EXTRACTION_AGGREGATOR, "-aggr");
        soGetOpt.addValidOption(MARF.F0, "-f0");
        soGetOpt.addValidOption(MARF.SEGMENTATION, "-segm");
        soGetOpt.addValidOption(MARF.CEPSTRAL, "-cepstral");

        // Classification
        soGetOpt.addValidOption(MARF.NEURAL_NETWORK, "-nn");
        soGetOpt.addValidOption(MARF.EUCLIDEAN_DISTANCE, "-eucl");
        soGetOpt.addValidOption(MARF.CHEBYSHEV_DISTANCE, "-cheb");
        soGetOpt.addValidOption(MARF.MINKOWSKI_DISTANCE, "-mink");
        soGetOpt.addValidOption(MARF.MAHALANOBIS_DISTANCE, "-mah");
        soGetOpt.addValidOption(MARF.RANDOM_CLASSIFICATION, "-randcl");
        soGetOpt.addValidOption(MARF.DIFF_DISTANCE, "-diff");
        soGetOpt.addValidOption(MARF.ZIPFS_LAW, "-zipf");
        soGetOpt.addValidOption(MARF.MARKOV, "-markov");
        soGetOpt.addValidOption(MARF.HAMMING_DISTANCE, "-hamming");
        soGetOpt.addValidOption(MARF.COSINE_SIMILARITY_MEASURE, "-cos");

        // Misc
        soGetOpt.addValidOption(OPT_SPECTROGRAM, "-spectrogram");
        soGetOpt.addValidOption(OPT_DEBUG, "-debug");
        soGetOpt.addValidOption(OPT_GRAPH, "-graph");
    }
    
    
    /**
     * Main body.
     * @param argv command-line arguments
     */
    public static final void main(String[] argv)
    {
        try
        {
            // Since some new API is always introduced...
            validateVersions();

            /*
             * Load the speakers database
             */
            soDBPerConfig.connect();
            soDBPerConfig.query();

            soDBPerSpeaker.connect();
            soDBPerSpeaker.query();

            soDBPerConfig.setFilename("config." + soDBPerConfig.getFilename());
            soDBPerSpeaker.setFilename("speaker." + soDBPerSpeaker.getFilename());

            setDefaultConfig();
            
            // Parse extra arguments
            int iValidOptions = soGetOpt.parse(argv);

            if(iValidOptions == 0)
            {
                throw new Exception("No valid options found: " + soGetOpt);
            }

            setCustomConfig();

            /*
             * If supplied in the command line, the system when classifying,
             * will output this ID next to the guessed one.
             */
            int iExpectedID = -1;

            switch(soGetOpt.getInvalidOptions().size())
            {
                // Unknown
                case 0:
                {
                    iExpectedID = -1;
                    break;
                }
                
                /*
                 * Extract and make active and valid option out of
                 * a filename and and an expected speaker ID. An
                 * assumption is that the expected speaker ID is
                 * always second on the command line somewhere after
                 * a file or directory name. Presence of the expected
                 * speaker ID always implies presence of the file
                 * or directory argument. 
                 */
                case 2:
                {
                    try
                    {
                        iExpectedID = Integer.parseInt(soGetOpt.getInvalidOptions().elementAt(2).toString());
                        soGetOpt.addActiveOption(OPT_EXPECTED_SPEAKER_ID, soGetOpt.getInvalidOptions().elementAt(2).toString());
                    }

                    /*
                     * May happened when trying to get expected ID,
                     * but the argument doesn't parse as an int.
                     */
                    catch(NumberFormatException e)
                    {
                        iExpectedID = -1;

                        System.err.println
                        (
                            "SpeakerIdentApp: WARNING: could not parse expected speaker ID ("
                            + e.getMessage() + "), ignoring..."
                        );
                    }

                    // No break required as the file or directory
                    // must always be present. Also, the clearance
                    // of the invalid options need to be postponed.
                }
                
                /*
                 * In the case of a single invalid option
                 * always assume it is either a filename
                 * or a directory name for the --ident
                 * or --train options.
                 */
                case 1:
                {
                    soGetOpt.addActiveOption(OPT_DIR_OR_FILE, soGetOpt.getInvalidOptions().firstElement().toString());
                    soGetOpt.getInvalidOptions().clear();
                    break;
                }

                default:
                {
                    throw new Exception("Unrecognized options found: " + soGetOpt.getInvalidOptions());
                }
            }

            // Set misc configuration
            MARF.setDumpSpectrogram(soGetOpt.isActiveOption(OPT_SPECTROGRAM));
            MARF.setDumpWaveGraph(soGetOpt.isActiveOption(OPT_GRAPH));
            Debug.enableDebug(soGetOpt.isActiveOption(OPT_DEBUG));

            Debug.debug("Option set: " + soGetOpt);

            int iMainOption = soGetOpt.getOption(argv[0]);

            switch(iMainOption)
            {
                /*
                 * --------------
                 * Identification
                 * --------------
                 */
                
                // Single case
                case OPT_IDENT:
                {
                    ident(getConfigString(argv), soGetOpt.getOption(OPT_DIR_OR_FILE), iExpectedID);
                    break;
                }

                // A directory with files for identification
                case OPT_BATCH_IDENT:
                {
                    // Store config and error/successes for that config
                    String strConfig = getConfigString(argv);

                    // Dir contents
                    File[] aoSampleFiles = new File(soGetOpt.getOption(OPT_DIR_OR_FILE)).listFiles();

                    for(int i = 0; i < aoSampleFiles.length; i++)
                    {
                        String strFileName = aoSampleFiles[i].getPath();

                        if(aoSampleFiles[i].isFile() && strFileName.toLowerCase().endsWith("." + DEFAULT_SAMPLE_FILE_EXTENSION))
                        {
                            ident(strConfig, strFileName, iExpectedID);
                        }
                    }

                    break;
                }

                /*
                 * --------
                 * Training
                 * --------
                 */

                // Add a single sample to the training set
                case OPT_SINGLE_TRAIN:
                {
                    train(soGetOpt.getOption(OPT_DIR_OR_FILE));
                    System.out.println("Done training with file \"" + soGetOpt.getOption(OPT_DIR_OR_FILE) + "\".");
                    break;
                }
                
                // Train on a directory of files
                case OPT_TRAIN:
                {
                    try
                    {
                        // Dir contents
                        File[] aoSampleFiles = new File(soGetOpt.getOption(OPT_DIR_OR_FILE)).listFiles();
    
                        Debug.debug("Files array: " + aoSampleFiles);
                        
                        if(Debug.isDebugOn())
                        {
                            System.getProperties().list(System.err);
                        }
                        
                        String strFileName = "";

                        // XXX: this loop has to be in MARF
                        for(int i = 0; i < aoSampleFiles.length; i++)
                        {
                            strFileName = aoSampleFiles[i].getPath();
    
                            if(aoSampleFiles[i].isFile() && strFileName.toLowerCase().endsWith("." + DEFAULT_SAMPLE_FILE_EXTENSION))
                            {
                                train(strFileName);
                            }
                        }
                    }
                    catch(NullPointerException e)
                    {
                        System.err.println("Folder \"" + soGetOpt.getOption(OPT_DIR_OR_FILE) + "\" not found.");
                        e.printStackTrace(System.err);
                        System.exit(-1);
                    }

                    System.out.println("Done training on folder \"" + soGetOpt.getOption(OPT_DIR_OR_FILE) + "\".");

                    break;
                }

                /*
                 * -----
                 * Stats
                 * -----
                 */
                case OPT_STATS:
                {
                    soDBPerConfig.restore();
                    soDBPerConfig.printStats();
                    soDBPerSpeaker.restore();
                    soDBPerSpeaker.printStats();
                    break;
                }

                /*
                 * Best Result with Stats
                 */
                case OPT_BEST_SCORE:
                {
                    soDBPerConfig.restore();
                    soDBPerConfig.printStats(true);
                    soDBPerSpeaker.restore();
                    soDBPerSpeaker.printStats(true);
                    break;
                }

                /*
                 * Reset Stats
                 */
                case OPT_RESET:
                {
                    soDBPerConfig.resetStats();
                    soDBPerSpeaker.resetStats();
                    System.out.println("SpeakerIdentApp: Statistics has been reset.");
                    break;
                }

                /*
                 * Versioning
                 */
                case OPT_VERSION:
                {
                    System.out.println("Text-Independent Speaker Identification Application, v." + getVersion());
                    System.out.println("Using MARF, v." + MARF.getVersion());
                    validateVersions();
                    break;
                }

                /*
                 * Help
                 */
                case OPT_HELP_SHORT:
                case OPT_HELP_LONG:
                {
                    usage();
                    break;
                }

                /*
                 * Invalid major option
                 */
                default:
                {
                    throw new Exception("Unrecognized option: " + argv[0]);
                }
            } // major option switch
        } // try

        /*
         * No arguments have been specified
         */
        catch(ArrayIndexOutOfBoundsException e)
        {
            System.err.println("No arguments have been specified.");

            //if(Debug.isDebugOn())
            {
                System.err.println(e.getMessage());
                e.printStackTrace(System.err);
            }

            usage();
        }

        /*
         * MARF-specific errors
         */
        catch(MARFException e)
        {
            System.err.println(e.getMessage());
            e.printStackTrace(System.err);
        }

        /*
         * Invalid option and/or option argument
         */
        catch(Exception e)
        {
            System.err.println(e.getMessage());
            e.printStackTrace(System.err);
            usage();
        }

        /*
         * Regardless whatever happens, close the db connection.
         */
        finally
        {
            try
            {
                Debug.debug("Closing DB connection...");
                soDBPerConfig.close();
                soDBPerSpeaker.close();
            }
            catch(Exception e)
            {
                Debug.debug("Closing DB connection failed: " + e.getMessage());
                e.printStackTrace(System.err);
                System.exit(-1);
            }
        }
    }

    /**
     * Identifies a speaker using MARF given configuration,
     * wave filename, and possibly expected speaker ID.
     *
     * @param pstrConfig configuration string for stats
     * @param pstrFilename name of the wave file with voice sample to identify
     * @param piExpectedID expected speaker ID; if -1 then no stats is kept
     *
     * @throws MARFException in case of any error while processing is in MARF
     * @since 0.3.0.5
     */
    public static final void ident(String pstrConfig, String pstrFilename, int piExpectedID)
    throws MARFException
    {
        /*
         * If no expected speaker present on the command line,
         * attempt to fetch it from the database by filename.
         */
        if(piExpectedID < 0)
        {
            piExpectedID = soDBPerConfig.getIDByFilename(pstrFilename, false);
        }

        int iIdentifiedID = -1;

        // Second best
        int iSecondClosestID = -1;

        long lTimeStart = System.currentTimeMillis();
        {
            MARF.setSampleFile(pstrFilename);
            MARF.recognize();
    
            // First guess
            iIdentifiedID = MARF.queryResultID();
    
            // Second best
            iSecondClosestID = MARF.getResultSet().getSecondClosestID();
        }
        long lTimeFinish = System.currentTimeMillis();

        // XXX: constify
        int iOneSecInMs  = 1000;
        int iOneMinInMs  = 60 * iOneSecInMs;
        int iOneHourInMs = 60 * iOneMinInMs; 
        int iOneDayInMs  = 24 * iOneHourInMs; 
        
        long lDuration = lTimeFinish - lTimeStart;

        long lDays = lDuration / iOneDayInMs;
        long lModTimeLeft = lDuration % iOneDayInMs;
        
        long lHours = lModTimeLeft / iOneHourInMs;
        lModTimeLeft %= iOneHourInMs;
        
        long lMinutes = lModTimeLeft / iOneMinInMs;
        lModTimeLeft %= iOneMinInMs;
        
        long lSeconds = lModTimeLeft / iOneSecInMs;
        lModTimeLeft %= iOneSecInMs;
        
        long lMilliSeconds = lModTimeLeft; 

        StringBuffer oDuration = new StringBuffer()
            .append(lDays).append("d:")
            .append(lHours).append("h:")
            .append(lMinutes).append("m:")
            .append(lSeconds).append("s:")
            .append(lMilliSeconds).append("ms:")
            .append(lDuration).append("ms")
            ;
        
        System.out.println("                 File: " + pstrFilename);
        System.out.println("               Config: " + pstrConfig);
        System.out.println("      Processing time: " + oDuration);
        System.out.println("         Speaker's ID: " + iIdentifiedID);
        System.out.println("   Speaker identified: " + soDBPerConfig.getName(iIdentifiedID));

        /*
         * Only collect stats if we have expected speaker
         */
        if(piExpectedID > 0)
        {
            System.out.println("Expected Speaker's ID: " + piExpectedID);
            System.out.println("     Expected Speaker: " + soDBPerConfig.getName(piExpectedID));

            soDBPerConfig.restore();
            soDBPerSpeaker.restore();
            {
                // 1st match
                soDBPerConfig.addStats(pstrConfig, (iIdentifiedID == piExpectedID));
                soDBPerSpeaker.addStats(soDBPerConfig.getName(piExpectedID), (iIdentifiedID == piExpectedID));

                // 2nd best: must be true if either 1st true or second true (or both :))
                boolean bSecondBest =
                    iIdentifiedID == piExpectedID
                    ||
                    iSecondClosestID == piExpectedID;

                soDBPerConfig.addStats(pstrConfig, bSecondBest, true);
                soDBPerSpeaker.addStats(soDBPerConfig.getName(piExpectedID), bSecondBest, true);
            }
            soDBPerConfig.dump();
            soDBPerSpeaker.dump();
        }

        System.out.println("       Second Best ID: " + iSecondClosestID);
        System.out.println("     Second Best Name: " + soDBPerConfig.getName(iSecondClosestID));
        System.out.println("            Date/time: " + new Date());
        System.out.println("----------------------------8<------------------------------");
    }

    /**
     * Updates training set with a new sample from a given file.
     *
     * @param pstrFilename name of the wave file with voice sample train the system on
     *
     * @throws MARFException in case of any error while processing is in MARF
     * @since 0.3.0.5
     */
    public static final void train(String pstrFilename)
    throws MARFException
    {
        MARF.setSampleFile(pstrFilename);
        
        int iID = soDBPerConfig.getIDByFilename(pstrFilename, true);

        if(iID == -1)
        {
            System.out.println("No speaker found for \"" + pstrFilename + "\" for training.");
        }
        else
        {
            MARF.setCurrentSubject(iID);
            MARF.train();
        }
    }

    /**
     * Displays application's usage information and exits.
     */
    private static final void usage()
    {
        System.out.println
        (
            "Usage:\n" +
            "  java SpeakerIdentApp --train <samples-dir> [options]        -- train mode\n" +
            "                       --single-train <sample> [options]      -- add a single sample to the training set\n" +
            "                       --ident <sample> [options]             -- identification mode\n" +
            "                       --batch-ident <samples-dir> [options]  -- batch identification mode\n" +
            "                       --gui                                  -- use GUI as a user interface\n" +
            "                       --stats=[per-config|per-speaker|both]  -- display stats (default is per-config)\n" +
            "                       --best-score                           -- display best classification result\n" +
            "                       --reset                                -- reset stats\n" +
            "                       --version                              -- display version info\n" +
            "                       --help | -h                            -- display this help and exit\n\n" +

            "Options (one or more of the following):\n\n" +

            "Loaders:\n\n" +
            "  -wav          - assume WAVE files loading (default)\n" +
            "  -text         - assume loading of text samples\n" +
            "\n" +

            "Preprocessing:\n\n" +
            "  -silence      - remove silence (can be combined with any of the below)\n" +
            "  -noise        - remove noise (can be combined with any of the below)\n" +
            "  -raw          - no preprocessing\n" +
            "  -norm         - use just normalization, no filtering\n" +
            "  -low          - use low-pass FFT filter\n" +
            "  -high         - use high-pass FFT filter\n" +
            "  -boost        - use high-frequency-boost FFT preprocessor\n" +
            "  -band         - use band-pass FFT filter\n" +
            "  -bandstop     - use band-stop FFT filter\n" +
            "  -endp         - use endpointing\n" +
            "  -lowcfe       - use low-pass CFE filter\n" +
            "  -highcfe      - use high-pass CFE filter\n" +
            "  -bandcfe      - use band-pass CFE filter\n" +
            "  -bandstopcfe  - use band-stop CFE filter\n" +
            "\n" +

            "Feature Extraction:\n\n" +
            "  -lpc          - use LPC\n" +
            "  -fft          - use FFT\n" +
            "  -minmax       - use Min/Max Amplitudes\n" +
            "  -randfe       - use random feature extraction\n" +
            "  -aggr         - use aggregated FFT+LPC feature extraction\n" +
            "  -f0           - use F0 (pitch, or fundamental frequency; NOT IMPLEMENTED)\n" +
            "  -segm         - use Segmentation (NOT IMPLEMENTED)\n" +
            "  -cepstral     - use Cepstral analysis (NOT IMPLEMENTED)\n" +
            "\n" +

            "Classification:\n\n" +
            "  -nn           - use Neural Network\n" +
            "  -cheb         - use Chebyshev Distance\n" +
            "  -eucl         - use Euclidean Distance\n" +
            "  -mink         - use Minkowski Distance\n" +
            "  -diff         - use Diff-Distance\n" +
            "  -zipf         - use Zipf's Law-based classifier\n" +
            "  -randcl       - use random classification\n" +
            "  -markov       - use Hidden Markov Models (NOT IMPLEMENTED)\n" +
            "  -hamming      - use Hamming Distance\n" +
            "  -cos          - use Cosine Similarity Measure\n" +
            "\n" +

            "Misc:\n\n" +
            "  -debug        - include verbose debug output\n" +
            "  -spectrogram  - dump spectrogram image after feature extraction\n" +
            "  -graph        - dump wave graph before preprocessing and after feature extraction\n" +
            "  <integer>     - expected speaker ID\n" +
            "\n"
        );

        System.exit(0);
    }

    /**
     * Retrieves String representation of the application's version.
     * @return version String
     */
    public static final String getVersion()
    {
        return MAJOR_VERSION + "." + MINOR_VERSION + "." + REVISION;
    }

    /**
     * Retrieves integer representation of the application's version.
     * @return integer version
     */
    public static final int getIntVersion()
    {
        return MAJOR_VERSION * 100 + MINOR_VERSION * 10 + REVISION;
    }

    /**
     * Makes sure the applications isn't run against older MARF version.
     * Exits with 1 if the MARF version is too old.
     */
    public static final void validateVersions()
    {
        if(MARF.getDoubleVersion() < (0 * 100 + 3 * 10 + 0 + .6))
        {
            System.err.println
            (
                "Your MARF version (" + MARF.getVersion() +
                ") is too old. This application requires 0.3.0.6 or above."
            );

            System.exit(1);
        }
    }

    /**
     * Composes the current configuration of in a string form.
     *
     * @param pstrArgv set of configuration options passed through the command line;
     * can be null or empty. If latter is the case, MARF itself is queried for its
     * numerical set up inside.
     *
     * @return the current configuration setup
     */
    public static final String getConfigString(String[] pstrArgv)
    {
        // Store config and error/successes for that config
        String strConfig = "";

        if(pstrArgv != null && pstrArgv.length > 2)
        {
            // Get config from the command line
            for(int i = 2; i < pstrArgv.length; i++)
            {
                strConfig += pstrArgv[i] + " ";
            }
        }
        else
        {
            // Query MARF for it's current config
            strConfig = MARF.getConfig();
        }
        
        return strConfig;
    }
    
    /**
     * Sets default MARF configuration parameters as normalization
     * for preprocessing, FFT for feature extraction, Euclidean
     * distance for training and classification with no spectrogram
     * dumps and no debug information, assuming WAVE file format.
     *
     * @throws MARFException
     * @since 0.3.0.5
     */
    public static final void setDefaultConfig()
    throws MARFException
    {
        /*
         * Default MARF setup
         */
        MARF.setPreprocessingMethod(MARF.DUMMY);
        MARF.setFeatureExtractionMethod(MARF.FFT);
        MARF.setClassificationMethod(MARF.EUCLIDEAN_DISTANCE);
        MARF.setDumpSpectrogram(false);
        MARF.setSampleFormat(MARF.WAV);
//        MARF.setSampleFormat(MARF.TEXT);

        Debug.enableDebug(false);
    }
    
    /**
     * Customizes MARF's configuration based on the options.
     * @throws MARFException if some options are out of range
     * @since 0.3.0.5
     */
    public static final void setCustomConfig()
    throws MARFException
    {
        ModuleParams oParams = new ModuleParams();
        
        for
        (
            int iPreprocessingMethod = MARF.MIN_PREPROCESSING_METHOD;
            iPreprocessingMethod <= MARF.MAX_PREPROCESSING_METHOD;
            iPreprocessingMethod++
        )
        {
            if(soGetOpt.isActiveOption(iPreprocessingMethod))
            {
                MARF.setPreprocessingMethod(iPreprocessingMethod);
                
                switch(iPreprocessingMethod)
                {
                    // Endpointing did not respond well to silence removal
                    // and the definition of RAW assumes none of that is done.
                    // Endpointing also has some extra params of its own.
                    case MARF.RAW:
                    case MARF.ENDPOINT:

                    case MARF.DUMMY:
                    case MARF.HIGH_FREQUENCY_BOOST_FFT_FILTER:
                    case MARF.HIGH_PASS_FFT_FILTER:
                    case MARF.LOW_PASS_FFT_FILTER:
                    case MARF.BANDPASS_FFT_FILTER:
                    case MARF.BAND_STOP_FFT_FILTER:
                    case MARF.HIGH_PASS_BOOST_FILTER:
                    case MARF.BAND_PASS_CFE_FILTER:
                    case MARF.BAND_STOP_CFE_FILTER:
                    case MARF.LOW_PASS_CFE_FILTER:
                    case MARF.HIGH_PASS_CFE_FILTER:
                    {
                        // Normalization and filters seem to respond better to silence removal.
                        // The setting of the third protocol parameter (silence threshold)
                        // is yet to be implemented here.
                        oParams.addPreprocessingParam(new Boolean(soGetOpt.isActiveOption(OPT_NOISE_REMOVAL)));
                        oParams.addPreprocessingParam(new Boolean(soGetOpt.isActiveOption(OPT_SILENCE_REMOVAL)));
                        break;
                    }
                    
                    default:
                    {
                        assert false : "Not implemented valid preprocessing configuration parameter: " + iPreprocessingMethod;
                    }
                } // switch

                break;
            }
        }

        for
        (
            int iFeatureExtractionMethod = MARF.MIN_FEATUREEXTRACTION_METHOD;
            iFeatureExtractionMethod <= MARF.MAX_FEATUREEXTRACTION_METHOD;
            iFeatureExtractionMethod++
        )
        {
            if(soGetOpt.isActiveOption(iFeatureExtractionMethod))
            {
                MARF.setFeatureExtractionMethod(iFeatureExtractionMethod);
                
                switch(iFeatureExtractionMethod)
                {
                    case MARF.FFT:
                    case MARF.LPC:
                    case MARF.RANDOM_FEATURE_EXTRACTION:
                    case MARF.MIN_MAX_AMPLITUDES:
                    case MARF.F0:
                    case MARF.CEPSTRAL:
                    case MARF.SEGMENTATION:
                        // For now do nothing; customize when these methods
                        // become parametrizable.
                        break;
                    
                    case MARF.FEATURE_EXTRACTION_AGGREGATOR:
                    {
                        // For now aggregate FFT followed by LPC until
                        // it becomes customizable
                        oParams.addFeatureExtractionParam(new Integer(MARF.FFT));
                        oParams.addFeatureExtractionParam(null);
                        oParams.addFeatureExtractionParam(new Integer(MARF.LPC));
                        oParams.addFeatureExtractionParam(null);
                        break;
                    }

                    default:
                        assert false;
                } // switch

                break;
            }
        }

        for
        (
            int iClassificationMethod = MARF.MIN_CLASSIFICATION_METHOD;
            iClassificationMethod <= MARF.MAX_CLASSIFICATION_METHOD;
            iClassificationMethod++
        )
        {
            if(soGetOpt.isActiveOption(iClassificationMethod))
            {
                MARF.setClassificationMethod(iClassificationMethod);
                
                switch(iClassificationMethod)
                {
                    case MARF.NEURAL_NETWORK:
                    {
                        // Dump/Restore Format of the TrainingSet
                        oParams.addClassificationParam(new Integer(TrainingSet.DUMP_GZIP_BINARY));
    
                        // Training Constant
                        oParams.addClassificationParam(new Double(0.5));
    
                        // Epoch number
                        oParams.addClassificationParam(new Integer(20));
    
                        // Min. error
                        oParams.addClassificationParam(new Double(0.1));
                        
                        break;
                    }

                    case MARF.HAMMING_DISTANCE:
                    {
                        // Dump/Restore Format
                        oParams.addClassificationParam(new Integer(TrainingSet.DUMP_GZIP_BINARY));
    
                        // Type of hamming comparison
                        oParams.addClassificationParam(new Integer(HammingDistance.STRICT_DOUBLE));

                        break;
                    }

                    case MARF.MINKOWSKI_DISTANCE:
                    {
                        // Dump/Restore Format
                        oParams.addClassificationParam(new Integer(TrainingSet.DUMP_GZIP_BINARY));
    
                        // Minkowski Factor
                        oParams.addClassificationParam(new Double(6.0));

                        break;
                    }

                    case MARF.EUCLIDEAN_DISTANCE:
                    case MARF.CHEBYSHEV_DISTANCE:
                    case MARF.MAHALANOBIS_DISTANCE:
                    case MARF.RANDOM_CLASSIFICATION:
                    case MARF.DIFF_DISTANCE:
                    case MARF.MARKOV:
                    case MARF.ZIPFS_LAW:
                    case MARF.COSINE_SIMILARITY_MEASURE:
                        // For now do nothing; customize when these methods
                        // become parametrizable.
                        break;
                    
                    default:
                        assert false : "Unrecognized classification module";
                } // switch
                
                // Method is found, break out of the look up loop
                break;
            }
        }

        // Assign meaningful params only
        if(oParams.size() > 0)
        {
            MARF.setModuleParams(oParams);
        }
    }
}

// EOF
\end{verbatim}
}
\hrule
\vspace{15pt}

\section{\texttt{SpeakersIdentDb.java}}

\vspace{15pt}
\hrule
{\scriptsize \begin{verbatim}
import java.awt.Point;
import java.io.BufferedReader;
import java.io.FileInputStream;
import java.io.FileNotFoundException;
import java.io.FileOutputStream;
import java.io.FileReader;
import java.io.IOException;
import java.io.ObjectInputStream;
import java.io.ObjectOutputStream;
import java.text.DecimalFormat;
import java.util.Enumeration;
import java.util.Hashtable;
import java.util.StringTokenizer;
import java.util.Vector;
import java.util.zip.GZIPInputStream;
import java.util.zip.GZIPOutputStream;

import marf.Storage.Database;
import marf.Storage.StorageException;
import marf.util.Arrays;
import marf.util.Debug;


/**
 * <p>Class SpeakersIdentDb manages database of speakers on the application level.</p>
 * <p>XXX: Move stats collection over to MARF.</p>
 *
 * $Id: SpeakersIdentDb.java,v 1.25 2007/02/19 03:28:13 mokhov Exp $
 *
 * @author Serguei Mokhov
 * @version $Revision: 1.25 $
 * @since 0.0.1
 */
public class SpeakersIdentDb
extends Database
{
    /**
     * Hashes "config string" -&gt; Vector(FirstMatchPoint(XSuccesses, YFailures),
     * SecondMatchPoint(XSuccesses, YFailures)).
     */
    private Hashtable oStatsPerConfig = null;

    /**
     * Array of sorted stats refs.
     */
    private transient Vector[] aoSortedStatsRefs = null;

    /**
     * A vector of vectors of speakers info pre-loded on <code>connect()</code>.
     * @see #connect()
     */
    private Hashtable oDB = null;

    /**
     * "Database connection".
     */
    private BufferedReader oConnection = null;

    /**
     * For serialization versioning.
     * @since 0.3.0.5
     */
    private static final long serialVersionUID = -7185805363856188810L;

    /**
     * Constructor.
     * @param pstrFileName filename of a CSV file with IDs and names of speakers
     */
    public SpeakersIdentDb(final String pstrFileName)
    {
        this.strFilename = pstrFileName;
        this.oDB = new Hashtable();
        this.oStatsPerConfig = new Hashtable();
    }

    /**
     * Retrieves Speaker's ID by a sample filename.
     * @param pstrFileName Name of a .wav file for which ID must be returned
     * @param pbTraining indicates whether the filename is a training (<code>true</code>) sample or testing (<code>false</code>)
     * @return int ID
     * @throws StorageException in case of an error in any I/O operation
     */
    public final int getIDByFilename(final String pstrFileName, final boolean pbTraining)
    throws StorageException
    {
        String strFilenameToLookup;

        // Extract actual file name without preceeding path (if any)
        if(pstrFileName.lastIndexOf('/') >= 0)
        {
            strFilenameToLookup = pstrFileName.substring(pstrFileName.lastIndexOf('/') + 1, pstrFileName.length());
        }
        else if(pstrFileName.lastIndexOf('\\') >= 0)
        {
            strFilenameToLookup = pstrFileName.substring(pstrFileName.lastIndexOf('\\') + 1, pstrFileName.length());
        }
        else
        {
            strFilenameToLookup = pstrFileName;
        }

        Enumeration oIDs = this.oDB.keys();

        // Traverse all the info vectors looking for sample filename
        while(oIDs.hasMoreElements())
        {
            Integer oID = (Integer)oIDs.nextElement();

            Debug.debug("File: " + pstrFileName + ", id = " + oID.intValue());

            Vector oSpeakerInfo = (Vector)this.oDB.get(oID);
            Vector oFilenames;

            if(pbTraining == true)
            {
                oFilenames = (Vector)oSpeakerInfo.elementAt(1);
            }
            else
            {
                oFilenames = (Vector)oSpeakerInfo.elementAt(2);
            }

            // Start from 1 because 0 is speaker's name
            for(int i = 0; i < oFilenames.size(); i++)
            {
                String strCurrentFilename = (String)oFilenames.elementAt(i);

                if(strCurrentFilename.equals(strFilenameToLookup))
                {
                    return oID.intValue();
                }
            }
        }

        return -1;
    }

    /**
     * Retrieves speaker's name by their ID.
     * @param piID ID of a person in the DB to return a name for
     * @return name string
     * @throws StorageException
     */
    public final String getName(final int piID)
    throws StorageException
    {
        //Debug.debug("getName() - ID = " + piID + ", db size: " + oDB.size());
        String strName;

        Vector oDBEntry = (Vector)this.oDB.get(new Integer(piID));

        if(oDBEntry == null)
        {
            strName = "Unknown Speaker (" + piID + ")";
        }
        else
        {
            strName = (String)oDBEntry.elementAt(0);
        }

        return strName;
    }

    /**
     * Connects to the "database" of speakers (opens the text file :-)).
     * @throws StorageException in case of any I/O error
     */
    public void connect()
    throws StorageException
    {
        // That's where we should establish file linkage and keep it until closed
        try
        {
            this.oConnection = new BufferedReader(new FileReader(this.strFilename));
            this.bConnected = true;
        }
        catch(IOException e)
        {
            throw new StorageException
            (
                "Error opening speaker DB: \"" + this.strFilename + "\": " +
                e.getMessage() + "."
            );
        }
    }

    /**
     * Retrieves speaker's data from the text file and populates
     * internal data structures. Uses StringTokenizer to parse
     * data read from the file.
     * @throws StorageException in case of any I/O error
     */
    public void query()
    throws StorageException
    {
        // That's where we should load db results into internal data structure

        String strLine;
        int iID = -1;

        try
        {
            strLine = this.oConnection.readLine();

            while(strLine != null)
            {
                StringTokenizer oTokenizer = new StringTokenizer(strLine, ",");
                Vector oSpeakerInfo = new Vector();

                // get ID
                if(oTokenizer.hasMoreTokens())
                {
                    iID = Integer.parseInt(oTokenizer.nextToken());
                }

                // speaker's name
                if(oTokenizer.hasMoreTokens())
                {
                    strLine = oTokenizer.nextToken();
                    oSpeakerInfo.add(strLine);
                }

                // training file names
                Vector oTrainingFilenames = new Vector();

                if(oTokenizer.hasMoreTokens())
                {
                    StringTokenizer oSTK = new StringTokenizer(oTokenizer.nextToken(), "|");

                    while(oSTK.hasMoreTokens())
                    {
                        strLine = oSTK.nextToken();
                        oTrainingFilenames.add(strLine);
                    }
                }

                oSpeakerInfo.add(oTrainingFilenames);

                // testing file names
                Vector oTestingFilenames = new Vector();

                if(oTokenizer.hasMoreTokens())
                {
                    StringTokenizer oSTK = new StringTokenizer(oTokenizer.nextToken(), "|");

                    while(oSTK.hasMoreTokens())
                    {
                        strLine = oSTK.nextToken();
                        oTestingFilenames.add(strLine);
                    }
                }

                oSpeakerInfo.add(oTestingFilenames);

                Debug.debug("Putting ID=" + iID + " along with info vector of size " + oSpeakerInfo.size());

                this.oDB.put(new Integer(iID), oSpeakerInfo);

                strLine = this.oConnection.readLine();
            }
        }
        catch(IOException e)
        {
            throw new StorageException
            (
                "Error reading from speaker DB: \"" + this.strFilename +
                "\": " + e.getMessage() + "."
            );
        }
    }

    /**
     * Closes (file) database connection.
     * @throws StorageException if not connected or fails to close inner reader
     */
    public void close()
    throws StorageException
    {
        // Close file
        if(this.bConnected == false)
        {
            throw new StorageException("SpeakersIdentDb.close() - not connected");
        }

        try
        {
            this.oConnection.close();
            this.bConnected = false;
        }
        catch(IOException e)
        {
            throw new StorageException(e.getMessage());
        }
    }

    /**
     * Adds one more classification statics entry.
     * @param pstrConfig String representation of the configuration the stats are for
     * @param pbSuccess <code>true</code> if classification was successful; <code>false</code> otherwise
     */
    public final void addStats(final String pstrConfig, final boolean pbSuccess)
    {
        addStats(pstrConfig, pbSuccess, false);
    }

    /**
     * Adds one more classification statics entry and accounts for the second best choice.
     * @param pstrConfig String representation of the configuration the stats are for
     * @param pbSuccess <code>true</code> if classification was successful; <code>false</code> otherwise
     * @param pbSecondBest <code>true</code> if classification was successful; <code>false</code> otherwise
     */
    public final void addStats(final String pstrConfig, final boolean pbSuccess, final boolean pbSecondBest)
    {
        Vector oMatches = (Vector)this.oStatsPerConfig.get(pstrConfig);
        Point oPoint = null;

        if(oMatches == null)
        {
            oMatches = new Vector(2);
            oMatches.add(new Point());
            oMatches.add(new Point());
            oMatches.add(pstrConfig);
        }
        else
        {
            if(pbSecondBest == false)
            {
                // First match
                oPoint = (Point)oMatches.elementAt(0);
            }
            else
            {
                // Second best match
                oPoint = (Point)oMatches.elementAt(1);
            }
        }

        int iSuccesses = 0; // # of successes
        int iFailures = 0; // # of failures

        if(oPoint == null) // Didn't exist yet; create new
        {
            if(pbSuccess == true)
            {
                iSuccesses = 1;
            }
            else
            {
                iFailures = 1;
            }

            oPoint = new Point(iSuccesses, iFailures);

            if(oPoint == null)
            {
                System.err.println("SpeakersIdentDb.addStats() - oPoint is null! Out of memory?");
                System.exit(-1);
            }

            if(oMatches == null)
            {
                System.err.println("SpeakersIdentDb.addStats() - oMatches is null! Out of memory?");
                System.exit(-1);
            }

            if(oMatches.size() == 0)
            {
                System.err.println("SpeakersIdentDb.addStats() - oMatches.size = 0");
                System.exit(-1);
            }

            if(pbSecondBest == false)
            {
                oMatches.setElementAt(oPoint, 0);
            }
            else
            {
                oMatches.setElementAt(oPoint, 1);
            }

            this.oStatsPerConfig.put(pstrConfig, oMatches);
        }

        else // There is an entry for this config; update
        {
            if(pbSuccess == true)
            {
                oPoint.x++;
            }
            else
            {
                oPoint.y++;
            }
        }
    }

    /**
     * Dumps all collected statistics to STDOUT.
     * @throws Exception
     */
    public final void printStats()
    throws Exception
    {
        printStats(false);
    }

    /**
     * Dumps collected statistics to STDOUT.
     * @param pbBestOnly <code>true</code> - print out only the best score number; <code>false</code> - all stats
     * @throws Exception
     */
    public final void printStats(boolean pbBestOnly)
    throws Exception
    {
        if(this.oStatsPerConfig.size() == 0)
        {
            System.err.println("SpeakerIdentDb: no statistics available. Did you run the recognizer yet?");
            return;
        }

        // First row is for the identified results, 2nd is for 2nd best ones.
        String[][] astrResults = new String[2][this.oStatsPerConfig.size()];

        this.aoSortedStatsRefs = (Vector[])oStatsPerConfig.values().toArray(new Vector[0]);
        Arrays.sort(this.aoSortedStatsRefs, new StatsPercentComparator(StatsPercentComparator.DESCENDING));

        int iResultNum = 0;

        System.out.println("guess,run,config,good,bad,%");

        for(int i = 0; i < this.aoSortedStatsRefs.length; i++)
        {
            String strConfig = (String)(this.aoSortedStatsRefs[i]).elementAt(2);

            for(int j = 0; j < 2; j++)
            {
                Point oGoodBadPoint = (Point)(this.aoSortedStatsRefs[i]).elementAt(j);
                String strGuess = (j == 0) ? "1st" : "2nd";
                String strRun = (iResultNum + 1) + "";
                DecimalFormat oFormat = new DecimalFormat("#,##0.00;(#,##0.00)");
                double dRate = ((double)oGoodBadPoint.x / (double)(oGoodBadPoint.x + oGoodBadPoint.y)) * 100;

                if(pbBestOnly == true)
                {
                    System.out.print(oFormat.format(dRate));
                    return;
                }

                astrResults[j][iResultNum] =
                    strGuess + "," +
                    strRun + "," +
                    strConfig + "," +
                    oGoodBadPoint.x + "," + // Good
                    oGoodBadPoint.y + "," + // Bad
                    oFormat.format(dRate);
            }

            iResultNum++;
        }

        // Print all of the 1st match
        for(int i = 0; i < astrResults[0].length; i++)
        {
            System.out.println(astrResults[0][i]);
        }

        // Print all of the 2nd match
        for(int i = 0; i < astrResults[1].length; i++)
        {
            System.out.println(astrResults[1][i]);
        }
    }

    /**
     * Resets in-memory and on-disk statistics.
     * @throws StorageException
     */
    public final void resetStats()
    throws StorageException
    {
        this.oStatsPerConfig.clear();
        dump();
    }

    /**
     * Dumps statistic's Hashtable object as gzipped binary to disk.
     * @throws StorageException
     */
    public void dump()
    throws StorageException
    {
        try
        {
            FileOutputStream oFOS = new FileOutputStream(this.strFilename + ".stats");
            GZIPOutputStream oGZOS = new GZIPOutputStream(oFOS);
            ObjectOutputStream oOOS = new ObjectOutputStream(oGZOS);

            oOOS.writeObject(this.oStatsPerConfig);
            oOOS.flush();
            oOOS.close();
        }
        catch(Exception e)
        {
            throw new StorageException(e);
        }
    }

    /**
     * Reloads statistic's Hashtable object from disk.
     * If the file did not exist, it creates a new one.
     * @throws StorageException
     */
    public void restore()
    throws StorageException 
    {
        try
        {
            FileInputStream oFIS = new FileInputStream(this.strFilename + ".stats");
            GZIPInputStream oGZIS = new GZIPInputStream(oFIS);
            ObjectInputStream oOIS = new ObjectInputStream(oGZIS);

            this.oStatsPerConfig = (Hashtable)oOIS.readObject();
            oOIS.close();
        }
        catch(FileNotFoundException e)
        {
            System.out.println
            (
                "NOTICE: File " + this.strFilename +
                ".stats does not seem to exist. Creating a new one...."
            );

            resetStats();
        }
        catch(ClassNotFoundException e)
        {
            throw new StorageException
            (
                "SpeakerIdentDb.retore() - ClassNotFoundException: " +
                e.getMessage()
            );
        }
        catch(Exception e)
        {
            throw new StorageException(e);
        }
    }
}

/**
 * <p>Used in sorting by percentage of the stats entries
 * in either ascending or descending order.</p>
 *
 * <p>TODO: To be moved to Stats.</p>
 *
 * @author Serguei Mokhov
 * @version $Revision: 1.25 $
 * @since 0.0.1 of MARF
 */
class StatsPercentComparator
extends marf.util.SortComparator
{
    /**
     * For serialization versioning.
     * @since 0.3.0.5
     */
    private static final long serialVersionUID = -7185805363856188810L;

    /**
     * Mimics parent's constructor.
     */
    public StatsPercentComparator()
    {
        super();
    }

    /**
     * Mimics parent's constructor.
     * @param piSortMode either DESCENDING or ASCENDING sort mode 
     */
    public StatsPercentComparator(final int piSortMode)
    {
        super(piSortMode);
    }

    /**
     * Implementation of the Comparator interface for the stats objects.
     * @see java.util.Comparator#compare(java.lang.Object, java.lang.Object)
     */
    public int compare(Object poStats1, Object poStats2)
    {
        Vector oStats1 = (Vector)poStats1;
        Vector oStats2 = (Vector)poStats2;

        Point oGoodBadPoint1 = (Point)(oStats1.elementAt(0));
        Point oGoodBadPoint2 = (Point)(oStats2.elementAt(0));

        double dRate1 = ((double)oGoodBadPoint1.x / (double)(oGoodBadPoint1.x + oGoodBadPoint1.y)) * 100;
        double dRate2 = ((double)oGoodBadPoint2.x / (double)(oGoodBadPoint2.x + oGoodBadPoint2.y)) * 100;

        switch(this.iSortMode)
        {
            case DESCENDING:
            {
                return (int)((dRate2 - dRate1) * 100);
            }

            case ASCENDING:
            default:
            {
                return (int)((dRate1 - dRate2) * 100);
            }
        }
    }
}

// EOF
\end{verbatim}
}
\hrule
\vspace{15pt}

\clearpage

%%%%%%%%%%%%%%%%%%%%%%%%%%%%%%%%%%%%%%%%%%%%%%%%%%

\chapter{TODO}
\label{appx:todo}

\vspace{15pt}
\hrule
{\scriptsize \begin{verbatim}
MARF TODO/Wishlist
------------------

$Header: /cvsroot/marf/marf/TODO,v 1.75 2007/08/05 18:36:40 mokhov Exp $

Legend:
-------

"-"  -- TODO
"*"  -- done (for historical records)
"+-" -- somewhat done ("+" is degree of completed work, "-" remaining).
"?"  -- unsure if feature is needed or how to proceed about it or it's potentially far away
{..} -- projected release (tentative)


THE APPS

- SpeakerIdentApp
  - GUI
  - Real-Time recording and recognition (does it belong here?)
    - Integrate fully marf.Storage.SampleRecorder of Jimmy
  - Move dir. read from the app to MARF in training section {0.3.0}
  - Enhance batch recognition (do not re-load training set per sample) {0.3.0}
  - Enhance Option Processing
    * Make use of OptionProcessor {0.3.0.5}
    - Enhance options with arguments, e.g. -fft=1024, -lpc=40, -lpc=40,256, keeping the existing defaults
    - Add options:
      * --batch-ident option {0.3.0.5}
      - -data-dir=DIRNAME other than default to specify a dir where to store training sets and stuff
      - -mink=r
      - -fft=x,y
      - -lpc=x,y
      * single file training option {0.3.0.5}
      - Dump stats: -classic -latex -csv
  +- Make binary/optimized distro
  - Generate data.tex off speakers.txt in the manual.
  +---- Convert testing.sh to Java
  - Open up -randcl with -nn
  * Finish testing.bat
  * Make executable .jar
  * Improve on javadoc
  * ChangeLog
  * Sort the stats
  * Add classification methods to training in testing.sh
  * Implement batch plan execution

- LangIdentApp
  * Integrate
  - Add GUI
  * Make executable .jar
  - Release.

- ProbabilisticParsingApp
  * Integrate
  - Add GUI
  * Make executable .jar
  - Release.

- ZipfLawApp
  * Integrate
  - Add GUI
  * Make executable .jar
  - Release.

- TestFilters
  - The application has already a Makefile and a JBuilder project file.
    We would want to add a NetBeans project as well.
  - Add GUI
  * Release.
  * It only tests HighFrequencyBoost, but we also have BandpassFilter,
    HighPassFilter, and LowPassFilter.
    These have to be tested.
  * The output of every filter will have to be stored in a expected
    (needs to be created) folder for future regression testing. Like
    with TestMath.
  * Option processing has to be done and standartized
    by using marf.util.OptionProcessor uniformly in all apps. But we can
    begin with this one. The options then would be: --high, --low, --band,
    and --boost that will correspond to the appropriate filters I mentioned.
  * The exception handling has to be augmented to print the error message
    and the stack trace to System.err.
  * Apply coding conventions to naming variables, etc.
  * Make executable .jar

- TestNN
  * Fix to work with new MARF API
  * Add use of OptionProcessor
  * Apply coding conventions
  - Add GUI
  * Make executable .jar
  - Release.

- TestLPC
  - Add GUI
  * Release.
  * Fix to work with new MARF API
  * Add use of OptionProcessor
  * Apply coding conventions
  * Make executable .jar

- TestFFT
  - Add GUI
  * Release.
  * Make executable .jar
  * Fix to work with new MARF API
  * Add use of OptionProcessor
  * Apply coding conventions

- MathTestApp
  - Add GUI
  * Release.
  * Make executable .jar

- TestWaveLoader
  - Add GUI
  * Release.
  * Fix to work with new MARF API
  * Apply coding conventions
  * Make executable .jar

- TestLoaders
  - Add GUI
  * Create a-la TestFilters with option processing.
  * Make executable .jar
  * Release.

- Regression
  - one script calls all the apps and compares new results vs. expected
  - Matrix ops
  +- Fix TestNN
  - Add GUI
  * Make executable .jar
  - Release.

- Graph % vs. # of samples used per method

- Release all apps as packages at Source Forge
  ++- Bundle
  ++- Individually

- SpeechRecognition
  - Define

- InstrumentIdentification
  - Define

- EmotionDetection
  - Define

- User Authentication


THE BUILD SYSTEM

  - Perhaps at some point we'd need make/project files for other Java IDEs, such as
    - IBM Visual Age
    - Ant
    - Windoze .bat file(s)?
    * Sun NetBeans
  * A Makefile per directory
  * Make utility detection test
  * global makefile in /marf
  * fix doc's global makefile
  * Global Makefile for apps
    (will descend to each dir and build all the apps.)
  * Build and package distrubution
    * MARF
    * App


DISTROS

  * Apps jars
  - Different JDKs (1.4/1.5/1.6)
  - rpm
    - FCs
    - Mandrake
    - RH
  - deb
  - dmg
  - iso


THE FRAMEWORK

- All/Optimization/Testing

  - Implement text file support/toString() method for
    all the modules for regression testing

  - Threading and Parallelism
    - PR and FE are safe to run in ||
    - Fix NNet for ArrayLists->Vector

  ++--- Adopt JUnit

  +- Use of StringBuffer for String concatenation

  * Implement getMARFSourceCodeRevision() method for CVS revision info.

  +- Distributed MARF
    - WAL
    - Replication


- Preprocessing

  ++- Make dump()/restore() to serialize filtered output {0.3.0}
  - Fix hardcoding of filter boundaries
  - Implement
    - Enable changing values of frequency boundaries and coeffs. in filters by an app.
    - "Compressor" [steve]
    - Methods:  {0.3.0.6}
      - removeNoise()
      * removeSilence()
      - cropAudio()
    - streamed normalization
    * Endpoint {0.3.0.5}
    * High-pass filter with high-frequency boost together {0.3.0}
    * Band-pass Filter {0.2.0}
  * Move BandpassFilter and HighFrequencyBoost under FFTFilter package with CVS comments
  * Tweak the filter values of HighPass and HighFrequencyBoost filters


- Feature Extraction

  - Make modules to dump their features for future use by NNet and maybe others {0.3.0}
  - Implement {1.0.0}
    - F0
      - fundamental frequency estimation,
        providing us with a few more features.
    - Cepstral Analysis
    - Segmentation
    * RandomFeatureExtraction {0.2.0}
  - Enhance
    - MinMaxAmplitudes to pick different features, not very close to each other

- Classification

  - Implement
    +++-- Mahalanobis Distance {0.3.0}
      - Learning Covariance Matrix
    - Stochastic [serge] {1.0.0}
      - Gaussian Mixture Models
      - Hidden Markov Models {1.0.0}
    - SimilarityClassifier {0.3.0}
      ? Boolean Model
      ? Vector Space Model
      ? sine similarity model
    * Minkowski's Distance {0.2.0}
    * RandomClassification {0.2.0}

  - Fully Integrate
    +- MaxProbabilityClassifier
      - Push StatisticalEstimator to Stochastic
    +- ZipfLaw

  - Fix and document NNet {0.*.*}
    - add % of correct/incorrect expected to train() {0.3.0}
    - ArrayList ---> Vector, because ArrayList is NOT thread-safe {0.3.0}
    - Second Best (by doubling # of output Neurons with the 2nd half being the 2nd ID)
    +- Epoch training
    * dump()/retore() {0.2.0}

  - Distance Classifiers
    - make distance() throw an exception maybe?
    * Move under Distance package
    * DiffDistance

- Sample Loaders

  - Create Loaders for Java-supported formats:
    +--- AIFC
    +--- AIFF
    +--- AU
    +--- SND
  +--- Add MIDI support
  * Create MARFAudioFileFormat extends AudioFileFormat
  * Enumerate all types in addition to ours from FileAudioFormat.Types

- marf.speech package
  - Recognition (stt)
  - Dictionaries
  - Generation (tts)


- NLP package
  ++++- Integrate
    +- Classification
    - Potter's Stemmer
    * Stats
    * Storage management
    * StatisticalEstimators
    * NLP.java
    * Parsing
    * Util
    * Collocations

? Integrate AIMA stuff


- Stats {0.3.0}

  - Move stats collection from the app and other places to StatsCollector
  - Timing
  - Batch progress report
  - Rank results
  - Results validation (that numbers add up properly e.g. sum of BAD and GOOD should be equal to total)


- Algos

  +- Algorithm decoupling to marf.algos or marf.algorithms or ... {0.4.0}
    * To marf.math.Algorithms {0.3.0}
  - marf.algos.Search
  - marf.util.DataStructures -- Node / Graph --- to be used by the networks and state machines
  * move out hamming() from FeatureExtraction


* marf.math

  * Integrate Vector operations
  * Matrix:
    * Add {0.3.0.3}
      * translate
      * rotate
      * scale
      * shear


- GUI {0.5.0}

  - Make them actual GUI components to be included into App
    - Spectrogram
      +----- Implement SpectrogramPanel
      - Draw spectrogram on the panel
      * Fix filename stuff (module_dirname to module_filename)
    - WaveGrapher
      +----- Implement WaveGrapherPanel
      - Draw waves on the panel
  - Fix WaveGrapher
    - Sometimes dumps files of 0 length
    - Make it actually ouput PPM or smth like that (configurable?)
    - Too huge files for samp output.
    - Have LPC understand it
  - Config tool
  - Web interface?


- MARF.java

  +- Concurrent use of modules of the same type
    - FFT and F0 can both be applied like normalization and filtering
  - Implement
    - streamedRecognition()
    +- train()
       - Add single file training
    - Inter-module compatibility (i.e. specific modules can only work
      with some other specific modules and not the others)
      - Module Compatibility Matrix
      - Module integer and String IDs
  ++++- Server Part {2.0.0}
  * enhance error reporting
  * Embed the NLP class


* MARF Exceptions Framework {0.3.0}
  * Propagate to NLP properly
    * NLPException
  * StorageException
  * Have all marf exceptions inherit from util.MARFException


- marf.util

  - complete and document
    +++- Matrix
    +++- FreeVector
    * Arrays
    * Debug
  ? PrintFactory
    - Move NeuralNetowork.indent()
  ? marf.util.upgrade
  * OptionProcessor
    * Integrate {0.3.0.2}
    * Add parsing of "name=value" options {0.3.0.3}
  * Add marf.util.Debug module. {0.3.0.2}
    * marf.util.Debug.debug()
    * Replicate MARF.debug() --> marf.util.Debug.debug()


- Storage

  - ModuleParams:
    - have Hashtables instead of Vectors
      to allow params in any order and in any number.
    - maybe use OptionProcessor or be its extension?
  - Keep all data files under marf.data dir, like training sets, XML, etc {0.3.0}
  - Implement
    - Schema (as in DB for exports)
      - Common attribute names for
        - SQL
        - XML
        - HTML
        - CSV
      - Metadata / DDL
    - Dump/Restore Types
      +++- DUMP_BINARY (w/o compression) {0.3.0}
      - DUMP_XML {?.?.?}
      - DUMP_CSV {?.?.?}
      - DUMP_HTML {?.?.?}
      - DUMP_SQL {?.?.?}
  +- Revise TrainingSet stuff
    ? Cluster mode vs. feature set mode
    - TrainingSet
      - upgradability {?.?.?}
      - convertability: gzbin <-> bin <-> csv <-> xml <-> html <-> sql
  +--- Add FeatureSet {0.3.0}
  - Revise dump/restore implementation to check for unnecessary
    file writes
  * Integrate IStorageManager
  * Move DUMP_* flags up to IStorageManager
  * Add re-implemented StorageManager and integrate it


- The MARF Language (MARFL)
  - A meta language to write MARF applications in a script/shell-like
    manner along the lines of:
      MARF:
      {
         use normalization;
         use FFT 1024;
         use NeuralNetwork;
         use WAVLoader 8000 1 mono;
         pipeline start on dir /var/data/samples;
         print stats paged 10;
      }
  - Fully define syntax
  - Complete compiler
    - grammar file
    - semantic analysis
    - code generator


* Clean up
  * CVS:
    ? Rename /marf/doc/sgml to /marf/doc/styles
    x Remove /marf/doc/styles/ref
    * Remove --x permissions introduced from windoze in files:
      * /marf/doc/src/tex/sampleloading.tex
      * /marf/doc/src/graphics/*.png
      * /marf/doc/src/graphics/arch/*.png
      * /marf/doc/src/graphics/fft_spectrograms/*.ppm
      * /marf/doc/src/graphics/lpc_spectrograms/*.ppm
      * /marf/doc/arch.mdl
      * /marf/src/marf/Classification/Distance/EuclideanDistance.java
      * /marf/src/marf/Preprocessing/FFTFilter.java
      * /apps/SpeakerIdentApp/SpeakerIdentApp.jpx
      * /apps/SpeakerIdentApp/testing-samples/*.wav
      * /apps/SpeakerIdentApp/testing-samples/*.wav
      * /apps/TestFilters/TestFilters.*
    * Add NLP revisions directly to the CVS (by SF staff)
      * Parsing
      * Stemming
      * Collocations
    * Move distance classifiers with CVS log
      to Distance
    * remove uneeded attics and corresponding dirs
      * "Ceptral"
      * Bogus samples


THE CODE

* Define coding standards
+++- Propagate them throughout the code
* Document


THE SAMPLE DATABASES

- /samples/ -- Move all wav and corpora files there from apps
  - WAV/
    - training-samples/
    - testing-samples/
    - README
    - speakers.txt
    - <training-sets>

  - corpora/
    - training/
      - en/
      - fr/
      - ru/
      - ...
    - testing/
      - en/
      - fr/
      - ru/
      - ...
    - heldout/
    - README
    - <training-sets>

- Make releases


THE TOOLS

* Add module

- Add tools:
  - marfindent
  * stats2latex
  * cvs2cl
  * cvs2html

- upgrade/
  - MARFUpgrade.java
    - an app or a marf module?
    - cmd-line?
    - GUI?
    - Interactive?
  - v012/TrainingSet.java
  - v020/TrainingSet.java
         FeatureSet.java


THE DOCS

- docs [s]
  - report.pdf -> manual.pdf
  - autosync from the report
    - history.tex -> HISTORY
    - legal.tex -> COPYRIGHT
    - installation.tex -> INSTALL
  - Arch Update [serge]
    +- gfx model (rr)
      - add util package
      - add math package
      - add nlp package
      * gui: add StorageManager
    * update doc
    * newer images
  - MARF-specific exceptions
  - Account for dangling .tex files
    - old-results.tex
    - output.tex
    * installation.tex
    * training.tex
    * sample-formats.tex
    * cvs.tex
    * history.tex
    * f0.tex
    * notation.tex
    * sources.tex
  +++- better doc format and formulas
  - Results:
    - Add modules params used, like r=6 in Minkowski, FFT input 1024, etc
    - Add time took
  * fix javadoc 1.4/1.5 warnings
  * fix Appendix
  * split-out bibliography
  * index
  * ChangeLog
  * report components [serge]

- web site
  - Publish
    * TODO
    +- ChanageLog
      * Raw
      - HTML
    - Manual
      - Add HTML
  * Add training sets
  * CVS
  * autoupdate from CVS

EOF
\end{verbatim}
}
\hrule
\vspace{15pt}

% EOF

	\printindex
\end{document}